%% file: dissertation.tex
\documentclass[12pt]{report}
\usepackage{uwmthesis,latexsym,graphics,epsfig,lscape,amsbsy,amsmath,amssymb}

\allowdisplaybreaks
\newcommand{\mA}{\mathrm{A}}%
\newcommand{\vecA}{\boldsymbol{A}}%
\newcommand{\vecAp}{\boldsymbol{A'}}%
\newcommand{\vecalph}{\boldsymbol{\alpha}}%
\newcommand{\vecB}{\boldsymbol{B}}%
\newcommand{\vecBp}{\boldsymbol{B'}}%
\newcommand{\mc}{\mathrm{c}}%
\newcommand{\dif}{\mathrm{d}}%
\newcommand{\vecdel}{\boldsymbol{\triangledown}}%
\newcommand{\vecdelp}{\boldsymbol{\triangledown'}}%
\newcommand{\vecE}{\boldsymbol{E}}%
\newcommand{\vecEp}{\boldsymbol{E'}}%
\newcommand{\me}{\mathrm{e}}%
\newcommand{\veceps}{\boldsymbol{\varepsilon}}%
\newcommand{\vecepspar}{\boldsymbol{{\varepsilon} _{_{||}}}}%
\newcommand{\vecepsparp}{\boldsymbol{{\varepsilon}' _{_{||}}}}%
\newcommand{\vecepsperp}{\boldsymbol{{\varepsilon} _{_\perp}}}%
\newcommand{\vecepsperpp}{\boldsymbol{{\varepsilon}' _{_\perp}}}%
\newcommand{\vecF}{\boldsymbol{F}}%
\newcommand{\mi}{\mathrm{i}}%
\newcommand{\vecJ}{\boldsymbol{J}}%
\newcommand{\veck}{\boldsymbol{k}}%
\newcommand{\veckappa}{\boldsymbol{\kappa}}%
\newcommand{\veckzero}{\boldsymbol{k_0}}%
\newcommand{\mL}{\mathrm{L}}%
\newcommand{\hatn}{\boldsymbol{\hat{n}}}%
\newcommand{\vecp}{\boldsymbol{p}}%
\newcommand{\mpi}{\mathrm{\pi}}%
\newcommand{\vecq}{\boldsymbol{q}}%
\newcommand{\vecqp}{\boldsymbol{q'}}%
\newcommand{\mR}{\mathrm{R}}%
\newcommand{\vecr}{\boldsymbol{r}}%
\newcommand{\hatrp}{\boldsymbol{\hat{r}'}}%
\newcommand{\vecrp}{\boldsymbol{r'}}%
\newcommand{\vecrpq}{\boldsymbol{{r_q}'}}%
\newcommand{\vecrbp}{\boldsymbol{\bar{r}'}}%
\newcommand{\ms}{\mathrm{s}}%
\newcommand{\vecS}{\boldsymbol{S}}%
\newcommand{\vecSp}{\boldsymbol{S'}}%
\newcommand{\hats}{\boldsymbol{\hat{s}}}%
\newcommand{\vecsig}{\boldsymbol{\sigma}}%
\newcommand{\mT}{\mathrm{T}}%
\newcommand{\vecT}{\boldsymbol{T}}%
\newcommand{\hatt}{\hat{t}}%
\newcommand{\vectau}{\boldsymbol{\tau}}%
\newcommand{\hattheta}{\boldsymbol{\hat{\theta}}}%
\newcommand{\vectheta}{\boldsymbol{\theta}}%
\newcommand{\vecv}{\boldsymbol{v}}%
\newcommand{\hatv}{\boldsymbol{\hat{v}}}%
\newcommand{\vecVg}{\boldsymbol{V_g}}%
\newcommand{\vecVp}{\boldsymbol{V_p}}%
\newcommand{\vecx}{\boldsymbol{x}}%
\newcommand{\hatx}{\boldsymbol{\hat{x}}}%
\newcommand{\mY}{\mathrm{Y}}%
\newcommand{\haty}{\boldsymbol{\hat{y}}}%
\newcommand{\hatyp}{\boldsymbol{\hat{y}'}}%
\newcommand{\hatz}{\boldsymbol{\hat{z}}}%
\newcommand{\hatzp}{\boldsymbol{\hat{z}'}}%
\newcommand{\veczero}{\boldsymbol{0}}%
\newcommand{\half}{\frac{1}{2}}%
\newcommand{\oort}{\frac{1}{\sqrt{2}}}%
\newcommand{\halfort}{\frac{1}{2\sqrt{2}}}%
\newcommand{\ootwopi}{\frac{1}{2\mpi}}%
\newcommand{\oofourpi}{\frac{1}{4\mpi}}%
\newcommand{\oortpi}{\frac{1}{\sqrt{2\mpi}}}%
\newcommand{\piot}{\frac{\mpi}{2}}%
\newcommand{\piof}{\frac{\mpi}{4}}%
\newcommand{\rpiot}{\sqrt{\frac{\mpi}{2}}}%
\newcommand{\toortpi}{\frac{2}{\sqrt{2\mpi}}}%
\newcommand{\tioortpi}{\frac{2\mi}{\sqrt{2\mpi}}}%

\begin{document}

%====================== UWM FORMAT ===========================

\title{AN ELECTROWEAK WEIZS\"{A}CKER-WILLIAMS METHOD}%
\author{Sean Ahern}%
\program{Physics}%
\majorprof{Major Professor}%
\degree{Doctor of Philosophy}%
\submitdate{December 2001}%
\copyrightyear{2001}%
\copyrighttrue%
\doctoratetrue%
\figurespagetrue%
\tablespagetrue%
\dblabstractfalse%
\multivolumesfalse%
\multiminorsfalse%

\input{abstract}
\beforepreface%
\prefacesection{Dedication}%
\input{Dedication}\prefacesection{Acknowledgments}%
\input{Acknowledgments}\afterpreface%

\pagenumbering{arabic}%

%====================== INPUT CHAPTERS =========================

\input{Introduction}%
\input{Chapter2}%
\input{Chapter3}%
\input{Chapter4}%
\input{Chapter5}%
\input{Summary}\clearpage%
\addcontentsline{toc}{chapter}{Appendix A: Electroweak 4-Currents}%
\input{AppendixA}\clearpage%
\addcontentsline{toc}{chapter}{Appendix B: Helicity and Chirality}%
\input{AppendixB}\clearpage%
\addcontentsline{toc}{chapter}{Bibliography}%
\input{Biblio}
\clearpage%
\input{Figures}\clearpage%

%===============================================================

%===================== VITA (required by UWM) ==================

%\addcontentsline{toc}{chapter}{Curriculum Vitae}%
%\birthplace{...}

%\begin{startvita}
%\end{startvita}

%\begin{publications}
%\end{publications}

%\finishvita

%===============================================================

\end{document}

%% file: abstract.tex
%%%%%%%%%%%%%%%%%%
% Abstract
%%%%%%%%%%%%%%%%%%

\Abstract{The Weizs\"{a}cker-Williams method is a semiclassical
approximation scheme used to analyze a wide variety of
electromagnetic interactions. It can greatly simplify calculations
that would otherwise be impractical or impossible to carry out
using the standard route of the Feynman rules. With a few
reasonable assumptions, the scope of the method was generalized so
as to accommodate weak, as well as the usual electromagnetic,
interactions. The results are shown to be in excellent agreement,
in the high energy limit of interest, with other methods, and the
generalized scheme is shown to still work in regimes of analysis
where those methods break down.}

%% file: Dedication.tex
%%%%%%%%%%%%%%%%%%
% Dedication
%%%%%%%%%%%%%%%%%%

\dblspace

\vspace*{4.5cm}

\begin{center}
To \\my parents, my sister\\ and\\ my fianc\'{e}e
\end{center}

%% file: Acknowledgments.tex
%%%%%%%%%%%%%%%%%%
% Acknowledgments
%%%%%%%%%%%%%%%%%%

\vspace*{0.3cm}

Support by the funding from the National Space Grant College and
Fellowship Program through the Wisconsin Space Grant Consortium
and by the University of Wisconsin-Milwaukee Graduate School is
gratefully acknowledged. Additional thanks to John Norbury and
Sudha Swaminathan for their continual encouragement to strive for
perfection.

%% file: Introduction.tex
%%%%%%%%%%%%%%%%%%%%%%%%%%%%
% INTRODUCTION
%%%%%%%%%%%%%%%%%%%%%%%%%%%

\dblspace

\chapter{Introduction} \label{sec:introduction} \indent

The physics of interacting charged particles can be quite
complicated. The situation simplifies considerably if the
particles are travelling at very nearly the speed of light. Due to
relativistic effects, the electric $\vecE$ and magnetic $\vecB$
fields of such a particle are (Lorentz) contracted into the plane
that is transverse to the direction of motion. At every point in
this plane, the $\vecE$ and $\vecB$ fields are of very nearly the
same magnitude and are transverse to one another, very much like
on the wavefront of a plane electromagnetic (EM) wave. In fact, to
an observer (viz, another particle) at rest some short distance
away from the passing particle, the effects of these fields are
practically indistinguishable from those of a passing EM wave. If
the particle's EM fields are approximated as EM plane waves, the
problem of analyzing a peripheral (near-miss) collision of two
ultrarelativistic (UR) charged particles thus simplifies to one of
analyzing the interaction between a passing EM wave and just one
particle.

This idea was first introduced in 1924 by E. Fermi, and extended
ten years later by C. Weizs\"{a}cker and E. Williams, and forms
the basis of what is now called the Weizs\"{a}cker-Williams method
(WWM), or Equivalent Photon Approximation (EPA)
\cite{ref:Jack1,ref:Dali}. Since then, much progress has been
made, including the development of a quantum mechanical version of
the method (QWWM) that is in very good agreement (in the UR limit)
with the original semiclassical version (SWWM)
\cite{ref:Dali,ref:Tera,ref:Jack2}. The most basic approximations
of both variants of the WWM are that (1) the colliding particles
are ultrarelativistic; (2) the particles follow straight-line
trajectories; and (3) the particles have no internal structure.
Various complications have been introduced throughout the years,
including the relaxation of approximations (2) and (3), but in its
simplest form, the WWM turns out to be an impressively accurate
approximation scheme. A brief informative survey of the history of
the WWM is presented in \cite{ref:Kim}. The success of the
Standard Model (SM) (viz, the electroweak sector) has brought
about the development of a weak interaction analog of the QWWM,
called the Effective-$W$ Method (EWM)
\cite{ref:Daws1,ref:Daws2,ref:Guni,ref:Kane,ref:Cahn1,ref:Alta}.
The only difference between the EWM and the QWWM is that the
particles mediating the interactions are $W$ and $Z$ bosons
instead of photons. Like the QWWM, the EWM is a very accurate
alternative to the more exact theory.

The purpose of this thesis is to generalize the scope of the
\emph{SWWM} so as to accommodate $W$ and $Z$ bosons, in addition
to photons, as the mediators of the particle collisions. The
motivation for this endeavor is threefold. First, the SM is
generally believed to only be a low energy approximation to a more
(as yet undiscovered) comprehensive theory; it is only reliable up
to interaction energies of about the TeV scale. For energies above
about a TeV, a different approach (one not based on perturbation
theory) is needed. Currently, the EWM is the only such viable
alternative \cite{ref:Guni}. However, the EWM assumes that the
mediating bosons are on-shell (``real"), so in principle it cannot
be used to investigate interactions in which the bosons are
necessary off-shell (``virtual"). An example of such an
interaction is the production of a sufficiently light-weight
resonance $R$ via $W$ or $Z$ boson fusion in fermion-fermion
scattering (cf. Fig. \ref{fig:Rprod}). The EWM can be used if the
two bosons are real and the mass of $R$ is greater than the sum of
their masses. However, if the mass of $R$ is less than the sum of
the on-shell values of the boson masses, conservation of energy
precludes the reaction from proceeding unless the bosons are
virtual. But, in that case, the EWM is not applicable because it
assumes that the bosons are real. So, the second goal of this
research is to formulate a scheme in which the mediating bosons
are not necessarily on-shell, with the hope that it might have a
greater scope of applicability than the EWM. The last factor is
pedagogical --- such a generalization has never been done before.
With the phenomenal success of the electroweak theory within the
past few decades, it seems timely to return to the classical
theory to try to formulate a classical (or, at least
semiclassical) description of the same phenomena. Perhaps
interesting parallels can be drawn between the two viewpoints.

There are usually two parts to the typical SWWM analysis --- a
semiclassical part and a quantum part. For concreteness in
understanding the implementation of the method, consider the
prototypical interaction shown in Fig. \ref{fig:Xprod}. An
incident particle $q$ undergoes a peripheral collision with a
target particle $P$ by way of a one-photon exchange. The
semiclassical part of the calculation is concerned with the
emission of EM energy $\gamma$ (a photon in the quantum viewpoint)
from $q$. The useful quantity is the number spectrum $N(E)$ of
photons --- the differential number d$n(E)$ of photons of energy
$E$ in $q$'s EM fields per unit photon energy d$E$:
\begin{equation}
N(E)=\frac{\dif n(E)}{\dif E}\qquad \mbox{(number spectrum).}\label{eq:N1}%
\end{equation}
Usually, the quantum part involves the description of the
interaction between the emitted photon and a target particle ($P$
in this case). The quantity of interest here is the (microscopic)
cross section $\sigma_{mic}$ for this photon-induced subprocess.
The (macroscopic) cross section $\sigma_{mac}$ for the overall
process is found by folding $N(E)$ with $\sigma_{mic}$:
\begin{equation}
\sigma_{mac}=\int \dif E\, N(E)\, \sigma_{mic},\label{eq:s1}
\end{equation}
where the integral runs over all allowable (by energy
conservation) photon energies. So, in a crude sense, $N(E)$ gives
the probability that $q$ emits the photon $\gamma$ at energy $E$,
and $\sigma_{mic}$ is the probability that $\gamma$ then collides
with $P$ and produces the final state $X$. A more interesting
application of the method is where \emph{two} bosons are
simultaneously exchanged, and collide to produce some final state
$X$. The microscopic cross section then describes the
two-boson-induced subprocess. An example of one such process is SM
Higgs boson production via boson-boson fusion (cf. Fig.
\ref{fig:Hprod}). The Higgs boson is the last remaining unverified
prediction of the SM that presumably endows all particles with
mass. For more information on this interesting particle, see
\cite{ref:Grif}, \cite{ref:Halz}, or practically any other
reference on modern particle physics. For such a two-boson
process, $\sigma_{mac}$ is of the form
\begin{equation}
\sigma_{mac}=\int \dif E_1 \int \dif E_2\;\, N(E_1)\, N(E_2)\,
\sigma_{mic}.\label{eq:s2}
\end{equation}
The expression for the macroscopic cross section in the SWWM, QWWM
and EWM all share this same form, but whereas $N(E)$ is calculated
using semiclassical considerations in the SWWM, it is treated
quantum mechanically in both the QWWM and EWM. The determination
of $\sigma_{mic}$ is a quantum calculation in all of these
schemes. The method developed here is formulated in much the same
spirit as in the (traditional) SWWM --- the $N(E)$ functions are
derived semiclassically and $\sigma_{mic}$ is derived using
quantum field theory. So, like the SWWM, this scheme is
semiclassical in nature.

This document is arranged as follows. Chapter \ref{sec:notation}
summarizes the notation and conventions that will be used, and the
overall scenario that is under consideration. In Chapter
\ref{sec:N}, the number spectrum function $N(E)$ is derived in a
more general way than is done in the traditional analysis ---
allowing for bosons with some (as yet unspecified) \emph{nonzero}
mass. In Chapter \ref{sec:Ncases}, $N(E)$ is shown to reduce to
the relevant functions (that appear in the other above-mentioned
methods) in the various limits of interest when a few extra
reasonable assumptions are made. And, finally, in Chapter
\ref{sec:othmethods}, the results obtained using the current
approach are compared to those of other studies.

%% file: Chapter2.tex
%%%%%%%%%%%%%%%%%%
% Chapter 2
%%%%%%%%%%%%%%%%%%

\chapter{Notation, Conventions and Overall Scenario} \label{sec:notation} \indent

Natural units (rationalized Heaviside-Lorentz units (where
$\varepsilon_0=\mu_0=1$) with $\hbar=c=1$) and the Einstein
summation convention (the repeated appearance of the same index in
any tensor equation implies a sum over all allowable values) are
used throughout. Greek indices $\mu,\, \nu,\, \xi,\, \ldots$ can
take on the values $0,\, 1,\, 2,\, 3,\mbox{ or }4$, and Latin
indices $i,\, j,\, k,\, \ldots$ can only take on the values $1,\,
2\mbox{ or }3$. The signature of the metric is taken to be $+\,
-\, -\, -$\,, so that
\begin{equation}
g^{\mu \nu}=g^{\nu \mu}=
\begin{pmatrix}
1 & \;\;\; 0 & \;\;\; 0 & \;\;\; 0\\
0 & -1 & \;\;\; 0 & \;\;\; 0\\
0 & \;\;\; 0 & -1 & \;\;\; 0\\
0 & \;\;\; 0 & \;\;\; 0 & -1
\end{pmatrix}
\qquad \mbox{(metric tensor).}\label{eq:MetMat}
\end{equation}
Of interest is a peripheral collision between an incident particle
$q$ and some target point $P$. $P$ can be thought of as another
particle or merely an interaction point; a prototypical Feynman
diagram that takes $P$ to be an actual particle appears in Fig.
\ref{fig:Xprod}. At the point of closest approach between $q$ and
$P$, a boson (either a photon or a $W$ or $Z$ boson) is emitted by
$q$ that subsequently interacts with some other particle at $P$.
It is convenient to proceed from the standpoint of an observer at
rest in the frame comoving with $P$, who views $q$ pass by with
some UR speed $v\simeq 1$. The basic scenario is depicted in Fig.
\ref{fig:LTWWM}. Frames $K$ (rest frame of $P$) and $K'$ (rest
frame of $q$) share a common $z$ axis and coincide at time
$t=t'=0$. $q$ is located at the origin of $K'$ and moves at
constant velocity $\vecv=\hatv v$ (with magnitude $v$ and
direction $\hatv$) past $P$ along the common $z/z'$ axis (the
identifications $\hatv =\hatz$ in $K$ and $\hatv =\hatzp$ in $K'$
can always be made). $P$ is located at a fixed perpendicular
distance $b$ (the ``impact parameter") from the $z/z'$ axis, along
the $x$ axis. $b$ is also the distance of closest approach between
$q$ and $P$, which occurs at time $t=t'=0$.

As the WWM formalism is (by construction) only valid at high
energies, the most frequent approximation made throughout this
thesis is that the colliding particles (both fermions \emph{and}
bosons) are travelling at UR speeds. This condition is
alternatively expressed as $v=|\vecp|/E\simeq 1$, $\gamma=E/m \gg
1$, etc., where
\begin{equation}
\gamma =\frac{1}{\sqrt{1-v^2}} \qquad \mbox{(Lorentz
factor)}\label{eq:LorFac}
\end{equation}
is the Lorentz factor. It will be assumed that any errors
introduced by this approximation are negligible.

Lastly, the symbols $\gamma$, $Z$ and $W$ are used to label
quantities corresponding to EM, neutral weak and charged weak
interactions, respectively. It should be clear from the context
whether $\gamma$ means $1/\sqrt{1-v^2}$ or is a label identifying
an EM quantity, and whether $Z$ means $z$ axis or $Z$ boson, and
so on.

%% file: Chapter3.tex
%%%%%%%%%%%%%%%%%%
% Chapter 3
%%%%%%%%%%%%%%%%%%

\chapter{Derivation of $N(E)$} \label{sec:N} \indent

The derivation of $N(E)$ consists of two separate analyses. One is
the specification of the scalar $\Phi$ and vector $\vecA$
potentials and the $\vecE$ and $\vecB$ fields of an UR charged
particle $q$. The other is the determination of the same set of
quantities for relevant plane waves of radiation in a vacuum. The
next step, which is the hallmark of the WWM, is to make
appropriate identifications between the two sets of quantities,
and thereby approximate the fields of $q$ as equivalent pulses of
plane wave radiation. This approximation is known to work in the
SWWM, and will be shown to be interconsistent as well in the
generalized version (GWWM) developed herein. The identification
allows for the interpretation of the fields as quantum mechanical
bosons, and hence makes plausible the subsequent quantum field
theoretic treatment of any boson-induced microscopic cross section
of interest.

As mentioned briefly in the Introduction, the function $N(E)$ is
the differential number of bosons of energy $E$ in the fields
surrounding $q$ per unit boson energy. The quantity of paramount
importance in the derivation of $N(E)$ is the Poynting vector
$\vecS(\vecr ,t)$, or energy flux, associated with the fields of
$q$. With appropriate factors of $\hbar$ taken into account,
$N(E)$ can be thought of as the Fourier transform (FT)
$\vecS(\vecr ,\omega)$ of $\vecS(\vecr ,t)$, integrated over the
(infinite) wavefront area of the equivalent pulse, and divided by
the energy $E=\omega$ of the bosons. The analysis thus begins with
a general (coordinate-independent) discussion of the Poynting
vector $\vecS$.

\section{Energy Flux}
\label{sec:SPulses} \indent

Insofar as energy transport by radiation is concerned, the most
descriptive dynamic quantity is the Poynting vector $\vecS$, which
represents the energy flux (energy per unit time per unit area)
carried by the wave. It is a familiar result from classical
electrodynamics (ED) that $\vecS=\vecE \times \vecB$ in a
source-free region. This short section sets the stage for future
calculations by stating the generalization of this formula to
cases where the fields have some nonzero mass $m$.

The $i^{th}$ component of $\vecS$ is the $i0^{th}$ component of
the stress-energy tensor $T^{\mu\nu}$: $S^i=T^{i0}$. More
generally, $T^{\mu\nu}=T^{\nu\mu}$ is the flow of the $\mu^{th}$
component of 4-momentum along the $\nu$ direction. The
generalization being sought can be derived using standard
variational techniques once the Lagrangian is known. For a massive
spin-1 field, the result is (see \cite{ref:Grei}, for example)
\begin{equation}
\vecS=\vecE \times \vecB +m^2\Phi \vecA -\rho \vecA,\label{eq:S1}
\end{equation}
where $\Phi$, $\vecA$, $\vecE$, $\vecB$ are the aforementioned
scalar and vector potentials, and electric and magnetic fields,
respectively, and $\rho$ is the charge density of the source. In
the case under consideration, the region of space at the
observation point, where $\vecS$ is to be evaluated, is
source-free. So $\rho=0$ and Eq. (\ref{eq:S1}) becomes
\begin{equation}
\vecS =\vecE \times \vecB + m^2\Phi \vecA.\label{eq:S2}
\end{equation}
Note that Eq. (\ref{eq:S2}) consistently reduces to the familiar
expression $\vecS=\vecE \times \vecB$ in the $m=0$ limit. It's
also important to note that in the $m=0$ limit, the functional
forms of $\Phi$ and $\vecA$ are inconsequential insofar as the
physically measurable $\vecS$ is concerned. This subtlety has its
roots in gauge field theory. It is a familiar result from
classical ED that the solutions to Maxwell's equations (ME) are
arbitrary up to a choice of gauge. More specifically, there is a
whole family of functions $\Phi$ and $\vecA$ that yield the same
$\vecE$ and $\vecB$ fields, which are the only quantities of any
importance in determining $\vecS$. The same cannot be said for the
massive field case --- if $m \ne 0$, there is no arbitrariness
allowed to any of these functions.

\section{Potentials and Fields of an UR Point Charge}
\label{sec:PFURParticle} \indent

The potentials and fields of $q$ are first solved for in $K'$, the
rest frame of $q$, and are then (Lorentz) transformed to frame
$K$, the rest frame of $P$, where they are evaluated at the
location of $P$.

\subsection{$\Phi'(\vecrp ,t')$ and $\vecAp(\vecrp ,t')$ of a Point
Charge at Rest} \label{sec:QatRest} \indent

The equations of motion for the four vector bosons (the radiation
fields in the semiclassical picture) appearing in the SM are
generally quite complex. The equations of motion for a given field
have vestiges of the other three bosons. These ``coupling terms"
arise mathematically because the SU(2) algebra describing the
interactions is nonabelian (noncommutative). More intuitively, the
terms correspond to interactions among the bosons. In particular,
interactions couple $W^+$ and $W^-$ bosons to each other and to
$Z$ bosons and photons. Photons and $Z$ bosons do not interact
because photons only couple to electrically charged particles, and
$Z$ bosons are electrically neutral. For the problem of interest
to this thesis, the coupling terms naturally disappear from the
equations because the interactions of interest only involve bosons
being emitted from fermions. In the language of quantum field
theory, the processes are always tree-level fermion-boson
interactions, so there are no further boson-boson couplings to
consider. For a given boson, one simply sets all other boson
fields to zero. A detailed presentation of this procedure, applied
to decoupling the $Z$ boson and photon equations of motion, is
given in \cite{ref:Carr}. The following analysis implicitly
assumes that such a step has already been carried out, so that the
field equations for the four bosons are uncoupled from one
another.

The functions $\Phi$ and $\vecA$, describing the radiation fields,
are combined into a 4-vector called the 4-potential,
\begin{equation}
A^\mu=(\Phi,\vecA) \qquad \mbox{(4-potential),}\label{eq:Amu1}
\end{equation}
and the charge $\rho$ and current $\vecJ$ densities, describing
the source charge, are combined into the 4-current:
\begin{equation}
J^\mu=(\rho,\vecJ) \qquad \mbox{(4-current).}\label{eq:Jmu1}
\end{equation}
For massless fields (photons), the equations of motion linking
$A^\mu$ and $J^\mu$ are ME:
\begin{equation}
\Box A^\mu =J^\mu \qquad \mbox{(Maxwell's
equations).}\label{eq:ME}
\end{equation}
For massive fields ($W$ and $Z$ bosons), the set of equations is
called the Proca equation (PE):
\begin{equation}
\Box A^\mu +m^2A^\mu =J^\mu \qquad \mbox{(Proca
equation).}\label{eq:PE}
\end{equation}
Both sets of field equations are shown here after the Lorentz
condition (LC), $\partial^\mu A_\mu =0$, has already been imposed.
The LC is an optional constraint in the former case, but is a
natural consequence of current conservation in the latter. $\Box$
is the D'Alembertian operator,
\begin{equation}
\Box =\partial^\mu \partial_\mu =\frac{\partial^2}{\partial
t^2}-\vecdel^2 \qquad \mbox{(D'Alembertian
operator),}\label{eq:dAlem}
\end{equation}
and the quantity $m$ appearing in Eq. (\ref{eq:PE}) is the mass of
the boson, which, for the usual application to on-shell $W$ and
$Z$ bosons, is set equal to $80.42$ and $91.19$ GeV, respectively
\cite{ref:RPP}. For the sake of generality, the analysis will be
confined entirely to the PE (note that ME are the $m=0$ limit of
the PE), and $m$ will be left unspecified for the time being.
Later, special cases of interest, where the appropriate value of
$m$ will be explicitly stated, will be considered. From time to
time, $m$ will be set to $0$ in the formulas to verify that they
correctly reduce to those of the SWWM.

The starting point is thus Eq. (\ref{eq:PE}). This equation will
first be solved for a point charge $q$ at rest (in frame $K'$).
${A^\mu}'$ does not \emph{explicitly} depend on time, as the
source is simply a static charge at rest, so $\partial {A^\mu}'
/\partial t'=0$, and Eq. (\ref{eq:PE}) reduces to
\begin{equation}
\vecdel^2 {A^\mu}' - m^2{A^\mu}' =-{J^\mu}'.\label{eq:PE1}
\end{equation}
The solution is a standard exercise in Fourier analysis (see
\cite{ref:Roln}, or the methods outlined in \cite{ref:Jack1}, for
example). In frame $K'$, ${A^\mu}'$ is found to be
\begin{equation}
{A^\mu}'(\vecrp ,t')=\oofourpi \, \int \dif^3 \vecrbp \,
\frac{{J^\mu}'(\vecrbp ,t')}{|\vecrp-\vecrbp|}\, \mbox{\Large
e}^{-m|\vecrp-\vecrbp|},\label{eq:PEsol1}
\end{equation}
where $\vecrp =\vecrp (t')$ and $\vecrbp =\vecrbp (t')$ are the
position vectors of $P$ and $\dif q$ (the differential source
charge element within $q$), respectively. From basic Lorentz
transformation (LT) equations, it is known that the position
vector of $P$ in $K'$ is $\vecrp (t')=(b,\, 0,\, -vt')$, and it
might be expected that $\vecrbp (t')=(0,\, 0,\, 0)$, as $q$ is
supposed to be a point charge located at the origin of $K'$.

After a moderate amount of work, a form for $J^\mu(\vecr ,t)$ that
is common to all three types of electroweak (EW) interactions of
\emph{point charges} can be found. The derivation is presented in
Appendix A. The result is
\begin{equation}
J^\mu(\vecr ,t)=\delta [\vecr (t)]\, q^\mu \qquad \mbox{(4-current
of a point charge).}\label{eq:Jmu2}
\end{equation}
$\delta [\vecr (t)]$ is the (3 dimensional) Dirac delta function,
that forces all $\dif q$ elements to be located at the same point,
$\vecr (t)=\veczero$. $q^\mu=(q^0,\vecq)$ is a new 4-vector
introduced here, called the ``4-charge". It can be expressed in
all cases of interest as a linear combination of the 4-velocity
$u^\mu$ and normalized 4-spin $s^\mu$ of the particle:
\begin{equation}
q^\mu=q_Vu^\mu + q_As^\mu \qquad \mbox{(4-charge).}\label{eq:qmu1}
\end{equation}
The subscripts $V$ and $A$ on the coefficients $q_V$ and $q_A$
refer to the fact that $u^\mu$ and $s^\mu$ are vector and
axial-vector quantities, respectively. $u^\mu$ can always be
written as
\begin{equation}
u^\mu=\gamma (1,\hatv v) \qquad \mbox{(4-velocity),}\label{eq:umu}
\end{equation}
and $s^\mu$ is expressed in a simple way in terms of the helicity
$\lambda$ of the particle, as
\begin{equation}
s^\mu=\gamma (\lambda v, \hats) \qquad \mbox{(normalized
4-spin).}\label{eq:smu1}
\end{equation}
$\hats$ is the normalized spin (3-vector) of the particle, which
is chosen to be measured along the $z$ axis so that $\hats=+\hatz$
for spin-up and $\hats=-\hatz$ for spin-down. Helicity is the
normalized projection of the particle's spin along the direction
of motion, and can be expressed in terms of $\hats$ as
\begin{equation}
\lambda=\hats \cdot \hatv \qquad \mbox{(helicity).}\label{eq:hel1}%
\end{equation}
See Appendix B for a more in depth discussion of helicity. Because
$\vecv$ is being chosen to be along the $z$ axis, $\hats$ can also
be written as
\begin{equation}
\hats=\hatz \lambda \qquad \mbox{(normalized 3-spin),}\label{eq:spinvec}%
\end{equation}
where $\lambda =\hats \cdot \hatz =+1(-1)$ for spin-up(down).
While the helicity of a massive particle is not Lorentz invariant
in general, it can be easily shown that for the simple Lorentz
boost that will be considered here (where $\hatv$ is fixed in
direction), $\lambda'=\lambda$ is constant. Plugging Eq.
(\ref{eq:spinvec}) into Eq. (\ref{eq:smu1}), and using $\hatv
=\hatz$, the general result
\begin{equation}
s^\mu=\gamma \lambda (v,\hatz)\label{eq:smu2}
\end{equation}
is found. $q^\mu$ can then be written
\begin{equation}
q^\mu=\gamma ((q_V+q_A \lambda v),\hatz (q_Vv+q_A
\lambda)).\label{eq:qmu2}
\end{equation}
For completeness, $u^\mu$, $s^\mu$, $q^\mu$, and $J^\mu$ are now
specified in two limiting cases of interest: the zero-velocity
limit (i.e., the rest frame $K'$ of the charge $q$ itself, where
$v=0$), and the UR limit (i.e., the rest frame $K$ of the observer
$P$, where $v=1$). In the former limit (with $\hatz=\hatzp$),
\begin{equation}
{u^\mu}'=(1,\veczero) \qquad \mbox{(rest frame),}\label{eq:umu3}
\end{equation}
\begin{equation}
{s^\mu}'=(0,\hatzp \lambda) \qquad \mbox{(rest
frame),}\label{eq:smu3}
\end{equation}
\begin{equation}
{q^\mu}'=(q_V,\hatzp q_A \lambda) \qquad \mbox{(rest
frame),}\label{eq:qmu3}
\end{equation}
and
\begin{equation}
{J^\mu}'(\vecrp,t')=\delta [\vecrp (t')]\, (q_V,\hatzp q_A
\lambda) \qquad \mbox{(rest frame).}\label{eq:Jmu3}
\end{equation}
In the UR limit,
\begin{equation}
u^\mu=\gamma (1,\hatz) \qquad \mbox{(UR limit),}\label{eq:umu4}
\end{equation}
\begin{equation}
s^\mu=\gamma \lambda (1,\hatz)=\lambda u^\mu \qquad \mbox{(UR
limit),}\label{eq:smu4}
\end{equation}
\begin{equation}
q^\mu=\gamma q_{VA}(1,\hatz)=q_{VA}u^\mu \qquad \mbox{(UR
limit),}\label{eq:qmu5}
\end{equation}
and
\begin{equation}
J^\mu (\vecr,t)=\delta [\vecr (t)]\, \gamma q_{VA}(1,\hatz)=\delta
[\vecr (t)]\, q_{VA}u^\mu \qquad \mbox{(UR limit).}\label{eq:Jmu4}
\end{equation}
The new quantity
\begin{equation}
q_{VA}\equiv q_V+q_A\lambda \qquad \mbox{(definition of VA
charge),} \label{eq:qeff}
\end{equation}
called the ``VA charge", has been introduced in the expressions
for $q^\mu$ and $J^\mu (\vecr,t)$.

While the form of $q^\mu$ is common to all interactions of
interest, the charges $q_V$ and $q_A$ differ among them. These
charges generally depend on the \emph{dimensionless} charge
quantum numbers $Q^\gamma$ and $T^3$. $Q^\gamma$ is the EM charge
quantum number, which is related to the usual electric charge
$q^\gamma$ according to $q^\gamma=Q^\gamma e$, where $e$ is the
magnitude of the charge on the electron. In the rationalized
Heaviside-Lorentz system of units being used here, $e=\sqrt{4\mpi
\alpha}=0.3028$ to four significant figures, where $\alpha
=7.297\times 10^{-3}\simeq 1/137$ is the fine structure constant
\cite{ref:RPP}. $T^3$ is the third component of the vector
$\boldsymbol{T}=(T^1,\, T^2,\, T^3)$ of weak isospin quantum
numbers of the fermion. $T^3$ is either $+1/2$ or $-1/2$ for
left-handed quarks and leptons (denoted with a subscript $L$), and
$0$ for their right-handed counterparts (denoted with a subscript
$R$). Note that $L$ and $R$ states are also eigenstates of the
helicity operator in the high energy limit, with respective
eigenvalues $-1$ and $+1$ (cf. Appendix B). So in that limit,
$\lambda$ is also an informative quantum number. An additional
quantum number---weak hypercharge $Y$---is introduced here as
well, as it will be referred to in a future section. Each
left-handed weak isospin doublet that appears in the SM is
assigned a unique value of $Y$; that is, each member of the pair
has that same quantum number. Also, each right-handed weak
isosinglet has a unique value of $Y$. So it is an informative
parameter because it differentiates $R$ singlet states from those
states that belong together in an $L$ doublet. $Y$ is easily
deduced from $Q^\gamma$ and $T^3$ by a weak interaction analog of
the Gell-Mann--Nishijima formula,
\begin{equation}
Y=2(Q^\gamma -T^3)\qquad \mbox{(Gell-Mann--Nishijima formula).}\label{eq:GMN}%
\end{equation}
Table \ref{tab:charges} summarizes the relevant charge quantum
number assignments for quarks, leptons, nucleons, and nuclei. The
right-handed neutrino states have been included for completeness,
although they do not appear in the minimal version of the SM
(\emph{all} neutrinos are left-handed). Protons, neutrons and
nuclei also do not appear in the SM, but can nevertheless be
parameterized by these quantum numbers as well. Technically
speaking, they form a more generic type of isospin doublet,
instead of a left-handed weak isospin doublet, and are not
assigned a weak hypercharge quantum number, but those subtleties
will be ignored here. Except for helicity, all charge quantum
numbers of an antiparticle state are simply the negatives of the
quantum numbers of the corresponding particle state. Helicity does
not change sign because neither spin nor velocity change sign
under the charge conjugation operation that transforms a particle
state into its antiparticle state (or vice versa); hence $\lambda
=\hats \cdot \hatv$ does not change sign either.

\renewcommand{\arraystretch}{1.4}
\begin{table}
\renewcommand{\tablename}{TABLE}
\caption{\label{tab:charges} CHARGE QUANTUM NUMBERS OF VARIOUS
FERMIONS}
\medskip
\begin{tabular*}{5in}{@{\extracolsep{\fill}}p{1.85in}cccc@{\hspace{1em}}@{\extracolsep{1em}}}
\hline\hline
Fermion & $\lambda$ & $Q^\gamma$ & $T^3$ & $Y$\\
\hline
$(\nu _e)_\mL $, $(\nu _\mu )_\mL $, $(\nu _\tau )_\mL $ & $-1$ & \hspace{4pt} $0$ & \hspace{4pt} $\half$ & $-1$\\
$e^-_\mL $, $\mu^-_\mL $, $\tau^-_\mL $ & $-1$ & $-1$ & $-\half$ & $-1$\\
$(\nu _e)_\mR $, $(\nu _\mu )_\mR $, $(\nu _\tau )_\mR $ & \hspace{4pt} $1$ & \hspace{4pt} $0$ & \hspace{4pt} $0$ & \hspace{4pt} $0$\\
$e^-_\mR $, $\mu^-_\mR $, $\tau^-_\mR $ & \hspace{4pt} $1$ & $-1$ & \hspace{4pt} $0$ & $-2$\\
$u_\mL $, $c_\mL $, $t_\mL $ & $-1$ & \hspace{4pt} $\frac{2}{3}$ & \hspace{4pt} $\half$ & \hspace{4pt} $\frac{1}{3}$\\
$d_\mL $, $s_\mL $, $b_\mL $ & $-1$ & $-\frac{1}{3}$ & $-\half$ & \hspace{4pt} $\frac{1}{3}$\\
$u_\mR $, $c_\mR $, $t_\mR $ & \hspace{4pt} $1$ & \hspace{4pt} $\frac{2}{3}$ & \hspace{4pt} $0$ & \hspace{4pt} $\frac{4}{3}$\\
$d_\mR $, $s_\mR $, $b_\mR $ & \hspace{4pt} $1$ & $-\frac{1}{3}$ & \hspace{4pt} $0$ & $-\frac{2}{3}$\\
proton, $p=uud$ & $\pm 1$ & \hspace{4pt} $1$ & \hspace{4pt} $\half$ & \hspace{4pt} $1$\\
neutron, $n=udd$ & $\pm 1$ & \hspace{4pt} $0$ & $-\half$ & \hspace{4pt} $1$\\
nucleus (with $Z$ protons and $N$ neutrons) & \raisebox{-1.5ex}[0pt]{$\pm 1$} & \raisebox{-1.5ex}[0pt]{$Z$} & \raisebox{-1.5ex}[0pt]{$\half (Z-N)$} & \hspace{4pt} \raisebox{-1.5ex}[0pt]{$Z+N$}\\
\hline\hline
\end{tabular*}
\end{table}

Having now specified the relevant charge quantum numbers, the
vector, axial-vector, and $VA$ charges are now enumerated for the
three types of interactions. For EM currents,
\begin{equation}
\left.
\begin{array}{r@{\, = \;}l}
q_V & Q^\gamma e\\
q_A & 0\\
q_{VA} & Q^\gamma e\\
\end{array}
\right\} \qquad \mbox{(charges to which the $\gamma$
couples),}\label{eq:chEM}
\end{equation}
For neutral weak currents,
\begin{equation}
\left.
\begin{array}{r@{\, = \;}l}
q_V & \half g_Z(T^3-2Q^\gamma \sin^2\theta_W)\\
q_A & -\half g_ZT^3\\
q_{VA} & \half g_Z[(1-\lambda )T^3-2Q^\gamma
\sin^2\theta_W]\\
\end{array}
\right\} \mbox{(charges to which the $Z$ boson
couples).}\label{eq:chNEW}
\end{equation}
The quantity $g_Z$ here is the neutral weak coupling constant. It
has a value $g_Z=e/\sin\theta_W\cos\theta_W=0.7183$ (to four
significant figures), where $\theta_W=28.74^\circ$ is the weak
mixing (or Weinberg) angle \cite{ref:RPP}. Charged weak currents
have these charges:
\begin{equation}
\left.
\begin{array}{r@{\, = \;}l}
q_V & \mbox{\large $\halfort$}g_W\\
q_A & \mp \mbox{\large $\halfort$}g_W\\
q_{VA} & \mbox{\large $\halfort$}g_W(1\mp \lambda )\\
\end{array}
\right\} \qquad \mbox{(charges to which the $W^\pm$ boson
couples),}\label{eq:chCEW}
\end{equation}
where $g_W=e/\sin\theta_W=0.6298$ is the charged weak coupling
constant \cite{ref:RPP}. The canonical (Lorentz invariant) charge,
as defined via the Noether prescription, for a given interaction
can be shown to be simply $q_V$:
\begin{equation}
\left.
\begin{array}{r@{\, = \;}l}
q^\gamma & Q^\gamma e\\
q^Z & \half g_Z(T^3-2Q^\gamma \sin^2\theta_W)\\
q^W & \mbox{\large $\halfort$}g_W\\
\end{array}
\right\} \qquad \mbox{(Noether charges).}\label{eq:chNoether}
\end{equation}
See also Appendix A for an in depth derivation of these charge
assignments, and for a short list of useful references that also
use these charge parameters.

Returning to the calculation of interest, it is found that Eq.
(\ref{eq:Jmu2}) helps to simplify Eq. (\ref{eq:PEsol1})
considerably. The delta function in $J^\mu$ kills the integral and
makes $q$ a \emph{point charge}, whose position vector is now
given by $\vecrpq (t')=(0,\, 0,\, 0)$. Upon plugging Eq.
(\ref{eq:Jmu2}) into Eq. (\ref{eq:PEsol1}), it is found that
\begin{subequations}
\begin{align}
A^\mu(\vecrp ,t')&=\oofourpi\, \frac{{q^\mu}'}{|\vecrp|}\,
\mbox{\Large e}^{-m|\vecrp|}\label{eq:PEsol2}\\%
&=\oofourpi\, \frac{{q^\mu}'}{r'}\, \mbox{\Large
e}^{-mr'},\label{eq:PEsol3}%
\end{align}
\end{subequations}
where
\begin{equation}
r'=r'(t')=\sqrt{b^2+(vt')^2}\label{eq:rp}%
\end{equation}
is the magnitude of the position vector $\vecrp(t')$ of $P$
relative to $q$ (in frame $K'$). The general form of the scalar
potential works out to be
\begin{subequations}
\begin{align}
\Phi'(\vecrp ,t')&=\oofourpi\, \frac{{q^0}'}{r'}\, \mbox{\Large
e}^{-mr'}\label{eq:Phip1}\\%
&=\oofourpi\, \frac{q_V}{r'}\, \mbox{\Large
e}^{-mr'},\label{eq:Phip2}%
\end{align}
\end{subequations}
and that of the vector potential is, similarly,
\begin{subequations}
\begin{align}
\vecAp(\vecrp ,t')&=\oofourpi\, \frac{\vecqp}{r'}\, \mbox{\Large e}^{-mr'}\label{eq:Ap1}\\%
&=\hatzp \oofourpi\, \frac{q_A\lambda}{r'}\, \mbox{\Large e}^{-mr'}.\label{eq:Ap2}%
\end{align}
\end{subequations}
For future analysis, it will be useful to express $\vecAp$ in
Cartesian components.
\begin{subequations}
\begin{align}
{A_x}'(\vecrp ,t')&=0\label{eq:Apx}\\%
{A_y}'(\vecrp ,t')&=0\label{eq:Apy}\\%
{A_z}'(\vecrp ,t')&=\oofourpi\, \frac{q_A\lambda}{r'}\, \mbox{\Large e}^{-mr'}.\label{eq:Apz}%
\end{align}
\end{subequations}
Note that Eqs. (\ref{eq:Phip2}) and (\ref{eq:Ap2}) reassuringly
reduce to the EM limit (to the familiar expressions in classical
ED) when $m$ is set to zero and $q_V$ and $q_A$ are set to the
values specified in Eq. (\ref{eq:chEM}):
\begin{equation}
\renewcommand{\arraystretch}{1.3}
\left.
\begin{array}{r@{\, = \;}l}
{\Phi^\gamma}'(\vecrp ,t') & \mbox{\Large $\oofourpi\,
\frac{q^{\mbox{\scriptsize $\gamma$}}}{r'}$}\\
{\boldsymbol{A}^\gamma}'(\vecrp ,t') & \veczero\\
\end{array}
\right\} \qquad \mbox{(EM limit).} \label{eq:AmuEM}
\end{equation}

\subsection{$\vecEp(\vecrp ,t')$ and $\vecBp(\vecr', t')$ of a Point
Charge at Rest} \label{sec:EandB} \indent

The generalized $\vecE$ and $\vecB$ fields are identified in the
same way that they are in classical ED --- as components of the
field strength tensor $F^{\mu \nu}$
\cite{ref:Carr,ref:Huan,ref:Frau}. In the absence of other fields,
$F^{\mu \nu}$ is defined (for either massless or massive fields)
as $F^{\mu \nu}=\partial^\mu A^\nu -\partial^\nu A^\mu$, and its
components are related to the components of $\vecE$ and $\vecB$ as
$F^{i0}=E^i$ and $F^{ij}=-{\epsilon^{ij}}_kB^k$, where
$\epsilon^{ijk}$ is the completely antisymmetric Levi-Civita
tensor density (with $\epsilon^{123}=+1$). In matrix notation,
\begin{equation}
F^{\mu \nu}=-F^{\nu \mu}=
\begin{pmatrix}
0 & -E_x & -E_y & -E_z\\
\,E_x & \;\; 0 & -B_z & \quad\! B_y\\
\,E_y & \quad\! B_z & \;\; 0 & -B_x\\
\,E_z & -B_y & \quad\! B_x & \;\; 0
\end{pmatrix}
.\label{eq:FSTMat}%
\end{equation}
It is easy to then verify that
\begin{subequations}
\begin{align}
E^i&=F^{i0}\label{eq:Ei1}\\%
&=\partial^iA^0-\partial^0A^i,\label{eq:Ei2}%
\end{align}
\end{subequations}
or
\begin{equation}
\vecE=-\vecdel \Phi-\frac{\partial \vecA}{\partial
t},\label{eq:Evec1}%
\end{equation}
and
\begin{subequations}
\begin{align}
B^i&=-\half {\epsilon^i}_{jk}F^{jk}\label{eq:Bi1}\\%
&=-\half
{\epsilon^i}_{jk}(\partial^jA^k-\partial^kA^j)\label{eq:Bi2}\\%
&=-{\epsilon^i}_{jk}\partial^jA^k,\label{eq:Bi3}%
\end{align}
\end{subequations}
or
\begin{equation}
\vecB=\vecdel \times \vecA.\label{eq:Bvec1}%
\end{equation}
The forms of Eqs. (\ref{eq:Evec1}) and (\ref{eq:Bvec1}) are the
same for massless and massive fields, but the explicit expressions
will be seen to differ because $\Phi$ and $\vecA$ differ for the
two cases (cf. Eqs. (\ref{eq:Phip2}) and (\ref{eq:Ap2})). The
explicit expressions for $\vecEp(\vecrp ,t')$ and $\vecBp(\vecrp
,t')$ are easily worked out by plugging Eqs. (\ref{eq:Phip2}) and
(\ref{eq:Ap2}) into Eqs. (\ref{eq:Evec1}) and (\ref{eq:Bvec1}).
\begin{subequations}
\begin{align}
\vecEp(\vecrp ,t')&=-\vecdelp \Phi'(\vecrp ,t')-\frac{\partial
\vecAp(\vecrp ,t')}{\partial t'}\label{eq:Ep1}\\%
&=-\hatrp\, \frac{\partial \Phi'(\vecrp ,t')}{\partial
r'}\label{eq:Ep2}\\%
&=\hatrp\, \oofourpi\, \frac{q_V}{{r'}^2}\, (1+mr')\, \mbox{\Large
e}^{-mr'},\label{eq:Ep3}%
\end{align}
\end{subequations}
where the static charge condition, $\partial \vecAp(\vecrp
,t')/\partial t'=0$, has been used in Eq. (\ref{eq:Ep1}). In
Cartesian components,
\begin{subequations}
\begin{align}
{E_x}'(\vecrp ,t')&=\oofourpi\, \frac{q_Vb}{{r'}^3}\, (1+mr')\,
\mbox{\Large e}^{-mr'}\label{eq:Epx}\\%
{E_y}'(\vecrp ,t')&=0\label{eq:Epy}\\%
{E_z}'(\vecrp ,t')&=-\oofourpi\, \frac{q_Vvt'}{{r'}^3}\, (1+mr')\,
\mbox{\Large e}^{-mr'}.\label{eq:Epz}%
\end{align}
\end{subequations}
And, for the $\vecB$ field,
\begin{subequations}
\begin{align}
\vecBp(\vecrp ,t')&=\vecdelp \times \vecAp(\vecrp ,t')\label{eq:Bp1}\\%
&=\vecdelp \times [\hatzp {A_z}'(\vecrp ,t')]\label{eq:Bp2}\\%
&=\hatzp \times [-\vecdelp {A_z}'(\vecrp ,t')]+[\vecdelp \times
\hatzp]\, {A_z}'(\vecrp ,t')\label{eq:Bp3}\\%
&=[\hatzp \times \hatrp]\, \oofourpi\, \frac{q_A\lambda}{{r'}^2}\,
(1+mr')\, \mbox{\Large e}^{-mr'}\label{eq:Bp4}\\%
&=\hatyp\, \oofourpi\, \frac{q_A \lambda b}{{r'}^3}\, (1+mr')\,
\mbox{\Large e}^{-mr'}.\label{eq:Bp5}%
\end{align}
\end{subequations}
Or, in Cartesian coordinates,
\begin{subequations}
\begin{align}
{B_x}'(\vecrp ,t')&=0\label{eq:Bpx}\\%
{B_y}'(\vecrp ,t')&=\oofourpi\, \frac{q_A\lambda b}{{r'}^3}\,
(1+mr')\, \mbox{\Large e}^{-mr'}\label{eq:Bpy}\\%
{B_z}'(\vecrp ,t')&=0.\label{eq:Bpz}%
\end{align}
\end{subequations}
As a double check, note that Eqs. (\ref{eq:Ep3}) and
(\ref{eq:Bp5}) reduce to the expected formulas in the EM limit:
\begin{equation}
\left.
\begin{array}{r@{\, = \;}l}
{\boldsymbol{E}^\gamma}'(\vecrp ,t') & \hatrp \mbox{\Large
$\oofourpi\, \frac{q^{\mbox{\scriptsize
$\gamma$}}}{{r'}^{\mbox{\scriptsize $2$}}}$}\\
{\boldsymbol{B}^\gamma}'(\vecrp ,t') & \veczero\\
\end{array}
\right\} \qquad \mbox{(EM limit).} \label{eq:FEM}
\end{equation}

\subsection{$A^\mu (b,t)$ and $F^{\mu\nu}(b,t)$ of an UR Point Charge}
\label{sec:QinMotion} \indent

The expressions for the potentials and fields evaluated at $P$ in
frame $K$ can be easily obtained from the above results using
standard LT equations. These formulas will first be obtained
without making any approximations, and then, at the end of this
section, the $v=1$ limit will be taken where possible so as to
arrive at a simpler set of equations.

The scalar potential in the observer's rest frame is found to be
\begin{subequations}
\begin{align}
\Phi&=\gamma (\Phi'+v{A_z}')\label{eq:Phi1}\\%
&=\oofourpi\, \frac{\gamma (q_V+q_A\lambda v)}{r'}\, \mbox{\Large
e}^{-mr'}.\label{eq:Phi2}%
\end{align}
\end{subequations}
This expression is to be evaluated at the location of the observer
$P$, in terms of the her rest frame $K'$ coordinates. It is known
from Eq. (\ref{eq:rp}) how $r'$ depends on the coordinates of
$K'$; this quantity needs to be reexpressed in terms of the
coordinates in frame $K$. $b$ is the same in the two frames, and a
LT equation can be used to transform $t'$:
\begin{subequations}
\begin{align}
t'&=\gamma (t-vz)\label{eq:tp1}\\%
&=\gamma t,\label{eq:tp2}%
\end{align}
\end{subequations}
where $z$ has been set to $0$ because the evaluation point $P$ has
coordinates $(b,\, 0,\, 0)$ in $K$. Thus, the quantity denoted
$r'$ in Eq. (\ref{eq:Phi2}) is to be now read as
$r'=r'(t)=\sqrt{b^2+(\gamma vt)^2}$. The prime notation shall
henceforth be dropped, and this quantity will simply be called
$r$:
\begin{equation}
r\equiv r'(t)=\sqrt{b^2+(\gamma vt)^2} \qquad \mbox{(definition of
r).}\label{eq:r}
\end{equation}
Thus, the observer $P$, who is monitoring the effects of the
passing charge $q$ at the fixed location $(b,\, 0,\, 0)$ in frame
$K$, will determine the magnitude of the scalar potential
$\Phi(b,t)$ to vary in time $t$ according to
\begin{equation}
\Phi(b,t)=\oofourpi\, \frac{\gamma (q_V+q_A\lambda v)}{r}\,
\mbox{\Large e}^{-mr},\label{eq:Phi3}%
\end{equation}
where $r$ is given by Eq. (\ref{eq:r}). It must be kept in mind,
however, that $r$ is not actually the magnitude of the position
vector of $P$ relative to $q$, as measured in frame $K$.
\emph{That} vector is $\vecr(t)=(b,\, 0,\, -vt)$, and its
magnitude is not given by Eq. (\ref{eq:r}), but by
$r(t)=|\vecr(t)|=\sqrt{b^2+(vt)^2}$.

The components of $\vecA(b,t)$ are also identified via LTs. As
with $\Phi(b,t)$, each component depends on components of
${A^\mu}'$, each of which is naturally expressed in terms of the
coordinates of $K'$, so $t'$ must be set to $\gamma t$ in all
final expressions.
\begin{subequations}
\begin{align}
A_x&={A_x}'=0\label{eq:Ax}\\%
A_y&={A_y}'=0\label{eq:Ay}\\%
A_z&=\gamma ({A_z}'+v\Phi')\label{eq:Az1}\\%
&=\left. \left[ \oofourpi\, \frac{\gamma (q_A\lambda +q_Vv)}{r'}\,
\mbox{\Large e}^{-mr'}\right] \right| _{t'=\gamma
t}\label{eq:Az2}\\%
&=\oofourpi\, \frac{\gamma (q_Vv+q_A\lambda )}{r}\, \mbox{\Large e}^{-mr}.\label{eq:Az3}%
\end{align}
\end{subequations}
Or,
\begin{equation}
\vecA(b,t)=\hatz \oofourpi\, \frac{\gamma (q_Vv+q_A\lambda)}{r}\, \mbox{\Large e}^{-mr}.\label{eq:Avec}%
\end{equation}
Recalling Eqs. (\ref{eq:Amu1}) and (\ref{eq:qmu2}), this equation
can be combined with Eq. (\ref{eq:Phi3}) into a very elegant
expression for the 4-potential:
\begin{subequations}
\begin{align}
A^\mu(b,t) &= \oofourpi\, \frac{q^\mu}{r}\, \mbox{\Large
e}^{-mr}\label{eq:Amu2}\\%
&= \oofourpi\, \frac{q_{VA}u^\mu}{r}\,
\mbox{\Large e}^{-mr}\qquad \mbox{(UR limit)},\label{eq:Amu3}%
\end{align}
\end{subequations}
where the last step follows from Eq. (\ref{eq:qmu5}). This result
is identical in form to the expression for the 4-potential of an
UR point charge in classical ED, with $q_{VA}\to q^\gamma$ and
$m\to 0$.

The components of the $\vecE$ and $\vecB$ fields are components of
a tensor instead of a 4-vector, so their transformation equations
differ from those corresponding to $\Phi$ and $\vecA$. The
components of $\vecE$ transform as
\begin{subequations}
\begin{align}
E_x&=\gamma ({E_x}'+v{B_y}')\label{eq:Ex1}\\%
&=\left. \left[ \oofourpi\, \frac{\gamma (q_V+q_A\lambda
v)b}{{r'}^3}\, (1+mr)\, \mbox{\Large e}^{-mr'}\right] \right|
_{t'=\gamma t}\label{eq:Ex2}\\%
&=\oofourpi\, \frac{\gamma (q_V +q_A\lambda v)b}{r^3}\, (1+mr)\,
\mbox{\Large e}^{-mr}\label{eq:Ex3}\\%
E_y&=\gamma ({E_y}'-v{B_x}')=0\label{eq:Ey}\\%
E_z&={E_z}'\label{eq:Ez1}\\%
&=\left. \left[ -\oofourpi\, \frac{q_Vvt'}{{r'}^3}\, (1+mr')\,
\mbox{\Large e}^{-mr'}\right] \right| _{t'=\gamma
t}\label{eq:Ez2}\\%
&=-\oofourpi\, \frac{\gamma q_Vvt}{r^3}\, (1+mr)\, \mbox{\Large
e}^{-mr},\label{eq:Ez3}%
\end{align}
\end{subequations}
and those of $\vecB$ as
\begin{subequations}
\begin{align}
B_x&=\gamma ({B_x}'-v{E_y}')=0\label{eq:Bx}\\%
B_y&=\gamma ({B_y}'+v{E_x}')\label{eq:By1}\\%
&=\left. \left[ \oofourpi\, \frac{\gamma (q_A\lambda
+q_Vv)b}{{r'}'3}\, (1+mr)\, \mbox{\Large e}^{-mr'}\right] \right|
_{t'=\gamma t}\label{eq:By2}\\%
&=\oofourpi\, \frac{\gamma (q_Vv+q_A\lambda )b}{r^3}\,
(1+mr)\, \mbox{\Large e}^{-mr}\label{eq:By3}\\%
B_z&={B_z}'=0.\label{eq:Bz}%
\end{align}
\end{subequations}
Collecting together these latest results,
\begin{subequations}
\begin{align}
E_x(b,t)&=\oofourpi\, \frac{\gamma (q_V+q_A\lambda v)b}{r^3}\,
(1+mr)\, \mbox{\Large e}^{-mr},\label{eq:EB1}\\%
E_z(b,t)&=-\oofourpi\, \frac{\gamma q_Vvt}{r^3}\, (1+mr)\,
\mbox{\Large e}^{-mr},\qquad \mbox{and}\label{eq:EB2}\\%
B_y(b,t)&=\oofourpi\, \frac{\gamma (q_Vv+q_A\lambda )b}{r^3}\,
(1+mr)\, \mbox{\Large e}^{-mr},\label{eq:EB3}%
\end{align}
\end{subequations}
with all other components vanishing.

The UR limit is now taken. Using Eq. (\ref{eq:qeff}), with $v=1$,
the nonvanishing components of $A^\mu(b,t)$ and $F^{\mu\nu}(b,t)$
are found to simplify to
\begin{equation}
\renewcommand{\arraystretch}{2}
\left.
\begin{array}{r@{\, = \;}l}
\Phi(b,t) & \mbox{\Large $\oofourpi\, \frac{\gamma
q_{\mbox{\scriptsize $VA$}}}{r}$}\, \mbox{\Large e}^{-mr}\\
A_z(b,t) & \mbox{\Large $\oofourpi\, \frac{\gamma
q_{\mbox{\scriptsize $VA$}}}{r}$}\,\mbox{\Large e}^{-mr}\\
E_x(b,t) & \mbox{\Large $\oofourpi\, \frac{\gamma
q_{\mbox{\scriptsize $VA$}}b}{r^{\mbox{\scriptsize $3$}}}$}\,
(1+mr)\, \mbox{\Large e}^{-mr}\\
E_z(b,t) & \mbox{\Large $-\oofourpi\, \frac{\gamma
q_{\mbox{\scriptsize $V$}}vt}{r^{\mbox{\scriptsize $3$}}}$}\,
(1+mr)\, \mbox{\Large e}^{-mr}\\
B_y(b,t) & \mbox{\Large $\oofourpi\, \frac{\gamma
q_{\mbox{\scriptsize $VA$}}b}{r^{\mbox{\scriptsize $3$}}}$}\,
(1+mr)\, \mbox{\Large e}^{-mr}\\
\end{array}
\right\} \qquad \mbox{(UR limit).}\label{eq:AFUR}
\end{equation}
In the EM limit, the following equations result:
\begin{equation}
\renewcommand{\arraystretch}{2}
\left.
\begin{array}{r@{\, = \;}l}
\Phi^\gamma (b,t) & \mbox{\Large $\oofourpi\, \frac{\gamma
q^{\mbox{\scriptsize $\gamma$}}}{r}$}\\
A^\gamma _z(b,t) & \mbox{\Large $\oofourpi\, \frac{\gamma
q^{\mbox{\scriptsize $\gamma$}}}{r}$}\\
E^\gamma _x(b,t) & \mbox{\Large $\oofourpi\, \frac{\gamma
q^{\mbox{\scriptsize $\gamma$}} b}{r^{\mbox{\scriptsize $3$}}}$}\\
E^\gamma _z(b,t) & \mbox{\Large $-\oofourpi\, \frac{\gamma
q^{\mbox{\scriptsize $\gamma$}}vt}{r^{\mbox{\scriptsize $3$}}}$}\\
B^\gamma _y(b,t) & \mbox{\Large $\oofourpi\, \frac{\gamma
q^{\mbox{\scriptsize $\gamma$}}b}{r^{\mbox{\scriptsize $3$}}}$}\\
\end{array}
\right\} \qquad \mbox{(EM limit),}\label{eq:AFUREM}
\end{equation}
in complete agreement with results from classical ED (cf.
\cite{ref:Jack1}).%

\section{Massive and Massless Plane Waves}
\label{sec:planewaves} \indent

Equally as important as the functional forms of the components of
$A^\mu$ and $F^{\mu\nu}$ (shown in Eq. (\ref{eq:AFUR})) are the
interrelationships among them. Most relevant to the SWWM is the
fact that the components $E_x$ and $B_y$ are almost exactly equal
in magnitude, and are oriented perpendicular to one another and to
the direction of motion, just like the $\vecE$ and $\vecB$ fields
on the wavefront of a plane EM wave. Furthermore, as in the SWWM,
the fields are only appreciable at time $t=0$, within a time
interval $\Delta t\simeq b/\gamma v$. Thus, the components
oriented longitudinal to the direction of motion (viz, $E_z$, with
$t \lesssim b/\gamma v$) are suppressed by a factor of $\gamma \gg
1$ compared to the components oriented transverse to the direction
of motion (viz, $E_x$ and $B_y \simeq E_x$). So, the $\vecE$ and
$\vecB$ fields are strongly contracted into the plane transverse
to the direction of motion, which then hints at the idea of
approximating these fields as freely propagating plane waves of
radiation. It is precisely this similarity that is used in the
SWWM to simplify the physics of complicated interactions between
particles. Any reaction induced by the fields of a passing UR
particle can be analyzed, to a good approximation, in terms of the
familiar equations of radiation theory. The assignment of a
nonzero mass to the fields complicates the picture, as the physics
of massive plane waves is not as well documented as that of
massless EM waves. It is very worthwhile, therefore, to work out
the intricacies of massive plane waves, without making any
reference to the results of the previous section. In the
subsequent section, similarities between the two descriptions will
be sought, and the appropriate identifications will be made.

\subsection{The Proca Equation in Vacuum}
\label{sec:PEV} \indent

Just as in Section \ref{sec:QatRest}, the starting point is the
PE, Eq. (\ref{eq:PE}). But, unlike the previous case, where the
potential was static ($\partial \vecA/\partial t=0$) and the
source was a point charge, here the solutions of interest
correspond to plane waves travelling through a vacuum --- an
entirely different problem. With the vacuum condition $J^\mu=0$
assumed, the PE reduces to
\begin{equation}
\Box A^\mu +m^2A^\mu =0 \qquad \mbox{(PE in vacuum).}\label{eq:PE2}%
\end{equation}
This analysis will be carried out in frame $K$, of course, because
it is only in that frame that the $\vecE$ and $\vecB$ fields are
contracted into the plane transverse to the direction of motion.
Using previous results and basic vector identities, the PE in
vacuum (PEV) can be recast into a set of four vector equations:
\begin{subequations}
\begin{align}
\vecdel \cdot \vecE &= -m^2\Phi & \mbox{(PEV 1) } \label{eq:PEvec1}\\%
\vecdel \cdot \vecB &= 0 & \mbox{(PEV 2) } \label{eq:PEvec2}\\%
\vecdel \times \vecE + \frac{\partial \vecB}{\partial t} &= \veczero & \mbox{(PEV 3) } \label{eq:PEvec3}\\%
\vecdel \times \vecB - \frac{\partial \vecE}{\partial t} &= -m^2\vecA & \mbox{(PEV 4).} \label{eq:PEvec4}%
\end{align}
\end{subequations}
These equations are the $m\neq 0$ generalization of the vector
form of ME in vacuum; they neatly reduce to the familiar set of
equations in the $m=0$ limit.

\subsection{Wave Modes and Wave Packets}
\label{sec:ModesPackets} \indent

From PEV 1\,--\,4, previous results and vector identities can now
be used to obtain the following decoupled wave equations:
\begin{subequations}
\begin{align}
\vecdel^2 \Phi -\frac{\partial^2 \Phi}{\partial t^2}&= m^2\Phi \label{eq:WEPhi}\\%
\vecdel^2 \vecA -\frac{\partial^2 \vecA}{\partial t^2}&= m^2\vecA \label{eq:WEA}\\%
\vecdel^2 \vecE -\frac{\partial^2 \vecE}{\partial t^2}&= m^2\vecE \label{eq:WEE}\\%
\vecdel^2 \vecB -\frac{\partial^2 \vecB}{\partial t^2}&= m^2\vecB. \label{eq:WEB}%
\end{align}
\end{subequations}
Eqs. (\ref{eq:WEPhi}) and (\ref{eq:WEA}) could, of course, have
been deduced immediately from Eq. (\ref{eq:PE2}).

These equations are all of the form
\begin{equation}
\vecdel^2 u -\frac{\partial^2 u}{\partial t^2}= m^2u.\label{eq:WEu}%
\end{equation}
The basic one-dimensional solutions are called wave modes, and are
of the form
\begin{equation}
u(z,t)= \mbox{\Large e}^{-\mi (\omega t-kz)} \qquad \mbox{(wave mode solution).}\label{eq:WEsolu}%
\end{equation}
Eq. (\ref{eq:WEsolu}) describes an infinitely-long plane wave (a
mode) of definite energy $E=\omega$ and 3-momentum
$\vecp=\veck=\hatz k$ propagating with \emph{phase} velocity
\begin{equation}
\vecVp = \hatz \frac{\omega}{k} \qquad \mbox{(phase velocity).}\label{eq:vphase}%
\end{equation}
Upon substituting Eq. (\ref{eq:WEsolu}) into Eq. (\ref{eq:WEu}),
the usual frequency\,--\,wave number\,--\,mass relation,
\begin{equation}
\omega ^2-k^2 = m^2\qquad \mbox{(equation of motion),}\label{eq:okm}%
\end{equation}
is recovered. The general solution, called a wave packet, is of
the form
\begin{equation}
U(z,t)= \oortpi \, \int ^\infty_{-\infty} \dif k\, A(k)\, \mbox{\Large e}^{-\mi (\omega t-kz)} \qquad \mbox{(wave packet solution).}\label{eq:WEsolU1}%
\end{equation}
Here $k$ is taken as an independent parameter and $\omega$ is
generally a function of $k$. The amplitude $A(k)$ describes the
properties of the linear superposition of different modes. It is
given by the FT of U(z,\,0):
\begin{equation}
A(k)= \oortpi \, \int ^\infty_{-\infty} \dif z\, U(z,0)\, \mbox{\Large e}^{-\mi kz} \qquad \mbox{(amplitude).}\label{eq:WEamp}%
\end{equation}
If $U(z,0)$ represents a finite wave train at time $t=0$, with a
length on the order of some $\Delta z$, $A(k)$ is a function
peaked at some $k \equiv k_0$, which is the dominant wave number
in the modulated wave $U(z,0)$, and has a breadth on the order of
some $\Delta k$. With $\Delta z$ and $\Delta k$ defined as rms
(root-mean-squared) deviations from the average values of $z$ and
$k$, they satisfy the Heisenberg uncertainty principle in the form
$\Delta z\,\Delta k \geqslant \half$ \cite{ref:Jack1}. If $\Delta
k$ is not very broad (i.e., $A(k)$ is fairly sharply peaked at
some $k \equiv k_0$), or $\omega$ depends only weakly on $k$,
\begin{equation}
\omega(k) \simeq \left. \omega _0+\frac{\dif \omega}{\dif k}\right| _{t=0}(k-k_0),\label{eq:oofk}%
\end{equation}
it can be shown (cf. \cite{ref:Jack1}, for example) that the
packet travels along undistorted in shape, with the approximate
waveform
\begin{equation}
U(z,t) \simeq U(z-V_gt,0)\, \mbox{\Large e}^{-\mi (\omega _0-k_0V_g)t},\label{eq:WEsolU2}%
\end{equation}
at a velocity (the \emph{group} velocity)
\begin{equation}
\vecVg =\left. \hatz \frac{\dif \omega}{\dif k}\right| _{t=0} \qquad \mbox{(group velocity).}\label{eq:vgroup1}%
\end{equation}
Unlike the well-defined energy and 3-momentum of an individual
wave mode, the energy $E$ and 3-momentum $\vecp$ of a wave packet
can only be defined to within uncertainties $\Delta E$ and $\Delta
\vecp$: $E=\omega _0\, \pm\, \Delta E$ and $\vecp =\veckzero\,
\pm\, \Delta \vecp$, where $\veckzero=\hatz k_0$. As in the above
discussion, it is being assumed that $\omega_0 \gg \Delta E$ and
$\veckzero \gg \Delta \vecp$, so that the packet has a fairly
well-defined energy $E=\omega \simeq \omega _0 $ and 3-momentum
$\vecp = \veck \simeq \veckzero$. The transport of 4-momentum by
the packet occurs at the group velocity. Taking all modes to have
roughly the same energy $\omega =\sqrt{k^2+m^2}$,
\begin{equation}
\vecVg \simeq \hatz \frac{k}{\sqrt{k^2+m^2}}=\hatz \frac{k}{\omega}=\hatz \frac{1}{V_p}.\label{eq:vgroup2}%
\end{equation}
Due to the factor of $m$ in the denominator, the group velocity of
a massive packet will be less than that of a massless packet,
which is identically the velocity of light. It will be assumed
that all components of $A^\mu$ and $F^{\mu\nu}$ are of the same
form as $U(z,t)$ shown in Eq. (\ref{eq:WEsolU2}). That is, with
each component is associated a wave packet (a ``pulse"), that is
described by a wave function
\begin{equation}
U(z,t) = U_0\, \mbox{\Large e}^{-\mi (\omega t-kz)} \qquad
\mbox{($U_0=$\,const)}.\label{eq:WEsolU3}%
\end{equation}
A given pulse has a well-defined energy $E=\omega$ and 3-momentum
$\vecp =\hatz k$, and travels at a subluminal group velocity
$\vecv =\hatz v=\hatz k/\omega$. The relation $z=vt$ also holds,
because these waves are comoving with $q$ and are being viewed by
$P$, which is at rest in frame $K$. Note, then, that the factor
$U(z-V_gt,0)$ in Eq. (\ref{eq:WEsolU2}), which is denoted $U_0$ in
Eq. (\ref{eq:WEsolU3}), is a constant.

So, the quantities of interest to the present analysis are as
follows:
\begin{subequations}
\begin{align}
\Phi &= \Phi _0\, \mbox{\Large e}^{-\mi (\omega t-kz)}, \qquad \Phi _0=\,\mbox{const}\label{eq:PhiPEV1}\\%
\vecA &= \vecA _0\, \mbox{\Large e}^{-\mi (\omega t-kz)}, \qquad \vecA _0=\,\hatx A_{0x}+\haty A_{0y}+\hatz A_{0z}=\,\mbox{const}\label{eq:APEV1}\\%
\vecE &= \vecE _0\, \mbox{\Large e}^{-\mi (\omega t-kz)}, \qquad \vecE _0=\,\hatx E_{0x}+\haty E_{0y}+\hatz E_{0z}=\,\mbox{const}\label{eq:EPEV1}\\%
\vecB &= \vecB _0\, \mbox{\Large e}^{-\mi (\omega t-kz)}, \qquad \vecB _0=\,\hatx B_{0x}+\haty B_{0y}+\hatz B_{0z}=\,\mbox{const}.\label{eq:BPEV1}%
\end{align}
\end{subequations}

\subsection{Polarization 4-Vector}
\label{sec:PolarizationFourVector} \indent

It will prove to be very useful to introduce some new notation at
this point. In quantum mechanics, $A^\mu$ is interpreted as the
wave function of the boson, and is expressed in the form
\begin{equation}
A^\mu = \varepsilon ^\mu\, \mbox{\large e}^{-\mi (\omega
t-kz)},\label{eq:APEV2}
\end{equation}
where $\varepsilon ^\mu =(\varepsilon ^0, \veceps)$ is a 4-vector
called the 4-polarization \cite{ref:Halz}. $\varepsilon ^\mu$ is
used to identify the three different possible helicity states of a
spin-1 boson, corresponding to projections of its angular momentum
parallel to ($\lambda =+1$), antiparallel to ($\lambda =-1$), and
perpendicular to ($\lambda =0$) the direction of propagation.
These transverse ($\lambda =\pm 1$) and longitudinal ($\lambda
=0$) states will be referred to a great deal in future sections,
so a brief digression is in order here to properly set forth a few
definitions.

Recalling that $E=\omega$ and $\vecp =\hatz k$ (so that $p^\mu
=(\omega ,\, 0,\, 0,\, k)$), the LC ($\partial ^\mu A_\mu =0$)
applied to Eq. (\ref{eq:APEV2}) yields $p^\mu \varepsilon _\mu=0$.
This equation, which is the LC in momentum-space, reduces the
number of independent components of $\varepsilon ^\mu$ from 4 to
3. It is convenient to split $\veceps$ into components
perpendicular to (denoted with the subscript $\perp$) and parallel
to (denoted with the subscript $||$) the direction of propagation;
thus $\veceps =\vecepsperp +\vecepspar$. Of course, $\vecepsperp$
will be some linear combination of $\hatx$ and $\haty$, and
$\vecepspar$ will be some multiple of $\hatz$. One can write
$\veceps =\vecepsperp +\hatz \varepsilon _z$. The LC in
momentum-space thus becomes
\begin{equation}
\omega \varepsilon ^0 = k\varepsilon _z\qquad \mbox{(LC in
momentum-space)} .\label{eq:LCpspace1}
\end{equation}

At this juncture, the explicit specifications of the components of
$\varepsilon ^\mu$ differ for the massive and massless cases. This
difference stems from the fact that after imposing the LC, the PE
cannot accommodate any further gauge transformations, while ME are
still invariant under additional gauge transformations of the form
$A^\mu \rightarrow {A^\mu }'=A^\mu +\partial ^\mu F$, where
$F=F(x^\mu )$ is an arbitrary function satisfying $\Box F =0$.
Choosing $F$ to be of the form $F=\mi\, a\, \mbox{\large e}^{-\mi
(\omega t-kz)}$, where $a$ is an arbitrary constant that one is
free to specify, this transformation is equivalent to $\varepsilon
^\mu \rightarrow {\varepsilon ^\mu }'= \varepsilon ^\mu +ap^\mu$.
For massless pulses, this freedom can be used to further reduce
the number of independent components of $\varepsilon ^\mu$ from 3
to 2; the $\varepsilon ^\mu$ for massive spin-1 fields, on the
other hand, always has exactly 3 independent components. By
choosing $a$ such that $\varepsilon ^0=0$, Eq.
(\ref{eq:LCpspace1}) is seen to reduce to $\varepsilon _z =\hatz
\cdot \veceps =0$, so that only 2 components of $\veceps$ are
actually independent (for massless pulses). These components are
in the $x-y$ plane, so that $\vecA$ is purely transverse: $\vecA
=\vecepsperp A_{\perp}$. This particular gauge is called the
Coulomb, or transverse, gauge. The more familiar pair of
characteristic equations for this gauge in classical ED, namely
$\Phi =0$ and $\vecdel \cdot \vecA =0$, can be shown to be
equivalent to $\varepsilon ^0=0$ and $\varepsilon _z=0$ (by way of
Eq. (\ref{eq:APEV2})). Because there is a whole family of possible
gauges that yield the same $\vecE$ and $\vecB$ fields (hence
$\vecS$) for the massless case, the 4-polarization is not uniquely
defined for massless pulses; it is, however, well-defined for
massive pulses.

It is conventional to use the following two linear combinations of
$\hatx$ and $\haty$ for $\vecepsperp$:
\begin{subequations}
\begin{align}
\vecepsperp &= \oort (-\hatx - \mi \haty)\qquad \mbox{(for $\lambda =+1$ states)}\label{eq:eR}\\%
\vecepsperp &= \oort (+\hatx - \mi \haty)\qquad \mbox{(for $\lambda =-1$ states)}.\label{eq:eL}%
\end{align}
\end{subequations}
These vectors are called circular polarization vectors, and can be
easily shown to be eigenvectors of the helicity operator, with
eigenvalues $+1$ and $-1$, respectively \cite{ref:Ryde}. They thus
correspond to photons with right and left circular polarizations,
respectively. These quantities can be generalized to 4-vectors. As
both are oriented in the plane transverse to the direction of
propagation, neither will be affected by a LT in the $\hatz$
direction. The corresponding \emph{transverse} polarization
4-vectors of interest (which shall be denoted with the subscript
$T$) are found via LT equations to be
\begin{equation}
\varepsilon ^\mu _\mT =\oort (0,\, \pm 1,\, -\mi ,\, 0)\qquad
\mbox{(for $\lambda =\pm 1$ states),}\label{eq:epstrans}
\end{equation}
in any Lorentz frame. These 4-vectors can be shown to be
eigenvectors of a 4-dimensional generalization of the helicity
operator,
\begin{equation}
\Lambda ^{\mu \nu}=
\begin{pmatrix}
0 & \;\;\; 0 & \;\;\; 0 & \;\;\; 0\\
0 & \;\;\; 0 & -\mi & \;\;\; 0\\
0 & \;\;\; \mi & \;\;\; 0 & \;\;\; 0\\
0 & \;\;\; 0 & \;\;\; 0 & \;\;\; 0
\end{pmatrix}
\qquad \mbox{(helicity operator),}\label{eq:HelMat}
\end{equation}
with respective eigenvalues $\lambda =\pm 1$
\cite{ref:Halz,ref:Ryde}. Note that these pairs of 4-vectors can
be used to describe transverse polarization states of both
massless \emph{and} massive pulses. See Appendix B for a more in
depth treatment of helicity.

In contrast, the \emph{longitudinal} polarization vector will be
different for the massless and massive cases. For massless pulses,
it was found above that the component of the polarization vector
that is oriented in the (longitudinal) $\hatz$ direction vanishes:
$\vecepspar =\hatz \varepsilon _z=\veczero$. A 4-vector
constructed from this 3-vector (to which the 4-vector reduces in
the observer's frame $K$) and the Coulomb gauge condition
$\varepsilon ^0=0$ is
\begin{equation}
\varepsilon ^\mu _\mL =(0,\, 0,\, 0,\, 0).\label{eq:epslongless1}
\end{equation}
The subscript $L$ has been introduced here to denote the
``longitudinal" ($\lambda =0$) helicity state. But this quantity
is not a valid polarization vector, as it is not normalized. In
fact, according to \cite{ref:Grei}, it is impossible to construct
such a third polarization vector for a massless vector field that
is both normalized and transverse (in four dimensions) to
$\varepsilon ^\mu _\mT $. Instead, the following 4-vector is used
for the longitudinal polarization vector:
\begin{equation}
\renewcommand{\arraystretch}{1.25}
\left.
\begin{array}{r@{\, = \;}l}
\vecepspar & \hatz\\
\varepsilon ^\mu _\mL  & (0,\, 0,\, 0,\, 1)\\
\end{array}
\right\} \qquad \mbox{(for \emph{massless} states with $\lambda
=0$).} \label{eq:epslongless2}
\end{equation}
This 4-vector is only defined in the special Lorentz frame in
which the pulse has 3-momentum $\vecp=\hatz k$. It is normalized
and transverse to $\varepsilon ^\mu _\mT $, as it should be. It is
also clearly \emph{not} formulated in the Coulomb gauge, where
$\varepsilon ^0=\varepsilon _z=0$, but that is of no significance
because, as mentioned above, the 4-polarization is not uniquely
defined for massless pulses. That $\varepsilon ^\mu _\mL $
corresponds to a helicity $\lambda =0$ state is easily verified by
applying the generalized helicity operator, Eq. (\ref{eq:HelMat}).
For massive pulses, a suitable longitudinal polarization 4-vector
may be constructed by first considering the longitudinal 3-vector
in the rest frame of the pulse, and then enact a LT. In the rest
frame (i.e., frame $K'$) of a massive pulse, the obvious choice
for $\vecepsparp$ is $\hatzp$, as it is normalized and orthogonal
to $\vecepsperpp$. The rest-frame polarization 4-vector is thus
${\varepsilon ^\mu_{_{||}}}'=(0,\, 0,\, 0,\, 1)$. A LT to frame
$K$ yields
\begin{equation}
\renewcommand{\arraystretch}{1.25}
\left.
\begin{array}{r@{\, = \;}l}
\vecepspar & \hatz \gamma\\
\varepsilon ^\mu _\mL  & \gamma (v,\, 0,\, 0,\, 1)\\
\end{array}
\right\} \qquad \mbox{(for \emph{massive} states with $\lambda
=0$).} \label{eq:epslongive}
\end{equation}
This $\varepsilon ^\mu _\mL $ is both normalized and transverse to
$\varepsilon ^\mu _\mT $. Furthermore, by using Eq.
(\ref{eq:HelMat}) as the helicity operator, it is easily seen to
be an eigenvector with eigenvalue $\lambda =0$. Thus, Eq.
(\ref{eq:epslongive}) is a suitable representation of the
4-polarization for longitudinal massive pulses.

In addition to expanding $A^\mu$ in terms of these new
4-polarizations, all 3-vectors of interest can be expressed in
terms of the pair of 3-vectors $\vecepsperp$ and $\vecepspar$.
But, it will prove to be less confusing (mostly because of the
factor of $\gamma$ appearing in Eq. (\ref{eq:epslongive})) in the
long run if these 3-vectors are, instead, expressed in terms of
$\vecepsperp$ and $\hatz$.
\begin{subequations}
\begin{align}
\vecA &= \vecA _{\perp} + \vecA _z = \vecepsperp A_{\perp} + \hatz A_z \label{eq:APEV3}\\%
\vecE &= \vecE _{\perp} + \vecE _z = \vecepsperp E_{\perp} + \hatz E_z \label{eq:EPEV3}\\%
\vecB &= \vecB _{\perp} + \vecB _z = \vecepsperp B_{\perp} + \hatz B_z. \label{eq:BPEV3}%
\end{align}
\end{subequations}

\subsection{Solution to the Proca Equation in Vacuum for Massive Pulses}
\label{sec:MassivePulses} \indent

In proceeding with the analysis of these pulses, the details of
pulses that are generally massive will be worked out first. At the
end of the analysis, analogous results for massless pulses will be
specified. A few of the $m=0$ results cannot be found by simply
setting $m=0$ in the equations describing massive pulses, but are
easily worked out nevertheless. A bit of subtlety, involving the
choice of additional gauge beyond the LC, is actually involved for
the massless case. If there is any ambiguity as to this choice of
additional gauge, it is to be assumed that it is the Coulomb gauge
(where $\Phi =0$ and $\vecdel \cdot \vecA =0$ for massless plane
waves). The following characteristics of plane waves are derived
from PEV 1\,--\,4 (Eqs. (\ref{eq:PEvec1})\,--\,(\ref{eq:PEvec4})),
assuming the potentials and fields are of the forms specified in
Eqs. (\ref{eq:PhiPEV1})\,--\,(\ref{eq:BPEV1}) and/or
(\ref{eq:APEV3})\,--\,(\ref{eq:BPEV3}), and using the notation of
polarization 3-vectors introduced in the previous section.

PEV 1 yields
\begin{subequations}
\begin{align}
\frac{\partial E_x}{\partial x} + \frac{\partial E_y}{\partial y} + \frac{\partial E_z}{\partial z} &= -m^2 \Phi \label{eq:PEV1sol1}\\%
\mi k E_z &= -m^2 \Phi \label{eq:PEV1sol2}\\%
E_z &= \mi\, \frac{m}{\gamma v}\, \Phi , \label{eq:PEV1sol3}%
\end{align}
\end{subequations}
where $k=\gamma m v$ was used in the last step. Eq.
(\ref{eq:PEV1sol3}) shows that the plane wave $\vecE$ field is not
purely transverse in general, as it is known to be in the massless
case.

From PEV 2, one finds
\begin{subequations}
\begin{align}
\frac{\partial B_x}{\partial x} + \frac{\partial B_y}{\partial y} + \frac{\partial B_z}{\partial z} &= 0 \label{eq:PEV2sol1}\\%
\mi k B_z &= 0 \label{eq:PEV2sol2}\\%
B_z &= 0 \label{eq:PEV2sol3}%
\end{align}
\end{subequations}
Thus, like in the massless case, the $\vecB$ field is always
purely transverse.

The solution to PEV 3 is
\begin{subequations}
\begin{align}
\left[ \hatx \left( -\frac{\partial E_y}{\partial z}\right) + \haty \left( \frac{\partial E_x}{\partial z}\right) \right] + (-\mi \omega \vecB) &= \veczero \label{eq:PEV3sol1}\\%
\mi k (-\hatx E_y + \haty E_x) &= \mi \omega \vecB \label{eq:PEV3sol2}\\%
v\hatz \times \vecE &= \vecB \label{eq:PEV3sol3}\\%
\vecB &= \vecv \times \vecE .\label{eq:PEV3sol4}%
\end{align}
\end{subequations}
Here it can be seen that, just like in the EM case, $\vecE$ and
$\vecB$ are always perpendicular to each another, and to the
direction of propagation.

Finally, the solution to PEV 4 is as follows:
\begin{subequations}
\begin{align}
\left[ \hatx \left( -\frac{\partial B_y}{\partial z}\right) + \haty \left( \frac{\partial B_x}{\partial z}\right) \right] - (-\mi \omega \vecE) &= -m^2 \vecA \label{eq:PEV4sol1}\\%
\mi k (-\hatx B_y + \haty B_x) + \mi \omega \vecE &= -m^2 \vecA \label{eq:PEV4sol2}\\%
\mi k \hatz \times \vecB + \mi \omega \vecE &= -m^2 \vecA \label{eq:PEV4sol3}\\%
\mi k v \hatz \times (\hatz \times \vecE) + \mi \omega \vecE &= -m^2 \vecA \qquad \mbox{via Eq. (\ref{eq:PEV3sol3})} \label{eq:PEV4sol4}\\%
\mi k v [\hatz (\hatz \cdot \vecE) - \vecE (\hatz \cdot \hatz)] + \mi \omega \vecE &= -m^2 \vecA \label{eq:PEV4sol5}\\%
-\mi k v (\vecE - \vecE _z) + \mi \omega \vecE &= -m^2 \vecA \label{eq:PEV4sol6}\\%
-\mi k v \vecE _{\perp} + \mi \omega (\vecE _{\perp} + \vecE _z) &= -m^2 (\vecA _{\perp} + \vecA _z). \label{eq:PEV4sol7}%
\end{align}
\end{subequations}
Equating $\perp$ and $z$ components,
\begin{subequations}
\begin{align}
\mi (\omega - k v) \vecE _{\perp} &= -m^2 \vecA _{\perp} \label{eq:PEV4sol8}\\%
\mi \frac{(\omega ^2 - k^2)}{\omega} \vecE _{\perp} &= -m^2 \vecA _{\perp} \label{eq:PEV4sol9}\\%
\vecE _{\perp} &= \mi\, \omega \vecA _{\perp}, \label{eq:PEV4sol10}%
\end{align}
\end{subequations}
where $\omega ^2 - k^2=m^2$ was used in the last step, and
\begin{subequations}
\begin{align}
\mi\, \omega \vecE _z &= -m^2 \vecA _z \label{eq:PEV4sol11}\\%
\vecE _z &= \mi \frac{m}{\gamma} \vecA _z, \label{eq:PEV4sol12}%
\end{align}
\end{subequations}
where $\omega =\gamma m$ was used.

By employing the LC, another useful relation can be obtained:
$\Phi = vA_z=\vecv \cdot \vecA _z$. This equation can also be
found, though only for the massive case, by comparing Eqs.
(\ref{eq:PEV1sol3}) and (\ref{eq:PEV4sol12}). In summary, for
massive pulses, the following relations have been established:
\begin{equation}
\left.
\begin{array}{r@{\, = \;}l}
\Phi & \vecv \cdot \vecA _z\\
\vecE _{\perp} & \mi\, \omega \vecA _{\perp}\\
\vecE _z & \mi \mbox{\Large $\frac{m^2}{\omega}$} \vecA _z\\
\vecB _{\perp} & \vecv \times \vecE _{\perp}\\
\vecB _z & \veczero\\
\end{array}
\right\} \qquad \mbox{(massive pulses).} \label{eq:PEViveresults}
\end{equation}

In contrast to the massive case, the Poynting vector $\vecS$ for
massless pulses depends only on the $\vecE$ and $\vecB$ fields, so
the exact forms of $\Phi$ and $\vecA$ are inconsequential insofar
as the physically measurable quantity $\vecS$ is concerned (cf.
Section \ref{sec:SPulses}). For completeness, the analog of the
above set of equations for massless pulses (in the Coulomb gauge)
is listed here as well.
\begin{equation}
\left.
\begin{array}{r@{\, = \;}l}
\Phi & 0\\
\vecA _z & \veczero\\
\vecE _{\perp} & \mi\, \omega \vecA _{\perp}\\
\vecE _z & \veczero\\
\vecB _{\perp} & \hatz \times \vecE _{\perp}\\
\vecB _z & \veczero\\
\end{array}
\right\} \qquad \mbox{(massless pulses in Coulomb gauge).}
\label{eq:PEVlessresults}
\end{equation}
The first two relations were obtained in the previous section, and
only hold in the Coulomb gauge. Though, as mentioned above, $\Phi
=A_z$ always holds in view of the LC and the fact that $v=1$ for
massless pulses. The third relation follows from $\vecE =-\vecdel
\Phi -\partial \vecA/\partial t$ and previous results; it does not
necessarily follow from PEV 4 (i.e., Eq. (\ref{eq:PEV4sol10})) by
setting $m=0$. The vanishing of $\vecE _z$ results from using
either $\Phi =0$ in Eq. (\ref{eq:PEV1sol3}) or $\vecA _z=\veczero$
in Eq. (\ref{eq:PEV4sol12}). And the last two equations are
consequences of Eqs. (\ref{eq:PEV2sol3}) and (\ref{eq:PEV3sol4})
(with $v=1$) found above.

\subsection{Transverse Pulses}
\label{sec:TransversePulses} \indent

Transverse pulses are associated with the two helicity states
$\lambda =\pm 1$. Such a pulse is defined in the same way that a
transverse helicity state of a spin-1 boson is identified in
quantum mechanics --- by the polarization vector $\vecepsperp$.
$\vecA _z=\hatz A_z$ is thus set equal to $\veczero$ so that the
pulse is purely ``transverse":
\begin{equation}
\vecA =\vecA _{\perp}\qquad \mbox{(definition of a transverse pulse).}\label{eq:Tcondition}%
\end{equation}
Eq. (\ref{eq:PEViveresults}) reduces to the following set of
relations:
\begin{equation}
\left.
\begin{array}{r@{\, = \;}l}
\Phi & 0\\
\vecA _z & \veczero\\
\vecE _{\perp} & \mi\, \omega \vecA _{\perp}\\
\vecE _z & \veczero\\
\vecB _{\perp} & \vecv \times \vecE _{\perp}\\
\vecB _z & \veczero\\
\end{array}
\right\} \qquad \mbox{(transverse pulses).} \label{eq:PEViveresultsT}%
\end{equation}
Note that Eq. (\ref{eq:PEVlessresults}), describing massless
pulses, is a special case of this equation, with $v=1$. Thus, the
familiar result has been obtained that massless pulses, such as EM
waves, are purely transverse!

The important quantity for the project is the Poynting vector
$\vecS$. Eq. (\ref{eq:S2}) reveals, quite generally, that $\vecS =
\vecE \times \vecB + m^2 \Phi \vecA$. For the transverse pulses of
interest here, $\Phi=0$, so the $\vecS$ corresponding to these
types of waves (which shall be denoted with the subscript $T$) is
$\vecS _\mT =\vecE \times \vecB$. Since $\vecE$ and $\vecB$ are
both oriented in the plane transverse to the direction of
propagation, $\vecS$ can be expressed as
\begin{equation}
\left.
\begin{array}{r@{\, = \;}l}
\vecS _\mT  & \vecE _{\perp} \times \vecB _{\perp}\\
& \vecv\, (E_{\perp})^2\\
\end{array}
\right\} \qquad \mbox{(transverse pulses).} \label{eq:STPulses}%
\end{equation}
The second line follows from the first by using Eq.
(\ref{eq:PEV3sol4}) and a basic vector identity. The important
point to make is that, for the purpose of calculating energy
transport, only $\vecE _{\perp}$ and $\vecB _{\perp}$ are needed.
Whether the pulse is massive or massless, the values of $\Phi$ and
$\vecA$ are inconsequential.

In summary, a transverse pulse has the following generic
properties. The wavefront is a plane transverse to the direction
of propagation, that is spanned by $\vecE$ and $\vecB$ field
lines, where $\vecB =\vecv \times \vecE$. On this wavefront, the
fields are constant in magnitude, and are perpendicular to each
other and to the direction of propagation. The energy flux $\vecS$
associated with the pulse is uniquely determined by the values of
these fields (it does not depend on $\Phi$ and $\vecA$), and is
given by Eq. (\ref{eq:STPulses}).

\subsection{Longitudinal Pulses}
\label{sec:LongitudinalPulses} \indent

Longitudinal pulses are associated with a helicity of $\lambda
=0$. Such a pulse is defined by setting $\vecA
_{\perp}=\vecepsperp A_{\perp}=\veczero$, so that
\begin{equation}
\vecA =\vecA _z\qquad \mbox{(definition of a longitudinal pulse).}\label{eq:Lcondition}%
\end{equation}
Eq. (\ref{eq:PEViveresults}) reduces to
\begin{equation}
\left.
\begin{array}{r@{\, = \;}l}
\Phi & \vecv \cdot \vecA _z\\
\vecA _{\perp} & \veczero\\
\vecE _{\perp} & \veczero\\
\vecE _z & \mi \mbox{\Large $\frac{m}{\gamma}$} \vecA _z\\
\vecB _{\perp} & \veczero\\
\vecB _z & \veczero\\
\end{array}
\right\} \qquad \mbox{(longitudinal pulses).}
\label{eq:PEViveresultsL}
\end{equation}
In view of the fact that $\vecB =\vecB _{\perp} +\vecB
_z=\veczero$, the Poynting vector $\vecS$ for such pulses (which
shall be denoted with the subscript $L$) reduces to
\begin{equation}
\left.
\begin{array}{r@{\, = \;}l}
\vecS _\mL  & m^2 \Phi \vecA _z\\
& \vecv\, (mA_z)^2\\
\end{array}
\right\} \qquad \mbox{(longitudinal pulses),} \label{eq:SLPulses}%
\end{equation}
where use has been made of the fact that $\Phi =vA_z$ in the
second line. Thus, for the purpose of calculating energy
transport, only $m$, $\Phi$ and $\vecA _z$ are needed. An
interesting special case of this formula is the $m=0$ limit:
$\vecS _\mL =\veczero$ for massless pulses --- a result consistent
with the point made in the previous section, that massless pulses
are purely transverse. Here it is seen that, indeed, there is no
field energy associated with the $\lambda =0$ state of such a
pulse. Another point worth noting is that, in contrast to the
massless case, it is the $\vecE$ and $\vecB$ fields, instead of
the potentials, that are inconsequential here. And lastly, $\vecS
_\mL $, like $\vecS _\mT $, points in the same direction as
$\vecv$. So, in the ideal case that is being considered here,
where there is no component of 3-momentum in the transverse plane,
all energy transported by a pulse, regardless of its helicity
state, is done so in the forward direction. It is quite common to
make the assumption (the ``forward scattering" approximation) that
the particles in a high energy collision follow approximately
straight-line trajectories. An excellent discussion of this issue
in the case of electron-electron scattering via photon exchange is
presented in \cite{ref:Tera}. See \cite{ref:Krau,ref:Bert} for
usage of this approximation in peripheral relativistic heavy-ion
collisions mediated by photons. See \cite{ref:Papa} for usage in
peripheral relativistic heavy-ion collisions mediated by
$Z$-bosons. And see
\cite{ref:Daws1,ref:Daws2,ref:Guni,ref:Kane,ref:Cahn1,ref:Alta}
for usage in quark-quark scattering mediated by $W$- and
$Z$-bosons. An actual probability distribution function for such
scattering angles is derived in a future section; it is found to
be sharply peaked at an average scattering angle of zero.

To summarize, a longitudinal pulse is characterized by the
following generic properties. The wavefront is a plane that is
transverse to the direction of propagation, and defined by a
longitudinal $\vecA$ field line configuration that is constant in
magnitude everywhere. The energy flux $\vecS$ associated with the
pulse is uniquely determined by $m$, $\Phi$ and $\vecA$ (it does
not depend on the $\vecE$ and $\vecB$ fields), where $\Phi =\vecv
\cdot \vecA$, and is given by Eq. (\ref{eq:SLPulses}).

\section{Equivalent Pulses} \label{sec:EquivPulses} \indent

The crux of the SWWM is to approximate the $\vecE$ and $\vecB$
fields of an UR charge as appropriate plane wave pulses
(``equivalent pulses") of EM radiation. The same types of
identifications are being sought for the generalized scheme being
developed here, but, because $\vecS$ depends generally on both the
fields \emph{and} potentials, $\Phi$ and $\vecA$ must somehow be
incorporated into this procedure. Having enumerated the potentials
and fields of an UR charge and the characteristic relationships
among these functions for both transverse and longitudinal pulses,
the identification proceeds in a very simple way.

The potentials and fields of an UR charged particle (evaluated by
an observer at point $P$ in frame $K$) are listed in Eq.
(\ref{eq:AFUR}). Three equivalent pulses can be constructed to
reproduce this set of quantities. The first two are transverse
pulses built up out of the three nonvanishing components of the
$\vecE$ and $\vecB$ fields, and are the ones appearing in the
SWWM. The third one is a new feature that is being introducing
into the formalism so as to incorporate modifications due to a
nonzero pulse mass. In the first part of this section, the
traditional SWWM scheme is reviewed, and the properties of these
two equivalent transverse pulses are defined. Then, it will be
shown why a third equivalent pulse is needed at all, and how it
should be constructed.

In the SWWM, Pulse 1 is a transverse EM wave that travels in the
$\hatz$ direction, and its wavefront is defined by the $\vecE _x$
and $\vecB _y$ fields specified in Eq. (\ref{eq:AFUR}). It was
noted in Sec. \ref{sec:TransversePulses} that massless transverse
waves are simply a special case of massive transverse waves (i.e.,
they have exactly the same properties). The only general
requirement is that $\vecB =\vecv \times \vecE$. As this relation
is satisfied by the $\vecE _x$ and $\vecB _y$ fields of the UR
charge discussed in Sec. \ref{sec:QinMotion}, whether $m=0$ or
not, the form of this pulse in the generalized scheme being
developed here is identical in nature to that of Pulse 1 in the
original method. Again, the only measurable quantity of interest
associated with these fields is the energy flux $\vecS _\mT =\vecE
_{\perp} \times \vecB _{\perp}$, which does not depend at all on
$\Phi$ and $\vecA$. So, the $\vecA _x$ field that must accompany
these $\vecE _x$ and $\vecB _y$ fields (cf. Eq.
(\ref{eq:PEViveresultsT})) need not actually exist. Such
fictitious quantities will be denoted with a $\sim$; thus this
vector potential construct is denoted $\tilde{\vecA _x}$. No
errors would be introduced if there were only transverse waves
propagating in the region, as $\vecS _\mT $ does not depend at all
on $\tilde{\vecA _x}$. But, one may wonder whether introducing
such an artificial quantity would contribute to an overestimation
of the actual total energy flux associated with
\emph{longitudinal} pulses propagating in the \emph{transverse}
plane, as $\vecS _\mL =\vecv (mA _y)^2$, where $\vecv$ is the
velocity of the pulse. Well, in reality, the potentials and fields
of the UR charge are only moving in the longitudinal direction, so
there is no physical motion in the transverse plane, and thus the
relevant $\vecv$ here vanishes. Hence, \emph{any} resulting energy
flux corresponding to this artificial quantity, whether associated
with a transverse \emph{or} longitudinal state, would vanish. The
Poynting vector $\vecS _1$ for Pulse 1 is given as
\begin{equation}
\left.
\begin{array}{r@{\, = \;}l}
\vecS _1 & \vecE _x \times \vecB _y\\
& \hatz (E_x)^2\\
\end{array}
\right\} \qquad \mbox{(Pulse 1 (transverse)),} \label{eq:SPulse1}%
\end{equation}
where $\vecB _y=\vecv \times \vecE _x$ has been used along with a
vector identity, and the approximation $\vecv =\hatz$ was made.
See Eq. (\ref{eq:AFUR}) for explicit expressions for $\vecE _x$
and $\vecB _y$.

Pulse 2 in the SWWM is composed of the $\vecE _z$ field specified
in Eq. (\ref{eq:AFUR}) and an artificial magnetic field
$\tilde{\vecB}$. In reality, $\tilde{\vecB}$ does not exist, so
the pulse is not actually realized. But, the effects of $\vecE _z$
felt by $P$ are real, so $\tilde{\vecB}$ is invented to simulate a
transverse pulse travelling from the charge $q$ to $P$ in the
$\hatx$ direction, with some UR velocity $\simeq \hatx$. In order
to properly construct the wavefront of this pulse, $\tilde{\vecB}$
must satisfy the defining equation for a transverse pulse:
$\tilde{\vecB} =\hatx \times \vecE _z=-\haty E_z$. Thus,
$\tilde{\vecB}$ points in the $-\haty$ direction and has the same
magnitude as $\vecE _z$. In the traditional SWWM (where all the
pulses are transverse), it can be shown that the introduction of
this new $\tilde{\vecB _y}$ introduces a negligible error to
$\vecS _\mT $ in the overall analysis \cite{ref:Jack1,ref:Eich}.
In the generalized analysis here, it must be shown that neither
$\tilde{\vecB _y}$ nor the additional artificial $\tilde{\vecA
_z}$ field appearing in Eq. (\ref{eq:AFUR}) introduces any
significant errors. First note that $\tilde{\vecB _y}$ will not
contribute to any $\vecS _\mL $ associated with longitudinal
pulses because the energy flux in that case only depends on the
potentials. As for the $\tilde{\vecA _z}$ field needed to complete
the picture, it can first be argued that for a \emph{peripheral}
interaction of any significance, the condition $1/\gamma \lesssim
mr \lesssim 1$ must hold; it can then be worked out that this new
$\tilde{\vecA _z}$ field is smaller than $\vecA _z$ (cf. Eq.
(\ref{eq:AFUR})) by a factor of $\gamma$. So, any errors
introduced by incorporating these two fictitious quantities into
the actual physics are negligible. The Poynting vector $\vecS _2$
for Pulse 2 is given as
\begin{equation}
\left.
\begin{array}{r@{\, = \;}l}
\vecS _2 & \vecE _z \times \tilde{\vecB} _y\\
& \hatx (E_z)^2\\
\end{array}
\right\} \qquad \mbox{(Pulse 2 (transverse)),} \label{eq:SPulse2}%
\end{equation}
where $\tilde{\vecB} _y=-\haty E _z$ is an artificial magnetic
field used with $\vecE _z$ to form a hypothetical transverse wave
propagating with velocity $\vecv =\hatx$ from $q$ to $P$. See Eq.
(\ref{eq:AFUR}) for the explicit expression for $\vecE _z$.

In the SWWM, the $\Phi$ and $\vecA _z$ potentials specified in Eq.
(\ref{eq:AFUR}) do not contribute in any way to the energy flux,
because the two pulses there are both transverse, and $\vecS$ for
such waves ($\vecS _\mT =\vecE _{\perp } \times \vecB _{\perp }$)
does not depend on these functions. In developing the GWWM, where
all pulses are generally massive, the total $\vecS$ was found in
Eq. (\ref{eq:S2}) to be given by $\vecS =\vecS _\mT +\vecS _\mL $,
where $\vecS _\mT =\vecE _{\perp } \times \vecB _{\perp }$, as
before, and $\vecS _\mL =m^2\Phi \vecA _z$. So, if $m \ne 0$,
there is an additional contribution $\vecS _\mL $ to consider when
determining the total energy flux associated with the particle's
potentials and fields. The $\Phi$ and $\vecA$ potentials
associated with the charge $q$ are thus no longer inconsequential
in terms of observable effects --- they evidently contribute to a
new energy flux term that is longitudinally polarized. In a
seeming miraculous way, these potentials are related in exactly
the way that they are expected to be for a longitudinal plane
wave: $\Phi =\vecv \cdot \vecA _z$ (cf. Eq.
(\ref{eq:PEViveresultsL}))! Therefore, the formalism generalizes
quite naturally. In addition to the two transverse pulses
appearing in the SWWM, a third (longitudinal) pulse -- ``Pulse 3"
-- is thus defined. The wavefront of this pulse is defined by the
longitudinal $\vecA _z$ field, much like the surface of a bed of
nails is defined by the array of nails or the wavefront of a
volley of arrows is defined by the arrows, themselves. According
to Eq. (\ref{eq:PEViveresultsL}), there is an (artificial)
$\tilde{\vecE _z}$ field that must be introduced in order to
complete the picture of a longitudinal wave. But, insofar as
$\vecS _\mL $ is concerned, this $\tilde{\vecE _z}$ need not be
real. As in the construction of Pulses 1 and 2, it must be shown
that the introduction of this fictitious field does not result in
any errors. First of all, any such $\tilde{\vecE _z}$ field would
not contribute to an energy flux associated with
\emph{longitudinal} waves, as $\vecS _\mL $ does not depend at all
on electric and magnetic fields. It can also be reasoned that any
contribution of $\tilde{\vecE _z}$ to the energy flux of a
\emph{transverse} pulse propagating in the transverse plane would
vanish on account of the fact that the velocity $\vecv$ of any
such pulse would vanish because there is nothing in reality
actually propagating in the transverse plane. Because $\vecv
=\veczero$, the fictitious magnetic field $\tilde{\vecB _y}=\vecv
\times \tilde{\vecE _z}$ associated with the pulse would vanish,
and hence so would the associated energy flux $\vecS _\mT
=\tilde{\vecE _z} \times \tilde{\vecB _y}$. In short, then, the
approximation of the charge's potentials as a longitudinal plane
wave pulse does not introduce any errors into the overall
analysis. The Poynting vector $\vecS _3$ for Pulse 3 is given as
\begin{equation}
\left.
\begin{array}{r@{\, = \;}l}
\vecS _3 & m^2\Phi \vecA _z\\
& \hatz (mA_z)^2\\
\end{array}
\right\} \qquad \mbox{(Pulse 3 (longitudinal)).} \label{eq:SPulse3}%
\end{equation}
$\Phi =\vecv \cdot \vecA _z$ has been used, and the approximation
$\vecv =\hatz$ has been made here. See Eq. (\ref{eq:AFUR}) for
explicit expressions for $\Phi$ and $\vecA _z$.

To complete this section, the issue of spatial and temporal
variations of the potentials and fields, as they sweep across the
observer's location, must be addressed. It is being assumed that
the collisions are non-contact (so that the uncertainty $\Delta x$
in the location of the interaction is $\ll b$), and the duration
$\Delta t$ of a typical interaction of interest is $\lesssim
b/\gamma v$ (so that $r \simeq b=\mbox{const}$ during the
encounter). Hence, the magnitudes of the potentials and fields
will not vary appreciably in space (across the target location)
and time during the interaction. Therefore, for the application of
interest, the magnitudes of these quantities can all be taken to
be approximately constant, just like they are on the plane waves
with which they are to be identified.

\section{Fourier Transform of the Energy Flux} \label{sec:FTofS} \indent

In this section, the FT of the energy flux is derived in a general
way, and follows fairly closely the method used in Section 14.5 of
\cite{ref:Jack1}. The differential amount of power $P(t)$ radiated
by $q$ into a differential solid angle element $\dif \Omega$ in
some direction $\hatn$ is given in frame $K$ as
\begin{equation}
\dif P(t)=\dif \Omega \, [r^2(\vecS \cdot \hatn)]_{ret}.\label{eq:dP}%
\end{equation}
$r=r(t)$ here is the relative distance between $q$ and $P$:
$r(t)=\sqrt{b^2+(v t)^2}$. The notation $[\quad]_{ret}$ indicates
that the time $t$ appearing in the term in square brackets is to
be evaluated at the retarded time $\hatt$, which, in the quantum
viewpoint, is the time when the boson was emitted from $q$. This
subtlety is needed to take into account the fact that $q$ cannot
affect $P$ instantaneously; there must be some time delay between
emission and absorption of energy. $t$ is related to $\hatt$ via
\begin{equation}
t = \hatt + \frac{r(\hatt )}{v},\label{eq:rett}%
\end{equation}
where $v$ is the speed of the boson. Thus, the present time $t$,
at which the boson is just influencing $P$, is equal to the
retarded time $\hatt$, at which the boson left $q$, plus the time
delay $r(\hatt )/v$ needed for the boson to travel from $q$ (at
time $\hatt$) to $P$ (at time $t$).

The power radiated per unit solid angle can generally be written
as
\begin{equation}
\frac{\dif P(t)}{\dif \Omega}=|\mathcal{A}(t)|^2,\label{eq:dPdOmega}%
\end{equation}
where $\mathcal{A}(t)$ is a function introduced here for
simplification. For a given pulse, $\mathcal{A}(t)$ is defined as
\begin{equation}
\mathcal{A}(t)=[r(t)\,\sqrt{S_n(t)}]_{ret}.\label{eq:calAt}%
\end{equation}
Note that $\mathcal{A}(t)$ is to be generally complex, so
$|\mathcal{A}(t)|^2$ means $\mathcal{A}(t) \mathcal{A}^*(t)$.
Also, $S_n(t)=\vecS (t) \cdot \hatn$, where $\hatn$ is the unit
vector pointing in the direction of propagation of the pulse.
Thus,
\begin{subequations}
\begin{align}
\mathcal{A}_1(t) &= [r(t)\,E_x(t)]_{ret} & \mbox{(Pulse 1)}\label{eq:calAt1}\\%
\mathcal{A}_2(t) &= [r(t)\,E_z(t)]_{ret} & \mbox{(Pulse 2)}\label{eq:calAt2}\\%
\mathcal{A}_3(t) &= [r(t)\,mA_z(t)]_{ret} & \mbox{(Pulse 3)}\label{eq:calAt3}%
\end{align}
\end{subequations}
for Pulses 1, 2 and 3, respectively. The total energy radiated per
unit solid angle is the integral over all time of Eq.
(\ref{eq:dPdOmega}):
\begin{equation}
\frac{\dif W}{\dif \Omega}=\int^\infty_{-\infty} \dif t\, |\mathcal{A}(t)|^2.\label{eq:dWdOmega1}%
\end{equation}
To reexpress this quantity as an integral over all frequencies,
and thereby provide a link to the FT of the energy flux, the FT
$\mathcal{A}(\omega )$ of $\mathcal{A}(t)$ is introduced:
\begin{equation}
\mathcal{A}(\omega )=\oortpi \, \int^\infty_{-\infty} \dif t\, \mathcal{A}(t)\, \mbox{\Large e}^{\mi \omega t}.\label{eq:FTofA}%
\end{equation}
The FT of this equation yields the inverse relation:
\begin{equation}
\mathcal{A}(t)=\oortpi \, \int^\infty_{-\infty} \dif \omega\, \mathcal{A}(\omega)\, \mbox{\Large e}^{-\mi \omega t}.\label{eq:FTofFTofA}%
\end{equation}
Using Eq. (\ref{eq:FTofFTofA}), Eq. (\ref{eq:dWdOmega1}) can be
written
\begin{subequations}
\begin{align}
\frac{\dif W}{\dif \Omega}&=\int^\infty_{-\infty} \dif t\, \mathcal{A}(t)\, \mathcal{A}^*(t)\label{eq:dWdOmega2}\\%
&=\ootwopi \, \int^\infty_{-\infty} \dif t\, \int^\infty_{-\infty} \dif \omega\, \int^\infty_{-\infty} \dif {\omega }'\, \mathcal{A}^*({\omega }')\, \mathcal{A}(\omega )\, \mbox{\Large e}^{\mi ({\omega }'-\omega)t}.\label{eq:dWdOmega3}%
\end{align}
\end{subequations}
The Fourier representation of the Dirac delta function,
\begin{equation}
\delta ({\omega}'-\omega)=\ootwopi \, \int^\infty_{-\infty} \dif t\, \mbox{\Large e}^{\mi ({\omega }'-\omega)t},\label{eq:FTdelta}%
\end{equation}
can be used to kill the $t$ and ${\omega}'$ integrals, leaving
\begin{equation}
\frac{\dif W}{\dif \Omega}=\int^\infty_{-\infty} \dif \omega\, |\mathcal{A}(\omega )|^2.\label{eq:dWdOmega4}%
\end{equation}
Since $\mathcal{A}(t)$ is purely real for all three pulses, it is
evident from Eq. (\ref{eq:FTofA}) that $\mathcal{A}(\omega
)=\mathcal{A}^*(-\omega )$, so that Eq. (\ref{eq:dWdOmega4}) can
be written as an integral over only positive frequencies.
\begin{equation}
\frac{\dif W}{\dif \Omega}=2 \int^\infty_0 \dif \omega\, |\mathcal{A}(\omega )|^2,\label{eq:dWdOmega5}%
\end{equation}
or
\begin{equation}
\frac{\dif W}{\dif \Omega}=\int^\infty_0 \dif \omega\, \frac{\dif^2I(\omega ,\hatn)}{\dif \omega\, \dif \Omega},\label{eq:dWdOmega6}%
\end{equation}
where
\begin{equation}
\frac{\dif^2I(\omega ,\hatn)}{\dif \omega\, \dif \Omega}=2\, |\mathcal{A}(\omega )|^2\label{eq:dIdwdOmega}%
\end{equation}
is a new quantity that represents the energy radiated in direction
$\hatn$ per unit solid angle per unit frequency. As
$\mathcal{A}(\omega )$ is simply $\mathcal{A}$ expressed as a
function of frequency instead of retarded time, the functional
form of $\mathcal{A}(\omega )$ should be the same as
$\mathcal{A}(t)$. Recalling Eq. (\ref{eq:calAt}),
$\mathcal{A}(\omega )$ can be written
\begin{equation}
\mathcal{A}(\omega )=r(\omega)\,\sqrt{S_n(\omega )}.\label{eq:calAomega}%
\end{equation}
The quantities $r(\omega )$ and $S_n(\omega )$ are the FTs of
$r(t)$ and $S_n(t)$, respectively. Using this equation, Eq.
(\ref{eq:dIdwdOmega}) can be converted into an expression for the
FT of the energy flux of a given pulse:
\begin{equation}
\frac{\dif^2I(\omega ,\hatn)}{\dif \omega\, \dif A}=2\, |S_n(\omega )|.\label{eq:dIdwdA}%
\end{equation}
$dA=r^2\dif \Omega$, where $r=r(\omega )$, is the differential
area element presented by the target to the incident pulse.
Recalling Eqs. (\ref{eq:SPulse1})\,--\,(\ref{eq:SPulse3}), the FTs
of the energy fluxes of the three pulses are found to be
\begin{subequations}
\begin{align}
\frac{\dif^2I_1(\omega ,\hatz)}{\dif \omega\, \dif A} &= 2\, |E_x(\omega )|^2 & \mbox{(Pulse 1)}\label{eq:dIdwdA1}\\%
\frac{\dif^2I_2(\omega ,\hatx)}{\dif \omega\, \dif A} &= 2\, |E_z(\omega )|^2 & \mbox{(Pulse 2)}\label{eq:dIdwdA2}\\%
\frac{\dif^2I_3(\omega ,\hatz)}{\dif \omega\, \dif A} &= 2\, m^2|A_z(\omega )|^2 & \mbox{(Pulse 3)}&.\label{eq:dIdwdA3}%
\end{align}
\end{subequations}
Here, $E_x(\omega )$, $E_z(\omega )$ and $A_z(\omega )$ are the
FTs of $E_x(t)$, $E_z(t)$ and $A_z(t)$, respectively. It remains
now to work out the explicit functional forms of these quantities.

\section{General Fourier Transform Integrals} \label{sec:GenFTInts} \indent

The transformations of the $E_x$, $E_z$ and $A_z$ fields from the
time to the frequency domains are accomplished by way of standard
FT integrals. In this section, the basic FT integrals (Fourier
sine and cosine transforms) that will be solved in subsequent
sections are set up. Denoting a general field in the time domain
as $\Psi (t)$, the corresponding FT $\Psi (\omega )$ is given as
\begin{equation}
\Psi (\omega )=\oortpi\, \int^\infty_{-\infty} \dif t\, \Psi (t)\, \mbox{\Large e}^{\mi \omega t}.\label{eq:FTPsi1}%
\end{equation}
Since $t$ is just a dummy index, this equation can also be written
as
\begin{equation}
\Psi (\omega )=\oortpi\, \int^\infty_{-\infty} \dif t'\, \Psi (t')\, \mbox{\Large e}^{\mi \omega t'},\label{eq:FTPsi2}%
\end{equation}
where $t'=-t$. Thus,
\begin{subequations}
\begin{align}
\Psi (\omega )&=-\oortpi\, \int^{t'\,=\,-t\,=\,+\infty}_{t'\,=\,-t\,=\,-\infty} \dif t\, \Psi (-t)\, \mbox{\Large e}^{-\mi \omega t}\label{eq:FTPsi3}\\%
&=\oortpi\, \int^{t\,=\,+\infty}_{t\,=\,-\infty} \dif t\, \Psi (-t)\, \mbox{\Large e}^{-\mi \omega t},\label{eq:FTPsi4}%
\end{align}
\end{subequations}
where the minus sign has been used to swap the limits of
integration.

If $\Psi (t)$ is an even functions of $t$, $\Psi (-t)=\Psi (t)$,
and hence
\begin{equation}
\Psi (\omega )=\oortpi\, \int^\infty_{-\infty} \dif t\, \Psi (t)\, \mbox{\Large e}^{-\mi \omega t}.\label{eq:FTPsi5}%
\end{equation}
Adding Eq. (\ref{eq:FTPsi1}) to Eq. (\ref{eq:FTPsi5}),
\begin{subequations}
\begin{align}
2\Psi (\omega )&=\oortpi\, \int^\infty_{-\infty} \dif t\, \Psi (t)\, \left( \mbox{\Large e}^{\mi \omega t} + \mbox{\Large e}^{-\mi \omega t} \right) \label{eq:FTPsi6}\\%
&=\toortpi\, \int^\infty_{-\infty} \dif t\, \Psi (t)\, \cos \omega t.\label{eq:FTPsi7}%
\end{align}
\end{subequations}
Or,
\begin{equation}
\Psi (\omega )=\oortpi\, \int^\infty_{-\infty} \dif t\, \Psi (t)\, \cos \omega t.\label{eq:FTPsi8}%
\end{equation}
Since both $\Psi (t)$ and $\cos \omega t$ are even functions of
$t$, their product (the integrand) is also an even function of
$t$, and an even simpler expression can be obtained:
\begin{equation}
\Psi (\omega )=\toortpi\, \int^\infty_0 \dif t\, \Psi (t)\, \cos \omega t\qquad \mbox{(for $\Psi (-t)=\Psi (t)$).}\label{eq:FTPsi9}%
\end{equation}

If, on the other hand, $\Psi (t)$ is an odd functions of $t$,
$\Psi (-t)=-\Psi (t)$, so that
\begin{equation}
\Psi (\omega )=-\oortpi\, \int^\infty_{-\infty} \dif t\, \Psi (t)\, \mbox{\Large e}^{-\mi \omega t}.\label{eq:FTPsi10}%
\end{equation}
Adding Eq. (\ref{eq:FTPsi1}) to Eq. (\ref{eq:FTPsi10}),
\begin{subequations}
\begin{align}
2\Psi (\omega )&=\oortpi\, \int^\infty_{-\infty} \dif t\, \Psi (t)\, \left( \mbox{\Large e}^{\mi \omega t} - \mbox{\Large e}^{-\mi \omega t} \right) \label{eq:FTPsi11}\\%
&=\tioortpi\, \int^\infty_{-\infty} \dif t\, \Psi (t)\, \sin \omega t.\label{eq:FTPsi12}%
\end{align}
\end{subequations}
Or,
\begin{equation}
\Psi (\omega )=\mi \oortpi\, \int^\infty_{-\infty} \dif t\, \Psi (t)\, \sin \omega t.\label{eq:FTPsi13}%
\end{equation}
$\Psi (t)$ and $\sin \omega t$ are both odd functions of $t$, so
their product (the integrand) is an even function of $t$. Thus
\begin{equation}
\Psi (\omega )=\mi \toortpi\, \int^\infty_0 \dif t\, \Psi (t)\, \sin \omega t\qquad \mbox{(for $\Psi (-t)=-\Psi (t)$).}\label{eq:FTPsi14}%
\end{equation}

\section{Fourier Transforms of Fields} \label{sec:FTsofFields} \indent

The three fields of interest are $E_x$ (for Pulse 1), $E_z$ (for
Pulse 2) and $A_z$ (for Pulse 3). In the time domain, they are
(recall Eqs. (\ref{eq:AFUR})):
\begin{subequations}
\begin{align}
E_x(t) &= \oofourpi\, \frac{\gamma q_{\mbox{\scriptsize $VA$}}b}{r^{\mbox{\scriptsize $3$}}}\, (1+mr)\, \mbox{\Large e}^{-mr} & \qquad \mbox{(Pulse 1)}\label{eq:P1field1}\\%
E_z(t) &= -\oofourpi\, \frac{\gamma q_{\mbox{\scriptsize $V$}}vt}{r^{\mbox{\scriptsize $3$}}}\, (1+mr)\, \mbox{\Large e}^{-mr} & \qquad \mbox{(Pulse 2)}\label{eq:P2field1}\\%
A_z(t) &= \oofourpi\, \frac{\gamma q_{\mbox{\scriptsize $VA$}}}{r}\,\mbox{\Large e}^{-mr} & \qquad \mbox{(Pulse 3)}&,\label{eq:P3field1}%
\end{align}
\end{subequations}
where (recall Eq. (\ref{eq:r})) $r\equiv r'(t)=\sqrt{b^2+(\gamma
vt)^2}$. It is immediately apparent that $E_x(t)$ and $A_z(t)$ are
even functions of $t$, and $E_z(t)$ is an odd function of $t$.
Because the derivation of $E_z(\omega )$ easily follows from a
knowledge of $E_x(\omega )$, and that of $E_x(\omega )$ depends on
$A_z(\omega )$, these quantities will be solved here in the
reverse order.

As mentioned above, $A_z(t)$ is an even function of time. So, the
Fourier cosine transform integral equation (Eq. (\ref{eq:FTPsi9}))
is used to find $A_z(\omega )$:
\begin{subequations}
\begin{align}
A_z (\omega )&=\toortpi\, \int^\infty_0 \dif t\, A_z(t)\, \cos \omega t\label{eq:FTAz1}\\%
&=\toortpi\, \int^\infty_0 \dif t\, \left[ \frac{\gamma
q_{\mbox{\scriptsize $VA$}}}{4\mpi}\, \frac{1}{\sqrt{b^2+(\gamma
vt)^2}}\, \mbox{\Large e}^{-m\sqrt{b^2+(\gamma vt)^2}} \right] \,
\cos \omega t \label{eq:FTAz2}\\%
&=\toortpi\, \left( \frac{\gamma q_{\mbox{\scriptsize
$VA$}}}{4\mpi} \right)\, \left[ \int^\infty_0 \dif t\,
\frac{1}{\sqrt{b^2+(\gamma vt)^2}}\, \mbox{\Large
e}^{-m\sqrt{b^2+(\gamma vt)^2}}\, \cos \omega t \right]
\label{eq:FTAz3}\\%
&=\frac{1}{(2\mpi )^{3/2}}\, \frac{q_{\mbox{\scriptsize
$VA$}}}{v}\, K_0(\xi),\label{eq:FTAz4}%
\end{align}
\end{subequations}
where
\begin{equation}
\xi \equiv b\, \sqrt{m^2+\left( \frac{\omega}{\gamma v}
\right)^2}\qquad \mbox{(definition of $\xi$).}\label{eq:xi}%
\end{equation}
The solution to the integral in Eq. (\ref{eq:FTAz3}) was found in
\cite{ref:Grad} (cf. Eq. (3.961.2) therein); $K_0(\xi)$ is a
modified Bessel function of the second kind order $0$.

The derivation of $A_z(\omega )$ was fairly straightforward; that
of $E_x(\omega )$ is much more complicated. As in the previous
analysis, $E_x(t)$ is an even function of $t$, so the Fourier
cosine integral transform equation is used:
\begin{subequations}
\begin{align}
E_x (\omega )&=\toortpi\, \int^\infty_0 \dif t\, E_x(t)\, \cos \omega t\label{eq:FTEx1}\\%
&=\toortpi\, \int^\infty_0 \dif t\, \left[ \oofourpi\,
\frac{\gamma q_{\mbox{\scriptsize $VA$}}b}{r^{\mbox{\scriptsize
$3$}}}\, (1+mr)\, \mbox{\Large e}^{-mr} \right] \, \cos
\omega t \label{eq:FTEx2}\\%
&=\toortpi\, \left( \oofourpi\, \gamma q_{\mbox{\scriptsize
$VA$}}b \right)\, \left[ \int^\infty_0 \dif t\, \left(
\frac{1}{r^3}+\frac{m}{r^2} \right) \, \mbox{\Large e}^{-mr}\,
\cos \omega t \right] \label{eq:FTEx3}\\%
&=\frac{1}{(2\mpi )^{3/2}}\, \gamma q_{\mbox{\scriptsize
$VA$}}b\, (I_1+I_2),\label{eq:FTEx4}%
\end{align}
\end{subequations}
where
\begin{equation}
I_1 \equiv \int^\infty_0 \dif t\, \frac{1}{r^3}\, \mbox{\Large e}^{-mr}\, \cos \omega t\label{eq:I1}%
\end{equation}
and
\begin{equation}
I_2 \equiv \int^\infty_0 \dif t\, \frac{m}{r^2}\, \mbox{\Large e}^{-mr}\, \cos \omega t.\label{eq:I2}%
\end{equation}
Note that
\begin{equation}
\frac{\partial I_1}{\partial m} = -\int^\infty_0 \dif t\, \frac{1}{r^2}\, \mbox{\Large e}^{-mr}\, \cos \omega t\label{eq:I1p}%
\end{equation}
and
\begin{subequations}
\begin{align}
\frac{\partial I_2}{\partial m} &= \int^\infty_0 \dif t\,
\frac{1}{r^2}\, \mbox{\Large e}^{-mr}\, \cos \omega t -
\int^\infty_0 \dif t\, \frac{m}{r}\, \mbox{\Large e}^{-mr}\, \cos
\omega t\label{eq:I2p1}\\%
&=-\frac{\partial I_1}{\partial m}-\frac{m}{\gamma v}K_0(\xi),\label{eq:I2p2}%
\end{align}
\end{subequations}
where the second integral in this equation was solved above, in
determining $A_z(\omega )$. Therefore,
\begin{subequations}
\begin{align}
\frac{\partial E_x(\omega )}{\partial m} &= \frac{1}{(2\mpi
)^{3/2}}\, \gamma q_{\mbox{\scriptsize
$VA$}}b\, \left( \frac{\partial I_1}{\partial m}+\frac{\partial I_2}{\partial m} \right) \label{eq:FTExp1}\\%
&=\frac{1}{(2\mpi )^{3/2}}\, \gamma q_{\mbox{\scriptsize
$VA$}}b\, \left[ -\frac{m}{\gamma v}K_0(\xi) \right].\label{eq:FTExp2}%
\end{align}
\end{subequations}
The parameter $m$ here is being taken as variable, while $\omega$
and $b$ are being treated as constants. Thus,
$\xi=\xi(m)=b\sqrt{m^2+(\omega/\gamma v)^2}$, as defined in Eq.
(\ref{eq:xi}). Because the argument of the Bessel function is
$\xi$, it is easier to reexpress this equation in terms of the
variable $\xi$, instead of $m$. First note that
\begin{equation}
\frac{\partial \xi}{\partial m} = \frac{b^2m}{\xi}.\label{eq:xip1}%
\end{equation}
Then $\partial E_x(\omega)/\partial \xi$ works out to be
\begin{subequations}
\begin{align}
\frac{\partial E_x(\omega )}{\partial \xi} &= \frac{\partial
E_x(\omega )}{\partial m}\, \frac{\partial m}{\partial
\xi}\label{eq:FTExp3}\\%
&=\left[ -\frac{1}{(2\mpi )^{3/2}}\, \frac{q_{\mbox{\scriptsize
$VA$}}bm}{v}K_0(\xi) \right] \left[ \frac{1}{\partial
\xi/\partial m} \right]\label{eq:FTExp4}\\%
&=\left[ \frac{1}{(2\mpi )^{3/2}}\, \frac{q_{\mbox{\scriptsize
$VA$}}}{bv} \right] \left[ -\xi K_0(\xi)
\right].\label{eq:FTExp5}%
\end{align}
\end{subequations}
The second factor in Eq. (\ref{eq:FTExp5}) can be expressed in an
alternative useful form, by using a standard recursion formula for
the derivatives of $K_\nu (\xi)$. It can be worked out that
\begin{equation}
-\xi K_0(\xi) = \frac{\partial [\xi K_1(\xi)]}{\partial \xi},\label{eq:recBes1}%
\end{equation}
where $K_1(\xi)$ is a modified Bessel function of the second kind
of order $1$. Then Eq. (\ref{eq:FTExp5}) becomes
\begin{equation}
\frac{\partial E_x(\omega )}{\partial \xi} = \left[
\frac{1}{(2\mpi )^{3/2}}\, \frac{q_{\mbox{\scriptsize $VA$}}}{bv}
\right] \frac{\partial [\xi K_1(\xi)]}{\partial
\xi}.\label{eq:FTExp6}%
\end{equation}
Since $b$ and $v$ (and of course $q_{\mbox{\scriptsize $VA$}}$)
are constants, this equation can easily be integrated with respect
to $\xi$. It is found that
\begin{equation}
E_x(\omega ) = const + \frac{1}{(2\mpi )^{3/2}}\,
\frac{q_{\mbox{\scriptsize $VA$}}}{bv} \xi
K_1(\xi),\label{eq:FTEx5}%
\end{equation}
where $const$ is any function that does not explicitly depend on
$\xi$. Since $\xi$ is simply proportional to $b$, $const$ is
independent of $b$, too. By demanding that $E_x(\omega )$ vanish
in the $b\to \infty$ limit, $const$ is found to be identically
zero. Note that $\lim_{m\to 0}\xi = \omega b/\gamma v$, so that in
the EM limit (where $m\to 0$ and $q_{\mbox{\scriptsize $VA$}}\to
q^\gamma$), the expression for $E_x(\omega )$ reassuringly reduces
to the familiar formula encountered in ED (see p. 625 of
\cite{ref:Jack1}, for example):
\begin{equation}
\left. \lim_{m\to 0} E_x(\omega )\right| _{q_{\mbox{\scriptsize
$VA$}}=q^\gamma} = \frac{1}{(2\mpi )^{3/2}}\,
\frac{q^\gamma}{bv}\, \left[ \frac{\omega b}{\gamma v} K_1\left(
\frac{\omega b}{\gamma v}\right) \right] \qquad \mbox{(EM
limit).}\label{eq:FTEx6}%
\end{equation}
The final form of $E_x(\omega )$ is thus
\begin{equation}
E_x(\omega ) = \frac{1}{(2\mpi )^{3/2}}\,
\frac{q_{\mbox{\scriptsize $VA$}}}{bv}\, \left[ \xi
K_1(\xi)\right] .\label{eq:FTEx7}%
\end{equation}

Having now solved for $E_x(\omega )$, the determination of
$E_z(\omega )$ is quite easy. Since $E_z(t)$ is an odd function of
$t$, the Fourier sine transform equation must be used.
\begin{subequations}
\begin{align}
E_z(\omega )&=\mi \toortpi\, \int^\infty_0 \dif t\, E_z(t)\, \sin
\omega t\label{eq:FTEz1}\\%
&=\mi \toortpi\, \int^\infty_0 \dif t\, \left[ -\oofourpi\,
\frac{\gamma q_{\mbox{\scriptsize $V$}}vt}{r^{\mbox{\scriptsize
$3$}}}\, (1+mr)\, \mbox{\Large e}^{-mr} \right] \, \sin
\omega t\label{eq:FTEz2}\\%
&=-\mi \toortpi\, \left( \oofourpi\, \gamma q_{\mbox{\scriptsize
$V$}}v \right) \, \int^\infty_0 \dif t\, \left[ \left(
\frac{1}{r^3}+\frac{m}{r^2} \right) \, \mbox{\Large e}^{-mr} \,
(t\sin \omega t) \right] \label{eq:FTEz3}\\%
&=\mi \frac{1}{(2\mpi )^{3/2}}\, \gamma q_{\mbox{\scriptsize
$V$}}v\, \frac{\partial}{\partial \omega} \left[ \int^\infty_0
\dif t\, \left( \frac{1}{r^3}+\frac{m}{r^2} \right)
\, \mbox{\Large e}^{-mr} \, \cos \omega t \right] .\label{eq:FTEz4}%
\end{align}
\end{subequations}
The term in square brackets is identically the term in square
brackets in Eq. (\ref{eq:FTEx3}), which was called $I_1+I_2$ in
Eq. (\ref{eq:FTEx4}). Comparing Eq. (\ref{eq:FTEx4}) to Eq.
(\ref{eq:FTEx7}), it can be gleaned that
\begin{equation}
I_1+I_2 = \frac{1}{\gamma b^2v}\, \xi K_1(\xi).\label{eq:I1plusI2}%
\end{equation}
Thus, Eq. (\ref{eq:FTEz4}) simplifies to
\begin{subequations}
\begin{align}
E_z(\omega )&= \mi \frac{1}{(2\mpi )^{3/2}}\, \gamma
q_{\mbox{\scriptsize $V$}}v\, \frac{\partial}{\partial \omega}
\left[ \frac{1}{\gamma b^2v}\, \xi K_1(\xi)
\right]\label{eq:FTEz5}\\%
&=\mi \frac{1}{(2\mpi )^{3/2}}\, \frac{q_{\mbox{\scriptsize
$V$}}}{b^2} \frac{\partial}{\partial \omega} \left[ \xi K_1(\xi)
\right]\label{eq:FTEz6}\\%
&=\mi \frac{1}{(2\mpi )^{3/2}}\, \frac{q_{\mbox{\scriptsize
$V$}}}{b^2} \frac{\partial \left[ \xi K_1(\xi)
\right]}{\partial \xi} \frac{\partial \xi}{\partial \omega}.\label{eq:FTEz7}%
\end{align}
\end{subequations}
Now, from Eq. (\ref{eq:recBes1}), it is known that
\begin{equation}
\frac{\partial [\xi K_1(\xi)]}{\partial \xi} = -\xi K_0(\xi),\label{eq:recBes2}%
\end{equation}
and it can easily be worked out that
\begin{equation}
\frac{\partial \xi}{\partial \omega} = \frac{b^2 \omega}{\xi (\gamma v)^2},\label{eq:xip2}%
\end{equation}
so that
\begin{subequations}
\begin{align}
E_z(\omega )&=\mi \frac{1}{(2\mpi )^{3/2}}\,
\frac{q_{\mbox{\scriptsize $V$}}}{b^2} [-\xi K_0(\xi)] \left[
\frac{b^2 \omega}{\xi (\gamma v)^2} \right] \label{eq:FTEz8}\\%
&=-\mi \frac{1}{(2\mpi )^{3/2}}\, \frac{q_{\mbox{\scriptsize $V$}}
\omega}{(\gamma v)^2} K_0(\xi) .\label{eq:FTEz9}%
\end{align}
\end{subequations}
Expressed in a different form,
\begin{equation}
E_z(\omega )=-\mi \frac{1}{(2\mpi )^{3/2}}\,
\frac{q_{\mbox{\scriptsize $V$}}}{\gamma vb}\, \left[ \sqrt{\xi
^2-(mb)^2} K_0(\xi)\right] , \label{eq:FTEz10}
\end{equation}
this equation is seen to reduce in the EM limit to a result
encountered in ED (cf. p. 625 of \cite{ref:Jack1}), as it should!
\begin{equation}
\left. \lim_{m\to 0} E_z(\omega )\right| _{q_{\mbox{\scriptsize
$V$}}=q^\gamma}=-\mi \frac{1}{(2\mpi )^{3/2}}\,
\frac{q^\gamma}{\gamma vb}\, \left[ \frac{\omega b}{\gamma v} K_0
\left( \frac{\omega b}{\gamma v} \right) \right] \qquad \mbox{(EM
limit).} \label{eq:FTEz11}
\end{equation}

\section{Frequency Spectra} \label{sec:FreqSpec} \indent

The results of the previous section can be used to find explicit
expressions for the frequency spectra (the FT of the energy flux)
of each pulse. Using Eqs. (\ref{eq:FTEx7}), (\ref{eq:FTEz10}) and
(\ref{eq:FTAz4}) in Eqs. (\ref{eq:dIdwdA1}), (\ref{eq:dIdwdA2})
and (\ref{eq:dIdwdA3}), respectively, the following formulas are
found.
\begin{subequations}
\begin{align}
\frac{\dif^2I_1(\omega ,\hatz)}{\dif \omega\, \dif A} &=
\frac{2}{(2\mpi )^3}\, \left( \frac{q_{\mbox{\scriptsize
$VA$}}}{bv}\right) ^2\,\left[ \xi ^2 K^2_1(\xi)\right] &
\mbox{(Pulse 1)}\label{eq:dIdwdA1exp1}\\%
\frac{\dif^2I_2(\omega ,\hatx)}{\dif \omega\, \dif A} &=
\frac{2}{(2\mpi )^3}\, \left( \frac{q_{\mbox{\scriptsize
$V$}}}{bv}\right) ^2\, \left\{ \frac{1}{\gamma ^2} \left[ \xi
^2-(mb)^2\right] K^2_0(\xi)\right\} & \mbox{(Pulse
2)}\label{eq:dIdwdA2exp1}\\%
\frac{\dif^2I_3(\omega ,\hatz)}{\dif \omega\, \dif A} &=
\frac{2}{(2\mpi )^3}\, \left( \frac{q_{\mbox{\scriptsize
$VA$}}}{bv} \right) ^2 \, \left[ (mb)^2 K^2_0(\xi) \right] &
\mbox{(Pulse 3)}&,\label{eq:dIdwdA3exp1}%
\end{align}
\end{subequations}
where
\begin{equation}
\xi=b\, \sqrt{m^2+\left( \frac{\omega}{\gamma v} \right) ^2},\label{eq:xiagain}%
\end{equation}
as before (cf. Eq. (\ref{eq:xi})).

These functions correspond to a UR charge in a definite helicity
state. The helicity $\lambda$ of the charge appears in the VA
charge $q_{VA}=q_V+q_A\lambda$ (recall Eq. (\ref{eq:qeff})). The
usual application of the WWM is to cases where the moving charges
are unpolarized, such as in the beam of particles in an
accelerator. So it is more useful to consider the average over all
helicity states of the above quantities. This averaging procedure
boils down to averaging $q_{VA}^2$ over all possible $\lambda$
($+1$ or $-1$):
\begin{subequations}
\begin{align}
\langle q_{VA}^2\rangle &= \langle (q_V+q_A\lambda )^2\rangle \label{eq:qva2ave1}\\%
&=q_V^2+2q_Vq_A\langle \lambda \rangle + q_A^2\langle \lambda ^2 \rangle \label{eq:qva2ave2}\\%
&=q_V^2+q_A^2.\label{eq:qva2ave3}%
\end{align}
\end{subequations}
The last step follows from the fact that $\lambda ^2=1$ and the
assumption that the spins of all the charges in the beam are
randomly oriented, so that $\langle \lambda \rangle =0$. The
helicity-averaged frequency spectra are thus
\begin{subequations}
\begin{align}
\left\langle \frac{\dif^2I_1(\omega ,\hatz)}{\dif \omega\, \dif A} \right\rangle &= \frac{2}{(2\mpi )^3}\, \frac{q_{\mbox{\scriptsize $V$}}^{\mbox{\scriptsize $2$}}+q_{\mbox{\scriptsize $A$}}^{\mbox{\scriptsize $2$}}}{b^{\mbox{\scriptsize $2$}}v^{\mbox{\scriptsize $2$}}}\,\left[ \xi ^2 K^2_1(\xi)\right] & \mbox{(Pulse 1)}\label{eq:dIdwdA1exp2}\\%
\left\langle \frac{\dif^2I_2(\omega ,\hatx)}{\dif \omega\, \dif A}
\right\rangle &= \frac{2}{(2\mpi )^3}\, \frac{q_{\mbox{\scriptsize
$V$}}^{\mbox{\scriptsize $2$}}}{b^{\mbox{\scriptsize
$2$}}v^{\mbox{\scriptsize $2$}}}\, \left\{ \frac{1}{\gamma ^2}
\left[ \xi ^2-(mb)^2\right] K^2_0(\xi)\right\} & \mbox{(Pulse
2)}\label{eq:dIdwdA2exp2}\\%
\left\langle \frac{\dif^2I_3(\omega ,\hatz)}{\dif \omega\, \dif A}
\right\rangle &= \frac{2}{(2\mpi )^3}\, \frac{q_{\mbox{\scriptsize
$V$}}^{\mbox{\scriptsize $2$}}+q_{\mbox{\scriptsize
$A$}}^{\mbox{\scriptsize $2$}}}{b^{\mbox{\scriptsize
$2$}}v^{\mbox{\scriptsize $2$}}}\, \left[ (mb)^2 K^2_0(\xi)
\right] &
\mbox{(Pulse 3)}&.\label{eq:dIdwdA3exp2}%
\end{align}
\end{subequations}

In the EM limit, these formulas agree with the expected results
(cf. \cite{ref:Jack1}):
\begin{subequations}
\begin{align}
\left\langle \frac{\dif^2I_1(\omega ,\hatz)}{\dif \omega\, \dif A}
\right\rangle &=
\frac{2}{(2\mpi )^3}\, \left( \frac{q^\gamma}{bv}\right) ^2\,\left[ \left( \frac{\omega b}{\gamma v} \right) ^2 K^2_1\left( \frac{\omega b}{\gamma v} \right) \right] \hspace{.15in} \mbox{(Pulse 1 - EM limit)}\label{eq:dIdwdA1exp3}\\%
\left\langle \frac{\dif^2I_2(\omega ,\hatx)}{\dif \omega\, \dif A}
\right\rangle &=
\frac{2}{(2\mpi )^3}\, \left( \frac{q^\gamma}{bv}\right) ^2\, \left[ \frac{1}{\gamma ^2}\left( \frac{\omega b}{\gamma v} \right) ^2 K^2_0\left( \frac{\omega b}{\gamma v} \right) \right] \nonumber \\ & \hspace{2.8in} \mbox{(Pulse 2 - EM limit)}\label{eq:dIdwdA2exp3}\\%
\left\langle \frac{\dif^2I_3(\omega ,\hatz)}{\dif \omega\, \dif A} \right\rangle &= 0 \hspace{2.502in} \mbox{(Pulse 3 - EM limit).}\label{eq:dIdwdA3exp3}%
\end{align}
\end{subequations}
In the traditional WWM (the SWWM), both Pulse 1 and Pulse 2
correspond to transversely polarized EM radiation. Pulse 3 does
not appear in that theory because it corresponds, by construction,
to longitudinally polarized EW radiation. As photons are always
only transversely polarized, no EM energy can ever be transported
by such a third pulse; that is why the FT of the energy flux for
Pulse 3 vanishes in the EM limit.

Recall that, for a given pulse, $\dif^2I(\omega,\, \hatn)/\dif
\omega\, \dif A$ is the differential energy carried by the pulse
per unit frequency per unit transverse area. The number spectrum
that is ultimately being sought here is easily derived from the
frequency spectrum of a given pulse integrated over the entire
wavefront area of the pulse, which is merely the differential
energy carried by the pulse per unit boson frequency. It must be
kept in mind, however, that the only types of collisions being
considered in this study are those in which the particles do not
come into contact with each other. The peripheral nature of these
collisions is characterized by $b_{min}$, the minimum impact
parameter. For values of the impact parameter $b$ greater than
$b_{min}$, the effects of the fields of the incident particle are
accurately represented in the method by equivalent pulses.
Collisions in which $b$ is less than $b_{min}$ are categorized as
contact collisions, and are not of interest to this study. The
exact specification and a more in depth discussion of $b_{min}$
will be put off until a later section. For now, suffice it to say
that the frequency spectrum integrated over the wavefront area
$\dif I(\omega,\, \hatn)/\dif \omega$ for a given pulse is
obtained from the frequency spectrum $\dif^2I(\omega,\,
\hatn)/\dif \omega\, \dif A$ according to the following formula.
\begin{equation}
\frac{\dif I(\omega ,\hatn)}{\dif \omega} = \int^{2\mpi }_0 \dif
\phi\, \int^\infty_{b_{min}} \dif b\, b\, \left[
\frac{\dif^2I(\omega,\, \hatn)}{\dif \omega\, \dif A}
\right].\label{eq:frsp1}%
\end{equation}
Each such expression corresponds to one pulse, which has fixed
values of $m$ and $\omega$, so these parameters are to be taken as
constants during the following procedures. Because $m$ and
$\omega$ are constants, $\xi =\xi (b)$ only, and the following
useful relations can be easily derived:
\begin{equation}
\dif b\, b = \dif \xi \, \frac{\xi}{\left[ m^2+\left( \mbox{\Large $\frac{\omega}{\gamma v}$} \right)^2 \right]}\label{eq:brel1}%
\end{equation}
and
\begin{equation}
\frac{\dif b}{b} = \frac{\dif \xi}{\xi}.\label{eq:brel2}%
\end{equation}
The minimum value of $\xi$ corresponding to the minimum value
$b_{min}$ of $b$ will be denoted $\chi$:
\begin{equation}
\chi \equiv \xi (b_{min}) = b_{min}\sqrt{m^2+\left(
\frac{\omega}{\gamma v} \right)^2}\qquad \mbox{(definition of
$\chi$);}\label{eq:brel3}%
\end{equation}
the corresponding upper limit is $\infty$. Expressions for the
frequency spectra integrated over the wavefront area will now be
derived for each of the three pulses.

For Pulse 1,
\begin{subequations}
\begin{align}
\frac{\dif I_1(\omega ,\hatz)}{\dif \omega} &= \int^{2\mpi }_0
\dif \phi\, \int^\infty_{b_{min}} \dif b\, b\, \left[
\frac{\dif^2I_1(\omega,\, \hatz)}{\dif \omega\, \dif A}
\right]\label{eq:frspP11}\\%
&= \int^{2\mpi }_0 \dif \phi\, \int^\infty_{b_{min}} \dif b\, b\,
\left\{ \frac{2}{(2\mpi )^3}\, \left( \frac{q_{\mbox{\scriptsize
$VA$}}}{bv}\right) ^2\,\left[ \xi ^2 K^2_1(\xi)\right] \right\} \label{eq:frspP12}\\%
&= (2\mpi )\, \frac{2}{(2\mpi )^3}\, \left(
\frac{q_{\mbox{\scriptsize $VA$}}}{v}\right) ^2\,
\int^\infty_{b_{min}} \frac{\dif b}{b}\, \left[ \xi ^2 K^2_1(\xi)\right] \label{eq:frspP13}\\%
&= \frac{2}{(2\mpi )^2}\, \left( \frac{q_{\mbox{\scriptsize
$VA$}}}{v}\right) ^2\, \int^\infty_{\chi} \frac{\dif \xi}{\xi}\,
\left[ \xi ^2 K^2_1(\xi)\right] \quad \mbox{(via Eqs.
(\ref{eq:brel2})
and (\ref{eq:brel3}))}\label{eq:frspP14}\\%
&= \frac{2}{(2\mpi )^2}\, \left( \frac{q_{\mbox{\scriptsize
$VA$}}}{v}\right) ^2\, \int^\infty_{\chi} \dif \xi\, \left[ \xi
K^2_1(\xi) \right] \label{eq:frspP15}\\%
&= \frac{2}{(2\mpi )^2}\, \left( \frac{q_{\mbox{\scriptsize
$VA$}}}{v}\right) ^2\, \left. \left\{ \half \xi ^2\left[
K^2_1(\xi) - K_0(\xi)K_2(\xi) \right] \right\}
\right|^\infty_{\chi} \nonumber \\ & \quad \mbox{(via Eq. (5.54.2)
in
\cite{ref:Grad})}\label{eq:frspP16}\\%
&= \frac{2}{(2\mpi )^2}\, \left( \frac{q_{\mbox{\scriptsize
$VA$}}}{v}\right) ^2\, \left( \lim_{\xi \to \infty} \left\{ \half
\xi ^2\left[ K^2_1(\xi) - K_0(\xi)K_2(\xi) \right] \right\} -
\right.
\\ & \quad \left. - \half \chi ^2\left[ K^2_1(\chi) - K_0(\chi)K_2(\chi) \right]
\right). \label{eq:frspP17}%
\end{align}
\end{subequations}
The term in curly brackets vanishes in view of the fact that
\begin{equation}
K_\nu (x) \to \sqrt{\frac{\mpi}{2x}}\,\mbox{\Large e}^{-x}\qquad
\mbox{for $x \gg 1$,}\label{eq:Klimit}
\end{equation}
for all values of $\nu \geqslant 0$. Thus,
\begin{subequations}
\begin{align}
\frac{\dif I_1(\omega ,\hatz)}{\dif \omega} &= \frac{2}{(2\mpi
)^2}\, \left( \frac{q_{\mbox{\scriptsize $VA$}}}{v}\right) ^2\,
\left\{ -\half \chi ^2\left[ K^2_1(\chi) -
K_0(\chi)K_2(\chi) \right] \right\}\label{eq:frspP18}\\%
&= \frac{2}{(2\mpi )^2}\, \left( \frac{q_{\mbox{\scriptsize
$VA$}}}{v}\right) ^2\, (-\half \chi ^2)\, \left\{ K^2_1(\chi) -
\right. \nonumber\\ & \quad \left. - K_0(\chi) \left[
K_0(\chi)+\frac{2}{\chi}K_1(\chi) \right] \right\}\qquad
\mbox{(via Eq. (8.486.10) in
\cite{ref:Grad})}\label{eq:frspP19}\\%
&= \frac{2}{(2\mpi )^2}\, \left( \frac{q_{\mbox{\scriptsize
$VA$}}}{v}\right) ^2\, \left\{ \chi K_0(\chi )K_1(\chi )-\half
\chi ^2 \left[ K^2_1(\chi)-K^2_0(\chi) \right]
\right\}.\label{eq:frspP110}%
\end{align}
\end{subequations}

For Pulse 2,
\begin{subequations}
\begin{align}
\frac{\dif I_2(\omega ,\hatx)}{\dif \omega} &= \int^{2\mpi }_0
\dif \phi\, \int^\infty_{b_{min}} \dif b\, b\, \left[
\frac{\dif^2I_2(\omega,\, \hatx)}{\dif \omega\, \dif A}
\right]\label{eq:frspP21}\\%
&= \int^{2\mpi }_0 \dif \phi\, \int^\infty_{b_{min}} \dif b\, b\,
\left( \frac{2}{(2\mpi )^3}\, \left( \frac{q_{\mbox{\scriptsize
$V$}}}{bv}\right) ^2\, \left\{ \frac{1}{\gamma ^2} \left[ \xi
^2-(mb)^2\right] K^2_0(\xi)\right\} \right) \label{eq:frspP22}\\%
&= (2\mpi )\, \frac{2}{(2\mpi )^3}\, \left(
\frac{q_{\mbox{\scriptsize $V$}}}{\gamma v}\right) ^2\,
\int^\infty_{b_{min}} \dif b\, b\, \left\{ \left[ m^2+\left(
\frac{\omega}{\gamma v} \right) ^2 \right]
K^2_0(\xi)\right\} \label{eq:frspP23}\\%
&= \frac{2}{(2\mpi )^2}\, \left( \frac{q_{\mbox{\scriptsize
$V$}}}{\gamma v}\right) ^2\, \int^\infty_{\chi} \dif \xi \,
\frac{\xi}{\left[ m^2+\left( \mbox{\Large $\frac{\omega}{\gamma
v}$} \right)^2 \right]}\, \left\{ \left[ m^2 + \left(
\frac{\omega}{\gamma v} \right) ^2 \right] K^2_0(\xi)\right\}
\nonumber\\ & \quad \mbox{(via Eqs. (\ref{eq:brel1}) and
(\ref{eq:brel3}))}\label{eq:frspP24}\\%
&= \frac{2}{(2\mpi )^2}\, \left( \frac{q_{\mbox{\scriptsize
$V$}}}{\gamma v}\right) ^2\, \int^\infty_{\chi} \dif \xi\, \left[
\xi K^2_0(\xi) \right] \label{eq:frspP25}\\%
&= \frac{2}{(2\mpi )^2}\, \left( \frac{q_{\mbox{\scriptsize
$V$}}}{\gamma v}\right) ^2\, \left. \left\{ \half \xi ^2\left[
K^2_0(\xi) - K^2_1(\xi) \right] \right\}
\right|^\infty_{\chi}\nonumber\\ & \quad \mbox{(via Eq. (5.54.2)
in
\cite{ref:Grad})}\label{eq:frspP26}\\%
&= \frac{2}{(2\mpi )^2}\, \left( \frac{q_{\mbox{\scriptsize
$V$}}}{\gamma v}\right) ^2\, \left( \lim_{\xi \to \infty} \left\{
\half \xi ^2\left[ K^2_0(\xi) - K^2_1(\xi) \right] \right\} -
\right.
\\ & \quad \left. - \half \chi ^2\left[ K^2_0(\chi) - K^2_1(\chi) \right]
\right). \label{eq:frspP27}%
\end{align}
\end{subequations}
As before, the term in curly brackets vanishes (via Eq.
(\ref{eq:Klimit})), thus yielding the final result
\begin{subequations}
\begin{align}
\frac{\dif I_2(\omega ,\hatx)}{\dif \omega} &= \frac{2}{(2\mpi
)^2}\, \left( \frac{q_{\mbox{\scriptsize $V$}}}{v}\right) ^2\,
\left\{ \frac{1}{2\gamma ^2} \chi ^2\left[ K^2_1(\chi) -
K^2_0(\chi) \right]
\right\}.\label{eq:frspP28}%
\end{align}
\end{subequations}

And, for Pulse 3,
\begin{subequations}
\begin{align}
\frac{\dif I_3(\omega ,\hatz)}{\dif \omega} &= \int^{2\mpi }_0
\dif \phi\, \int^\infty_{b_{min}} \dif b\, b\, \left[
\frac{\dif^2I_3(\omega,\, \hatz)}{\dif \omega\, \dif A}
\right]\label{eq:frspP31}\\%
&= \int^{2\mpi }_0 \dif \phi\, \int^\infty_{b_{min}} \dif b\, b\,
\left\{ \frac{2}{(2\mpi )^3}\, \left( \frac{q_{\mbox{\scriptsize
$VA$}}}{bv} \right) ^2 \, \left[ (mb)^2 K^2_0(\xi) \right]
\right\} \label{eq:frspP32}\\%
&= (2\mpi )\, \frac{2}{(2\mpi )^3}\, \left(
\frac{q_{\mbox{\scriptsize $VA$}}}{v} \right) ^2\,
\int^\infty_{b_{min}} \dif b\, b\, \left[ (m)^2 K^2_0(\xi) \right]
\label{eq:frspP33}\\%
&= \frac{2}{(2\mpi )^2}\, \left( \frac{q_{\mbox{\scriptsize
$VA$}}}{v} \right) ^2\, \int^\infty_{\chi} \dif \xi \,
\frac{\xi}{\left[ m^2+\left( \mbox{\Large $\frac{\omega}{\gamma
v}$} \right)^2 \right]}\, \left[ (m)^2 K^2_0(\xi)
\right]\nonumber\\ & \quad \mbox{(via Eqs. (\ref{eq:brel1}) and
(\ref{eq:brel3}))}\label{eq:frspP34}\\%
&= \frac{2}{(2\mpi )^2}\, \left( \frac{q_{\mbox{\scriptsize
$VA$}}}{v} \right) ^2\, \frac{(mb_{min})^2}{\left\{
b^2_{min}\left[ m^2+\left( \mbox{\Large $\frac{\omega}{\gamma v}$}
\right)^2 \right] \right\} }\, \int^\infty_{\chi} \dif \xi\,
\left[ \xi K^2_0(\xi) \right] \label{eq:frspP35}\\%
&= \frac{2}{(2\mpi )^2}\, \left( \frac{q_{\mbox{\scriptsize
$VA$}}}{v} \right) ^2\, \frac{(mb_{min})^2}{\chi ^2}\, \left.
\left\{ \half \xi ^2\left[ K^2_0(\xi) - K^2_1(\xi) \right]
\right\} \right|^\infty_{\chi}\nonumber\\ & \quad \mbox{(via Eq.
(\ref{eq:brel3}) and Eq. (5.54.2) in
\cite{ref:Grad})}\label{eq:frspP36}\\%
&= \frac{2}{(2\mpi )^2}\, \left( \frac{q_{\mbox{\scriptsize
$VA$}}}{v} \right) ^2\, \frac{(mb_{min})^2}{\chi ^2}\, \left(
\lim_{\xi \to \infty} \left\{ \half \xi ^2\left[ K^2_0(\xi) -
K^2_1(\xi) \right] \right\} - \right. \notag \\ & \quad \left. -
\half \chi ^2\left[ K^2_0(\chi) - K^2_1(\chi) \right]
\right)\label{eq:frspP37}\\%
&= \frac{2}{(2\mpi )^2}\, \left( \frac{q_{\mbox{\scriptsize
$VA$}}}{v} \right) ^2\, \left\{ \half (mb_{min})^2 \left[
K^2_1(\chi) - K^2_0(\chi) \right]
\right\},\label{eq:frspP38}%
\end{align}
\end{subequations}
where the last step follows from Eq. (\ref{eq:Klimit}).

The corresponding helicity-averaged quantities are then found to
be
\begin{subequations}
\begin{align}
\left\langle \frac{\dif I_1(\omega ,\hatz)}{\dif \omega}
\right\rangle &= \frac{1}{2\mpi ^2}\, \frac{q^2_{\mbox{\scriptsize
$V$}}+q^2_{\mbox{\scriptsize $A$}}}{v^2}\, \left\{ \chi K_0(\chi
)K_1(\chi )-\half \chi ^2 \left[ K^2_1(\chi)-K^2_0(\chi) \right]
\right\} \nonumber \\ &\hspace{3.75in} \mbox{(Pulse 1)}\label{eq:avefrspP1}\\%
\left\langle \frac{\dif I_2(\omega ,\hatx)}{\dif \omega}
\right\rangle &= \frac{1}{2\mpi ^2}\, \frac{q^2_{\mbox{\scriptsize
$V$}}}{v^2}\, \left\{ \frac{1}{2\gamma ^2} \chi ^2\left[
K^2_1(\chi) - K^2_0(\chi) \right] \right\}\hspace{0.995in}
\mbox{(Pulse
2)}\label{eq:avefrspP2}\\%
\left\langle \frac{\dif I_3(\omega ,\hatz)}{\dif \omega}
\right\rangle &= \frac{1}{2\mpi ^2}\, \frac{q^2_{\mbox{\scriptsize
$V$}}+q^2_{\mbox{\scriptsize $A$}}}{v^2}\, \left\{ \half
(mb_{min})^2 \left[ K^2_1(\chi) - K^2_0(\chi) \right] \right\}
\hspace{0.33in} \mbox{(Pulse 3)}.\label{eq:avefrspP3}%
\end{align}
\end{subequations}
For simplicity in terminology, the term frequency spectrum will
henceforth refer to what has been here referred to as the
frequency spectrum integrated over the wavefront area of a pulse.

\section{Transverse and Longitudinal Frequency Spectra} \label{sec:TandLFreqSpec} \indent

These three quantities can be regrouped as (helicity-averaged)
frequency spectra for transverse and longitudinal boson states. It
was stated previously that Pulses 1 and 2 correspond to transverse
helicity states, and Pulse 3 corresponds to longitudinal helicity
states. The total frequency spectrum for transverse states is thus
the sum of Eqs. (\ref{eq:avefrspP1}) and (\ref{eq:avefrspP2}).
Using a slightly simpler notation, the helicity-averaged
transverse (T) frequency spectrum takes the form
\begin{equation}
\begin{split}
\left\langle \frac{\dif I(\omega )}{\dif \omega} \right\rangle
_\mT &= \frac{1}{2\mpi ^2}\, \frac{q^2_{\mbox{\scriptsize
$V$}}+q^2_{\mbox{\scriptsize $A$}}}{v^2}\, \left\{ \chi K_0(\chi
)K_1(\chi )-\half v^2 \chi ^2 \left[ K^2_1(\chi)-K^2_0(\chi)
\right] \right\} - \\ &\quad -\frac{1}{2\mpi ^2}\,
\frac{q^2_{\mbox{\scriptsize $A$}}}{v^2}\, \left\{
\frac{1}{2\gamma ^2} \chi ^2\left[ K^2_1(\chi) - K^2_0(\chi)
\right] \right\}.\label{eq:avefrspT1}%
\end{split}
\end{equation}
The term proportional to $q^2_A$ is utterly negligible compared to
term proportional to $q^2_V+q^2_A$, because of the factor of
$\gamma ^2$ in the denominator of the latter. As a realistic
simplifying approximation, this term is henceforth discarded,
yielding
\begin{equation}
\left\langle \frac{\dif I(\omega )}{\dif \omega} \right\rangle
_\mT = \frac{1}{2\mpi ^2}\, \frac{q^2_{\mbox{\scriptsize
$V$}}+q^2_{\mbox{\scriptsize $A$}}}{v^2}\, \left\{ \chi K_0(\chi
)K_1(\chi )-\half v^2 \chi ^2 \left[ K^2_1(\chi)-K^2_0(\chi)
\right] \right\}.\label{eq:avefrspT2}%
\end{equation}
The helicity-averaged frequency spectrum for longitudinal boson
states is simply Eq. (\ref{eq:avefrspP3}), which corresponds to
the only longitudinally polarized pulse in the method.
\begin{equation}
\left\langle \frac{\dif I(\omega )}{\dif \omega} \right\rangle
_\mL = \frac{1}{2\mpi ^2}\, \frac{q^2_{\mbox{\scriptsize
$V$}}+q^2_{\mbox{\scriptsize $A$}}}{v^2}\, \left\{ \half
(mb_{min})^2 \left[ K^2_1(\chi) - K^2_0(\chi) \right]
\right\}.\label{eq:avefrspL}%
\end{equation}
In the EM limit, where $\chi=b_{min}\omega /\gamma v$, these
expressions reduce to the expected results (cf. \cite{ref:Jack1}):
\begin{equation}
\renewcommand{\arraystretch}{2}
\left.
\begin{array}{r@{\, = \;}l}
\left\langle \mbox{\Large $\frac{\dif I(\omega )}{\dif \omega}$}
\right\rangle ^\gamma_\mT  & \mbox{\Large $\frac{1}{2\mpi
^{\mbox{\scriptsize $2$}}}\, \frac{q^{\mbox{\scriptsize
$\gamma$}}}{v^{\mbox{\scriptsize $2$}}}$}\, \left\{ \chi K_0(\chi
)K_1(\chi )-\mbox{\Large $\half$} v^2 \chi ^2 \left[
K^2_1(\chi)-K^2_0(\chi)
\right] \right\}\\%
\left\langle \mbox{\Large $\frac{\dif I(\omega )}{\dif \omega}$}
\right\rangle ^\gamma_\mL  & 0\\%
\end{array}
\right\} \quad \mbox{(EM limit).} \label{eq:avefrspEM}
\end{equation}

\section{Transverse and Longitudinal Number Spectra} \label{sec:TandLNumSpec} \indent

It is just one small step now to arrive at the long sought after
number spectra formulas. For a given boson helicity state, the
frequency spectrum function $\langle \dif I(\omega)/\dif \omega
\rangle$ represents the differential energy per unit frequency
contained in the radiation fields surrounding the charged
particle. The number spectrum function $N(E)$ is the differential
number of such bosons per unit boson energy $E$. The relation
between these two quantities is simply
\begin{equation}
N_\Lambda (E)=\frac{1}{E}\left\langle \frac{\dif I(\omega )}{\dif
\omega} \right\rangle _\Lambda,\label{eq:numsp1}
\end{equation}
where $\Lambda =T\mbox{ or }L$ is the helicity of the boson. Thus,
\begin{subequations}
\begin{align}
N_\mT (E) &= \frac{N_0}{E}\left\{ \chi K_0(\chi )K_1(\chi )-\half
v^2 \chi ^2 \left[ K^2_1(\chi)-K^2_0(\chi) \right]
\right\}\label{eq:numspT}\\%
N_\mL (E) &= \frac{N_0}{E}\left\{ \half (mb_{min})^2 \left[
K^2_1(\chi) - K^2_0(\chi) \right]
\right\},\label{eq:numspL}%
\end{align}
\end{subequations}
where
\begin{equation}
N_0 \equiv \frac{1}{2\mpi ^2}\, \frac{q^2_{\mbox{\scriptsize
$V$}}+q^2_{\mbox{\scriptsize $A$}}}{v^2}=const\label{eq:N0}%
\end{equation}
and
\begin{equation}
\chi=b_{min}\, \sqrt{m^2+\left( \frac{E}{\gamma v} \right) ^2},\label{eq:chi}%
\end{equation}
as before (cf. Eq. (\ref{eq:brel3})).

%% file: Chapter4.tex
%%%%%%%%%%%%%%%%%
%  CHAPTER 4
%%%%%%%%%%%%%%%%%

\chapter{Special Cases of $N(E)$} \label{sec:Ncases} \indent

In order to implement these functions in the standard ways (cf.
Eqs. (\ref{eq:s1}) and (\ref{eq:s2})), the mass $m$ of the boson
and the minimum impact parameter $b_{min}$ of the collision must
be specified. As the discussion of these assignments necessarily
involves the fermions emitting the bosons and the bosons,
themselves, some new notation is introduced for clarity at this
point. A quantity associated with a fermion will be denoted with a
subscript $f$, and a quantity associated with a boson will be
denoted with a subscript $b$. Written in this new notation, the
recently derived results (Eqs. (\ref{eq:avefrspT2}) and
(\ref{eq:avefrspL}), and Eqs. (\ref{eq:numspT}) and
(\ref{eq:numspL})) thus become
\begin{subequations}
\begin{align}
\left\langle \frac{\dif I(\omega _b)}{\dif \omega _b}
\right\rangle _\mT  &= N_0\, \left\{ \chi K_0(\chi )K_1(\chi
)-\half v_f^2 \chi ^2 \left[ K^2_1(\chi)-K^2_0(\chi) \right]
\right\}\label{eq:avefrspTnew}\\%
\left\langle \frac{\dif I(\omega _b)}{\dif \omega _b}
\right\rangle _\mL  &= N_0\, \left\{ \half (m_bb_{min})^2 \left[
K^2_1(\chi) - K^2_0(\chi) \right]
\right\},\label{eq:avefrspLnew}%
\end{align}
\end{subequations}
and
\begin{subequations}
\begin{align}
N_\mT (E_b) &= \frac{N_0}{E_b}\left\{ \chi K_0(\chi )K_1(\chi
)-\half v_f^2 \chi ^2 \left[ K^2_1(\chi)-K^2_0(\chi) \right]
\right\}\label{eq:numspTnew}\\%
N_\mL (E_b) &= \frac{N_0}{E_b}\left\{ \half (m_bb_{min})^2 \left[
K^2_1(\chi) - K^2_0(\chi) \right]
\right\},\label{eq:numspLnew}%
\end{align}
\end{subequations}
where
\begin{equation}
N_0 \equiv \frac{1}{2\mpi ^2}\, \frac{q^2_{\mbox{\scriptsize
$V$}}+q^2_{\mbox{\scriptsize $A$}}}{v_f^2}=const\label{eq:N0new}%
\end{equation}
and
\begin{equation}
\chi=b_{min}\, \sqrt{m_b^2+\left( \frac{E_b}{\gamma _fv_f} \right) ^2}.\label{eq:chinew}%
\end{equation}

\section{Boson Mass} \label{sec:BosonMass} \indent

The SM makes definite predictions about the masses of the photon
and the $W$ and $Z$ bosons. According to that theory, of the four
bosons mediating EW interactions, one should be massless, one
should have a mass of about $91$ GeV, and the two remaining bosons
should each have a mass of about $80$ GeV. Of course, the
corresponding particles are identified as the photon, the $Z$
boson and the $W^\pm$ bosons, respectively. In 1983, the
predictions of the masses of the $W$ and $Z$ bosons were verified
with spectacular success at CERN (the European Laboratory for
Particle Physics), in Geneva, Switzerland.

\subsection{Real vs. Virtual Particles} \label{sec:RealvsVirt} \indent

The mass values quoted above are actually the masses of the bosons
when they are ``real". Most simply put, a real particle is one
whose properties can be directly detected; such particle states
are represented by external lines in Feynman diagrams. In
contrast, the vast majority of all particle interactions involve
particles that cannot be observed directly. Such particles are
called ``virtual", and are represented in Feynman diagrams by
internal lines. Properties of virtual particles can only be
inferred, at best. A classic example of such an inference is the
calculation of a correction to the Lamb Shift of hydrogen. The
experimental verification of the correction, which is due entirely
to the presence of virtual particles, provided a great impetus to
the early development of quantum field theory \cite{ref:Aitc}. The
distinction between real and virtual particles is usually made in
the context of the laws of conservation of energy and 3-momentum,
and is alternatively phrased in terms of either the Heisenberg
Uncertainty Principle or the on-shell condition.

The Heisenberg Uncertainty Principle is a restriction on the
values that various pairs of dynamic quantities (called
canonically conjugate observables) can assume. Most relevant to
this thesis are the pairs $E$ and $t$, and $p^i$ and $x^i$
($i=1,\,2,\,3$). For \emph{real} particles, the principle sets a
limit on the degree of accuracy with which the two quantities in a
given pair of observables can be simultaneously measured. Let the
symbol $\Delta$ denote an rms deviation from an average value of
an observable $O$:
\begin{equation}
\Delta O \equiv \sqrt{\left \langle (O-\langle O \rangle )^2
\right \rangle}\qquad \mbox{(definition of $\Delta$).}\label{eq:defdelO1}%
\end{equation}
After a few lines of algebra, a useful related equation is found:
\begin{equation}
\Delta O^2 = O_{rms}^2-\langle O \rangle ^2,\label{eq:defdelO2}%
\end{equation}
where $O_{rms}\equiv \sqrt{\langle O^2\rangle }$ is the
root-mean-squared average of $O$. The Heisenberg relations of
interest then interrelate $\Delta E$ and $\Delta t$, and $\Delta
p^i$ and $\Delta x^i$ ($i=1,\,2,\,3$):
\begin{equation}
\renewcommand{\arraystretch}{2}
\left.
\begin{array}{r@{\, \geqslant \;}l}
\Delta E \Delta t & \mbox{\Large $\frac{\hbar}{2}$}\\%
\Delta p^i \Delta x^i & \mbox{\Large $\frac{\hbar}{2}$}\qquad \mbox{($i=1,\,2,\,3$)}\\%
\end{array}
\right\} \qquad \mbox{(Heisenberg relations).} \label{eq:Heisrel}
\end{equation}
Thus, if a dynamical state exists only for a time on the order of
$\Delta t$, the energy of the state cannot be measured to a
precision better than about $\hbar/\Delta t$. Similarly, if the
location of such a state is known to an accuracy of some $\Delta
x$, then the state's 3-momentum cannot be specified any more
precisely than about $\hbar/\Delta x$. This principle is often
rephrased by stating that the conservation of energy and
3-momentum can be violated so long as $\Delta t \lesssim
\hbar/\Delta E$ and $\Delta x^i \lesssim \hbar/\Delta p^i$
($i=1,\,2,\,3$). Processes that occur on these length and time
scales are not ``observable", and can therefore (supposedly)
violate the conservation of energy and 3-momentum. They are
mediated by so-called virtual particles. In this picture, then,
the (invisible) virtual particles can have \emph{any} values of
$E$ and $\vecp$ whatsoever, so long as 4-momentum is conserved in
the overall macroscopic (observable) process. In summary, if the
Heisenberg relations hold for any intermediate particle state, the
particle can be either real \emph{or} virtual. But, if these
relations are violated, the particle can only be \emph{virtual}.
It will be shown later that the WW number spectrum functions
$N_\Lambda (E_b)$ are strongly suppressed if $\Delta p_{b\perp }
b_{min} \gtrsim \hbar$, where $\Delta p_{b\perp }$ is the
uncertainty in the transverse component of $\vecp _b$. This
behavior seems to imply that the only significant contributions to
$N_\Lambda (E_b)$ come from bosons that are virtual.
Interestingly, the SWWM is also called the Weizs\"{a}cker-Williams
Method of Virtual Quanta \cite{ref:Jack1}.

The on-shell condition is an equation interrelating the mass $m$,
energy $E$ and 3-momentum $\vecp$ of a particle. It reads
\begin{equation}
m^2=E^2-\vecp ^2\qquad \mbox{(on-shell condition).}\label{eq:onshell}%
\end{equation}
If this equation is satisfied, the particle is called ``real"; if
not, the particle is called ``virtual". So, real (virtual)
particles are also oftentimes referred to as on-shell (off-shell).
In the case of real particles, the interpretation of this equation
is straightforward: $E$ and $\vecp$ are the observable energy and
3-momentum, respectively, of the particle, and $m$ is the
particle's mass, which is a fixed value. If this equation is not
satisfied, the meanings of these variables are usually
reinterpreted in a different way. In two closely-related
standpoints, $m$ is taken to be the familiar fixed (on-shell)
value associated with the intermediate particle. One
interpretation then assumes that energy is conserved in the
intermediate process, but 3-momentum is not; the other is just the
reverse: 3-momentum is conserved, but energy is not. The problem
with these interpretations is that they are not covariant. That
is, they do not treat all the components of $p^\mu=(E,\,\vecp)$ on
an equal footing. A third interpretation that \emph{is} covariant
assumes that both energy and 3-momentum are \emph{always}
conserved, but the value of $m$ does not equal the familiar
on-shell value. A stark illustration of this interpretation is
provided by nuclear beta decay, wherein a nucleus of atomic mass
$A$ and atomic number $Z$ ``beta-decays" into a nucleus of atomic
mass $A$ and atomic number $Z+1$, with an electron and an
anti-neutrino being emitted in the process. According to the EW
theory, the process is mediated by a $W^-$ boson. But, experiments
imply that Eq. (\ref{eq:onshell}) is violated quite dramatically.
In the context of the third interpretation discussed above, the
mass of the (highly-virtual) $W^-$ boson has to be on the order of
a few MeV, which is far removed from the on-shell value of $80.42$
GeV \cite{ref:Aitc,ref:RPP}! In this same sense, the weak bosons
mediating the interactions of interest in this study will be shown
to be necessarily far off their mass shells (i.e., highly
virtual). The interpretation that shall be adopted will be the
third viewpoint discussed above. However, it will turn out that a
boson's energy $E_b$ and the longitudinal component ${\vecp
_{b||}}$ of its 3-momentum $\vecp _b$ are well-defined, while the
transverse component ${\vecp _{b\perp}}$ of $\vecp _b$ is not
well-defined. The term well-defined here means that the average
value of the quantity is much greater than its associated
uncertainty. Thus, $\langle E_b \rangle \gg \Delta E_b$ and
$\langle {\vecp _{b||}} \rangle \gg \Delta {\vecp _{b||}}$, but
$\langle {\vecp _b}_{\perp}\rangle \lesssim \Delta {\vecp
_{b\perp}}$. In fact, it will be shown that $\langle {\vecp
_{b\perp}}\rangle = \veczero$, but $\Delta {\vecp _{b\perp}} >
\veczero$, so that the bosons can always be taken to be travelling
nearly collinearly with the parent fermion.

\subsection{General Considerations} \label{sec:mbGeneral} \indent

To begin with, it must be realized that the pulses appearing in
the method are not bona fide freely propagating bosons. The plane
wave pulse construct is merely an approximation to the radiation
fields that are carried along with the UR charge. The main
similarity is the plane wave geometry of both the pulses and
fields --- that was the historical motivation for the original
method. In the UR limit, this similarity is also realized in the
generalization to weak interactions developed here. Another
similarity is the charge to which these quantities couple. As
electric charge couples to $\vecE$ and $\vecB$ fields, it makes no
difference whether these fields are in the form of freely
propagating EM plane waves or the EM fields of an UR charge. A
nearby charge will respond in the same way to both quantities ---
that was another similarity that made the original formulation
plausible. The same similarity holds in the generalization to weak
interactions, as well. A third similarity that was realized in the
SWWM was the mass of both pulses and fields. A given set of EM
fields are carried along with their associated charge at an UR
velocity $\vecv _f$, so that their velocity $\vecv _b\simeq \vecv
_f$. If an energy $E_b$, 3-momentum $\vecp _b$ and mass $m_b$ are
associated with these fields, the UR condition is equivalent to
$|\vecv _b|= |\vecp _b|/E_b \simeq 1$, or $m_b \ll E_b \simeq
|\vecp _b|$. Thus, taking $m_b \simeq 0$ is a realistic
approximation. Upon identifying these fields with a swarm of
photons, then, it is completely reasonable to take those photons
to be on their mass shells, i.e., massless. The generalization to
massive vector bosons (viz, $W$ and $Z$ bosons) is troublesome in
this regard, as the on-shell values of the masses of these
mediators is not zero; they are, in fact, much greater than the
masses of typical light nuclei. An easy solution is to state that
the weak fields must be identified with \emph{UR} on-shell bosons,
so that the boson energies must be much greater than about $100$
GeV. That is precisely what is done in the EWM (cf.
\cite{ref:Daws1,ref:Daws2,ref:Guni,ref:Kane,ref:Cahn1,ref:Alta}).
It can be done here as well (that is, set $m_b$ equal to the
relevant on-shell value), and the resulting cross sections agree
with those found via other methods, but only for collision
energies above a certain threshold value. It is of interest to
devise the mass scheme so as to allow for a wider range of
possible collision energies. Such a formulation is developed in
this section. It amounts to stating that the quantities (called
``equivalent bosons") that mediate these weak interactions are not
on-shell UR bosons, but something analogous to highly virtual
bosons. Unlike in the EWM, the boson energies $E_b$ here are
allowed to assume values anywhere from $0$ up to the kinetic
energy $E_f-m_f$ of the parent fermion (the upper limit being set
by conservation of energy). The semiclassical method thus
constructed, then, has a greater scope of applicability than its
(inherently more accurate) quantum-mechanical counterpart. The
value of $m_b$ that is adopted must be consistent with several key
assumptions. One condition that shall be required is that energy
and 3-momentum always be conserved during any subprocess. Another
restriction on $m_b$ is that it be Lorentz invariant; that is,
$m_b^2$ must be the square of a 4-vector. A third restriction is
that the concept of causality must be preserved; that is, the
boson that transmits the 4-momentum must travel at a subluminal
velocity.

\subsection{The Boson Mass $m_b$} \label{sec:mbwave1} \indent

As a first step towards specifying a well-defined value for $m_b$,
consider one of the plane-wave wave packets that are approximating
the potentials and fields of the UR charge $q$. Let this wave
packet travel in the $+\hatz$ direction. This choice limits the
analysis to Pulses 1 and 3, which both travel in that direction,
but it can easily be applied to Pulse 2, which is the hypothetical
pulse that propagates in the $\hatx$ direction. In this section,
the boson $b$ will manifest itself as a wave-like disturbance
(i.e., the wave packet) that propagates through the potentials and
fields from $q$ to $P$. In a future section, the equations will be
reinterpreted in such a way that a particle manifestation of $b$
becomes apparent. In either case, the basic process under scrutiny
is the emission of $b$ from $q$, as shown in Fig. \ref{fig:fbf}.
An incident fermion $f$ emits $b$ into an angle $\theta _b$ with
respect to the original direction $\hatz$. Because 3-momentum is
conserved in the process, the final state fermion $f'$
consequently recoils into an angle ${\theta _f}'$, also measured
with respect to $\hatz$. With an energy $\omega _b$ and a
3-momentum $\veck _b$, the boson's 4-momentum $k_b^\mu$ is
expressed as $k_b^\mu=(\omega _b,\, \veck _b)$. In this section,
three guiding principles will be used to derive several important
quantities. They are the Lorentz condition, conservation of
4-momentum, and causality. The quantities derived are the square
$k_b^2$ of the boson's 4-momentum, the transverse $\veck
_{b\perp}$ and longitudinal $k _{bz}$ components of the boson's
3-momentum, and the boson's speed $v_b$. At the end of the
section, a mass $m_b$ of the boson is identified. The results of
this section were somewhat unexpected: for a given energy $\omega
_b$, $m_b$ is found to be uniquely defined in terms of the set of
charge quantum numbers of the parent fermion!

It is vital for the method to express $\veck _b$ in terms of
components perpendicular to (denoted with the subscript $\perp$),
and parallel to (denoted with the subscript $||$), the direction
of motion, which, in frame $K$, is $\hatz$. Thus,
\begin{equation}
\veck _b={\veck _{b\perp}}+{\veck _{b||}},\quad \mbox{where\,
$\veck _{b||}=\hatz\, k_{bz}$.} \label{eq:pb1}
\end{equation}
So $k_b^\mu=(\omega _b,\, \veck _{b\perp},\, k_{bz})$. As the
waves approximating the potentials and fields are all travelling
in the $\hatz$ direction, the transverse component of 3-momentum
must vanish on average. The nonvanishing of this quantity is
interpreted as being due to inherent statistical fluctuations in
the potentials and fields surrounding $q$. The actual transfer of
4-momentum in the transverse direction is thus accomplished by
means of a mere fluctuation. The equation of motion for the wave
packet was encountered in a previous section (Section
\ref{sec:ModesPackets}).
\begin{subequations}
\begin{align}
k_b^2 &= \omega _b^2-\veck _b^2\label{eq:mb1}\\%
&= \omega _b^2-\veck^2_{b\perp}-k_{bz}^2.\label{eq:mb2}%
\end{align}
\end{subequations}
The LC in momentum-space,
\begin{equation}
k_b^\mu A_\mu=0\qquad \mbox{(LC in
momentum-space)}\label{eq:LCpspace2}
\end{equation}
(cf. also Eq. (\ref{eq:LCpspace1})), can be used to simplify.
Recalling Eq. (\ref{eq:Amu2}),
\begin{equation}
A^\mu(b,t) = q^\mu\, \left[ \oofourpi\, \frac{1}{r}\, \mbox{\Large
e}^{-m_br} \right],\label{eq:Amu4}%
\end{equation}
Eq. (\ref{eq:LCpspace2}) becomes
\begin{equation}
k_b^\mu q_\mu \left[ \oofourpi\, \frac{1}{r}\, \mbox{\Large
e}^{-m_br} \right]=0,\label{eq:LCpspace4}%
\end{equation}
and thus
\begin{subequations}
\begin{align}
0 &= k_b^\mu q_\mu \label{eq:LCpspace5}\\%
&= \gamma _f(\omega _bq^0-\veck _b \cdot
\vecq)\label{eq:LCpspace6}\\%
&= \gamma _f[\omega _b(q_V+q_A \lambda _f v_f)-k_{bz}(q_Vv_f+q_A
\lambda _f)]\qquad \mbox{(via Eq.
(\ref{eq:qmu2}))}.\label{eq:LCpspace7}%
\end{align}
\end{subequations}
Therefore,
\begin{subequations}
\begin{align}
k_{bz} &= \omega _b\, \frac{(q_V+q_A \lambda _f v_f)}{(q_Vv_f+q_A
\lambda _f)}\label{eq:LCpspace8}\\%
&= \frac{\omega _b}{v_f}\, \frac{[(q_Vv_f+q_A \lambda _f )-q_A
\lambda _f /\gamma _f^2 ]}{(q_Vv_f+q_A
\lambda _f)}\label{eq:LCpspace9}\\%
&= \frac{\omega _b}{v_f}\, \left[ 1- \frac{1}{\gamma _f^2}\,
\frac{q_A \lambda _f }{(q_Vv_f+q_A \lambda _f)} \right]
\label{eq:LCpspace10}\\%
&= \frac{\omega _b}{v_f}\, (1- \varepsilon ),\quad
\mbox{where}\label{eq:LCpspace11}\\%
\varepsilon &\equiv \frac{\alpha}{\gamma _f^2}\qquad \mbox{(definition of $\varepsilon$), and}\label{eq:defvarth1}\\%
\alpha &\equiv \frac{q_A \lambda _f}{(q_Vv_f+q_A \lambda _f)}\qquad \mbox{(definition of $\alpha$).}\label{eq:defalpha}%
\end{align}
\end{subequations}
Plugging Eq. (\ref{eq:LCpspace11}) into Eq. (\ref{eq:mb2}) yields
\begin{subequations}
\begin{align}
k_b^2 &= \omega _b^2-\veck^2_{b\perp}-\left( \frac{\omega
_b}{v_f}\right) ^2(1-\varepsilon)^2\label{eq:mb3}\\%
&=-\left( \frac{\omega _b}{\gamma _fv_f}\right) ^2\left[ 1-\gamma
_f^2\varepsilon (2-\varepsilon)\right] -\veck
^2_{b\perp}.\label{eq:mb4}%
\end{align}
\end{subequations}
For a given value of $\omega _b$, Eq. (\ref{eq:mb4}) involves two
unknown parameters --- $k_b$ and ${\veck _{b\perp}}$. To solve for
$k_b$, additional information about $\veck _{b\perp}$ is needed.
Towards this end, the conservation of 3-momentum can be invoked at
the point of boson emission to relate $\veck _{b\perp}$ to ${\vecp
_{f\perp}}'$, the transverse component of 3-momentum of the final
state fermion $f'$ (cf. Fig. \ref{fig:fbf}):
\begin{equation}
\vecp _{f\perp} = {\vecp _{f\perp}}' + {\veck
_{b\perp}}.\label{eq:pcons1}%
\end{equation}
Then, the Cartesian coordinate system can be oriented in such a
way that $\vecp _{f\perp}=\veczero$, so that
\begin{equation}
\veck _{b\perp} = -{\vecp _{f\perp}}'.\label{eq:pcons2}%
\end{equation}
\emph{After} the emission,
\begin{subequations}
\begin{align}
m_f^2 &= ({{p_f}'})^\mu ({{p_f}'})_\mu \label{eq:pcons3}\\%
&= {{E_f}'}^2 - {{\vecp _f}'}^2\label{eq:pcons4}\\%
&= {{E_f}'}^2 - {\vecp ^2_{f\perp}}'  - {{p _{fz}}'}^2\label{eq:pcons5}\\%
&= {{E_f}'}^2 - {\vecp ^2_{f\perp}}'  - {{v_{fz}}'}^2{{E_f}'}^2\label{eq:pcons6}\\%
&= {{E_f}'}^2(1-{{v_{fz}}'}^2) - {{\vecp _{f\perp}}'}^2\label{eq:pcons7}\\%
&= \left( \frac{{E_f}'}{{\gamma _{fz}}'}\right) ^2 - {{\vecp
_{f\perp}}' }^2,\label{eq:pcons8}%
\end{align}
\end{subequations}
where
\begin{equation}
{v_{fz}}' = {\vecv _f}' \cdot \hatz = {v_f}'\cos{{\theta
_f}'}\label{eq:defvfpz}%
\end{equation}
(cf. Fig. \ref{fig:fbf}) and
\begin{equation}
{\gamma _{fz}}' \equiv \frac{1}{\sqrt{1-{{v_{fz}}'}^2}}\qquad
\mbox{(definition of ${\gamma _{fz}}'$)}.\label{eq:defgamfz}%
\end{equation}
Solving for ${\vecp ^2_{f\perp}}' $,
\begin{equation}
{\vecp ^2_{f\perp}}' = -m_f^2 + \left( \frac{{E_f}'}{{\gamma
_{fz}}'}\right) ^2,\label{eq:pcons9}%
\end{equation}
and using Eq. (\ref{eq:pcons2}) then yields the useful relation
\begin{equation}
\veck ^2_{b\perp} = m_f^2 - \left( \frac{{E_f}'}{{\gamma
_{fz}}'}\right) ^2.\label{eq:pcons10}%
\end{equation}
Now, ${E_f}'$ can be expressed in terms of the energy $E_f$ of the
incident fermion and the independent variable $\omega _b$, as
${E_f}'=E_f-\omega _b$ (which follows from conservation of
energy), but ${\gamma _{fz}}'$ is unknown at this point. Introduce
the new parameter $R$ by the following definition:
\begin{equation}
R \equiv \frac{\gamma _f}{{\gamma _{fz}}'} =
\sqrt{\frac{1-{{{v_{fz}}'}^2}}{1-{v_f}^2}}\qquad \mbox{(definition
of R).}\label{eq:defR}
\end{equation}
Also, the following parameter (the Feynman scaling variable) is
introduced for convenience:
\begin{equation}
x \equiv \frac{\omega _b}{E_f}\qquad \mbox{(definition of
$x$).}\label{eq:defx}
\end{equation}
By conservation of energy, $0 \le \omega _b \le E_f-m_f$, so that
$0 \le x \le 1-1/\gamma _f \simeq 1$. In terms of these new
parameters, $\veck ^2_{\perp}$ can be written
\begin{equation}
\veck ^2_{\perp} = -m_f^2\, \left[1 - R^2\,(1-x)^2 \right]
.\label{eq:pcons11}%
\end{equation}
Then, by Eq. (\ref{eq:mb4}), $k_b^2$ becomes
\begin{subequations}
\begin{align}
k_b^2 &= -\veck ^2_{\perp} - m_f^2\, \left( \frac{x}{v_f} \right)
^2\, \left[ 1-\gamma _f^2\varepsilon
(2-\varepsilon)\right] \label{eq:mbsq1}\\%
&= m_f^2 - m_f^2\, R^2\, (1-x)^2 - m_f^2\, \left( \frac{x}{v_f}
\right) ^2\left[ 1-\gamma _f^2\varepsilon (2-\varepsilon)\right]
\label{eq:mbsq2}\\%
&= m_f^2\, \left[ 1 - \left( \frac{x}{v_f} \right) ^2 - R^2\,
(1-x)^2 + \left( \frac{\gamma _fx}{v_f} \right)
^2\varepsilon(2-\varepsilon)\right] .\label{eq:mbsq3}%
\end{align}
\end{subequations}

To simplify this equation, $R^2(1-x)^2$ will now be reexpressed in
terms of the other parameters in the equation.
\begin{subequations}
\begin{align}
R^2 &= \gamma _f^2\, \left( 1-{{{v_{fz}}'}^2} \right)\label{eq:R1}\\%
&= \gamma _f^2\, \left( 1-\frac{{{p_{fz}}'}^2}{{{E_f}'}^2}\right)
\label{eq:R2}\\%
&= \gamma _f^2\, \left[ 1-\frac{(p_f-p_{bz})^2}{E_f^2(1-x)^2}
\right] \label{eq:R3}\\%
&= \gamma _f^2\, \left\{ 1- \frac{[v_fE_f-xE_f(1-
\varepsilon)/v_f]^2}{E_f^2(1-x)^2} \right\}
\label{eq:R4}\\%
&= \frac{\gamma _f^2}{v_f^2(1-x)^2}\, \left\{ v_f^2(1-x)^2- \left[
v_f^2-x(1-\varepsilon) \right] ^2
\right\}.\label{eq:R5}%
\end{align}
\end{subequations}
After a bit of work, this equation is found to lead to
\begin{equation}
R^2(1-x)^2 = 1 - \left( \frac{x}{v_f} \right) ^2 + \left(
\frac{\gamma _fx}{v_f} \right) ^2\varepsilon (2-\varepsilon ) -
(2\gamma _f^2\varepsilon )x.\label{eq:R6}%
\end{equation}
$k_b^2$ then becomes (via Eq. (\ref{eq:mbsq3}))
\begin{equation}
k_b^2=2\alpha m_f^2x.\label{eq:kbsq1}%
\end{equation}
Similarly, Eq. (\ref{eq:pcons11}) yields
\begin{equation}
\veck _{b\perp}^2=\delta M_{\delta}^2-k_b^2,\label{eq:veckbsq1}%
\end{equation}
where
\begin{equation}
\delta \equiv \gamma _f^2\varepsilon(2-\varepsilon)-1=\alpha \left( 2-\frac{\alpha }{\gamma _f^2}\right) -1 \qquad \mbox{(definition of $\delta$)}\label{eq:defdelta1}%
\end{equation}
and
\begin{equation}
M_{\delta}^2 \equiv \left( \frac{\omega _b}{\gamma _fv_f} \right) ^2 = m_f^2\, \left( \frac{x}{v_f} \right) ^2 \qquad \mbox{(definition of $M_{\delta}$)}.\label{eq:defMdel}%
\end{equation}
The symbol $M$ is used because it turns out that the quantity must
be an invariant with the units of energy (hence $M$ for mass).

A caveat in this analysis is causality. The above formulas are all
valid as long as the boson that they describe travels at a
subluminal velocity. To check for this condition, identify the
(magnitude of the) group velocity in the usual way (recall Eq.
(\ref{eq:vgroup2})):
\begin{equation}
v_b=\frac{\dif \omega _b}{\dif |\veck _b|}=\frac{|\veck _b|}{\omega _b}\qquad \mbox{(boson speed).}\label{eq:vb}%
\end{equation}
Then recall Eq. (\ref{eq:mb1}):
\begin{subequations}
\begin{align}
k_b^2 &= \omega _b^2-\veck _b^2\label{eq:vbsq1}\\%
&= \omega _b^2\left( 1-\frac{\veck _b^2}{\omega _b^2}\right) \label{eq:vbsq2}\\%
&= \omega _b^2( 1-v_b^2). \label{eq:vbsq3}%
\end{align}
\end{subequations}
Thus,
\begin{subequations}
\begin{align}
v_b^2 &= 1-\frac{k_b^2}{\omega _b^2}\label{eq:vbsq4}\\%
&= 1-\frac{2\alpha m_f^2x}{E_f^2x^2}\label{eq:vbsq5}\\%
&= 1-\frac{x_C}{x},\quad \mbox{where} \label{eq:vbsq6}\\%
x_C &\equiv 2\varepsilon = \frac{2\alpha }{\gamma _f^2} \qquad \mbox{(definition of $x_C$).} \label{eq:vbsq7}\\%
\end{align}
\end{subequations}
From this equation, the $v_b^2 \le 1$ condition is easily seen to
be equivalent to $x_C \ge 0$, or (by the definition of $x_C$),
\begin{equation}
\alpha \ge 0\qquad \mbox{(causality condition \#1a).}\label{eq:caus1a}%
\end{equation}
Another informative variation of this condition is the
corresponding restriction on $k_b^2$. By way of Eq.
(\ref{eq:kbsq1}),
\begin{equation}
k_b^2 \ge 0\qquad \mbox{(causality condition \#1b).}\label{eq:caus1b}%
\end{equation}
Note that $v_b=1$ corresponds exactly to $k_b^2=0$, as it should.
The condition $v_b^2 \ge 0$ is also easily found to be equivalent
to
\begin{equation}
x \ge x_C\qquad \mbox{(causality condition \#2).}\label{eq:caus2}%
\end{equation}
Causality condition \#1 places a restriction on the helicity state
$\lambda _f$ of the fermion in order for a boson to be emitted
(recall Eq. (\ref{eq:defalpha})). Causality condition \#2 reveals
that the above expressions for $k_b^2$ and $\veck _{b\perp}^2$ are
only valid for values of $x \ge x_C$, or, equivalently, $\omega _b
\ge 2\varepsilon E_f = 2\alpha m_f/\gamma_f$. Bosons with energies
$\omega _b$ lower than $x_CE_f$ are simply never emitted, because
they would necessarily have to travel faster than light.

It would seem that a viable mass scheme has been devised. With the
boson's 4-momentum identified as $k_b^\mu =(\omega _b,\, \veck
_{b\perp },\, k_{bz})$, where $\veck _{b\perp }$ and $k_{bz}$ are
specified in Eqs. (\ref{eq:veckbsq1}) and (\ref{eq:LCpspace11}),
respectively, a natural choice for the boson mass is
$k_b=m_f\sqrt{2\alpha x}$. While this choice is perfectly
reasonable for a boson in a definite helicity state $\lambda _f$,
of interest to this study are helicity-averaged quantities (recall
Sec. \ref{sec:FreqSpec}). Therefore, the above analysis must be
modified somewhat so as to apply to a boson emitted from a
particle in an arbitrary helicity state, as, for example, in an
accelerator beam. That is to say, the kinematic variables of
interest must somehow be averaged over all possible helicity
states. As all of the above analysis was based on the Proca
equation (cf. Eq. (\ref{eq:PE}), consider the helicity-averaged
Proca equation:
\begin{equation}
\Box \langle A^\mu \rangle +\langle m^2\rangle \langle A^\mu
\rangle =\langle J^\mu \rangle  \qquad \mbox{(helicity-averaged
Proca equation),}\label{eq:PEhel}
\end{equation}
where $\langle \rangle$ represents an average over all possible
helicity states, as in Sec. \ref{sec:FreqSpec}. Note that $m^2$
and $A^\mu$ can vary independently of one another, so are
uncorrelated in the averaging procedure. It is $\langle m^2\rangle
$ that is the quantity to be identified as $m_b^2$, the
mass-squared of the boson. Recalling Eq. (\ref{eq:Amu4}), $\langle
A^\mu \rangle $ is given as
\begin{equation}
\langle A^\mu(b,t)\rangle  = \langle q^\mu \rangle \, \left[
\oofourpi\, \frac{1}{r}\, \mbox{\Large
e}^{-m_br} \right] \qquad \mbox{(helicity-averaged 4-potential),}\label{eq:Amu4hel1}%
\end{equation}
and $\langle J^\mu \rangle $ is found from Eq. (\ref{eq:Jmu2}) to
be
\begin{equation}
\langle J^\mu(\vecr ,t)\rangle =\delta [\vecr (t)]\, \langle q^\mu
\rangle \qquad \mbox{(helicity-averaged
4-current),}\label{eq:Jmu2hel}
\end{equation}
where
\begin{equation}
\langle q^\mu \rangle =q_V\langle u^\mu \rangle + q_A\langle s^\mu
\rangle \qquad \mbox{(helicity-averaged
4-charge).}\label{eq:qmu1hel}
\end{equation}
Since $u^\mu$ only depends on $\vecv _f$, which is independent of
$\lambda _f$, $\langle u^\mu \rangle =u^\mu$. The helicity-average
of $s^\mu$ is found from Eq. (\ref{eq:smu1}) to be
\begin{subequations}
\begin{align}
\langle s^\mu \rangle &=\gamma _f (\langle \lambda _f\rangle  v_f,\, \langle \hats \rangle )\label{eq:smu1hel1}\\%
&=(0,\, \veczero)\label{eq:smu1hel2}%
\end{align}
\end{subequations}
since $\langle \lambda _f\rangle =0$ and $\langle \hats \rangle
=\veczero$. Thus $\langle q^\mu \rangle =q_Vu^\mu$ and hence
\begin{equation}
\langle A^\mu(b,t)\rangle  = \oofourpi\, \frac{q_Vu^\mu}{r}\,
\mbox{\Large
e}^{-m_br}.\label{eq:Amu4hel2}%
\end{equation}
The helicity-averaged momentum-space LC is (recall Eq.
(\ref{eq:LCpspace2}))
\begin{subequations}
\begin{align}
0&=\langle k_b^\mu \rangle \langle A_\mu \rangle\label{eq:LCpspace2hel1}\\%
&=\langle k_b^\mu \rangle \left[ \oofourpi\, \frac{q_Vu_\mu}{r}\, \mbox{\Large e}^{-m_br} \right] \label{eq:LCpspace2hel2}%
\end{align}
\end{subequations}
and thus
\begin{subequations}
\begin{align}
0 &= \langle k_b^\mu \rangle u_\mu \label{eq:LCpspace5hel1}\\%
&= \gamma _f(\langle \omega _b \rangle -\langle \veck _b \rangle
\cdot
\vecv _f)\label{eq:LCpspace6hel2}\\%
&= \gamma _f(\langle \omega _b \rangle -\langle k_{bz} \rangle v_f).\label{eq:LCpspace7hel3}%
\end{align}
\end{subequations}
Therefore,
\begin{equation}
\langle k_{bz} \rangle = \frac{\langle \omega _b \rangle }{v_f}.\label{eq:LCpspace8hel1}%
\end{equation}
Since $\omega _b$ is serving as the independent parameter in the
analysis, it can be taken as independent of $\lambda _f$, so that
$\langle \omega _b \rangle =\omega _b$. Consequently $k_{bz}$ is a
function of both $\omega _b$ and $\lambda _f$ (cf. Eq.
(\ref{eq:LCpspace11})); $\langle k_{bz}\rangle$ is then only a
function of $\langle \omega _b \rangle$ (cf. Eq.
(\ref{eq:LCpspace8hel1})). For notational clarity, introduce the
following new variables:
\begin{subequations}
\begin{align}
E_b &\equiv \langle \omega _b \rangle = \omega _b \label{eq:Eb}\\%
p_{bz} &\equiv \langle k_{bz}\rangle = \frac{E_b}{v_f}. \label{eq:pbL}%
\end{align}
\end{subequations}
$E_b$ is the helicity-averaged boson energy and $p_{bz}$ is the
helicity-averaged longitudinal component of the boson's
3-momentum. With $k_{bz}=p_{bz}(1-\varepsilon)$ now (recall Eq.
(\ref{eq:LCpspace11})), the equation of motion (Eq.
(\ref{eq:mb2})) can be written
\begin{equation}
k_b^2 = E_b^2-\veck^2_{b\perp}-p_{bz}^2(1-\varepsilon)^2.\label{eq:mb2hel1}%
\end{equation}

This equation can be rearranged to yield a more suitable mass
scheme. Such a scheme should describe a boson with energy $E_b$
and longitudinal component of 3-momentum $p_{bz}$. That is to say,
the mass-squared $m_b^2$ of the boson should be identified with
the square of the 4-momentum $p_b^\mu =(E_b,\vecp _{b\perp },
p_{bz})$, whose components are $E_b$, $p_{bz}$, and some $\vecp
_{b\perp }$ to be determined. As it turns out, $\vecp _{b\perp }$
is found to be the rms deviation of specific values of $\vecp
_{b\perp }$ from the average value of $\vecp _{b\perp }$, which is
identically zero, as might be desired (see a later section).
$m_b^2$ should also be independent of $\lambda _f$ and Lorentz
invariant. Towards these ends, rearrange Eq. (\ref{eq:mb2hel1}) as
follows.
\begin{subequations}
\begin{align}
k_b^2 &= E_b^2-\veck^2_{b\perp}-\left( \frac{E_b}{v_f} \right) ^2(1-2\varepsilon +\varepsilon ^2)\label{eq:eqmothel1}\\%
&= E_b^2-\veck^2_{b\perp}-\left( \frac{E_b}{v_f} \right) ^2+\left( \frac{E_b}{v_f} \right) ^2[\varepsilon (2-\varepsilon )]\label{eq:eqmothel2}\\%
&= E_b^2-\veck^2_{b\perp}-p_{bz}^2+\left( \frac{E_b}{\gamma _fv_f} \right) ^2[\gamma _f^2\varepsilon (2-\varepsilon )]\label{eq:eqmothel3}\\%
&= E_b^2-\veck^2_{b\perp}-p_{bz}^2+(1+\delta)M_{\delta}^2\qquad \mbox{(cf. Eqs. (\ref{eq:defdelta1}) and (\ref{eq:defMdel})).}\label{eq:eqmothel4}%
\end{align}
\end{subequations}
Then, with
\begin{equation}
M_b^2 \equiv k_b^2-(1+\delta)M_{\delta}^2, \label{eq:eqmothel5}%
\end{equation}
the equation of motion becomes
\begin{equation}
M_b^2 = E_b^2-\veck^2_{b\perp}-p_{bz}^2.\label{eq:eqmothel6}%
\end{equation}
By a slight ($\varepsilon$ is typically $\ll 1$) rearrangement, a
more suitable equation of motion is now taking shape. $M_b^2$ can
be identified with the square of a 4-vector whose time component
is $E_b$ and whose spatial components are $\veck _{b\perp}$ and
$p_{bz}$. To identify it as a viable mass-squared for a boson in
the method, the equation must be further adjusted so that $M_b^2$
is Lorentz invariant and independent of $\lambda _f$.

$M_b^2$ is easily seen to \emph{not} be Lorentz invariant. Recall
Eq. (\ref{eq:veckbsq1}), and note that both $\veck _{b\perp}^2$
and $k_b^2$ are inherently Lorentz invariant. Thus $\delta
M_{\delta}^2$ is also Lorentz invariant. Unless $\delta$ is
invariant (which it generally is not), $M_{\delta}^2$ is not
invariant. So $\delta M_{\delta}^2$ \emph{is}, but $M_{\delta}^2$
by itself \emph{is not}, an invariant quantity; hence $(1+\delta
)M_{\delta}^2$ is not invariant. Therefore, $M_b^2 =
k_b^2-(1+\delta)M_{\delta}^2$ is also not invariant. Another
(fine-tuning) reparameterization must thus be made to render
$M_b^2$ invariant. Of interest to this project is the $v_f \to 1$
limit. Introduce a new parameter $\alpha _0$, defined to be the
$v_f \to 1$ limit of $\alpha$.
\begin{equation}
\alpha _0 \equiv \lim_{v_f \to 1} \alpha = \frac{q_A\lambda _f}{q_V+q_A\lambda _f},\qquad \mbox{where $\lambda _f=\pm 1$.} \label{eq:defalpha0}%
\end{equation}
It is a familiar result, from studies of the Dirac equation, that
$\lambda _f=\pm 1$ in the $v_f \to 1$ limit (see practically any
textbook on quantum field theory). Define another related
quantity, $\delta _0$, by way of Eq. (\ref{eq:defdelta1}):
\begin{equation}
\delta _0 \equiv \lim_{v_f \to 1} \delta = 2\alpha _0-1.\label{eq:defdelta0}%
\end{equation}
Both $\alpha _0$ and $\delta _0$ are invariant in this limit.
Since $\delta M_{\delta}^2$ is always invariant, it thus follows
that $M_{\delta}^2$ is also invariant in the $v_f \to 1$ limit.
Therefore, $M_b^2 = k_b^2-(1+\delta)M_{\delta}^2$ is invariant, as
well, \emph{in this limit}. Now, introduce the new parameter
${(M_b)}_0^2$, defined to be the $v_f \to 1$ limit of $M_b^2$.
\begin{subequations}
\begin{align}
{(M_b)}_0^2 &\equiv \lim_{v_f \to 1} M_b^2\label{eq:defMb0sq1}\\%
&= \lim_{v_f \to 1} \left[ k_b^2-(1+\delta)M_{\delta}^2 \right] \label{eq:defMb0sq2}\\%
&= \lim_{v_f \to 1} \left[ 2\alpha m_f^2x-(1+\delta)M_{\delta}^2 \right] \label{eq:defMb0sq3}\\%
&= 2\alpha _0m_f^2x-(1+\delta _0 )M_{\delta _0}^2\label{eq:defMb0sq4}\\%
&= 2\alpha _0m_f^2x(1-x)\qquad \mbox{(via Eqs. (\ref{eq:defdelta0}) and (\ref{eq:defMdel})).}\label{eq:defMb0sq5}%
\end{align}
\end{subequations}
Note that causality is violated for the $k_b^2$ formula, hence
here as well, when $x < x_C$ (recall Eq. (\ref{eq:caus2})), but
that subtlety is being put aside for the moment. A new set of
causality conditions will be derived once a viable mass scheme is
set forth. Also, $M_{\delta}$ has been replaced here by its $v_f
\to 1$ limit (cf. Eq. (\ref{eq:defMdel}), with $v_f=1$). In
summary, the $M_b^2$ of Eq. (\ref{eq:eqmothel5}) has been replaced
by a more suitable mass-squared, ${(M_b)}_0^2$, which is Lorentz
invariant in the $v_f \to 1$ limit. With ${(M_b)}_0^2$ defined in
the above way (in the high energy limit), it is still unclear
whether or not it is Lorentz invariant \emph{in general}. It can
be shown that it is not, but can be easily made so by another
slight modification. To see that ${(M_b)}_0^2$ is not invariant,
recall the Lorentz transformation equations for $E_b$ and
$p_{bz}$. Denoting quantities in a frame (the ``rest frame" of the
pulse) comoving with the parent fermion with a prime,
\begin{subequations}
\begin{align}
{E_b}' &= \gamma _f(E_b-p_{bz}v_f)\label{eq:LTEb1}\\%
{p_{bz}}' &= \gamma _f(p_{bz}-E_bv_f).\label{eq:LTpbz1}%
\end{align}
\end{subequations}
Since the WW pulses are characterized by $p_{bz}=E_b/v_f$, it is
deduced from these equations that
\begin{subequations}
\begin{align}
{E_b}' &= 0\label{eq:LTEb2}\\%
{p_{bz}}' &= \frac{E_b}{\gamma _fv_f}.\label{eq:LTpbz2}%
\end{align}
\end{subequations}
As ${p_{bz}}'$ is the rest-frame value of $p_{bz}$, it is a
Lorentz invariant; thus so is $M_{\delta}$:
\begin{equation}
M_{\delta} \equiv \frac{E_b}{\gamma _fv_f}=m_f\left( \frac{x}{v_f} \right) ={p_{bz}}'\quad \mbox{is Lorentz invariant.}\label{eq:MdeltaisLI}%
\end{equation}
Clearly, then, it is the quantity $x/v_f$, rather than merely $x$,
that is an invariant. Returning to Eq. (\ref{eq:defMb0sq5}), then,
a proper expression, that is  Lorentz invariant in general, is
\begin{equation}
\mu _b^2 \equiv 2\alpha _0m_f^2\, \left( \frac{x}{v_f}\right) \left[ 1-\left( \frac{x}{v_f}\right) \right].\label{eq:defmubsq}%
\end{equation}
This quantity should be used in place of $M_b^2$ as the
mass-squared in Eq. (\ref{eq:eqmothel6}). The ``value" of the
boson mass is necessarily expressed in this complicated way (i.e.,
not as a simple number) because each virtual boson has a different
energy; because there is no unique energy for a given value of
$\gamma _f$, there is no corresponding unique value of $\mu _b$.
It is precisely the boson number spectra that are being devised
here that give the distributions of the boson energies. All of
these functions (cf. Eqs. (\ref{eq:numspT}) and (\ref{eq:numspL}))
are seen to be sharply peaked at $E_b=0$, which means that the
boson energies are typically $\ll E_f$. Hence, $x$ is typically
$\ll 1$, and so $\mu _b^2$ is typically $\ll m_f^2$. At best,
these distributed masses can only be expressed as a function of
the boson energies, in much the same way as one might write
$m=E/\gamma$. The equation of motion now becomes
\begin{equation}
\mu _b^2 = E_b^2-\veckappa ^2_{b\perp}-p_{bz}^2,\label{eq:eqmothel7}%
\end{equation}
where
\begin{equation}
\veckappa ^2_{b\perp} \equiv \veck^2_{b\perp}+M_b^2-\mu _b^2\qquad \mbox{(definition of $\veckappa ^2_{b\perp}$)} \label{eq:defkappa}%
\end{equation}
is now identified as the square of the transverse component of
boson's 3-momentum. The explicit dependence of $\veckappa
^2_{b\perp}$ on $x$ is complicated, and, at this point,
inconsequential. A simple expression will be found later, after an
average of the equation of motion over all fermion helicity states
is taken.

The dependence of $\mu ^2_b$ on $\lambda _f$ is contained in the
parameter $\alpha _0$ (cf. Eq. (\ref{eq:defalpha0})). The explicit
values of $\mu ^2_b$ (\emph{before} averaging over all fermion
helicity states) for the three types of bosons are now solved for.
For clarity, let $(\mu _b)^2_{\lambda _f}$ denote the mass of the
boson $b$ that is emitted from a fermion $f$ in helicity state
$\lambda _f$. So $(\mu _b)^2_{\pm 1}$ represents the mass of a
boson that is emitted from a fermion in a $\lambda _f=\pm 1$
state, respectively. For photons, $q_A=0$ (recall Eq.
(\ref{eq:chEM})), so $\alpha _0=0$ in turn (via Eq.
(\ref{eq:defalpha0})). Therefore,
\begin{equation}
(\mu _{\gamma })^2_{\pm 1} = 0 \qquad \mbox{(photon emitted from a $\lambda _f=\pm 1$ state).}\label{eq:muphotonsq}%
\end{equation}
It is important to note that this mass scheme \emph{predicts} that
$\mu ^2_{\gamma }$ is always zero, and does not depend on $\lambda
_f$. In contrast to the photon case, the mass-squared of a $Z$
boson in this method \emph{does} depend on $\lambda _f$. First
consider cases where $\lambda _f=+1$. Recall from Sec.
\ref{sec:QatRest} that UR particles in such states are
right-handed, and have $T^3=0$ (cf. Table \ref{tab:charges}).
Therefore, $q_A=0$ and $\alpha _0=0$ as well (cf. Eqs.
(\ref{eq:chNEW}) and (\ref{eq:defalpha0})). Finally, then, by Eq.
(\ref{eq:defmubsq}), the mass-squared of the $Z$ boson emitted
from such a state is found to vanish:
\begin{equation}
(\mu _Z)^2_{+1} = 0 \qquad \mbox{(for $Z$ boson emitted from a $\lambda _f=+1$ state).}\label{eq:muZRsq}%
\end{equation}
Now consider cases where $\lambda _f=-1$. By plugging Eq.
(\ref{eq:chNEW}) into Eq. (\ref{eq:defalpha0}) with $\lambda _f$
set to $-1$, $\alpha _0$ is found to reduce to
\begin{subequations}
\begin{align}
\alpha _0 &= \frac{-(-\half g_ZT^3)}{[\half g_Z(T^3-2Q^\gamma \sin^2\theta_W)-(-\half g_ZT^3)]}\label{eq:alpha0ZL1}\\%
&= \frac{T^3}{[(T^3-2Q^\gamma \sin^2\theta_W)+T^3]}\label{eq:alpha0ZL2}\\%
&= \frac{T^3}{2(T^3-Q^\gamma \sin^2\theta_W)}.\label{eq:alpha0ZL3}%
\end{align}
\end{subequations}
This expression is somewhat misleading, because $T^3$ is different
for the two chiral states of a given fermion: $T^3=\pm 1/2$ for
$L$ particle states, and $T^3=0$ for $R$ particle states (cf.
Table \ref{tab:charges}); the values of $Q^\gamma$ are the same
for the two states, however. To clarify this ambiguity, let
$T^3_\mL $ denote the quantity $T^3$ corresponding to the $L$
\emph{particle} state, and $T^3_\mR $ denote the quantity $T^3$
corresponding to the $R$ \emph{particle} state. The above equation
can then be written in an even \emph{more} concrete way by
defining the new parameter $\alpha _Z$, as
\begin{equation}
\alpha _Z \equiv \frac{T^3_\mL }{2(T^3_\mL -Q^\gamma \sin^2\theta_W)}\qquad \mbox{(definition of $\alpha _Z$),}\label{eq:defalphaZ}%
\end{equation}
which is simply a different name for $\alpha _0$ when it is
evaluated at $\lambda _f=-1$. $\alpha _Z$ is always well-defined,
because $\sin^2\theta_W$ will never equal $T^3_\mL /Q^\gamma$,
since it is not possible to express $\sin^2\theta_W$ as a quotient
of such simple integers (cf. Table \ref{tab:charges}). Upon
replacing $\alpha _0$ by $\alpha _Z$ in Eq. (\ref{eq:defmubsq}),
the following result is found.
\begin{subequations}
\begin{align}
(\mu _Z)^2_{-1} &= 2\alpha _Zm^2_f \left( \frac{x}{v_f}\right) \left[ 1-\left( \frac{x}{v_f}\right) \right] \nonumber \\ &\hspace{1.8in} \mbox{($Z$ boson emitted from a $\lambda _f=-1$ state).}\label{eq:muZLsq}%
\end{align}
\end{subequations}
For $W^{\pm}$ bosons, $q_A$ is always $\mp q_V$, respectively
(recall Eq. (\ref{eq:chCEW})), so that
\begin{equation}
\alpha _0 = \frac{\mp \lambda _f}{1 \mp \lambda _f}, \label{eq:alpha0Wpm1}%
\end{equation}
respectively (via Eq. (\ref{eq:defalpha0})). Since in this $v_f
\to 1$ limit, $\lambda _f$ will be either $+1$ or $-1$, the
denominator of this equation will be either $0$ or $2$. In terms
of the way the equation is written, the $\lambda _f=\pm 1$
possibility is ruled out, as it corresponds to $\alpha _0=\infty$,
and hence $k_b^2=\infty$ (by way of Eq. (\ref{eq:kbsq1})) --- an
absurd result. Therefore, $\lambda _f$ must necessarily be equal
to $\mp 1$ for a $W^{\pm}$ boson, respectively. In short,
\begin{equation}
\alpha _0 = \half, \label{eq:alpha0Wpm2}%
\end{equation}
for \emph{either} a $W^+$ \emph{or} a $W^-$ boson. According to
this scheme, then, a $W^{\pm}$ boson will only be emitted when
$\lambda _f$ is precisely $\mp 1$, respectively. Consequently, the
two bosons have \emph{exactly} the same mass, as in the Standard
Model! $\mu ^2_W$ is found here to depend on the boson energy
(i.e., $x$) in the following way.
\begin{subequations}
\begin{align}
\mu ^2_W &= m^2_f \left( \frac{x}{v_f}\right) \left[ 1-\left( \frac{x}{v_f}\right) \right] \nonumber \\ &\hspace{1.8in} \mbox{($W^{\pm}$ boson; $\lambda _f$ must equal $\mp 1$, respectively).} \label{eq:muWpmsq}%
\end{align}
\end{subequations}
As in the photon case, an important point to note is that $\mu
^2_W$ does not depend on $\lambda _f$ at all! However, a $W^{\pm}$
boson can only be emitted from a fermion if the fermion is in a
$\lambda _f=\mp 1$ state, respectively. This scheme is consistent
with the known fact that the charged weak current (i.e., $W^-$
boson absorption or $W^+$ emission) couples only left-handed
\emph{particle} states (or right-handed \emph{antiparticle}
states). Equivalently (in the high energy limit), a $\lambda
_f=-1$ \emph{particle} state always only couples to another
$\lambda _f=-1$ \emph{particle} state by emitting a $W^+$ boson;
or, a $\lambda_f =+1$ \emph{antiparticle} state always only
couples to another $\lambda _f=+1$ \emph{antiparticle} state by
emitting a $W^-$ boson. This latter restatement (in terms of
antiparticle states) follows from the fact that, to a very good
approximation, weak interactions conserve the combined CP (parity
followed by charge conjugation) operation \cite{ref:Aitc}. P
changes $\lambda _f=\pm 1$ to $\lambda _f=\mp 1$, and C changes a
particle state into its corresponding antiparticle state, so that,
as far as the equations are concerned, an $L$ \emph{particle}
state is equivalent to an $R$ \emph{antiparticle} state.

Having elucidated the values of $\mu ^2_b$ corresponding to bosons
emitted from fermions in particular helicity states, a canonical
expression for $m^2_b$ is now set forth. As stated above, the
canonical $m_b^2$ to be used in the method is to be identified
with the average of $(\mu _b)^2_{\lambda _f}$ over \emph{all
possible} fermion helicity states:
\begin{equation}
m_b^2 \equiv \langle (\mu _b)^2_{\lambda _f}\rangle = \half [(\mu _b)^2_{+1}+(\mu _b)^2_{-1}] \qquad \mbox{(canonical definition of $m_b^2$).}\label{eq:candefmbsq}%
\end{equation}
The corresponding equation of motion is then the average of Eq.
(\ref{eq:eqmothel7}) over all possible $\lambda _f$:
\begin{subequations}
\begin{align}
m_b^2 &\equiv \langle \mu _b^2\rangle \label{eq:eqmothe18}\\%
&= \langle E_b^2 \rangle -\langle \veckappa ^2_{b\perp}\rangle -\langle p_{bz}^2\rangle \label{eq:eqmothel9}\\%
&= E_b^2 -\langle \veckappa ^2_{b\perp}\rangle -p_{bz}^2 \label{eq:eqmothe20}\\%
&= E_b^2-\vecp ^2_{b\perp}-p_{bz}^2,\label{eq:eqmothel21}%
\end{align}
\end{subequations}
where
\begin{equation}
\vecp ^2_{b\perp} \equiv \langle \veck^2_{b\perp}\rangle \qquad \mbox{(definition of $\vecp ^2_{b\perp}$).}\label{eq:defpbperpsq}%
\end{equation}
Note that $E_b$ and $p_{bz}$ are simply constants as far as an
average over all helicity states is concerned, because they
already represent helicity-averaged quantities. An explicit
expression for $\vecp ^2_{b\perp}$ in terms of $x$ will be derived
once explicit expressions for the different $m_b^2$s are
specified. According to the canonical prescription set forth in
Eq. (\ref{eq:candefmbsq}), the mass-squared of the photon is found
(via Eq. (\ref{eq:muphotonsq})) to be
\begin{subequations}
\begin{align}
m^2_{\gamma} &= \half [(\mu _{\gamma})^2_{+1}+(\mu _{\gamma})^2_{-1}]\label{eq:mphoton1}\\%
&= \half (0+0)\label{eq:mphoton2}\\%
&= 0.\label{eq:mphoton3}%
\end{align}
\end{subequations}
The mass-squared of the $Z$ boson is found from Eqs.
(\ref{eq:muZRsq}) and (\ref{eq:muZLsq}) to be
\begin{subequations}
\begin{align}
m^2_Z &= \half [(\mu _Z)^2_{+1}+(\mu _Z)^2_{-1}]\label{eq:mZ1}\\%
&= \half \left\{ 0+2\alpha _Zm^2_f\, \left( \frac{x}{v_f}\right) \left[ 1-\left( \frac{x}{v_f}\right) \right] \right\} \label{eq:mZ2}\\%
&= \alpha _Zm^2_f\, \left( \frac{x}{v_f}\right) \left[ 1-\left( \frac{x}{v_f}\right) \right] .\label{eq:mZ3}%
\end{align}
\end{subequations}
As found above, the fermion that emits a $W^{\pm}$ boson is
necessarily in one particular helicity state. If a $W^+$ boson is
emitted, the fermion must be in a $\lambda _f=-1$ \emph{particle}
state, and if a $W^-$ boson is emitted, the fermion must be in a
$\lambda _f=+1$ \emph{antiparticle} state. So the averaging
procedure is not carried out when specifying the masses of $W$
bosons. The canonical mass-squared $m^2_W$ of the $W^{\pm}$ boson
is simply the $\mu ^2_W$ of Eq. (\ref{eq:muWpmsq}):
\begin{equation}
m^2_W = m^2_f\, \left( \frac{x}{v_f}\right) \left[ 1-\left( \frac{x}{v_f}\right) \right]. \label{eq:mW}%
\end{equation}
These three formulas are all of the general form
\begin{equation}
m^2_b = \alpha _bm^2_f\, \left( \frac{x}{v_f}\right) \left[ 1-\left( \frac{x}{v_f}\right) \right] , \label{eq:mbsqcaus}%
\end{equation}
where $\alpha _b$ differentiates one boson from another.

This general expression is Lorentz invariant and independent of
$\lambda _f$, but causality has not yet been checked. The check
can be done in the same way it was done for the bosons with
mass-squared values $k^2_b$. The analysis involves the square of
the 3-momentum. In the present case, $\vecp ^2_b$ can be
identified as $\vecp ^2_b=\vecp ^2_{b\perp}+p_{bz}^2$, so that the
equation of motion (Eq. (\ref{eq:eqmothel21})) can be rewritten as
\begin{subequations}
\begin{align}
m_b^2 &= E_b^2-\vecp ^2_b\label{eq:eqmothel22}\\%
&= E_b^2\left( 1-\frac{\vecp ^2_b}{E_b^2}\right) \label{eq:eqmothel23}\\%
&= E_b^2\left( 1-v_b^2\right) ,\label{eq:eqmothel24}%
\end{align}
\end{subequations}
where
\begin{equation}
v_b=\frac{|\vecp _b|}{E_b}\label{eq:veb}%
\end{equation}
is the speed of the boson, defined in terms of the boson's energy
and 3-momentum as it was previously. Solving for $\vecv ^2_b$,
\begin{subequations}
\begin{align}
v_b^2 &= 1-\frac{m_b^2}{E_b^2}\label{eq:vebsq1}\\%
&= 1-\frac{\alpha _bm^2_f}{E_f^2x^2}\left( \frac{x}{v_f}\right) \left[ 1-\left( \frac{x}{v_f}\right) \right] \qquad \mbox{(via Eq. (\ref{eq:mbsqcaus}))}\label{eq:vebsq2}\\%
&= 1-\frac{\alpha _bm^2_f}{E_f^2v_f^2(x/v_f)^2}\left( \frac{x}{v_f}\right) \left[ 1-\left( \frac{x}{v_f}\right) \right] \label{eq:vebsq3}\\%
&= 1-\frac{\alpha _b}{\gamma _f^2v_f^2}\left[ \frac{1}{(x/v_f)}-1\right] .\label{eq:vebsq4}%
\end{align}
\end{subequations}
Enforcing $v_b^2 \ge 0$ yields
\begin{subequations}
\begin{align}
0 &\le 1-\frac{\alpha _b}{\gamma _f^2v_f^2}\left[ \frac{1}{(x/v_f)}-1\right] \label{eq:CC11}\\%
\frac{\alpha _b}{\gamma _f^2v_f^2}\left[ \frac{1}{(x/v_f)}-1\right] &\le 1\label{eq:CC12}\\%
\left[ \frac{1}{(x/v_f)}-1\right] &\le \frac{1}{(\alpha _b/\gamma _f^2v_f^2)}\label{eq:CC13}\\%
\frac{1}{(x/v_f)} &\le 1+\frac{1}{(\alpha _b/\gamma _f^2v_f^2)}\label{eq:CC14}\\%
\frac{1}{(x/v_f)} &\le \frac{1+(\alpha _b/\gamma _f^2v_f^2)}{(\alpha _b/\gamma _f^2v_f^2)}\label{eq:CC15}\\%
\frac{(\alpha _b/\gamma _f^2v_f^2)}{1+(\alpha _b/\gamma _f^2v_f^2)} &\le \frac{x}{v_f}.\label{eq:CC16}%
\end{align}
\end{subequations}
Therefore, one causality condition on these bosons is a lower
limit on the possible values of $x$:
\begin{equation}
x \ge \frac{\varepsilon _bv_f}{v_f^2+\varepsilon _b},\qquad \mbox{(causality restriction \#1 on $x$).}\label{eq:ecaus1}%
\end{equation}
where
\begin{equation}
\varepsilon _b \equiv \frac{\alpha _b}{\gamma _f^2}\qquad \mbox{(definition of $\varepsilon _b$).} \label{eq:defvarepsb}%
\end{equation}
Enforcing $v_b^2 \le 1$ yields
\begin{subequations}
\begin{align}
1 &\ge 1-\frac{\alpha _b}{\gamma _f^2v_f^2}\left[ \frac{1}{(x/v_f)}-1\right] \label{eq:CC21}\\%
\frac{\alpha _b}{\gamma _f^2v_f^2}\left[ \frac{1}{(x/v_f)}-1\right] &\ge 0.\label{eq:CC22}%
\end{align}
\end{subequations}
As found previously (cf. Eq. (\ref{eq:caus1a})), $\alpha$ is
always $\ge 0$; thus $\alpha _b\ge 0$ here as well. Therefore,
\begin{equation}
\left[ \frac{1}{(x/v_f)}-1\right] \ge 0,\label{eq:CC23}%
\end{equation}
and thus (after a few steps of algebra) a second restriction on
$x$ is found:
\begin{equation}
x \le v_f\qquad \mbox{(causality restriction \#2 on $x$).}\label{eq:ecaus2}%
\end{equation}
This condition is automatically satisfied if it is assumed that
energy is conserved (which it was in the above analysis) at the
vertex. Recall the sentence following the definition of $x$ (cf.
Eq. (\ref{eq:defx})) --- energy conservation enforces $x \le
1-1/\gamma _f$. It can easily be verified that this energy
conservation requirement is a more stringent one than causality
restriction \#2 on x. That is,
\begin{equation}
x \le 1-\frac{1}{\gamma _f} \le v_f.\label{eq:xmax}%
\end{equation}
So, in practice, the upper bound on $x$ is set by energy
conservation instead of causality.

To summarize, then, the mass $m_b$ of an equivalent boson in the
GWWM is
\begin{equation}
m_b = m_f\sqrt{ \alpha _b\left( \frac{x}{v_f}\right) \left[ 1-\left( \frac{x}{v_f}\right) \right] },\label{eq:mbspec1}%
\end{equation}
where $x$ is bounded within the range
\begin{equation}
x_{min} \le x \le x_{max},\label{eq:xbounds}%
\end{equation}
where
\begin{subequations}
\begin{align}
x_{min} &\equiv \frac{\varepsilon _bv_f}{v_f^2+\varepsilon _b} \qquad \mbox{(definition of $x_{min}$)}\label{eq:defxmin}\\%
x_{max} &\equiv 1-\frac{1}{\gamma _f} \qquad \mbox{(definition of $x_{max}$)},\label{eq:defxmax}%
\end{align}
\end{subequations}
by causality and energy conservation, and the parameter $\alpha
_b$ is defined according to
\begin{equation}
\alpha _b \equiv \left\{
\begin{array}{l}
\mbox{\Large $\frac{q_{\mbox{\scriptsize $A$}}^{\mbox{\scriptsize $2$}}}{q_{\mbox{\scriptsize $A$}}^{\mbox{\scriptsize $2$}}-q_{\mbox{\scriptsize $V$}}^{\mbox{\scriptsize $2$}}}$}=0 \quad \mbox{for the photon}\\
\mbox{\Large $\frac{q_{\mbox{\scriptsize $A$}}}{q_{\mbox{\scriptsize $A$}}-q_{\mbox{\scriptsize $V$}}}$}=\mbox{\Large $\frac{T^{\mbox{\scriptsize $3$}}_{\mbox{\scriptsize $L$}}}{2(T^{\mbox{\scriptsize $3$}}_{\mbox{\scriptsize $L$}}-Q^{\mbox{\scriptsize $\gamma$}} \sin^{\mbox{\scriptsize $2$}}\theta_{\mbox{\scriptsize $W$}})}$} \quad \mbox{for the $Z$ boson}\\
\mbox{\Large $\frac{2q_{\mbox{\scriptsize $A$}}}{q_{\mbox{\scriptsize $A$}}\mp q_{\mbox{\scriptsize $V$}}}$}=1 \quad \mbox{for the $W^{\pm}$ bosons}\\
\end{array}
\right. .\label{eq:alphabspec}
\end{equation}
An equivalent expression for $m_b$, that is simpler (and
oftentimes more useful) than the one stated above, can be found by
recalling Eq. (\ref{eq:MdeltaisLI}):
\begin{equation}
m_b = \sqrt{ \alpha _bM_{\delta}(m_f-M_{\delta})}.\label{eq:mbspec2}%
\end{equation}
Table \ref{tab:alphab} summarizes the values of $\alpha _b$
(specified to four significant figures) for the three types of
bosons emitted from various fermions. Note that $\sin^2\theta_W =
0.2312$ to four significant figures \cite{ref:RPP}. Obviously
$\alpha _{\gamma}$ is always $0$, $\alpha _W$ is always $1$, and
the value of $\alpha _Z$ depends on the fermion that emitted the
boson. The table shows that $\alpha _Z$ is always less than $1$
for quarks and leptons. It can be worked out that, except for
nuclei with values of $Z$ and $N$ such that $2N \lesssim Z
\lesssim 13N$, $\alpha _Z$ is also always less than $1$. A
detailed analysis reveals the following results. If $Z \ge
N/(1-4\sin^2\theta_W)$ (i.e., $Z \ge 13.28N$), then
$1/2(1-2\sin^2\theta_W) \le \alpha _Z \le 1$ (i.e., $0.9300 \le
\alpha _Z \le 1$). If $N/(1-2\sin^2\theta_W) < Z <
N/(1-4\sin^2\theta_W)$ (i.e., $1.860N < Z < 13.28N$), then $1 <
\alpha _Z < \infty$. If $N < Z < N/(1-2\sin^2\theta_W)$ (i.e., $N
< Z < 1.860N$), then $-\infty < \alpha _Z < 0$; this possibility
must be ruled out because it violates causality (recall Eq.
(\ref{eq:caus1a})). If $0 \le Z \le N$ (i.e., $0 \le Z \le N$),
then $0 \le \alpha _Z \le 0.500$, with equivalent $Z$ bosons
emitted from isoscalar nuclei (where $Z=N$) being massless (with
$\alpha _Z = 0$).

\renewcommand{\arraystretch}{1.4}
\begin{table}
\renewcommand{\tablename}{TABLE}
\caption{\label{tab:alphab} {\Large $\alpha _b$} FOR BOSONS
EMITTED FROM VARIOUS FERMIONS}
\medskip
\begin{tabular*}{6in}{@{\extracolsep{\fill}}p{1.85in}ccccc@{\hspace{1em}}@{\extracolsep{1em}}}
\hline\hline
Fermion & $Q^\gamma$ & $T^3_\mL $ & $\alpha _{\gamma}$ &  $\alpha _Z$ & $\alpha _W$\\
\hline
$\nu _e$, $\nu _\mu$, $\nu _\tau$ & \hspace{4pt} $0$ & \hspace{4pt} $\half$ & $0$ & $0.5000$ & $1$\\
$e^-$, $\mu^-$, $\tau^-$ & $-1$ & $-\half$ & $0$ & $0.9300$ & 1\\
$u$, $c$, $t$ & \hspace{4pt} $\frac{2}{3}$ & \hspace{4pt} $\half$ & $0$ & $0.7228$ & $1$\\
$d$, $s$, $b$ & $-\frac{1}{3}$ & $-\half$ & $0$ & $0.5911$ & $1$\\
proton, $p=uud$ & \hspace{4pt} $1$ & \hspace{4pt} $\half$ & $0$ & $0.9300$ & $1$\\
neutron, $n=udd$ & \hspace{4pt} $0$ & $-\half$ & $0$ & $0.5000$ & $1$\\
nucleus (with $Z$ protons and $N$ neutrons) & \raisebox{-1.5ex}[0pt]{$Z$} & \raisebox{-1.5ex}[0pt]{$\half (Z-N)$} & \raisebox{-1.5ex}[0pt]{$0$} & \raisebox{-1.5ex}[0pt]{\Large $\frac{1}{2[1-2Z\sin^{\mbox{\scriptsize $2$}}\theta_{\mbox{\scriptsize $W$}}/(Z-N)]}$} & \raisebox{-1.5ex}[0pt]{$1$}\\

\hline\hline
\end{tabular*}
\end{table}

For convenience, the masses of the three types of bosons in the
method are now specified explicitly. The photon is exactly
massless:
\begin{equation}
m_{\gamma} = 0\qquad \mbox{(mass of photon).}\label{eq:mphoton}%
\end{equation}
The $Z$ boson has the mass
\begin{equation}
m_Z = \sqrt{ \alpha _Z\left( \frac{E_Z}{\gamma _fv_f}\right) \left[ m_f-\left( \frac{E_Z}{\gamma _fv_f}\right) \right] }\qquad \mbox{(mass of $Z$ boson),}\label{eq:mZboson}%
\end{equation}
where $\alpha _Z = T^3_\mL /2(T^3_\mL -Q^{\gamma}
\sin^2\theta_W)$, which is typically $\sim 1$. And the $W^+$ and
$W^-$ bosons have the mass
\begin{equation}
m_W = \sqrt{ \left( \frac{E_W}{\gamma _fv_f}\right) \left[ m_f-\left( \frac{E_W}{\gamma _fv_f}\right) \right] }\qquad \mbox{(mass of $W$ boson).}\label{eq:mWboson}%
\end{equation}

There are two interesting points to be made about the boson mass
values. One is that, except for a $Z$ boson being emitted from a
nucleus with $Z \simeq 1.860N$ (in which case $m_Z \gg m_f$), the
mass $m_b$ of any given boson is on the order of or less than
$m_f$. This result can be seen from Eq. (\ref{eq:mbspec1}): the
maximum of $m_b$ with respect to $x$ occurs at $x=v_f/2$, so that
the greatest possible value of $m_b$ for any given $m_f$ and $v_f$
is $m_f\sqrt{\alpha _b}/2$. Then, since $\alpha _b$ is typically
$\lesssim 1$, it follows that $m_b \lesssim m_f$ as well (except
for certain nuclei). It may seem that this strange result
conflicts with the predictions of the SM, but that is not the
case. The SM makes no prediction whatsoever about the masses of
the \emph{virtual} particle(s) that mediate particle interactions.
The bosons appearing in this semiclassical generalized WWM are
entirely different entities than the bosons appearing in quantum
field theoretic methods. Most importantly, they have definite
energies and momenta, and well-defined trajectories. In contrast,
the mediating bosons in a Feynman diagram analysis are some sort
of an average over \emph{all} contributing virtual mediating
states. However reminiscent of SM bosons, the bosons in the GWWM
should really be thought of as ``equivalent bosons", constructs
that are tailor-made to fit the semiclassical WW formalism.

The other notable point is that, unless the fermion is a nucleus
with either $Z \simeq N$ (in which case $m_Z \ll m_W$) or $Z
\simeq 1.860N$ (in which case $m_Z \gg m_W$), the boson masses are
in roughly the same ratios as they are in the Standard Model.
There, the photon is massless, the two $W$ bosons have identically
equal masses, and are \emph{less massive} than the $Z$ boson by a
factor $\cos\theta_W\simeq 0.8768$ (equivalent to an $\alpha _Z$
here of about $1.3$). For comparison, the photon in the scheme
developed here is massless and the two $W$ bosons have exactly the
same mass, but the $W$ bosons appear to always be slightly
\emph{more massive} than the $Z$ boson (except in the case of
nuclei with $2N \lesssim Z \lesssim 13N$). A parameter frequently
encountered in the literature relating the $W$ and $Z$ boson
masses is $\rho$, defined as
\begin{equation}
\rho \equiv \frac{m^2_W}{m^2_Z\cos^2\theta_W}\qquad \mbox{(definition of $\rho$).}\label{eq:defrho}%
\end{equation}
The SM makes the definite prediction that (for on-shell bosons)
$\rho$ is exactly equal to $1$, and experimental data show that
$\rho =1$ to within a very small error \cite{ref:Halz,ref:RPP}. In
comparing the $\rho$ of the SM to the same parameter in the mass
scheme developed here, some care must be exercised. In the SM, the
mass-squared $m^2_W$ is the square of the 4-momentum transferred
in a charged current interaction, which only couples $L$
\emph{particle} states to other $L$ \emph{particle} states (or $R$
\emph{antiparticle} states to other $R$ \emph{antiparticle}
states). Unlike in the SM, the mass of a $Z$ boson in the scheme
being developed here depends on the fermion from which it was
emitted. It makes sense, then, when determining $\rho$ in this
scheme to restrict the analysis to $W$ and $Z$ bosons being
emitted from the same type of fermion --- $L$ \emph{particle}
states ($R$ \emph{antiparticle} states). By a seeming fluke of
parameters, the $\rho$ here is not even well-defined for $R$
\emph{particle} states. This is so for two reasons. One is that
$\lambda _f$ must equal $-1$ for the $m^2_W$ defined for a
\emph{particle} state ($\lambda _f$ must equal $+1$ for an
\emph{antiparticle} state); in other words $m^2_W$ is only defined
if $\lambda _f=-1$. The other is based on the result found above
(cf. Eq. (\ref{eq:muZRsq})), that $m_Z=0$ (which would appear in
the denominator of $\rho$) for all $\lambda _f=+1$ states. In
short, the helicity-averaged mass values specified in Eqs.
(\ref{eq:mbspec1}) and (\ref{eq:alphabspec}) should not be used to
determine $\rho$. Instead, the masses $(\mu _W)_{-1}$ and $(\mu
_Z)_{-1}$, as specified in Eqs. (\ref{eq:muWpmsq}) and
(\ref{eq:muZLsq}), respectively, should be used. In the above mass
scheme, then, $\rho$ works out to be
\begin{subequations}
\begin{align}
\rho &= \frac{(\mu _W)^2_{-1}}{(\mu _Z)^2_{-1}\cos^2\theta_W}\label{eq:GWWMrho1}\\%
&= \frac{m^2_f(x/v_f)[1-(x/v_f)]}{2\alpha _Zm^2_f(x/v_f)[1-(x/v_f)]\cos^2\theta_W}\label{eq:GWWMrho2}\\%
&= \frac{1}{2\alpha _Z\cos^2\theta_W}\label{eq:GWWMrho3}\\%
&= \frac{1}{2T^3_\mL \cos^2\theta_W/2(T^3_\mL -Q^\gamma \sin^2\theta_W)}\label{eq:GWWMrho4}\\%
&= \frac{(T^3_\mL -Q^\gamma \sin^2\theta_W)}{T^3_\mL \cos^2\theta_W}\label{eq:GWWMrho5}\\%
&= \sec^2\theta_W-\left( \frac{Q^\gamma }{T^3_\mL } \right) \tan^2\theta_W \label{eq:GWWMrho6}\\%
&= 1+\left[ \frac{(T^3_\mL -Q^\gamma )}{T^3_\mL } \right] \tan^2\theta_W \label{eq:GWWMrho7}\\%
&= 1-\left( \frac{Y}{2T^3_\mL } \right) \tan^2\theta_W.\label{eq:GWWMrho8}%
\end{align}
\end{subequations}
The weak hypercharge quantum number $Y$ (cf. Table
\ref{tab:charges}) has been used in the last step to simplify.
Since $Y$ never vanishes for any of the SM particles, $\rho$ is
never exactly equal to $1$ for any one particular particle. As an
interesting footnote, though, the \emph{average} of $\rho$ for two
members of \emph{any} weak isospin doublet can easily be seen to
be exactly $1$. This result follows from the fact that the two
members of a doublet have the same value of $Y$, but values of
$T^3_\mL $ that differ by a minus sign. So the average of
$Y/T^3_\mL $ for two members is always $[Y/(1/2)+Y/(-1/2)]/2=0$.
Table \ref{tab:rho} summarizes the values of $\rho$ (to four
significant figures) in this scheme. Note that $\tan^2\theta_W =
0.3007$ to four significant figures \cite{ref:RPP}. In any case,
it is interesting that while the $W$ and $Z$ boson masses could
have turned out to be vastly different from one another, they are
found to always be roughly equal, as in the SM.

\renewcommand{\arraystretch}{1.4}
\begin{table}
\renewcommand{\tablename}{TABLE}
\caption{\label{tab:rho} {\Large $\rho$} AND {\Large $\langle \rho
\rangle$} FOR VARIOUS FERMIONS}
\medskip
\begin{tabular*}{4.5in}{@{\extracolsep{\fill}}p{1.8in}cc@{\hspace{1.6em}}@{\extracolsep{1em}}}
\hline\hline
Fermion & $\rho$ & $\langle \rho \rangle$\\
\hline
$\nu _e$, $\nu _\mu$, $\nu _\tau$ & $1+\tan^2\theta_W=1.301$ & ---\\
$e^-$, $\mu^-$, $\tau^-$ & $1-\tan^2\theta_W=0.6993$ & ---\\
$(\nu _e,e^-)$, $(\nu _\mu,\mu^-)$, $(\nu _\tau,\tau^-)$ & --- & $1$\\
$u$, $c$, $t$ & $1-\frac{1}{3}\tan^2\theta_W=0.8998$ & ---\\
$d$, $s$, $b$ & $1+\frac{1}{3}\tan^2\theta_W=1.100$ & ---\\
$(u,d)$, $(c,s)$, $(t,b)$ & --- & $1$\\
proton, $p=uud$ & $1-\tan^2\theta_W=0.6993$ & ---\\
neutron, $n=udd$ & $1+\tan^2\theta_W=1.301$ & ---\\
$(p,n)$ & --- & $1$\\
nucleus (with $Z$ protons and $N$ neutrons) &
\raisebox{-1.5ex}[0pt]{$1-\left( \frac{Z+N}{Z-N}\right)
\tan^2\theta_W$} & \raisebox{-1.5ex}[0pt]{---}\\
\hline\hline
\end{tabular*}
\end{table}

To end this section, the number spectrum functions (cf. Eqs.
(\ref{eq:numspTnew}) -- (\ref{eq:chinew})) are listed again, this
time with the explicit form for $m_b$ found in this section.
\begin{subequations}
\begin{align}
N_\mT (E_b) &= \frac{N_0}{E_b}\left\{ \chi _b K_0(\chi )K_1(\chi
_b)-\half v_f^2 \chi _b^2 \left[ K^2_1(\chi _b)-K^2_0(\chi _b)
\right]
\right\}\label{eq:numspTmb}\\%
N_\mL (E_b) &= \frac{N_0}{E_b}\left\{ \half (m_bb_{min})^2 \left[
K^2_1(\chi _b) - K^2_0(\chi _b) \right]
\right\},\label{eq:numspLmb}%
\end{align}
\end{subequations}
where
\begin{equation}
N_0 \equiv \frac{1}{2\mpi ^2}\, \frac{q^2_{\mbox{\scriptsize
$V$}}+q^2_{\mbox{\scriptsize $A$}}}{v_f^2}=const\label{eq:N0mb}%
\end{equation}
and
\begin{subequations}
\begin{align}
\chi _b&= b_{min}\, \sqrt{\alpha _bM_{\delta}(m_f-M_{\delta})+M_{\delta}^2}\label{eq:chimb1}\\%
&= b_{min}\, \sqrt{M_{\delta}[\alpha _bm_f+(1-\alpha _b)M_{\delta}]}\label{eq:chimb2}\\%
&= b_{min}\, m_f\sqrt{\left( \frac{x}{v_f}\right) \left[ \alpha _b+(1-\alpha _b)\left( \frac{x}{v_f}\right) \right] }.\label{eq:chimb3}%
\end{align}
\end{subequations}
This new expression for $\chi _b$ was obtained by using the
formulas for $m_b$ given by Eqs. (\ref{eq:mbspec1}) and
(\ref{eq:mbspec2}). Explicitly,
\begin{equation}
\chi _{\gamma} = \frac{E_{\gamma}b_{min}}{\gamma _fv_f} \label{eq:chiphoton}%
\end{equation}
for the photon;
\begin{equation}
\chi _Z = b_{min}\, \sqrt{ \left( \frac{E_Z}{\gamma _fv_f}\right) \left[ \alpha _Zm_f+(1-\alpha _Z)\left( \frac{E_Z}{\gamma _fv_f}\right) \right] } \label{eq:chiZboson}%
\end{equation}
for the $Z$ boson, where again $\alpha _Z = T^3_\mL /2(T^3_\mL
-Q^{\gamma} \sin^2\theta_W)$, and is typically $\simeq 1$; and
\begin{equation}
\chi _W = b_{min}\, \sqrt{ \frac{E_Wm_f}{\gamma _fv_f} } \label{eq:chiWboson}%
\end{equation}
for the $W$ boson.

\subsection{Imaginary Transverse Momentum} \label{sec:mbTransMom} \indent

Causality imposed the restriction that $\alpha _b \ge 0$ (recall
Eq. (\ref{eq:caus1a})), so that (by Eq. (\ref{eq:mbsqcaus}))
$m_b^2\ge 0$ as well. By the equation of motion (Eq.
(\ref{eq:eqmothel21})), this latter restriction yields a curious
result --- that the square of the transverse component of the
boson's 3-momentum, $\vecp ^2_{b\perp}$, must be negative!
Referring back to Eq. (\ref{eq:eqmothel21}), the value of $\vecp
^2_{b\perp}$ works out as follows
\begin{subequations}
\begin{align}
E_b^2-\vecp ^2_{b\perp}-p_{bz}^2 &= m_b^2\label{eq:eqmothel25}\\%
E_b^2-\vecp ^2_{b\perp}-\left( \frac{E_b}{v_f} \right) ^2 &= m_b^2\qquad \mbox{(via Eq. (\ref{eq:pbL}))} \label{eq:eqmothel26}\\%
-\vecp ^2_{b\perp}-\left( \frac{E_b}{\gamma _fv_f} \right) ^2 &= m_b^2\label{eq:eqmothel27}\\%
-\vecp ^2_{b\perp}-M_{\delta}^2 &= m_b^2\qquad \mbox{(via Eqs. (\ref{eq:defMdel}) and (\ref{eq:Eb})),} \label{eq:eqmothel28}%
\end{align}
\end{subequations}
so that
\begin{equation}
\vecp ^2_{b\perp} = -(m_b^2+M_{\delta}^2),\label{eq:pbTsqmM}%
\end{equation}
or, as a function of $x$,
\begin{subequations}
\begin{align}
\vecp ^2_{b\perp} &= -\left\{ \alpha _bm^2_f\, \left( \frac{x}{v_f}\right) \left[ 1-\left( \frac{x}{v_f}\right) \right] + m_f^2\, \left( \frac{x}{v_f} \right) ^2 \right\}  \nonumber \\ &\hspace{3in} \mbox{(via Eqs. (\ref{eq:mbsqcaus}) and (\ref{eq:defMdel}))}\label{eq:pbTsqx1}\\%
&= -m^2_f\, \left( \frac{x}{v_f} \right) \left[ \alpha _b +\left( \frac{x}{v_f} \right) (1-\alpha _b) \right] .\label{eq:pbTsqx2}%
\end{align}
\end{subequations}
As both $m_b^2$ and $M_{\delta}^2$ are nonnegative quantities, it
is apparent that $\vecp ^2_{b\perp}$ is \emph{always} less than or
equal to zero. The idea of a negative value for $\vecp _{b\perp}$
seems nonsensical at first. It equivalently means that ${\vecp
_b}_{\perp}$ is a purely imaginary quantity, which is not only
hard to grasp, but seems to contradict one of the basic
assumptions of Section \ref{sec:ModesPackets} --- that the plane
waves that are approximating the radiation fields only have a
longitudinal component of 3-momentum. Two main results of the
previous section are that
\begin{equation}
{\vecp _{b||}} \equiv \hatz\,{p _{b||}}=\hatz\,\frac{E_b}{v_f}\label{eq:pbL1}%
\end{equation}
and (for the scenario depicted in Fig. \ref{fig:LTWWM})
\begin{subequations}
\begin{align}
{\vecp _{b\perp}} &\equiv \pm \, \mi \, \hatx \,
{p_{b\perp}},\quad \mbox{where}\label{eq:pbT1}\\%
p_{b\perp}&\equiv \sqrt{m_b^2+\left( \frac{E_b}{\gamma
_fv_f}\right) ^2}.\label{eq:pbT2}%
\end{align}
\end{subequations}
The first result shows that the magnitude of ${\vecp _{b||}}$ is
known to be $E_b/v_f$. Since $E_b$ is the independent variable in
the analysis, it must be a well-defined quantity: $E_b \gg \Delta
E_b$. By the above equation, it then also follows that ${\vecp
_b}_{||} \gg \Delta {\vecp _{b||}}$. Thus, like $E_b$, the
longitudinal component ${\vecp _{b||}}$ of the 3-momentum of the
boson is well-defined. The second result indicates that taking
${\vecp _{b\perp}}=\veczero$ is somehow an oversimplification of
what is really going on. A clearer interpretation of the quantity
${\vecp _{b\perp}}$ needs to be established. It will be shown in
this section that the ${\vecp _{b\perp}}$ stated above is actually
a representation of the \emph{uncertainty} in the average value of
${\vecp _{b\perp}}$, which is the quantity that vanishes. The
method presented here closely follows that outlined in
Frauenfelder and Henley's {\it Subatomic Physics} \cite{ref:Frau}
for finding the uncertainty in the energy of a resonance (unstable
particle); a short but informative discussion is given in
\cite{ref:Pesk1} for how this formula generalizes in relativistic
quantum mechanics.

The wave functions of the plane-wave wave packets sweeping past
$P$ are generally of the form (recall Eq. (\ref{eq:WEsolU3}) and
see Fig. \ref{fig:LTWWM})
\begin{equation}
U(\vecx,t) = \Theta (x)\, U_0\, \mbox{\Large e}^{-\mi (E_bt-\vecp
_b \cdot \vecx)} \qquad \mbox{($U_0=$\,const),}\label{eq:WEsolU4}%
\end{equation}
where $\vecx =(x,\,y,\,z)$, $\vecp _b=(p_{bx},\,p_{by},\,p_{bz})$
and $\Theta (x)$ is the step function, defined as
\begin{equation}
\Theta (x) \equiv \left\{
\begin{array}{ll}
0 & \mbox{if}\quad x<0\\
1 & \mbox{if}\quad x>0\\
\end{array}
\right. \qquad \mbox{(definition of step function).}\label{eq:defstepfunc}%
\end{equation}
Of course, the trajectory of the wave function that actually
strikes $P$ is described by $\vecx =(b,\,0,\,v_ft)$, plus or minus
some uncertainty $\Delta \vecx$. Using Eqs. (\ref{eq:pbL1}) and
(\ref{eq:pbT2}), Eq. (\ref{eq:WEsolU4}) simplifies to
\begin{subequations}
\begin{align}
U(\vecx,t) &= \Theta (x)\, U_0\, \mbox{\Large e}^{-\mi [E_bt-(\pm
\mi {p_{b\perp}}x)-({p _{b||}}z)]}\label{eq:WEsolU5}\\%
&= \Theta (x)\, U_0\, \mbox{\Large e}^{-\mi E_bt}\, \mbox{\Large
e}^{-(\pm {p_{b\perp}}x)}\, \mbox{\Large e}^{\mi
{p _{b||}}z}.\label{eq:WEsolU6}%
\end{align}
\end{subequations}
Demanding that $U(\vecx,t)<\infty$ at $x=\infty$ forces the choice
of the $+$ sign in the specification of ${\vecp _{b\perp}}$:
\begin{equation}
\vecp _{b\perp}=+\mi \, \hatx \, {p_{b\perp}}.\label{eq:pbT3}
\end{equation}
Thus,
\begin{equation}
U(\vecx,t) = \Theta (x)\, U_0\, \mbox{\Large e}^{-\mi E_bt}\,
\mbox{\Large e}^{-{p_{b\perp}}x}\, \mbox{\Large e}^{\mi
{p _{b||}}z}.\label{eq:WEsolU7}%
\end{equation}
$U(\vecx,t)$ can be Fourier-expanded into its component modes as
follows (recall Eq. (\ref{eq:WEsolU1})):
\begin{equation}
U(\vecx,t)= \frac{1}{(2\mpi )^{3/2}} \, \int ^\infty_{-\infty}
\dif ^3 \vecp _b\, A(\vecp _b)\, \mbox{\Large e}^{-\mi (E_bt-\vecp
_b \cdot \vecx)},\label{eq:U3d1}%
\end{equation}
where
\begin{equation}
A(\vecp _b)= \frac{1}{(2\mpi )^{3/2}} \, \int ^\infty_{-\infty}
\dif ^3 \vecx \, U(\vecx,0)\, \mbox{\Large e}^{-\mi \vecp _b \cdot
\vecx}.\label{eq:amp3d1}%
\end{equation}
Evaluating Eq. (\ref{eq:WEsolU7}) at $t=0$ yields
\begin{equation}
U(\vecx,0) = \Theta (x)\, U_0\, \mbox{\Large e}^{-{p_{b\perp}}x}\,
\mbox{\Large e}^{\mi
{p _{b||}}z},\label{eq:WEsolU8}%
\end{equation}
so that Eq. (\ref{eq:amp3d1}) then becomes
\begin{subequations}
\begin{align}
A(\vecp _b) &= \frac{1}{(2\mpi )^{3/2}} \, \int ^\infty_{-\infty}
\dif ^3 \vecx \, \left[ \Theta (x)\, U_0\, \mbox{\Large
e}^{-{p_{b\perp}}x}\, \mbox{\Large e}^{\mi {p _{b||}}z} \right] \,
\mbox{\Large e}^{-\mi \vecp _b \cdot
\vecx}\label{eq:amp3d2}\\%
&= \frac{1}{(2\mpi )^{3/2}} \, U_0\, \left[ \int ^\infty_{-\infty}
\dif x \, \Theta (x)\, \mbox{\Large e}^{-\mi (p_{bx}-\mi
{p_{b\perp}})x} \right] \, \times \, \left[ \int ^\infty_{-\infty}
\dif y \, \mbox{\Large e}^{-\mi p_{by}y} \right] \, \times
\nonumber \\ & \quad \times \left[ \int ^\infty_{-\infty} \dif z
\, \mbox{\Large e}^{-\mi
(p_{bz}-{p _{b||}})z} \right]\label{eq:amp3d3}\\%
&= U_0\,I_x\,I_y\,I_z.\label{eq:amp3d4}%
\end{align}
\end{subequations}
The quantities $I_x$, $I_y$ and $I_z$ are defined, and work out to
be, as follows.
\begin{subequations}
\begin{align}
I_x &\equiv \oortpi\, \int ^\infty_{-\infty} \dif x \, \Theta
(x)\, \mbox{\Large e}^{-\mi \varrho x},\quad \varrho \equiv
p_{bx}-\mi
{p_{b\perp}}\label{eq:Ix1}\\%
&= \oortpi\, \int ^\infty_0 \dif x \, \mbox{\Large e}^{-\mi
\varrho x}\label{eq:Ix2}\\%
&= \oortpi\, \left. \left[ \frac{1}{(-\mi \varrho)}\, \mbox{\Large
e}^{-\mi \varrho x} \right] \right|^\infty_0\label{eq:Ix3}\\%
&= \oortpi\, \left( \frac{\mi}{\varrho} \right)\, \left[
\lim_{x\to \infty} \left( \mbox{\Large e}^{-\mi p_{bx}x}\,
\mbox{\Large
e}^{-{p_{b\perp}}x}\right) -1 \right] \label{eq:Ix4}\\%
&= \oortpi\, \frac{-\mi}{(p_{bx}-\mi
{p_{b\perp}})}.\label{eq:Ix5}\\%
I_y &\equiv \oortpi\, \int ^\infty_{-\infty} \dif y \,
\mbox{\Large
e}^{-\mi p_{by}y}\label{eq:Iy1}\\%
&= \sqrt{2\mpi}\, \delta (p_{by}).\label{eq:Iy2}\\%
I_z &\equiv \oortpi\, \int ^\infty_{-\infty} \dif y \,
\mbox{\Large
e}^{-\mi (p_{bz}-{p _{b||}})z}\label{eq:Iz1}\\%
&= \sqrt{2\mpi}\, \delta (p_{bz}-{p _{b||}}).\label{eq:Iz2}%
\end{align}
\end{subequations}
Therefore,
\begin{subequations}
\begin{align}
A(\vecp _b) &= U_0\, \left[ \oortpi\, \frac{-\mi}{(p_{bx}-\mi
{p_{b\perp}})}\right]\, \left[ \sqrt{2\mpi}\, \delta (p_{by})
\right]\, \left[ \sqrt{2\mpi}\, \delta (p_{bz}-{p _{b||}})
\right] \label{eq:amp3d5}\\%
&= -\mi \, \frac{\sqrt{2\mpi}\, U_0}{(p_{bx}-\mi {p_{b\perp}})}\,
\delta (p_{by})\, \delta (p_{bz}-{p _{b||}}) \label{eq:amp3d6}%
\end{align}
\end{subequations}
Returning to Eq. (\ref{eq:U3d1}),
\begin{subequations}
\begin{align}
U(\vecx,t) &= -\mi\, \frac{U_0}{2\mpi} \, \int ^\infty_{-\infty}
\dif p_{bx}\, \int ^\infty_{-\infty} \dif p_{by}\, \int
^\infty_{-\infty} \dif p_{bz}\, \delta (p_{by})\, \delta
(p_{bz}-{p _{b||}})\, \frac{\mbox{\Large e}^{-\mi (E_bt-\vecp _b
\cdot \vecx)}}{(p_{bx}-\mi
{p_{b\perp}})}\label{eq:U3d2}\\%
&= \oortpi\, \int ^\infty_{-\infty} \dif p_{bx}\, A(p_{bx})\,
\mbox{\Large e}^{-\mi (E_bt-\vecp _b \cdot \vecx)},\label{eq:U3d3}%
\end{align}
\end{subequations}
where $p_{by}=0$, $p_{bz}={p _{b||}}$, and
\begin{equation}
A(p_{bx}) \equiv -\mi\, \frac{U_0}{\sqrt{2\mpi}}\,
\frac{1}{(p_{bx}-\mi {p_{b\perp}})}.\label{eq:amp3d7}%
\end{equation}
The problem has now been reduced to one dimension. The probability
density $P(p_{bx})$ of finding the boson with a certain value of
$p_{bx}$ of the $x$-component of $\vecp _b$ is proportional to
$|A(p_{bx})|^2 \equiv A^*(p_{bx})A(p_{bx})$:
\begin{subequations}
\begin{align}
P(p_{bx}) &= const\, A^*(p_{bx})A(p_{bx})\label{eq:probpx1}\\%
&= \frac{const}{2\mpi}\,
\frac{|U_0|^2}{(p_{bx}-i{p_{b\perp}})(p_{bx}+i{p_{b\perp}})}\label{eq:probpx2}\\%
&= \frac{const}{2\mpi}\,
\frac{|U_0|^2}{(p_{bx}^2+{p_b}^2_{\perp})}.\label{eq:probpx3}%
\end{align}
\end{subequations}
The condition $\int ^\infty_{-\infty} \dif p_{bx}\, P(p_{bx})=1$
yields
\begin{subequations}
\begin{align}
\frac{const}{2\mpi}\, |U_0|^2\, \left[ \int ^\infty_{-\infty} \dif
p_{bx}\, \frac{1}{(p_{bx}^2+{p_b}^2_{\perp})}\right] &= 1 \label{eq:probpx4}\\%
\frac{const}{2\mpi}\, |U_0|^2\, \left[
\frac{\mpi}{{p_{b\perp}}}\right] &= 1 \label{eq:probpx5}\\%
const &= \frac{2{p_{b\perp}}}{|U_0|^2}.\label{eq:probpx6}%
\end{align}
\end{subequations}
Thus
\begin{equation}
P(p_{bx}) =
\frac{{p_{b\perp}}/\mpi}{(p_{bx}^2+{p_b}^2_{\perp})}.\label{eq:probpx7}%
\end{equation}
With $P(p_{bx})$ plotted against $p_{bx}$, ${p_{b\perp}}$ (as
defined in Eq. (\ref{eq:pbT2})) is the full width at half maximum
(FWHM) of a Lorentzian, or Breit-Wigner, curve that is peaked at
$p_{bx}=0$ with maximum $[P(p_{bx})]_{max}=1/\mpi {p_{b\perp}}$.
See Fig. \ref{fig:BWcurve} for a picture and equation of a more
general Breit-Wigner curve. Thus $p_{bx}=0$ to within an
uncertainty $\Delta p_{bx}=2p_{b\perp}$; that is to say,
$-p_{b\perp} \lesssim p_x \lesssim +p_{b\perp}$. Of course, the
$x$ axis is arbitrary, so the result holds for any transverse
component of 3-momentum. Make the following slight change of
notation: $p_x \to p_{b\perp}$ (to generalize $p_x$ to \emph{any}
transverse component) and $p_{b\perp} \to \Delta p_{b\perp}/2$ (to
avoid using the same symbol $p_{b\perp}$ two for different
quantities). Then (recalling Eq. (\ref{eq:pbT2}))
\begin{subequations}
\begin{align}
\Delta p_{b\perp} &\equiv 2\, \sqrt{m_b^2+M_{\delta}^2}\qquad
\mbox{(definition of uncertainty in ${p_{b\perp}}$)}\label{eq:defDeltapbT1}\\%
&= 2\, \sqrt{m_b^2+\left(
\frac{E_b}{\gamma _fv_f} \right) ^2}\qquad \mbox{(via Eq. (\ref{eq:defMdel}))}\label{eq:defDeltapbT2}\\%
&= 2\, m_f \sqrt{ \left( \frac{x}{v_f} \right) \left[ \alpha _b +\left( \frac{x}{v_f} \right) (1-\alpha _b) \right] }\qquad \mbox{(via Eq. (\ref{eq:pbTsqx2}))}\label{eq:defDeltapbT3}%
\end{align}
\end{subequations}
defines the uncertainty in $p_{b\perp}$. Thus $-\vecepsperp
(\Delta p_{b\perp}/2) \lesssim \vecepsperp p_{b\perp} \lesssim
+\vecepsperp (\Delta p_{b\perp}/2)$, with the probability of
finding $p_{b\perp}$ with values outside this range being strongly
suppressed. So, unlike the well-defined quantities $E_b$ and
${\vecp _{b||}}$, where $E_b \gg \Delta E_b$ and ${\vecp _{b||}}
\gg \Delta {\vecp _{b||}}$, the exact value of $p_{b\perp}$ is
always quite uncertain.

Explicitly, the uncertainty in the photon's $\vecp _{\gamma
\perp}$ is
\begin{equation}
\Delta p_{\gamma \perp} = \frac{2E_{\gamma}}{\gamma _fv_f} ;\label{eq:DpphotonT}%
\end{equation}
the uncertainty in the $Z$ boson's $\vecp _{Z \perp}$ is
\begin{equation}
\Delta p_{Z \perp} = 2\, \sqrt{ \left( \frac{E_Z}{\gamma _fv_f}\right) \left[ \alpha _Zm_f+(1-\alpha _Z)\left( \frac{E_Z}{\gamma _fv_f}\right) \right] } ,\label{eq:DpZbosonT}%
\end{equation}
where $\alpha _Z = T^3_\mL /2(T^3_\mL -Q^{\gamma}
\sin^2\theta_W)$, and is typically $\simeq 1$, as before; and the
uncertainty in the $W$ boson's $\vecp _{W \perp}$ is
\begin{equation}
\Delta p_{W \perp} = 2\, \sqrt{ \frac{E_Wm_f}{\gamma _fv_f} } .\label{eq:DpWbosonT}%
\end{equation}

This discussion can also be put in the context of the uncertainty
principle, which will be useful later (for the specification of
the minimum impact parameter). Recalling Eq. (\ref{eq:WEsolU7}),
the wave function for any of the three pulses in the method is
seen to only be appreciable for values of $p_{b\perp}$ (in the new
notation) and $x$ satisfying $|\vecp _{b\perp}|x\lesssim 1$. The
trajectory of the cloud of virtual bosons surrounding the UR
charge $q$ (cf. Fig. \ref{fig:LTWWM}) can be described as a
cylindrical beam with some radius $\Delta x$, which can be defined
by the relation $|\vecp _{b\perp}|\Delta x \equiv 1$. Thus, the
potentials and fields are appreciable for values of $x\lesssim
\Delta x$, and are relatively insignificant for values of
$x\gtrsim \Delta x$. As $|\vecp _{b\perp}| \lesssim \Delta
p_{b\perp}/2$, the following general relation is obtained:
\begin{equation}
\Delta p_{b\perp} \Delta x \gtrsim 2\qquad \mbox{(WWM uncertainty
relation).}\label{eq:WWMuncrel}%
\end{equation}
In other words, if an interaction occurs that localizes the
mediating boson to within some uncertainty $\Delta x$, the
transverse component of its 3-momentum cannot be simultaneously
specified to within an accuracy of better than about $\Delta
p_{b\perp}$.

In summary, a purely imaginary value of ${\vecp _{b\perp}}$
corresponds to nonzero probabilities of finding the boson with
$\vecp _{b\perp} \ne \veczero$. As an example, in both the current
generalized method and the SWWM, all photons have $m_{\gamma}=0$,
and thus (by Eq. (\ref{eq:pbT2})) $p_{\gamma \perp}=E_b/\gamma
_fv_f$. For an UR fermion of energy $E_f \gg m_f$, where $m_f$ is
the mass of the fermion, $E_b$ can range from $0$ to $E_f - m_f
\simeq E_f$ (by conservation of energy), so $0 \le p_{\gamma
\perp} \lesssim E_f/\gamma _fv_f \simeq m_f$. At most, then, the
virtual photons can carry a transverse component of 3-momentum
$p_{\gamma \perp} \simeq m_f$. Since $m_f \ll E_f \simeq {p
_{b||}}$, this transverse component can be safely neglected, and
the photons can be viewed as travelling collinearly with the
fermion.

\subsection{Other Imaginary Quantities} \label{sec:otherimquan} \indent

Other informative quantities turn out to be purely imaginary in
this scheme as well. Consider, for instance, the scattering angle
$\vectheta _b$ of the boson (cf. Fig. \ref{fig:fbf}). By
definition,
\begin{subequations}
\begin{align}
|\vectheta _b| &= \arctan \left( \frac{|\vecp _{b\perp}|}{|\vecp
_{b||}|} \right)\label{eq:theta1}\\%
&= \arctan \left( \mi\, \frac{p_{b\perp}}{p_{b||}}
\right),\label{eq:theta2}%
\end{align}
\end{subequations}
where Eq. (\ref{eq:pbT3}) has been used. Then, by a trigonometric
identity,
\begin{equation}
|\vectheta _b| = \mi\, \mathrm{arctanh} \left(
\frac{p_{b\perp}}{p_{b||}}
\right).\label{eq:theta3}%
\end{equation}
Just as $\vecp _{b\perp}$ can be written $\vecp _{b\perp}=+\mi \,
\hatx \, p_{b\perp}$, where $p_{b\perp}$ is purely real,
$\vectheta _b$ can be expressed in terms of a purely real quantity
$\theta _b$:
\begin{equation}
\vectheta _b = +\mi \,
\hattheta \, \theta _b.\label{eq:theta4}%
\end{equation}
By the previous equation, it is easily seen that
\begin{equation}
\theta _b = \mathrm{arctanh} \left( \frac{p_{b\perp}}{p_{b||}}
\right).\label{eq:theta5}%
\end{equation}
The actual boson emission angle $|\vectheta _b|$ is evidently
purely imaginary. To understand this conclusion, recall the
analysis of the previous section. Note that Eq. (\ref{eq:probpx7})
conveys the idea that the probability $prob$ of finding the
boson's x-component of 3-momentum between $p_{bx}$ and
$p_{bx}+\dif p_{bx}$ is
\begin{equation}
prob = \dif
{p_{bx}}\,\frac{p_{b\perp}/\mpi}{p_{bx}^2+{p_b}^2_{\perp}}.\label{eq:theta6}%
\end{equation}
This equation can be divided through by $p_{b||}^2$ (which is
constant insofar as the differential operator $\dif$ is concerned)
to obtain the probability of finding the boson's angle $\theta _b$
of emission:
\begin{subequations}
\begin{align}
prob &= \dif \left( \frac{p_x}{p_{b||}} \right)\,
\frac{(p_{b\perp}/p_{b||}/\mpi)}{(p_{bx}/p_{b||})^2+(p_{b\perp}/p_{b||})^2}\label{eq:theta7}\\%
&= \dif (\tanh \theta )\, \frac{\tanh \theta _b/\mpi}{\tanh
^2\theta +\tanh ^2\theta _b}.\label{eq:theta8}%
\end{align}
\end{subequations}
The variable $\theta$ here is defined as $\theta \equiv
\mathrm{arctanh} (p_{bx}/p_{b||})$. In words, Eq.
(\ref{eq:theta6}) gives the probability that the state of the
boson is such that $-p_{b\perp}\le p_{bx} \le p_{b\perp}$. Eq.
(\ref{eq:theta8}) expresses the fact that there is an identical
probability that $\tanh \theta$ falls within $-\tanh \theta _b \le
\tanh \theta \le \tanh \theta _b$; the probability function has
simply been reparameterized. Eq. (\ref{eq:theta8}) is thus
interpreted to mean that on average, the state of the system will
be found with $\tanh \theta _b=0$ (i.e., $\theta _b=0$, or
emission of the boson in the forward direction), with a narrow
spread of values $\Delta \tanh \theta _b\simeq 2\,\tanh \theta
_b$. Or, since $\Delta \tanh \theta _b \simeq \Delta \theta _b\,
\mathrm{sech} ^2\theta _b$, the angle $\theta _b$ will be $0$ to
within an uncertainty
\begin{equation}
\Delta \theta _b \simeq 2\,\frac{\tanh \theta
_b}{\mathrm{sech}^2\theta _b} \simeq \sinh 2\theta _b\qquad
\mbox{(uncertainty in emission angle).}\label{eq:theta9}%
\end{equation}
Because $\theta _b$ is purely imaginary, it can be easily shown
that
\begin{subequations}
\begin{align}
p_{b||} &= |\vecp _b|\,\cosh \theta _b\label{eq:theta10}\\%
p_{b\perp} &= |\vecp _b|\,\sinh \theta _b.\label{eq:theta11}%
\end{align}
\end{subequations}
Eq. (\ref{eq:theta9}) then translates to
\begin{equation}
\Delta \theta _b \simeq 2\,\frac{p_{b||}\,p_{b\perp}}{p_b^2}\qquad
\mbox{(uncertainty in boson emission angle).}\label{eq:theta12}%
\end{equation}
After a bit of algebra, this expression can be recast into the
following form.
\begin{equation}
\Delta \theta _b \simeq 2\, \frac{\sqrt{1-(v_f^2+\varepsilon_b)(1-x_{min}/x)}} {(v_f^2+\varepsilon_b)(1-x_{min}/x)}.\label{eq:theta13}%
\end{equation}
In the limit where $x\to x_{min}$ (i.e., $x\to 1/\gamma _f^2$,
$E_b\to m_f/\gamma _f$, etc.), the uncertainty in $\Delta \theta
_b$ can become quite great. For values of $x\not \simeq x_{min}$,
this uncertainty is given by the approximate relation
\begin{equation}
\Delta \theta _b \simeq \frac{2}{\gamma _f},\label{eq:theta14}%
\end{equation}
where it was also assumed that $\varepsilon _b \ll v_f^2 \simeq
1$, which is almost always the case. Thus, the uncertainty in the
boson emission angle is typically $\ll 1$, which is another way of
seeing that the bosons travel more or less collinearly with the
fermion. With $\sinh 2\theta _b \simeq \Delta \theta _b \ll 1$
(recall Eq. (\ref{eq:theta9})), $\theta _b$ is thus found to be
roughly $\theta _b \simeq 1/\gamma _f$, since $\sinh 2\theta _b
\simeq 2\theta _b$ in this limit. Then, the probability density
function describing the distribution of boson emission angles (cf.
Eq. (\ref{eq:theta8})) is found to reduce to
\begin{subequations}
\begin{align}
prob &= \dif (\theta )\, \frac{\theta _b/\mpi}{\theta ^2+\theta _b^2}\label{eq:theta15}\\%
&= \dif (\theta )\, \frac{(1/\gamma _f)/\mpi}{\theta ^2+(1/\gamma _f)^2}.\label{eq:theta16}%
\end{align}
\end{subequations}
This form of this function, which is a Breit-Wigner curve sharply
peaked at $\theta =0$, is in good agreement with a result
presented in \cite{ref:Tera}. They showed that the cross section
for the reaction $e^{\pm}+e^-\to e^{\pm}+e^-+\gamma +\gamma \to
e^{\pm}+e^-+X$, where $X$ may be a lepton or neutral $C=+1$ hadron
state, is proportional to precisely this function. In other words,
the probability for the process to occur is relatively negligible
unless the photons are emitted at angles $\theta \lesssim 1/\gamma
_e$.

Two other important quantities are described in a similar way. One
is the transverse component ${p_{f\perp}}'$ of the 3-momentum of
the final-state fermion $f'$. It was shown in the last section
that, by conservation of 3-momentum, ${\vecp _{f\perp}}' = -\vecp
_{b\perp}$. In a way similar to how a purely imaginary value of
$\vecp _{b\perp}$ was interpreted, it is expected that
$-\vecepsperp (\Delta {p_{f\perp}}'/2) \lesssim \vecepsperp
{p_{f\perp}}' \lesssim +\vecepsperp (\Delta {p_{f\perp}}'/2)$,
where
\begin{equation}
\Delta {p_{f\perp}}' \equiv 2\, \sqrt{m_b^2+\left(
\frac{E_b}{\gamma _fv_f} \right) ^2}\qquad \mbox{(definition of
uncertainty in ${p_{f\perp}}'$)}\label{eq:defDeltapfT}
\end{equation}
is the uncertainty in ${p_{f\perp}}'$. The probability of finding
${p_{f\perp}}'$ with values outside this range is strongly
suppressed. The other important quantity is the scattering angle
${\theta _f}'$ of the outgoing fermion. Just as with $\theta _b$,
the values of ${\theta _f}'$ are bounded by $-(\Delta {\theta
_f}'/2) \lesssim {\theta _f}' \lesssim +(\Delta {\theta _f}'/2)$,
where
\begin{equation}
\Delta {\theta _f}' \simeq
2\,\frac{{p_{f||}}'\,{p_{f\perp}}'}{{{p_f}'}^2}\qquad
\mbox{(uncertainty in fermion scattering angle).}\label{eq:thetaf1}%
\end{equation}
After a page or so of algebra, one finds
\begin{equation}
\Delta {\theta _f}' \simeq \frac{2}{(v_f^2/x-1)}\, \frac {\sqrt{1-(v_f^2+\varepsilon_b)(1-x_{min}/x)}} {\{1-[1-(v_f^2+\varepsilon_b)(1-x_{min}/x)]/(v_f^2/x-1)^2\}}.\label{eq:thetaf2}%
\end{equation}
For values of $x\not \simeq x_{min}$ and $x\not \simeq
x_{max}\simeq v_f^2$, this formula simplifies to
\begin{equation}
\Delta {\theta _f}' \simeq \frac{2}{\gamma _f}.\label{eq:thetaf3}%
\end{equation}
The corresponding probability density function for the
distribution of fermion scattering angles is identical in form to
Eq. (\ref{eq:theta16}), which is another result in good agreement
with the study presented in \cite{ref:Tera}. They pointed out that
the dependence on the \emph{electron} scattering angle of the
cross section for the above-mentioned reaction is also
proportional to a Breit-Wigner function, with a FWHM of $x/\gamma
_e(1-x)$. For values of $x$ $\not \simeq x_{min}$ and $\not \simeq
x_{max}$, this is of more or less the same form as that which
follows from Eq. (\ref{eq:thetaf3}) --- a Breit-Wigner curve with
$\Delta {\theta _f}' \simeq 2/\gamma _f$.

This section is concluded with a respecification of the number
spectrum functions, this time in terms of $\Delta {p_{b\perp}}$,
as given in Eq. (\ref{eq:defDeltapbT1}).
\begin{subequations}
\begin{align}
N_\mT (E_b) &= \frac{N_0}{E_b}\left\{ \chi _bK_0(\chi _b)K_1(\chi
_b)-\half v_f^2 \chi _b^2 \left[ K^2_1(\chi _b)-K^2_0(\chi _b)
\right]
\right\}\label{eq:numspTpbT}\\%
N_\mL (E_b) &= \frac{N_0}{E_b}\left\{ \half (m_bb_{min})^2 \left[
K^2_1(\chi _b) - K^2_0(\chi _b) \right]
\right\},\label{eq:numspLpbT}%
\end{align}
\end{subequations}
where
\begin{equation}
N_0 \equiv \frac{1}{2\mpi ^2}\, \frac{q^2_{\mbox{\scriptsize
$V$}}+q^2_{\mbox{\scriptsize $A$}}}{v_f^2}=const\label{eq:N0pbT}%
\end{equation}
and
\begin{equation}
\chi _b = \half\, b_{min}\, \Delta p_{b\perp},\label{eq:chipbT1}%
\end{equation}
where $\Delta {p_{b\perp}}$ was defined in Eq.
(\ref{eq:defDeltapbT1}), as
\begin{subequations}
\begin{align}
\Delta p_{b\perp} &\equiv 2\, \sqrt{m_b^2+M_{\delta}^2}\label{eq:chipbT2}\\%
&= 2\, \sqrt{m_b^2+\left( \frac{E_b}{\gamma _fv_f} \right) ^2}\label{eq:chipbT3}\\%
&=2\, m_f \sqrt{ \left( \frac{x}{v_f} \right) \left[ \alpha _b +\left( \frac{x}{v_f} \right) (1-\alpha _b) \right] }.\label{eq:chipbT4}%
\end{align}
\end{subequations}
The mass parameter $m_b$ is
\begin{equation}
m_b = m_f\sqrt{ \alpha _b\left( \frac{x}{v_f}\right) \left[ 1-\left( \frac{x}{v_f}\right) \right] },\label{eq:mbpbT}%
\end{equation}
where $x$ is bounded within the range
\begin{equation}
x_{min} \le x \le x_{max},\label{eq:xboundspbT}%
\end{equation}
where
\begin{subequations}
\begin{align}
x_{min} &\equiv \frac{\varepsilon _bv_f}{v_f^2+\varepsilon _b}\label{eq:defxminpbT}\\%
x_{max} &\equiv 1-\frac{1}{\gamma _f}.\label{eq:defxmaxpbT}%
\end{align}
\end{subequations}

See also Eqs. (\ref{eq:chiphoton})--(\ref{eq:chiWboson}), Eqs.
(\ref{eq:DpphotonT})--(\ref{eq:DpWbosonT}), and Eqs.
(\ref{eq:mphoton})--(\ref{eq:mWboson}) for explicit forms for
$\chi _b$, $\Delta {p_{b\perp}}$, and $m_b$, respectively, for the
three types of bosons. The Bessel functions appearing in the above
formula for the number spectra are strongly peaked at values of
$\chi \ll 1$. Since $N_{\lambda _b}(E_b)$ represents the number of
bosons $b$ in helicity state $\lambda _b$ with energy $E_b$, there
are apparently a relatively insignificant number of such bosons
for values of $\chi \gtrsim 1$. Equivalently, there are a
relatively insignificant number of such bosons for values of
$b_{min}$ and $\Delta {p_{b\perp}}$ satisfying $b_{min}\Delta
{p_{b\perp}} \gtrsim 2$, by Eq. (\ref{eq:chipbT1}). If one recalls
the ``WWM uncertainty relation" set forth in Eq.
(\ref{eq:WWMuncrel}), and identifies $b_{min}$ as the minimum
allowable uncertainty $\Delta x$ in the locations of the bosons,
then the equations can be interpreted in the following way (as
discussed in Sec. \ref{sec:RealvsVirt}). The greatest
contributions to the number spectrum functions come from bosons
that are ``virtual", in the sense that the uncertainties in their
positions and momenta in the transverse plane violate the
uncertainty principle. While nothing prevents the bosons from
being ``real", the number densities of such states are strongly
suppressed relative to those of virtual states. The vast majority
of bosons in the swarm of bosons that surround the fermion thus
pop in and out of existence, and are incapable of being detected.

\section{Minimum Impact Parameter} \label{sec:Bmin} \indent

One last step in developing a generalization of the WWM is the
specification of the minimum impact parameter, $b_{min}$. The
value of $b_{min}$ represents the closest distance of approach
between two colliding particles before they actually come into
contact with each another. In other words, if the impact parameter
$b$ of a given collision is less than $b_{min}$, the collision is
considered to be a head-on collision, while for values of
$b>b_{min}$, the collision is termed ``peripheral." The parameter
is of course very important to the WWM, because of interest here
is the interaction between an ``incident" particle (particle $P$
in Fig. \ref{fig:LTWWM}) and the virtual bosons of a ``target"
particle (particle $q$ in Fig. \ref{fig:LTWWM}); the two particles
are, by construction, never in contact with each another. The
value chosen for $b_{min}$ depends a great deal on the process
under consideration. An excellent reference for these matters is
Jackson's {\it Classical Electrodynamics}
\cite{ref:Jack1,ref:Jack2}.

One way of categorizing particle collisions is by the types of
interacting particles --- they can be either pointlike or
composite. The difference between the two is best quantified by
the probability density function, $\rho (\vecr ,t)$, which
represents the differential probability of finding the particle
within a given differential volume element. In the rest frame of a
particle $f$, this function is independent of time, and related to
the charge density function $J_f^0(\vecr )$ by the equation
$J_f^0(\vecr )=\rho (\vecr )\, q_f^0=\gamma _f\, \rho (\vecr )\,
q_V$ (cf. Sec. \ref{sec:PFURParticle}). For an ideal point charge,
$\rho (\vecr )$ is equal to $\delta (\vecr )$ --- the usual Dirac
delta function, which is normalized to unity and vanishes
everywhere except at $\vecr =\veczero$. Throughout this thesis,
all interacting particles are assumed for simplicity to be
pointlike; the above assignment for the probability density
function was made in Eq. (\ref{eq:Jmu2}) in Sec.
\ref{sec:PFURParticle}. When trying to describe a composite
particle with such a function, an assumption must be made about
the distribution of charge within it. It is commonplace in such
studies to introduce the \emph{form factor}, $F(\vecq ^2)$, where
$\vecq ^2=({\vecp _f}'-\vecp _f)^2$ is the square of the relevant
3-momentum transfer, which is the Fourier transform of $\rho
(\vecr )$. A good reference on form factors is Frauenfelder's {\it
Subatomic Physics} \cite{ref:Frau}. The value of $F(\vecq ^2)$ at
zero momentum transfer, $F(0)$, is usually normalized to unity for
a charged particle and zero for a neutral one. Whatever the exact
form of $F(\vecq ^2)$ is, it is apparent from the Fourier
transform equation linking $\rho(\vecr )$ to $F(\vecq ^2)$ that
$\rho (\vecr )\to \delta (\vecr )$ in the limit where $\vecq \to
\veczero$. Thus, a point particle can equivalently be described by
either of the assignments $\rho (\vecr )=\delta (\vecr )$ or
$F(\vecq ^2)=1$. Values of $\rho(\vecr )$ and $F(\vecq ^2)$
differing from those values describe a composite particle. For
example, if the probability density is assumed to be of a Gaussian
form, say $\rho (\vecr )=\rho _0\mbox{\Large{$\me$}}^{-(r/b)^2}$,
where $r_0$ is some constant, the form factor is consequently of
the form $F(\vecq ^2)=\mbox{\Large{$\me$}}^{-\vecq ^2r_0^2/4}$,
and the associated mean-square radius of the particle is $\langle
r^2 \rangle =3r_0^2/2$ \cite{ref:Frau}. In either the limit where
$\vecq ^2\to \veczero$ or $r_0\to 0$ (or, equivalently,
$\sqrt{\langle r^2 \rangle }\to 0$), it is clear that $F(\vecq
^2)\to 1$. Thus, in either of those limits, the composite particle
can be treated as a point charge. So, besides the obvious examples
of the quarks and leptons appearing in the SM, pointlike particles
could also be composite particles, such as the nuclei of atoms, as
long as the mediating bosons are not energetic enough to resolve
any internal structure.

Another way of categorizing particle collisions is by the energy
transfer mechanism. In a given collision, a particle can lose
energy in two different ways --- by collisional energy loss or by
radiation. In the former case, its kinetic energy can be
transferred to the other particle, or go into producing one or
more other particles in the local region of space surrounding the
two particles (cf. Figs. \ref{fig:Rprod} and \ref{fig:Hprod}). If
the particle transfers any amount of transverse momentum in the
collision, regardless of the exact process, it will necessarily be
deflected. The amount of deflection depends on its mass in the
usual way (by Newton's 2nd law) --- for the same Coulomb force, a
relatively light particle will be deflected more significantly
(i.e., accelerate more) than a heavier particle. When a charged
particle is accelerated in this manner, it is known to emit
radiation, which is the second mode of energy transfer mentioned
above. This particular kind of radiation is called
\emph{bremsstrahlung}, which means ``braking radiation" in German,
because it was first observed in experiments where high energy
electrons were stopped in a thick metallic target
\cite{ref:Jack1,ref:Jack2,ref:Kran}. For \emph{nonrelativistic} EM
processes, energy loss by bremsstrahlung is negligible compared to
collisional energy loss, but can be the dominant mode of energy
loss in \emph{relativistic} EM collisions
\cite{ref:Jack1,ref:Jack2}. Similar results are found in the case
of weak force processes. Because it is beyond the scope of this
study to present \emph{those} results, weak force bremsstrahlung
will be the focus of a future paper. The prototypical example of a
bremsstrahlung process is the scattering of a fast light particle,
such as an electron, by an atom; electrons are the best radiators
of bremsstrahlung because they are the lightest of all charged
particles and are thus best scattered in the force field of an
atomic nucleus. In a collision with an atom, the particle can
interact with either the orbiting electrons or the nucleus. If the
incident particle is an electron, both the 4-momentum loss and
deflection arise predominantly from interactions with the atom's
electrons. If the particle is heavier than an electron, the the
4-momentum loss and deflection are due to different interactions.
Because the atomic electrons are substantially lighter than the
nucleus, they tend to absorb the bulk of the 4-momentum, but have
little influence on deflecting the incident particle. In contrast,
the nucleus does not absorb any significant amount of 4-momentum,
but, because it has a greater charge, it is more effective than
any of the electrons at scattering the incident particle. So, for
the most part, the electrons absorb the particle's energy and
momentum, while the nucleus is the source of the particle's
deflection. In summary, the energy transfer in a collision between
two particles can be either due to collisional or radiative energy
loss. As discussed below, each of these possibilities has
associated with it a different value of $b_{min}$.

When analyzing a \emph{generic} collision (i.e., with or without a
significant amount of bremsstrahlung emitted) between two
\emph{pointlike} particles, one might wonder if the choice
$b_{min}=0$ can be made. If the particles are truly pointlike, in
the sense of having zero dimension, it would seem that \emph{any}
collision between them should still be considered ``peripheral".
This choice for $b_{min}$ has two inherent drawbacks. The
practical one is that the number spectrum functions are found to
approach infinity in the limit where $b_{min}$ approaches zero! A
theoretical drawback is that the Heisenberg uncertainty principle
imposes lower bounds on quantities such as particle size, so the
notion of a particle with exactly zero dimension is quite
unrealistic. Quantum mechanics gives approximate meaning to the
concept of particle size through the use of a wave packet. If the
particle has a magnitude of 3-momentum $p_f$, then the uncertainty
$\Delta p_f$ in the $p_f$ of the wave packet representing the
moving particle must be $\lesssim p_f$ in order for $p_f$ to still
be well-defined. The uncertainty principle then states that the
particle's position cannot be specified to an accuracy better than
$\Delta x\simeq 1/\Delta p_f \gtrsim 1/p_f$. It might be said that
the particle's wave function is smeared out (or the particle's
position is completely uncertain) within this distance scale. The
classical idea of a smooth particle trajectory thus loses its
meaning at distance scales smaller than $\Delta x$. In short,
then, for values of $b < \Delta x \lesssim 1/p_f$, the simple
classical description (as adopted in this project) of the
particle's path should not be expected to be valid any longer. In
a frame in which $p_f=\gamma _fm_fv_f$, a form for $b_{min}$ that
correctly accounts for these quantum mechanical effects is
\begin{equation}
b_{min}=\frac{\eta _f}{(\Delta p_f)_{max}}=\frac{\eta _f}{\gamma _fm_fv_f} \qquad \mbox{(quantum mechanical formula \#1),}\label{eq:bminQM1}%
\end{equation}
where $\eta _f$ is a constant of order unity. This expression is
not Lorentz invariant, so some caution (as to which frame of
reference to use) must be exercised when applying the formula.
Also, for collisions of two particles of unequal masses, the
correct $b_{min}$ to use is the one corresponding to the
\emph{lighter} of the two particles, as the limiting uncertainty
is determined by the \emph{smaller} of the two masses. As a final
point to make about this formula, it might be noted that neither
it nor its derivation makes any reference to the boson
characterizing the interaction. The concept of a boson can be
incorporated into this scheme as follows. Consider the same idea
presented above, but as applied to a boson propagating away from
the fermion, as envisioned in the previous subsection when
developing the mass scheme for the GWWM. That is to say, view the
boson as a particle travelling along a classical trajectory from
the fermion to some interaction point (see Fig. \ref{fig:fbf}).
Just as in the fermion case above, the uncertainty principle sets
a limit on the distance scale within which the trajectory can no
longer be considered to be classical, according to $\Delta
x\gtrsim 1/(\Delta p_b)_{max}$, where $p_b$ is the magnitude of
the boson's \emph{total} 3-momentum. Now, since
$p_b=\sqrt{E_b^2-m_b^2}\lesssim E_b$, it might be expected that
$\Delta p_b\lesssim \Delta E_b$ as well. Or, $(\Delta
p_b)_{max}\simeq \Delta E_b$, and thus $\Delta x\gtrsim 1/\Delta
E_b$. By Eq. (\ref{eq:MdeltaisLI}), $E_b$ is related to
${p_{bz}}'$ (which is $p_{bz}$ as seen in a frame comoving with
$f$) as $E_b=\gamma _f({p_{bz}}')v_f$, so (taking rms
uncertainties of both sides of the equation) $\Delta E_b\simeq
\gamma _f({\Delta p_{bz}}')v_f$, and thus $\Delta x\gtrsim
1/\gamma _f({\Delta p_{bz}}')v_f$. The task now is to solve for
${\Delta p_{bz}}'$. Well, one would expect that (in a frame
comoving with $f$) ${\Delta p_{bz}}'={\Delta p_{b\perp}}'$
identically, since in that frame there is no relative motion of
the boson in \emph{any particular} direction. Then, since $\Delta
p_{b\perp}$ is an invariant, ${\Delta p_{b\perp}}'$ has the same
value, for a given $E_b$, as the value of $\Delta p_{b\perp}$ in
the frame moving relative to $f$ with speed $v_f$ (cf. Eq.
(\ref{eq:defDeltapbT2})). Consequently, the identification
${\Delta p_{bz}}'=\Delta p_{b\perp}$ can also be made. With this
identification, the inequality $\Delta x\gtrsim 1/\gamma _f(\Delta
p_{b\perp})v_f$ is finally arrived at. Identifying $b_{min}$ as
the minimum value of $\Delta x$ specified here, the following
variation of Eq. (\ref{eq:bminQM1}) is obtained:
\begin{equation}
b_{min}=\frac{\eta _b}{(\Delta p_b)_{max}}=\frac{\eta _b}{\gamma _f(\Delta p_{b\perp})v_f} \qquad \mbox{(quantum mechanical formula \#2),}\label{eq:bminQM21}%
\end{equation}
where, like $\eta _f$, $\eta _b$ must be $\sim 1$. Explicit forms
for this choice of $b_{min}$ for the three types of interactions
are as follows; refer to Eqs.
(\ref{eq:DpphotonT})--(\ref{eq:DpWbosonT}) for the relevant values
of $\Delta p_{b\perp}$ to use. For EM interactions,
\begin{equation}
b_{min}=\frac{\eta _{\gamma}}{2E_{\gamma }}\qquad \mbox{(EM interactions between point particles),}\label{eq:bminQM2photon}%
\end{equation}
where $\eta _{\gamma} \sim 1$. For neutral weak interactions,
\begin{subequations}
\begin{align}
b_{min}&=\frac{\eta _Z}{2\sqrt{E_Z[\alpha _Zp_f+(1-\alpha
_Z)E_Z]}} \nonumber \\&\hspace{1in} \mbox{(neutral weak
interactions between point particles),}\label{eq:bminQM2Zboson}%
\end{align}
\end{subequations}
where $\eta _Z \sim 1$, $p_f=\gamma _fm_fv_f$, and $\alpha _Z =
T^3_\mL /2(T^3_\mL -Q^{\gamma} \sin^2\theta_W)$, which is
typically $\simeq 1$ as before. And for charged weak interactions,
\begin{equation}
b_{min}=\frac{\eta _W}{2\sqrt{E_Wp_f}}\qquad \mbox{(charged weak interactions between point particles),}\label{eq:bminQM2Wboson}%
\end{equation}
where $\eta _W \sim 1$ and $p_f=\gamma _fm_fv_f$ as above. To
recover quantum mechanical formula $\#1$ (Eq. (\ref{eq:bminQM1}))
from Eq. (\ref{eq:bminQM21}), note that quantum mechanical formula
$\#1$ makes no reference to any parameters characterizing the
boson mediating the interaction. That ambiguity can be interpreted
as arising from the ambiguity in the exact nature of the boson,
i.e., as being a distinct particle travelling along a well-defined
classical trajectory. It might be argued that quantum mechanical
formula $\#1$ is suitable to cases where the boson's energy and
other parameters are completely uncertain. Equivalently (for such
a fuzzy scenario), $\Delta E_b\simeq E_b$ can be set equal to
$E_f$ in order of magnitude, so that it is never possible to
specify an exact value of $E_b$ with any certainty. Then, $x\equiv
E_b/E_f\simeq \Delta E_b/E_f\simeq 1$. By Eq.
(\ref{eq:defDeltapbT3}), $\Delta p_{b\perp}\simeq 2m_f$, and thus
$b_{min}=\eta _b/2\gamma _fm_fv_f$
--- which is equal to Eq. (\ref{eq:bminQM1}) in order of magnitude.
This revised analysis may seem a bit contrived and irrelevant, but
this new form for $b_{min}$ has two advantages over quantum
mechanical formula \#1. One is that it yields a formula for
$b_{min}$ that gets perfect agreement between the number spectrum
functions (for transversely-polarized photons) of the SWWM and the
QWWM in the low boson energy ($E_b\to 0$) limit, which is the only
regime where those functions have any appreciable magnitude
\cite{ref:Dali,ref:Jack2}. Note that for EM interactions, (where
$\alpha _{\gamma}=0$) Eq. (\ref{eq:bminQM21}) reduces to
$b_{min}=\eta _{\gamma }/\gamma _f(E_b/\gamma _fv_f)v_f=\eta
_{\gamma }/E_{\gamma }$. An interesting discussion of this
agreement is presented by Dalitz et al., in \cite{ref:Dali}. They
consider an electron as the source of the virtual photon field,
and also cite another more detailed study that arrived at the same
conclusion --- that $b_{min}=1/E_{\gamma}$ should be used in the
number spectrum function instead of $b_{min}=1/E_f$ (note that
$E_f\simeq p_f$ in the $v_f\to 1$ limit) when comparing the SWWM
to the QWWM. The invalidity of Eq. (\ref{eq:bminQM1}) is explained
as being ``... due to the fact that the Weizs\"{a}cker-Williams
calculation comprises the contributions of the matrix elements [of
the fermion current operator] transverse to the incident
direction. Since the momentum transfers actually make some angle
with this direction in general, some contributions corresponding
to matrix elements longitudinal to the momentum transfer are
consequently included in the semiclassical calculation, but are
omitted in the [quantum] calculation ..." \cite{ref:Dali} The
other advantage to using Eq. (\ref{eq:bminQM21}) instead of Eq.
(\ref{eq:bminQM1}) is that it is found to also get perfect
agreement between the number spectrum functions (for both
transversely- \emph{and} longitudinally-polarized $W$ and $Z$
bosons) of the GWWM and the EWM in the low boson energy limit,
which will be shown below. For weak interactions, Eq.
(\ref{eq:bminQM21}) simplifies to $b_{min}=\eta _b
/\sqrt{E_fE_b\alpha_b}$. The values of $b_{min}$ for these two
types of interactions can differ appreciably (esp in the $E_b\to
0$ limit) from that given in Eq. (\ref{eq:bminQM1}), but are the
only forms for that parameter that are able to simultaneously get
agreement with the \emph{all} quantum formulations of the WWM
number spectrum functions. To sum up, it will be Eq.
(\ref{eq:bminQM21}) that will be the form for $b_{min}$ adopted in
this project, at least for applications to collisions between
\emph{point particles}, when comparing results to those of other
theories.

When analyzing a \emph{generic} collision between a
\emph{composite} particle and any another particle, whether
pointlike or not, one obvious form for $b_{min}$ to consider is
the sum of the particle radii; another form for $b_{min}$ that
should not be overlooked for these types of interactions will be
discussed below. Examples of composite particles are atoms or the
nuclei of atoms. Atomic and nuclear radii are well documented. For
an atom with $Z$ protons, a first-guess estimate of the radius $a$
can be found using the Bohr model: $a\simeq a_0/Z$, where
$a_0=1/\alpha m_e\simeq 5.292 \times 10^4$ fm is the Bohr radius,
and $\alpha=e^2/4\mpi \simeq 1/137$ ($e$ being the magnitude of
the charge on the electron) is the usual fine structure constant
\cite{ref:Jack1,ref:Jack2,ref:Bran}. A more refined analysis
requires assumptions about the distribution of charge within the
atom. In the Fermi-Thomas model, where the scalar potential is
approximated by the form $\Phi (r)\simeq
(Ze/r)\mbox{\Large{$\me$}}^{-r/a}$ instead of the usual $Ze/r$ of
the Bohr model, $a$ is found to be $a\simeq 1.4a_0/Z^{1/3}$
\cite{ref:Jack1,ref:Jack2}. This latter form for $a$ should be
used in place of the previous one in cases where screening effects
of the atomic electrons are important. An example of where either
of these forms for $b_{min}$ should be used can be found in the
problem that Fermi considered when he developed the original
version of the method: a swiftly moving charged particle $f$
collides with a hydrogen-like atom with radius $a$ and one
electron $e$. The collisions can be of two types: close
collisions, in which $f$ passes ``through" the atom ($b<a$), and
distant collisions, where $f$ passes by outside the atom ($b>a$).
The latter case is the one where $b_{min}$ should be taken to be
the atomic radius.
\begin{subequations}
\begin{align}
b_{min} = a &= \mbox{radius of atom} \nonumber \\ &\hspace{1.2in} \mbox{(\emph{distant} Coulomb collision of $f$ with $e$ in atom).}\label{eq:bmindCoul}%
\end{align}
\end{subequations}
This result holds for all three types of electroweak Coulomb
interactions. The close Coulomb collision problem is a bit more
intricate, and will be treated in a separate paragraph below.
Calculations of the \emph{nuclear} charge radius can vary widely
in technique, but all tend to yield the same value. For a nucleus
with $A$ nucleons, the nuclear radius is always found to be
$R=R_0A^{1/3}$, where $R_0=1.2-1.25$ fm; see, for example, Refs.
\cite{ref:Bert} and \cite{ref:Kran}--\cite{ref:Baur3}. Therefore,
a reasonable expression to use for $b_{min}$ for a collision
between two nuclei (denoted with subscripts $1$ and $2$) is
\begin{equation}
b_{min} = R_1+R_2 = 1.2\,\left( A_1^{1/3}+A_2^{1/3}\right) \mbox{ fm} \qquad \mbox{(collision between two nuclei),}\label{eq:bminnuclei1}%
\end{equation}
where $A_1$ and $A_2$ are the atomic masses of the two nuclei. It
should be pointed out that $b_{min}$ is an important parameter in
the study of relativistic nuclear collisions because it can be
used to differentiate electromagnetic interactions from those that
are dominated by the strong force. If the two nuclei come within
about $1$ fm (the size of a typical nucleus) of one another, the
strong force is the dominant of the four forces, and all other
interactions are swamped by its effects. If the colliding
particles never come that close to one another, the dynamics are
most significantly governed by the EM force. So for these
applications, the use of the sum of the nuclear radii for
$b_{min}$ is a way of triggering against strong (and weak) force
interactions, in the sense that the resulting calculations only
convey information about the EM effects of interest
\cite{ref:Norb2}. Another point worth noting is that the actual
choice of the functional form for the \emph{nuclear} form factor
$F(\vecq ^2)$ (see above discussion) is not very important insofar
as the final results are concerned, besides influencing the chosen
value of $b_{min}$, as long as $b$ is greater than the sum of the
nuclear radii \cite{ref:Bert,ref:Baur3}. This is because, by
Gauss's law, the fields and potentials of the target particle at
the location of the incident particle only depend on the total
charge of the target particle, and do not depend at all on the
distribution of charge within the chosen Gaussian surface. Eq.
(\ref{eq:bminnuclei1}) is a very intuitively appealing formula to
keep in mind for applications of the WWM to nucleus-nucleus
collisions. Another distance scale that must not be overlooked in
choosing an appropriate $b_{min}$ for these types of reactions is
the range of the relevant interaction. In the usual way, the range
of a given force $\sim 1/m_b$, where $m_b$ is the mass of the
mediator. For EM processes, which are mediated by massless
photons, the range is infinite. But weak force processes are known
to only occur on subnuclear distance scales, on account of the
massiveness of the mediators. The choice for $b_{min}$ listed in
Eq. (\ref{eq:bminnuclei1}) for weak force interactions is thus
problematic because it exceeds (sometimes to a great extent) the
actual range of the force involved! So Eq. (\ref{eq:bminnuclei1})
is a reasonable value to use for applications to collisions of
composite particles mediated by \emph{photons}, but a reassessment
is in order for those collisions mediated by $W$ and $Z$ bosons.
Towards this end, reconsider the quantum mechanical formula for
$b_{min}$ discussed above, except note that this time there is
greater uncertainty as to exactly where the boson originated than
Eq. (\ref{eq:bminQM21}) would seem to indicate. To see this,
recall the derivation of that equation. The parent particle $f$
was taken to be truly pointlike, and the identification $(\Delta
p_b)_{max}\simeq \gamma _f(\Delta p_{b\perp})v_f$ was believable.
If the parent particle is instead a composite nucleus, such a
precise specification of $(\Delta p_b)_{max}$ is unrealistic.
Compared to the immediate vicinity of an ideal point particle, a
nucleus is a region of very complicated goings-on, and there are
plenty of sources of additional uncertainties in $(\Delta
p_b)_{max}$. At best, one might expect to only be able to trust
the Heisenberg uncertainty principle, which, in this context,
reads $\Delta p_{b\perp} \Delta x \gtrsim 2$ (recall Eq.
(\ref{eq:WWMuncrel})). Solving for $\Delta x$, $\Delta x \gtrsim
2/\Delta p_{b\perp}$. The corresponding $b_{min}$ is hence
\begin{equation}
b_{min}=\frac{\eta _b}{(\Delta p_{b\perp}/2)} \qquad \mbox{(quantum mechanical formula \#3),}\label{eq:bminQM3}%
\end{equation}
where $\eta _b$ is a constant that should be expected to be $\sim
1$. This form for $b_{min}$ will be the one used when analyzing
collisions between composite particles mediated by either $W$ or
$Z$ bosons. Note from Eqs. (\ref{eq:chipbT4}) and (\ref{eq:mbpbT})
that $\Delta p_{b\perp}/2$ is always $\ge m_b$, so that
$b_{min}\simeq 1/(\Delta p_{b\perp}/2)$ is always within the
expected range of the force. It is for this reason that this
choice for $b_{min}$ is better than $b_{min}=R_1+R_2$ (which is
typically $\gtrsim 1/m_b$) for weak force interactions. A subtle
detail has been overlooked here in refining the choice of
$b_{min}$ for these weak force cases. Of interest to this study
are interactions between two nuclei that remain intact during the
collision. In the usual application of the traditional WWM to
photon-mediated nucleus-nucleus collisions, this condition is
enforced by demanding that $b_{min}$ be $\ge R_1+R_2$, as
discussed above. For applications to $W$ and $Z$ boson-mediated
nucleus-nucleus collisions, it has been found here that $b_{min}$
should be chosen as $1/(\Delta p_{b\perp}/2)$, which is typically
$\lesssim 1/m_b \lesssim R_1+R_2$. So the restriction that the
nuclear spheres do not ever overlap is being somewhat relaxed
here; the actual numbers work out that this smaller $b_{min}$ is
still typically $\simeq$ fm. To end this paragraph, the
interesting application of the WWM to electron-positron pair
production in relativistic heavy-ion collisions is mentioned.
Here, one must not neglect to take into account quantum mechanical
effects associated with the produced electrons. The localization
of an electron to within a distance less than its rest frame
Compton wavelength, $(\lambda _C)_e=1/m_e=386.2$ fm, requires the
use of unphysical negative energy states \cite{ref:Eich,ref:RPP}.
As this distance is more than an order of magnitude greater than
the sum of any two nuclear radii, it is this value that sets the
minimum impact parameter for such reactions.
\begin{subequations}
\begin{align}
b_{min} &= \frac{1}{m_e} = 386.2 \mbox{ fm} \nonumber \\ &\hspace{1.2in} \mbox{($e^+$--$e^-$ pair production in nucleus-nucleus collisions).}\label{eq:bmineepair}%
\end{align}
\end{subequations}
These types of processes are interesting in their own right for
various reasons, and quite a number of papers have been published
on the subject. See, for example, references
\cite{ref:Eich,ref:Bert} and \cite{ref:Baur4}--\cite{ref:Alsc}. Of
course, the created leptons do not have to be electrons and
positrons. However, the Compton wavelengths of the muon and tau
are less than typical nuclear radii, so the usual formula
(\ref{eq:bminnuclei1}) can be used for those applications.

The discussion so far has dealt with choosing a suitable form for
$b_{min}$ for collisions between two particles in which there is
no significant 4-momentum lost in the form of bremsstrahlung. For
the prototypical bremsstrahlung process, where a light fermion $f$
is scattered by an atomic nucleus, one possible form for $b_{min}$
to keep in mind is the nuclear radius $R$. However, another factor
to take into account is the form of $b_{min}$ determined by the
uncertainty principle: $b_{min}\simeq 1/Q_{max}$, where $Q=|{\vecp
_f}'-{\vecp _f}|$ is the magnitude of 3-momentum transferred in
the collision. In the end, the correct $b_{min}$ to use is the
greater of these two possibilities. The greatest amount of
momentum transfer in a \emph{nonrelativistic} collision between
$f$ and a nucleus occurs when the collision is elastic, and is
given by $Q_{max}=2p_f=2m_fv_f$ (as calculated in the rest frame
of the (heavier) nucleus). So the quantum $b_{min}$ for such
interactions is $b_{min}=1/2m_fv_f$. As $f$ is assumed to be much
lighter than the nucleus, and $v_f\ll 1$ for these cases, this
value of $b_{min}$ will always be greater than the nuclear radius,
which is typically a few fm. Therefore, the correct $b_{min}$ to
use for a \emph{nonrelativistic} collision between $f$ and a
nucleus is
\begin{equation}
b_{min}=\frac{1}{Q_{max}}=\frac{1}{2m_fv_f}.\qquad \mbox{(\emph{nonrelativistic} bremsstrahlung formula).}\label{eq:bminNRbrem}%
\end{equation}
This value is entirely kinematic in nature, so holds for all three
types of electroweak bosons. A detailed analysis of EM
bremsstrahlung shows that the intensity of radiation is greatest
in the low frequency regime (in the limit where $\omega
_{\gamma}\to 0$), and (in that limit) increases linearly with
$Q^2$ up to a value of $Q=2m_f$, at which point it levels off to a
constant value, independent of $Q$ \cite{ref:Jack1,ref:Jack2}.
Thus, in the \emph{relativistic} limit of interest, where $v_f\to
1$, there is already an effective maximum value of $Q$ built into
the equations: $Q_{max}=2m_f$. A generalization of the EM
bremsstrahlung analysis to an electroweak formalism, to be
published in a future paper, arrives at similar equations, with
input parameters $b_{min}$, $m_b$, and the $q_V$ and $q_A$ charges
of the accelerating particle. Since the $Q_{max}=2m_f$ result for
EM bremsstrahlung was obtained in the $\omega _{\gamma}\to 0$
limit, the same limit should be taken for the massive boson
analysis to find a similar $Q_{max}$. The generalization is
trivial, because all boson masses vanish in the $\omega _b\to 0$
limit (cf. Eqs. (\ref{eq:defx}) and (\ref{eq:mbspec1}))! Since all
of the electroweak bosons are massless in this limit, the
equations derived in the EM bremsstrahlung analysis are identical
to those of the electroweak generalization. That is, except for
the values of $q_V$ and $q_A$. The $q_V$ and $q_A$ charges appear
in the form of an overall constant that premultiplies the radiated
intensity function, so do not affect the overall dependence on
$Q$. Because the overall behavior is the same as in the EM limit,
the maximum allowable value of $Q$ in the relativistic limit is
also the same. Therefore, for \emph{any} of the three types of
electroweak bremsstrahlung, the correct $b_{min}$ to use in the
relativistic limit is
\begin{equation}
b_{min}=\frac{1}{Q_{max}}=\frac{1}{2m_f}.\qquad \mbox{(\emph{relativistic} bremsstrahlung formula).}\label{eq:bminRbrem}%
\end{equation}
Note that this form for $b_{min}$, which is $\simeq 193.1$ fm if
the incident particle is an electron, is greater than the nuclear
radius, so it is the correct one to use for these types of
processes.

As a summary so far, for most applications, $b_{min}$ is either
identified with the the sum of the radii of the two colliding
particles (if such quantities can be identified), or determined by
using the Heisenberg uncertainty principle: $b_{min}\simeq
1/Q_{max}$. In practice, except for the cases of massive bosons
radiating from composite particles, the correct $b_{min}$ to use
is the \emph{greater} of these two possibilities. An exception to
these simple approaches can be found in the \emph{close} Coulomb
collision problem that Fermi originally considered, as discussed
above. The scenario for this interaction is as follows. A fast
particle $f$ of charge $Ze$ and velocity $v_f$ passes ``through" a
hydrogen-like atom with radius $a$ and one electron $e$ of charge
$-e$ and mass $m_e$. The derivation of a suitable $b_{min}$
proceeds as follows. The total momentum $\Delta \vecp$ transferred
from $f$ to $e$ is found via $\Delta \vecp =\int
_{-\infty}^{\infty}\, \vecF (t)\, dt$, where $\vecF =e\vecE$ is
the Coulomb force between $f$ and $e$. After a bit of work,
$\Delta \vecp$ is found to depend on the impact parameter $b$
according to $\Delta \vecp=\vecepsperp (2Z\alpha / bv_f)$, where
$\alpha = e^2/4\mpi \simeq 1/137$ is the usual fine structure
constant. Assuming $f$ is not deflected appreciably, the energy
$\Delta E$ transferred to the electron is given by $\Delta
E=(\Delta \vecp)^2/2m_e=(Z\alpha )^2/b^2(m_ev_f^2/2)$. Solving
this equation for $b$ yields $b=Z\alpha /\sqrt{\Delta
E(m_ev_f^2/2)}$. $b_{min}$ is then found by considering when this
equation is minimized, which clearly occurs when $\Delta E$ is set
equal to its maximum allowable value, $(\Delta E)_{max}$. Well,
$(\Delta E)_{max}=(\Delta \vecp)_{max}^2/2m_e$, where $(\Delta
\vecp)_{max}$ is the maximum allowable momentum transfer, which
occurs when the collision is elastic. In the rest frame of $f$,
$|(\Delta \vecp)_{max}|$ is found to be $|(\Delta
\vecp)_{max}|=2p_e=2\gamma _fm_ev_f$. Therefore, $(\Delta
E)_{max}=(2\gamma _fm_ev_f)^2/2m_e=2\gamma _f^2m_ev_f^2$. Hence,
$b_{min}$ for such interactions works out to be $b_{min}=Z\alpha
/\gamma _fm_ev_f^2$. The mathematics of the generalization to the
same reaction mediated by \emph{any} of the electroweak bosons is
quite involved, and will not be presented here. The upshot is that
\begin{equation}
b_{min}=\oofourpi \frac{q_fq_e}{\gamma _fm_ev_f^2}\qquad \mbox{(\emph{close} Coulomb collision of $f$ with $e$ in atom),}\label{eq:bmincCoul}%
\end{equation}
where $q_f=\sqrt{(q_V)_f^2+(q_A)_f^2v_f^2}$ and
$q_e=\sqrt{(q_V)_e^2+(q_A)_e^2v_f^2}$ are effective charges of $f$
and $e$, that reduce to $Ze$ and $e$, respectively, in the EM
limit. At the end of the analysis, the correct $b_{min}$ to use is
then the greater of this value and the one specified in Eq.
(\ref{eq:bminQM1}) \cite{ref:Jack1,ref:Jack2,ref:Dali}. It can be
shown that, except for certain \emph{nonrelativistic} collisions
involving particles with $Z\gg 1$, the quantum formula is the one
to use.

This section is concluded with a summary of the correct forms for
$b_{min}$ to use in the different instances discussed above. For a
\emph{distant Coulomb collision} between a fast moving particle
and an electron in an atom in which no bremsstrahlung is emitted,
$b_{min}$ should be taken to be the atomic radius, $a$. For a
\emph{close Coulomb collision} between a fast moving particle $f$
and an electron $e$ in an atom (where $b<a$) in which no
bremsstrahlung is emitted, $b_{min}=(q_fq_e/4\mpi)/\gamma
_fm_ev_f^2$, where $q_f$ and $q_e$ are defined in the previous
paragraph. For most other \emph{photon-mediated} collisions of
interest between two particles in which no bremsstrahlung is
emitted, $b_{min}$ should be chosen to be the greater of the sum
$R_1+R_2$ of the radii of the two particles and $\eta _f/\gamma
_fm_fv_f$, where $\eta _f$ is a constant of order unity and $m_f$
is the mass of the lighter of the two particles; see above for a
discussion of nuclear radii. Examples of this class of collisions
include all types of nucleus-nucleus collisions, and
electron-nucleus collisions accompanied by either particle
production or nuclear photo-disintegration. An important exception
to the use of this $b_{min}$ is when comparing results of the GWWM
being developed here to the QWWM, for collisions between
\emph{point particles}. For those applications, the correct value
of $b_{min}$ to use is $\eta _{\gamma}/\gamma _f(\Delta p_{\gamma
\perp})v_f$, where $\Delta p_{\gamma \perp}$ is the uncertainty in
the transverse component of the photon's 3-momentum and $\eta
_{\gamma}$ is a constant of order unity. Applications of the
method to similar collisions mediated by either $W$ or $Z$ bosons
should use similar formulas; that is, $b_{min}=\eta _b/\gamma
_f(\Delta p_{b\perp})v_f$. And applications of the method to weak
force collisions between \emph{composite particles}, in which no
bremsstrahlung is emitted, should use $b_{min}=\eta _b/(\Delta
p_{b\perp}/2)$. It has gone without mention that the $b_{min}$ for
these applications should really be chosen as the greater of $\eta
_b/(\Delta p_{b\perp}/2)$ and $\eta _f/\gamma _fm_fv_f$, like the
procedure for the EM force case. But Eq. (\ref{eq:chipbT4}) shows
that $\Delta p_{b\perp}/2$ is always $\lesssim m_f$, so it is
always the case (since $\gamma _f \gg 1$) that $\eta _b/(\Delta
p_{b\perp}/2) \gg \eta _f/\gamma _fm_fv_f$. It will be these forms
for $b_{min}$ that will be used in the remainder of this report.
Another exception is when applying the WWM to electron-positron
pair creation in relativistic heavy ion collisions. There, one
must use $b_{min}=1/m_e=386.2$ fm. Finally, for \emph{any}
collision (i.e., relativistic or not) between a light particle $f$
and the nucleus of an atom in which there is a significant amount
of \emph{bremsstrahlung} emitted, $b_{min}$ should be set equal to
$1/2m_fv_f$.

\section{Limiting Forms of the GWWM Number Spectra} \label{sec:limNTNLGWWM} \indent

Due to the wide variety of possible values of $b_{min}$, the value
of $\chi _b$ that appears in the number spectrum formulas (cf.
Eqs. (\ref{eq:numspTpbT}) and (\ref{eq:numspLpbT})) can vary
greatly as well. It is sometimes convenient to have available the
$\chi _b\to 0$ and $\chi _b\to \infty$ limiting forms of these
functions. Consider first the former limit. The $\chi _b\to 0$
limiting forms of the $K_0(\chi _b)$ and $K_1(\chi _b)$ functions
that appear in these equations can be found in any good reference
on mathematical functions. They are
\begin{subequations}
\begin{align}
\lim _{\chi _b\to 0}K_0(\chi _b) &= \ln \left( \frac{2\mbox{\large e} ^{-\gamma}}{\chi _b} \right)\label{eq:K0ll1}\\%
\lim _{\chi _b\to 0}K_1(\chi _b) &= \frac{1}{\chi _b},\label{eq:K1ll1}%
\end{align}
\end{subequations}
where $\gamma$ (not to be confused with the Lorentz factor or
``photon") is the Euler-Mascheroni constant (or simply Euler's
constant), which is $\gamma = 0.5772$ to four significant figures
\cite{ref:Arfk}. Using these limiting expressions, $N_\mT (E_b)$
simplifies to
\begin{subequations}
\begin{align}
\lim _{\chi _b\to 0}N_\mT (E_b) &= \frac{N_0}{E_b}\left\{ \chi _b\left[ \ln \left( \frac{2\mbox{\large e} ^{-\gamma}}{\chi _b} \right) \right] \left[ \frac{1}{\chi _b}\right] -\half v_f^2 \chi _b^2 \left[ \frac{1}{\chi _b^2} - \ln ^2\left( \frac{2\mbox{\large e} ^{-\gamma}}{\chi _b} \right) \right] \right\} \label{eq:NTGWWM2}\\%
&\to \frac{N_0}{E_b}\left[ \ln \left( \frac{2\mbox{\large e} ^{-\gamma}}{\chi _b} \right) -\half v_f^2 \right] \label{eq:NTGWWM3}\\%
&= \frac{N_0}{E_b}\, \ln \left[ \frac{2\mbox{\large e} ^{-(\gamma +v_f^2/2)}}{\chi _b} \right]\label{eq:NTGWWM4}\\%
&= \frac{N_0}{E_b}\, \ln \left[ \frac{4\mbox{\large e} ^{-(\gamma +v_f^2/2)}}{b_{min}\Delta p_{b\perp}} \right] \qquad \mbox{(via Eq. (\ref{eq:chipbT1}))}\label{eq:NTGWWM5}\\%
&= \frac{N_0}{E_b}\, \ln \left[ \frac{2\mbox{\large e} ^{-(\gamma +v_f^2/2)} \gamma _fv_f}{b_{min}\sqrt{E_b[\alpha _bp_f+E_b(1-\alpha _b)]}} \right] \qquad \mbox{(via Eq. (\ref{eq:chipbT4})).}\label{eq:NTGWWM6}%
\end{align}
\end{subequations}
For the UR collisions in this study, where $v_f\simeq 1$, the
constant in the numerator of the argument of the logarithm in Eq.
(\ref{eq:NTGWWM6}) has a magnitude of about $0.6811$. Noting that
$\alpha _{\gamma }=0$ and $\alpha _W=1$, while $\alpha _Z$ is a
bit more complicated (though is typically $\simeq 1$), the
limiting form for $N_\mT (E_b)$ for the three types of bosons can
easily be specified.
\begin{subequations}
\begin{align}
\lim _{\chi _{\gamma }\to 0}N_\mT (E_{\gamma }) &= \frac{N_0}{E_{\gamma }}\, \ln \left[ \frac{0.6811\gamma _fv_f}{b_{min}E_{\gamma }} \right] \label{eq:NTph0}\\%
\lim _{\chi _Z\to 0}N_\mT (E_Z) &= \frac{N_0}{E_Z}\, \ln \left[ \frac{0.6811\gamma _fv_f}{b_{min}\sqrt{E_Z[\alpha _Zp_f+E_Z(1-\alpha _Z)]}} \right] \label{eq:NTZb0}\\%
\lim _{\chi _W\to 0}N_\mT (E_W) &= \frac{N_0}{E_W}\, \ln \left[ \frac{0.6811}{b_{min}\sqrt{E_Wm_f/\gamma _fv_f}} \right] .\label{eq:NTWb0}%
\end{align}
\end{subequations}
Eq. (\ref{eq:NTph0}) agrees exactly with the expected result
\cite{ref:Jack1}. In a similar way, $N_\mL (E_b)$ simplifies to
\begin{subequations}
\begin{align}
\lim _{\chi _b\to 0}N_\mL (E_b) &= \frac{N_0}{E_b}\left\{ \half \left( \frac{2m_b}{\Delta p_{b\perp}} \right) ^2\chi _b^2 \left[ \frac{1}{\chi _b^2} - \ln ^2\left( \frac{2\mbox{\large e} ^{-\gamma}}{\chi _b} \right) \right] \right\}  \nonumber \\ &\hspace{3.2in} \mbox{(via Eq. (\ref{eq:chipbT1}))}\label{eq:NLGWWM2}\\%
&\to \frac{2N_0}{E_b}\, \frac{m_b^2}{(\Delta p_{b\perp})^2} \label{eq:NLGWWM3}\\%
&= \frac{2N_0}{E_b}\, \frac{\alpha _b(E_b/\gamma _fv_f)[m_f-(E_b/\gamma _fv_f)]}{4(E_b/\gamma _fv_f)[\alpha _bm_f+(E_b/\gamma _fv_f)(1-\alpha _b)]} \nonumber\\ &\mbox{(via Eqs. (\ref{eq:mbspec1}) and (\ref{eq:chipbT4}))}\label{eq:NLGWWM4}\\%
&= \frac{N_0}{2E_b}\, \frac{\alpha _b(p_f-E_b)}{[\alpha _bp_f+E_b(1-\alpha _b)]} .\label{eq:NLGWWM5}%
\end{align}
\end{subequations}
The limiting values for the three types of interactions of
interest are as follows.
\begin{subequations}
\begin{align}
\lim _{\chi _{\gamma }\to 0}N_\mL (E_{\gamma }) &= 0 \label{eq:NLph0}\\%
\lim _{\chi _Z\to 0}N_\mL (E_Z) &= \frac{N_0}{2E_Z}\, \frac{\alpha _Z(p_f-E_Z)}{[\alpha _Zp_f+E_Z(1-\alpha _Z)]} \label{eq:NLZb0}\\%
\lim _{\chi _W\to 0}N_\mL (E_W) &= \frac{N_0}{2E_W}\, \left( 1-\frac{E_W}{p_f} \right) .\label{eq:NLWb0}%
\end{align}
\end{subequations}

The $\chi _b\to \infty$ limiting forms of $K_0(\chi _b)$ and
$K_1(\chi _b)$ are
\begin{subequations}
\begin{align}
\lim _{\chi _b\to \infty}K_0(\chi _b) &= \sqrt{\frac{\mpi}{2\chi _b}}\, \mbox{\Large e} ^{-\chi _b}\, \left( 1-\frac{1}{8\chi _b} \right) \label{eq:K0gg1}\\%
\lim _{\chi _b\to \infty}K_1(\chi _b) &= \sqrt{\frac{\mpi}{2\chi _b}}\, \mbox{\Large e} ^{-\chi _b}\, \left( 1+\frac{3}{8\chi _b} \right) \label{eq:K1gg1}%
\end{align}
\end{subequations}
(see, e.g., \cite{ref:Arfk} again). The corresponding $\chi _b\to
\infty$ limiting form of $N_\mT (E_b)$ is
\begin{subequations}
\begin{align}
\lim _{\chi _b\to \infty}N_\mT (E_b) &= \frac{N_0}{E_b}\Bigg\{ \chi _b\, \left( \frac{\mpi}{2\chi _b}\, \mbox{\Large e} ^{-2\chi _b}\right) \, \left( 1-\frac{1}{8\chi _b} \right) \left( 1+\frac{3}{8\chi _b} \right) - \nonumber\\ & - \half v_f^2 \chi _b^2\, \left( \frac{\mpi}{2\chi _b}\, \mbox{\Large e} ^{-2\chi _b}\right) \, \left[ \left( 1+\frac{3}{8\chi _b} \right) ^2- \left( 1-\frac{1}{8\chi _b} \right) ^2\right] \Bigg\} \label{eq:NTGWWM7}\\%
&= \frac{\mpi N_0}{2E_b}\, \Bigg\{ \left( 1+\frac{1}{4\chi _b}-\frac{3}{64\chi _b^2} \right) - \half v_f^2 \chi _b\, \Bigg[ \left( 1+\frac{3}{4\chi _b} + \frac{9}{64\chi _b^2} \right) - \nonumber\\ & - \left( 1-\frac{1}{4\chi _b} +\frac{1}{64\chi _b^2} \right) \Bigg] \Bigg\} \, \mbox{\Large e} ^{-2\chi _b} \label{eq:NTGWWM8}\\%
&= \frac{\mpi N_0}{2E_b}\, \left[ \left( 1+\frac{1}{4\chi _b}-\frac{3}{64\chi _b^2} \right) -\half v_f^2 \chi _b\, \left( \frac{1}{\chi _b} + \frac{1}{8\chi _b^2} \right) \right] \, \mbox{\Large e} ^{-2\chi _b} \label{eq:NTGWWM9}\\%
&= \frac{\mpi N_0}{2E_b}\, \left[1-\half v_f^2+\frac{1}{4\chi _b}\left( 1-\frac{1}{4}v_f^2 \right) -\frac{3}{64\chi _b^2} \right] \, \mbox{\Large e} ^{-2\chi _b}\label{eq:NTGWWM10}\\%
&\to \frac{\mpi N_0}{2E_b}\, \left( 1-\half v_f^2\right) \, \mbox{\Large e} ^{-2\chi _b}\label{eq:NTGWWM11}\\%
&\simeq \frac{\mpi N_0}{4E_b}\, \mbox{\Large e} ^{-2\chi _b}\qquad \mbox{(via $v_f\simeq 1$ )}\label{eq:NTGWWM12}%
\end{align}
\end{subequations}
This formula shows that $N_\mT (E_b)$ is only appreciable for
values of $\chi _b\lesssim 1/2$, which is a result referred to in
the paragraph following the Heisenberg relations, Eq.
(\ref{eq:Heisrel}). The special cases of this limiting value
corresponding to the three different types of electroweak
interactions of interest are found by simply replacing the
parameter $\chi _b$ in the exponential functional with the
relevant expression (cf. Eqs.
(\ref{eq:chiphoton})--(\ref{eq:chiWboson})). The $\chi _b\to
\infty$ limiting form of $N_\mL (E_b)$ is found from Eq.
(\ref{eq:numspLpbT}) to be
\begin{subequations}
\begin{align}
\lim _{\chi _b\to \infty}N_\mL (E_b) &= \frac{N_0}{E_b}\Bigg\{
\half \left( \frac{2m_b}{\Delta p_{b\perp }} \right) ^2\, \chi
_b^2\, \Bigg[ \left( \frac{\mpi}{2\chi _b} \right) \, \mbox{\Large
e} ^{-2\chi _b}\, \left( 1+\frac{3}{8\chi _b} \right) ^2 -
\nonumber\\ & - \left( \frac{\mpi}{2\chi _b} \right) \,
\mbox{\Large e} ^{-2\chi _b}\, \left( 1-\frac{1}{8\chi _b} \right)
^2\Bigg]
\Bigg\} \qquad \mbox{(via Eq. (\ref{eq:chipbT1}))}\label{eq:NLGWWM6}\\%
&= \frac{\mpi N_0}{E_b}\Bigg\{ \frac{m_b^2}{(\Delta p_{b\perp
})^2}\, \chi _b\, \Bigg[ \left( 1 + \frac{3}{4\chi _b} +
\frac{9}{64\chi _b^2} \right) - \nonumber\\ & - \left(
1-\frac{1}{4\chi _b} + \frac{1}{64\chi _b^2} \right) \Bigg]
\Bigg\} \, \mbox{\Large e} ^{-2\chi _b}\label{eq:NLGWWM7}\\%
&= \frac{\mpi N_0}{E_b}\left[ \frac{m_b^2}{(\Delta p_{b\perp })^2}\, \chi _b\, \left( \frac{1}{\chi _b} + \frac{1}{8\chi _b^2} \right) \right] \, \mbox{\Large e} ^{-2\chi _b}\label{eq:NLGWWM8}\\%
&= \frac{\mpi N_0}{E_b}\, \frac{m_b^2}{(\Delta p_{b\perp })^2}\, \left( 1 + \frac{1}{8\chi _b} \right) \, \mbox{\Large e} ^{-2\chi _b}\label{eq:NLGWWM9}\\%
&\to \frac{\mpi N_0}{E_b}\, \frac{m_b^2}{(\Delta p_{b\perp })^2}\, \mbox{\Large e} ^{-2\chi _b}\label{eq:NLGWWM10}\\%
&= \frac{\mpi N_0}{E_b}\, \frac{\alpha _b(E_b/\gamma _fv_f)[m_f-(E_b/\gamma _fv_f)]}{4(E_b/\gamma _fv_f)[\alpha _bm_f+(E_b/\gamma _fv_f)(1-\alpha _b)]}\, \mbox{\Large e} ^{-2\chi _b} \nonumber\\ & \mbox{(via Eqs. (\ref{eq:mbspec1}) and (\ref{eq:chipbT4}))}\label{eq:NLGWWM11}\\%
&= \frac{\mpi N_0}{4E_b}\frac{\alpha _b(p_f-E_b)}{[\alpha _bp_f+E_b(1-\alpha _b)]}\, \mbox{\Large e} ^{-2\chi _b}.\label{eq:NLGWWM12}%
\end{align}
\end{subequations}
As with the $N_\mT (E_b)$ limiting expression, this formula is
found to decay exponentially with $\chi _b$, which means that the
number spectrum of longitudinally-polarized bosons is also
strongly suppressed for values of $\chi _b\gtrsim 1/2$. The
limiting forms of $N_\mL (E_b)$ for the three types of
interactions of interest are as follows.
\begin{subequations}
\begin{align}
\lim _{\chi _{\gamma }\to \infty}N_\mL (E_{\gamma }) &= 0 \label{eq:NLphinfty}\\%
\lim _{\chi _Z\to \infty}N_\mL (E_Z) &= \frac{\mpi N_0}{4E_Z}\, \frac{\alpha _Z(p_f-E_Z)}{[\alpha _Zp_f+E_Z(1-\alpha _Z)]}\, \mbox{\Large e} ^{-2\chi _Z} \label{eq:NLZbinfty}\\%
\lim _{\chi _W\to \infty}N_\mL (E_W) &= \frac{\mpi N_0}{4E_W}\, \left( 1-\frac{E_W}{p_f}\right) \, \mbox{\Large e} ^{-2\chi _W},\label{eq:NLWbinfty}%
\end{align}
\end{subequations}
where $\chi _Z$ and $\chi _W$ are specified in Eqs.
(\ref{eq:chiZboson}) and (\ref{eq:chiWboson}), respectively.

%% file: Chapter5.tex
%%%%%%%%%%%%%%%%%
%  CHAPTER 5
%%%%%%%%%%%%%%%%%

\chapter{Comparison with Other Methods} \label{sec:othmethods} \indent

In this section, the number spectrum functions $N_\mT (E_b)$ and
$N_\mL (E_b)$ of the GWWM developed here are compared to the same
functions appearing in other theories. The EM number spectra,
corresponding to massless photons, are compared to the same
functions in both the traditional semiclassical WWM (the SWWM), as
originally devised by Fermi \cite{ref:Jack1}, and the quantum WWM
(the QWWM) \cite{ref:Dali,ref:Tera,ref:Jack2}. The weak force
number spectra, corresponding to massive $W$ and $Z$ bosons, are
compared to the same functions appearing in the Effective-$W$
Method (EWM)
\cite{ref:Daws1,ref:Daws2,ref:Guni,ref:Kane,ref:Cahn1,ref:Alta}.
Any two number spectrum functions (from different theories)
describing the same boson state are found to differ in general.
Fortunately, though, they are generally in good agreement in the
low boson energy limit (where $E_b\to 0$, or $x\to 0$), which is
the regime in which they contribute most significantly to cross
sections. So the criterion that will test the accuracy of a given
GWWM number spectrum function is the agreement with its quantum
counterpart \emph{in the $x\to 0$ limit}. It is instructive to
\emph{briefly} review the GWWM and the other theories to which it
will be compared, and list the transverse $N_\mT (E_b)$ and
longitudinal $N_\mL (E_b)$ number spectrum functions (in the
notation used in this report).

\section{The GWWM Number Spectra} \label{sec:NTNLGWWM} \indent

The GWWM is a semiclassical generalization of the traditional
SWWM. In both of these schemes, the particles are assumed to
travel at UR speeds along classical straight-line trajectories,
and the equivalent bosons are identified as infinitesimal elements
within the plane waves that represent the Lorentz contracted
fields and potentials. This picture facilitates the development of
the general form of the number spectra, but fails to allow for any
precise specification of a minimum impact parameter $b_{min}$ and
a nonvanishing boson mass $m_b$. $b_{min}$ remains a free
parameter in both the SWWM and GWWM. $m_b$ is assumed to be
exactly zero in the SWWM, so it is not a free parameter there. In
the GWWM, it \emph{is} a free parameter, and its value for a given
type of interaction is uniquely determined based on 4-momentum and
causality considerations. An equivalent boson in the GWWM is
envisioned as a pointlike entity travelling along a classical,
straight-line trajectory, just like the fermion from which it was
emitted. The boson's transverse component $\vecp _{b\perp}$ of
3-momentum is naturally found to vanish on average, and its
4-momentum is reparameterized so that its energy $E_b$ and
longitudinal component $p_{bz}$ of 3-momentum are guaranteed to be
independent of any one particular helicity state of the parent
fermion; that is to say, $E_b$ and $p_{bz}$ are helicity-averaged
quantities. In this way, the equivalent bosons in the GWWM can be
taken to be travelling collinearly with the parent fermion. The
nonzero value of $\vecp _{b\perp}$ that is needed to propel the
boson from the parent fermion to the interaction point is
explained as being due to a mere fluctuation of the fields and
potentials.

The general forms of the number spectrum functions in the GWWM are
listed in Eqs. (\ref{eq:numspTpbT})--(\ref{eq:chipbT4}), and,
along with definitions of other relevant parameters, are reviewed
once again here for convenience.
\begin{subequations}
\begin{align}
N_\mT (E_b) &= \frac{N_0}{E_b}\left\{ \chi _bK_0(\chi _b)K_1(\chi
_b)-\half v_f^2 \chi _b^2 \left[ K^2_1(\chi _b)-K^2_0(\chi _b)
\right]
\right\}\label{eq:NTGWWM1}\\%
N_\mL (E_b) &= \frac{N_0}{E_b}\left\{ \half (m_bb_{min})^2 \left[
K^2_1(\chi _b) - K^2_0(\chi _b) \right]
\right\},\label{eq:NLGWWM1}%
\end{align}
\end{subequations}
where
\begin{equation}
N_0 \equiv \frac{1}{2\mpi ^2}\, \frac{q_V^2+q_A^2}{v_f^2}=const\label{eq:N0GWWM}%
\end{equation}
and
\begin{equation}
\chi _b = \half\, b_{min}\, \Delta {p_{b\perp}}.\label{eq:chibGWWM}%
\end{equation}
The charges to which the photon couples are $q_V = Q^\gamma e$,
where $e=\sqrt{4\mpi \alpha}=0.3028$ (with $\alpha =7.297\times
10^{-3}\simeq 1/137$), and $q_A = 0$. For $Z$ boson mediated
processes, $q_V = g_Z(T^3-2Q^\gamma \sin^2\theta_W)/2$ and $q_A =
-g_ZT^3/2$, where $g_Z=e/\sin\theta_W\cos\theta_W=0.7183$ (with
$\theta_W=28.74^\circ$). And, for $W^{\pm}$ boson mediated
processes, $q_V = g_W/2\sqrt{2}$ and $q_A = \mp g_W/2\sqrt{2}$,
where $g_W=e/\sin\theta_W=0.6298$. As discussed in Section
\ref{sec:Bmin}, the correct value of $b_{min}$ to use when
comparing the GWWM to other theories is
\begin{equation}
b_{min} = \frac{\eta _b}{\gamma _f(\Delta p_{b\perp})v_f}, \label{eq:bminGWWM1}%
\end{equation}
where $\eta _b\sim 1$, if the colliding particles are
\emph{pointlike}. If the colliding particles are \emph{nuclei} and
the mediating bosons are photons,
\begin{equation}
b_{min} = \mbox{the greater of $R_1+R_2$ and } \frac{\eta _f}{\gamma _fm_fv_f}. \label{eq:bminGWWM21}%
\end{equation}
$R_1$ and $R_2$ here are the radii of the two nuclei (typically
$R_1=R_2$) and $\eta _f\sim 1$. And, if the colliding particles
are \emph{nuclei} and the mediating bosons are $W$ or $Z$ bosons,
\begin{equation}
b_{min} = \frac{\eta _b}{(\Delta p_{b\perp}/2)}, \label{eq:bminGWWM22}%
\end{equation}
where $\eta _b\sim 1$. $\Delta {p_{b\perp}}$ is defined as
\begin{equation}
\Delta p_{b\perp} = 2\, \sqrt{ m_b^2 + \left( \frac{E_b}{\gamma _fv_f} \right) ^2},\label{eq:DlbT2GWWM}%
\end{equation}
where
\begin{equation}
m_b = \sqrt{ \alpha _b\left( \frac{E_b}{\gamma _fv_f}\right)
\left[ m_f-\left( \frac{E_b}{\gamma _fv_f}\right) \right]
},\label{eq:mbGWWM}%
\end{equation}
and the boson energy $E_b$ is bounded (by 4-momentum and causality
considerations) within the range
\begin{equation}
(E_b)_{min} \le E_b \le (E_b)_{max},\label{eq:EboundsGWWM}%
\end{equation}
where
\begin{subequations}
\begin{align}
(E_b)_{min} &\equiv \frac{\alpha _bp_f}{\gamma _f^2v_f^2+\alpha _b}\simeq \frac{\alpha _bm_f}{\gamma _f} \label{eq:defxminGWWM}\\%
(E_b)_{max} &\equiv E_f-m_f\simeq E_f.\label{eq:defxmaxGWWM}%
\end{align}
\end{subequations}
The values of $\alpha _b$ for the three different bosons of
interest are
\begin{equation}
\alpha _b = \left\{
\begin{array}{l}
\mbox{\Large $\frac{q_{\mbox{\scriptsize $A$}}^{\mbox{\scriptsize $2$}}}{q_{\mbox{\scriptsize $A$}}^{\mbox{\scriptsize $2$}}-q_{\mbox{\scriptsize $V$}}^{\mbox{\scriptsize $2$}}}$}=0 \quad \mbox{for the photon}\\
\mbox{\Large $\frac{q_{\mbox{\scriptsize $A$}}}{q_{\mbox{\scriptsize $A$}}-q_{\mbox{\scriptsize $V$}}}$}=\mbox{\Large $\frac{T^{\mbox{\scriptsize $3$}}_{\mbox{\scriptsize $L$}}}{2(T^{\mbox{\scriptsize $3$}}_{\mbox{\scriptsize $L$}}-Q^{\mbox{\scriptsize $\gamma$}} \sin^{\mbox{\scriptsize $2$}}\theta_{\mbox{\scriptsize $W$}})}$} \quad \mbox{for the $Z$ boson}\\
\mbox{\Large $\frac{2q_{\mbox{\scriptsize $A$}}}{q_{\mbox{\scriptsize $A$}}\mp q_{\mbox{\scriptsize $V$}}}$}=1 \quad \mbox{for the $W^{\pm}$ bosons}\\
\end{array}
\right. .\label{eq:alphabGWWM}
\end{equation}
A list of values of $T^3_\mL $ and $Q_{\gamma}$ for the various
fermions of interest are shown in Table \ref{tab:charges}, and a
list of values of $\alpha _b$ appears in Table \ref{tab:alphab}.
Eqs. (\ref{eq:chiphoton})--(\ref{eq:chiWboson}),
(\ref{eq:DpphotonT})--(\ref{eq:DpWbosonT}), and
(\ref{eq:mphoton})--(\ref{eq:mWboson}), which give explicit forms
for $\chi _b$, $\Delta p_{b\perp}$, and $m_b$, respectively, for
the three different types of bosons are perhaps more useful.

\section{The SWWM Number Spectra} \label{sec:NTNLSWWM} \indent

The first of the other theories whose formulas will be summarized
is the SWWM. The SWWM formalism is identical to that of the GWWM,
except for the boson mass assignment scheme. In the SWWM, the only
mediator of interest is the photon, and it is \emph{assumed to be}
massless. The equivalent pulses of EM radiation are found to be
purely transversely-polarized, so that only the transverse number
spectrum $N_\mT (E_{\gamma })$ is nonvanishing. As has been
pointed out previously, the $N_\mT (E_{\gamma })$ function in the
SWWM is exactly the $m_b\to 0$, $q_V\to Q^{\gamma}e$, and $q_A\to
0$ limit of the $N_\mT (E_b)$ function of the GWWM. The number
spectrum functions are thus
\begin{subequations}
\begin{align}
[N_\mT (E_{\gamma })]_{SWWM} &= \frac{N_0}{E_{\gamma }}\left\{
\chi _{\gamma }K_0(\chi _{\gamma })K_1(\chi _{\gamma })-\half
v_f^2 \chi _{\gamma }^2 \left[ K^2_1(\chi _{\gamma })-K^2_0(\chi
_{\gamma }) \right]
\right\}\label{eq:NTSWWM1}\\
[N_\mL (E_{\gamma })]_{SWWM} &= 0,\label{eq:NLSWWM1}%
\end{align}
\end{subequations}
where
\begin{equation}
N_0 \equiv \frac{2}{\mpi }\, \frac{(Q^{\gamma })^2\alpha}{v_f^2}=const\label{eq:N0SWWM}%
\end{equation}
and
\begin{equation}
\chi _{\gamma } = \frac{E_{\gamma }b_{min}}{\gamma _fv_f}.\label{eq:chiSWWM}%
\end{equation}
The parameter $Q^{\gamma }$ in Eq. (\ref{eq:N0SWWM}) is the
dimensionless electric charge of the fermion (cf. Table
\ref{tab:charges}), and $\alpha=e^2/4\mpi \simeq 1/137$ is the
usual fine structure constant. The correct value of $b_{min}$ to
use is
\begin{equation}
b_{min} = \frac{\eta _{\gamma }}{\gamma _f(\Delta p_{{\gamma }\perp})v_f} = \frac{\eta _{\gamma}}{E_\gamma }, \label{eq:bminSWWM1}%
\end{equation}
if the colliding particles are \emph{pointlike}, and
\begin{equation}
b_{min} = \mbox{the greater of $R_1+R_2$ and } \frac{\eta _{\gamma}}{\gamma _fm_fv_f}, \label{eq:bminSWWM2}%
\end{equation}
if the colliding particles are \emph{nuclei}. As discussed
previously, $R_1$ and $R_2$ are the radii of the two colliding
particles (typically $R_1=R_2$), $\eta _{\gamma }\sim 1$, and
$m_f$ is the mass of the lighter of the two colliding particles.
Also, the above formulas are only valid so long as $E_{\gamma }$
is bounded (by 4-momentum conservation and causality) within the
range
\begin{equation}
(E_{\gamma })_{min} \le E_{\gamma } \le (E_{\gamma })_{max},\label{eq:EboundsSWWM}%
\end{equation}
where
\begin{subequations}
\begin{align}
(E_{\gamma })_{min} &\equiv 0\label{eq:defxminSWWM}\\%
(E_{\gamma })_{max} &\equiv E_f-m_f\simeq E_f.\label{eq:defxmaxSWWM}%
\end{align}
\end{subequations}
The $\chi _{\gamma }\to 0$ and $\chi _{\gamma } \to \infty$
limiting forms of the SWWM number spectra are listed in Sec.
\ref{sec:limNTNLGWWM}.

A different semiclassical approach to photon distribution
functions was worked out by J\"{a}ckle and Pilkuhn \cite{ref:JaPi}
(see also \cite{ref:Bert}). They considered photons radiating from
projectiles moving at \emph{arbitrary} velocities in the eikonal
approximation (i.e., along straight-line trajectories). With
various additional kinematical assumptions, they deduced a set of
number spectrum functions, one for each of all the possible
electric and magnetic multipolarity states (labelled by different
values of $\ell$, the orbital angular momentum quantum number) of
the photons. Previously, it had been worked out that the E1
(electric $\ell=1$ multipolarity) number spectrum is the one
considered in the SWWM, and that all multipolarity number spectrum
functions are equal in the $v_f\to 1$ limit \cite{ref:Bert}.
Taking the projectile to be a point particle, the E1 number
spectra found in their analysis are
\begin{subequations}
\begin{align}
[N_\mT (E_{\gamma })]_{JP} &= \frac{N_0v_f^2}{2E_{\gamma }}\,
\Bigg\{ \chi _{\gamma}^2\Bigl[ K_0( \chi _{\gamma })K_2(\chi
_{\gamma })-K_1^2(\chi _{\gamma })-2K_0( \phi _{\gamma } )\bigl(
K_2\{\chi _{\gamma }\}- \nonumber\\ & -K_0\{\chi _{\gamma
}\}\bigr) \Big] + \frac{\chi _{\gamma }^2}{\gamma _f^2}\, \left[
K_1^2(\chi _{\gamma })-K_0^2(\chi _{\gamma }) \right]+4\phi
_{\gamma } K_0(\phi _{\gamma })K_1(\phi _{\gamma }) \Bigg\}
\label{eq:NTJP}\\
[N_\mL (E_{\gamma })]_{JP} &= 0,\label{eq:NLJP}%
\end{align}
\end{subequations}
where $\chi _{\gamma }=E_{\gamma }b_{min}/\gamma _fv_f$ (as
usual), $\phi _{\gamma }\equiv \gamma _f\chi _{\gamma }=E_{\gamma
}b_{min}/v_f$, and the $K_{\nu}$s are modified Bessel functions of
the second kind, of order $\nu =0,\, 1,\, 2$. Eq. (\ref{eq:NTJP})
does not reduce to the $N_\mT (E_{\gamma })$ of the SWWM (Eq.
(\ref{eq:NTSWWM1})) in general, but is \emph{very nearly} the same
in the $v_f\to 1$ limit. The differences in that limit are
attributed to the small kinematic corrections used by J\"{a}ckle
and Pilkuhn \cite{ref:Bert}. In short, then, Eq.
(\ref{eq:NTSWWM1}) correctly represents the number spectrum of
photons in any semiclassical model in which the projectile is
moving ultrarelativistically, which is the case of interest in
this study.

\section{The QWWM Number Spectra} \label{sec:NTNLQWWM} \indent

The QWWM is the quantum mechanical version of the SWWM. Like in
the SWWM, the photons are treated as on-shell, so only $N_\mT
(E_{\gamma })$ is nonvanishing. Unlike in the SWWM, $N_\mT
(E_{\gamma })$ is derived in the QWWM by applying the Feynman
rules to a specific interaction. For example, Dalitz and Yennie
considered pion production via one-photon exchange in
electron-nucleon scattering: $e^- + p\to e^- + \gamma + p\to n +
\mpi ^+ + e^-$ \cite{ref:Dali}. Terazawa considered a similar
process, as well as a more complicated one: particle production
via \emph{two-photon} exchange in electron-positron (or
electron-electron) scattering, $e^{\pm} + e^-\to e^{\pm} + \gamma
+ e^- + \gamma \to e^{\pm} + e^- + X$, where $X$ may be a lepton
or neutral $C=+1$ hadron state \cite{ref:Tera}. These references
are by no means the only good ones on quantum derivations of the
WWM number spectra. Detailed Feynman diagram analyses have been
presented by Kwang-Je Kim and Yung-Su Tsai, Bonneau and Martin,
and Vidovi\'{c} et al., to merely name a few
\cite{ref:KiTs,ref:BoMa,ref:Vido}; citations to other useful
references can be found in all of these papers. It is not the
purpose of this section to survey the literature on the subject
--- only to list the relevant formulas to which to compare the
GWWM number spectra. It is noteworthy, however, to point out two
common simplifying approximations used in all of these analyses,
which are really at ``...the heart of the equivalent photon
approach" \cite{ref:Vido}. One is the approximation that the
photons are on-shell, or $k_{\gamma }^2\simeq 0$, in the language
used here. The other is the neglect of the matrix elements
corresponding to longitudinally-polarized photons, which are
always found to be suppressed by a factor of $1/\gamma _f^2$
relative to the matrix elements corresponding to purely transverse
photon states. Dalitz and Yennie arrived at the same formula for
the number spectra that Terazawa found, and it is the one stated
in the third edition of Jackson's {\it Classical Electrodynamics}
\cite{ref:Jack2}.
\begin{subequations}
\begin{align}
[N_\mT (E_{\gamma })]_{QWWM} &= \frac{N_0}{2E_{\gamma }}\Bigg\{ \frac{E_f^2+{{E_f}'}^2}{p_f^2}\, \ln \left[ \frac{E_f{E_f}'+p_f{p_f}'+m_f^2}{m_fE_{\gamma }} \right] - \nonumber \\ & - \frac{(E_f+{E_f}')^2}{2p_f^2}\, \ln \left( \frac{p_f+{p_f}'}{p_f-{p_f}'} \right) -\frac{{p_f}'}{p_f} \Bigg\} \label{eq:NTQWWM1}\\
[N_\mL (E_{\gamma })]_{QWWM} &= 0,\label{eq:NLQWWM1}%
\end{align}
\end{subequations}
where ${E_f}'=E_f-E_{\gamma}$. As in the previous section, the
$N_0$ here is
\begin{equation}
N_0 \equiv \frac{2}{\mpi }\, \frac{(Q^{\gamma })^2\alpha}{v_f^2}=const.\label{eq:N0QWWM}%
\end{equation}
These functions were derived for the case of photons radiating
from UR electrons, but are more generally applicable to photons
radiating from \emph{any} UR spin-1/2 pointlike particle. The
$E_{\gamma }\to 0$ limiting forms of these formulas are
\begin{subequations}
\begin{align}
\lim _{E_{\gamma }\to 0}[N_\mT (E_{\gamma })]_{QWWM} &= \frac{N_0}{E_{\gamma }}\, \ln \left( 0.6065\, \gamma _f \right) \label{eq:NTQWWM0}\\%
\lim _{E_{\gamma }\to 0}[N_\mL (E_{\gamma })]_{QWWM} &= 0,\label{eq:NLQWWM0}%
\end{align}
\end{subequations}
where the constant $0.6065$ in Eq. (\ref{eq:NTQWWM0}) is more
precisely $\mbox{\large e} ^{-1/2}$.

\section{The EWM Number Spectra} \label{sec:NTNLEWM} \indent

The EWM is the weak force analog of the QWWM. The mediating bosons
are treated as on-shell partons within the parent fermion. Since
the $W$ and $Z$ bosons are all massive, effects of longitudinal
boson polarization states are no longer insignificant. Just as in
the QWWM, the number spectrum functions are derived by applying
the Feynman rules to a specific interaction. Almost all of the
references consider processes of the type $f_1 + f_2 \to {f_1}' +
b_1 + {f_2}' + b_2 \to {f_1}' + {f_2}' + R$, where $f$ is a quark
(more generally, a light fermion), $b$ is any of the weak force
vector bosons ($b=W^+$, $W^-$ or $Z$), and $R$ is some resonant
particle state, such as a Higgs boson
\cite{ref:Daws1,ref:Daws2,ref:Guni,ref:Kane,ref:Cahn1,ref:Alta};
see Fig. \ref{fig:Rprod} for the Feynman diagram. So the two
colliding quarks exchange a pair of massive bosons, which
subsequently fuse and form the resonant state $R$. According to
Cahn, it is not possible to calculate the full cross section for
this reaction analytically, so various auxiliary assumptions must
be made \cite{ref:Cahn1}. Besides the on-shell approximation, the
quarks and bosons are assumed to be ultrarelativistic, so that
their masses are always negligible compared to their energies, the
quark scattering angles are assumed to be small, and the
interference between the transverse and longitudinal states of the
bosons is assumed to be insignificant
\cite{ref:Daws1,ref:Daws2,ref:Guni,ref:Kane,ref:Cahn1,ref:Alta}.
It is important to keep in mind that the EWM is only applicable
provided that the fermion energies $E_f$ are greater than or equal
to the boson mass $m_b$: $E_f\ge m_b$. This restriction follows
from conservation of energy, and implies (by conservation of
energy again) that only particles $R$ whose mass $m_\mR $ is
greater than or equal to twice the boson mass, $m_\mR  \ge 2m_b$,
can be produced. Perhaps the simplest and most informative forms
for the number spectrum functions are presented in
\cite{ref:Kane}. In the language used in this thesis, these
functions are as follows.
\begin{subequations}
\begin{align}
[N_\mT (E_b)]_{EWM} &= \frac{N_0v_f^2}{4E_b}\, \left[ 1+\left(
1-\frac{E_b}{E_f}\right) ^2\right] \, \ln \left[
\frac{(p_{b\perp})_{max}^2+(1-E_b/E_f)m_b^2}{(1-E_b/E_f)m_b^2}
\right] \label{eq:NTEWM1}\\
[N_\mL (E_b)]_{EWM} &= \frac{N_0v_f^2}{2E_b}\, \left(
1-\frac{E_b}{E_f}\right) \,
\frac{(p_{b\perp})_{max}^2}{(p_{b\perp})_{max}^2+(1-E_b/E_f)m_b^2}
.\label{eq:NLEWM1}%
\end{align}
\end{subequations}
The $N_0$ in these equations is like the one in the GWWM:
\begin{equation}
N_0 \equiv \frac{1}{2\mpi ^2}\, \frac{q_V^2+q_A^2}{v_f^2}=const.\label{eq:N0EWM}%
\end{equation}
$(p_{b\perp})_{max}$ is the magnitude of the maximum transverse
component of the boson's 3-momentum. Unlike in the GWWM, $m_b$ is
the \emph{on-shell} value of the boson mass. So, $m_W=80.42$ GeV
(for $b=W$ bosons) and $m_Z=91.19$ GeV (for $b=Z$ bosons)
\cite{ref:RPP}. Two limiting forms are of interest: the
$(p_{b\perp})_{max} \gg m_b$ limit and the $(p _{b\perp})_{max}
\ll m_b$ limit. All of the above-mentioned studies use the former
limit to simplify their calculations. They all arrive at the same
limiting functions:
\begin{subequations}
\begin{align}
[N_\mT (E_b)]_{EWM} \bigg| _{(p_{b\perp})_{max} \gg m_b} &= \frac{N_0}{2E_b}\, \left[ 1+\left( 1-\frac{E_b}{E_f}\right) ^2\right] \, \ln \left[ \frac{(p _{b\perp})_{max}}{m_b} \right] \label{eq:NTEWM01}\\
[N_\mL (E_b)]_{EWM} \bigg| _{(p_{b\perp})_{max} \gg m_b} &= \frac{N_0}{2E_b}\, \left( 1-\frac{E_b}{E_f}\right) .\label{eq:NLEWM01}%
\end{align}
\end{subequations}
If, furthermore, the $E_b\to 0$ limit is taken, these functions
reduce to
\begin{subequations}
\begin{align}
\lim _{E_b\to 0}\left\{ [N_\mT (E_b)]_{EWM}\right\} \bigg| _{(p_{b\perp})_{max} \gg m_b} &= \frac{N_0}{E_b}\, \left( 1-\frac{E_b}{E_f}\right) \, \ln \left[ \frac{(p _{b\perp})_{max}}{m_b} \right] \label{eq:NTEWM02}\\
\lim _{E_b\to 0}\left\{ [N_\mL (E_b)]_{EWM}\right\} \bigg| _{(p_{b\perp})_{max} \gg m_b} &= \frac{N_0}{2E_b}\, \left( 1-\frac{E_b}{E_f}\right) .\label{eq:NLEWM02}%
\end{align}
\end{subequations}
In the $(p_{b\perp})_{max} \ll m_b$ limit, the number spectrum
functions become
\begin{subequations}
\begin{align}
[N_\mT (E_b)]_{EWM} \bigg| _{(p_{b\perp})_{max} \ll m_b} &= \frac{N_0}{4E_b}\, \left[ 1+\left( 1-\frac{E_b}{E_f}\right) ^2\right] \, \ln \left[ 1 + \frac{(p_{b\perp})_{max}^2}{m_b^2} \right] \label{eq:NTEWMinf1}\\
[N_\mL (E_b)]_{EWM} \bigg| _{(p_{b\perp})_{max} \ll m_b} &= \frac{N_0}{2E_b}\, \left( 1-\frac{E_b}{E_f}\right) \, \frac{1}{1+m_b^2/(p_{b\perp})_{max}^2} .\label{eq:NLEWMinf1}%
\end{align}
\end{subequations}
In the $E_b\to 0$ limit, these functions reduce to
\begin{subequations}
\begin{align}
\lim _{E_b\to 0}\left\{ [N_\mT (E_b)]_{EWM}\right\} \bigg| _{(p_{b\perp})_{max} \ll m_b} &= \frac{N_0}{2E_b}\, \left( 1-\frac{E_b}{E_f}\right) \, \ln \left[ 1 + \frac{(p_{b\perp})_{max}^2}{m_b^2} \right] \label{eq:NTEWMinf2}\\
\lim _{E_b\to 0}\left\{ [N_\mL (E_b)]_{EWM}\right\} \bigg| _{(p _{b\perp})_{max} \ll m_b} &= \frac{N_0}{2E_b}\, \left( 1-\frac{E_b}{E_f}\right) \,  \frac{1}{1+m_b^2/(p_{b\perp})_{max}^2} .\label{eq:NLEWMinf2}%
\end{align}
\end{subequations}
This latter set of limits was used by Papageorgiu in examining
Higgs boson production via vector boson fusion in coherent
relativistic heavy-ion collisions, ``coherent" meaning that the
nuclei do not break up in the process \cite{ref:Papa}. She also
identified the minimum impact parameter $b_{min}$ with $1/(p
_{b\perp})_{max}$, and with the nuclear radius. These assignments
are similar to the ones made in Sec. \ref{sec:Bmin} (cf. Eqs.
(\ref{eq:bminQM21}) and (\ref{eq:bminnuclei1})). So the above
expressions can be written in a form that better facilitates
comparing with the GWWM, which is formulated in terms of $b_{min}$
instead of $(p _{b\perp})_{max}$.
\begin{subequations}
\begin{align}
\lim _{E_b\to 0}\left\{ [N_\mT (E_b)]_{EWM}\right\} \bigg| _{m_bb_{min} \ll 1} &= \frac{N_0}{E_b}\, \left( 1-\frac{E_b}{E_f}\right) \, \ln \left( \frac{1}{m_bb_{min}} \right) \label{eq:NTEWM03}\\
\lim _{E_b\to 0}\left\{ [N_\mL (E_b)]_{EWM}\right\} \bigg| _{m_bb_{min} \ll 1} &= \frac{N_0}{2E_b}\, \left( 1-\frac{E_b}{E_f}\right) .\label{eq:NLEWM03}%
\end{align}
\end{subequations}
And
\begin{subequations}
\begin{align}
\lim _{E_b\to 0}\left\{ [N_\mT (E_b)]_{EWM}\right\} \bigg| _{m_bb_{min} \gg 1} &= \frac{N_0}{2E_b}\, \left( 1-\frac{E_b}{E_f}\right) \, \ln \left( 1 + \frac{1}{m_b^2b_{min}^2} \right) \label{eq:NTEWMinf3}\\
\lim _{E_b\to 0}\left\{ [N_\mL (E_b)]_{EWM}\right\} \bigg| _{m_bb_{min} \gg 1} &= \frac{N_0}{2E_b}\, \left( 1-\frac{E_b}{E_f}\right) \,  \frac{1}{1+m_b^2b_{min}^2} .\label{eq:NLEWMinf3}%
\end{align}
\end{subequations}

\section{Comparisons for Point Particles} \label{sec:PointParticles} \indent

If the colliding particles are pointlike, the correct form for the
minimum impact parameter in the GWWM is given in Eq.
(\ref{eq:bminGWWM1}). For clarity, parameters (viz, boson mass and
minimum impact parameter) in theories other than the GWWM will be
denoted with capital letters in this section and the next. The
same parameters in the GWWM will be denoted with lower case
letters, in the usual way. Thus,
\begin{equation}
b_{min}=\frac{\eta _b}{\gamma _f(\Delta p_{b\perp})v_f},\label{eq:bminQM22}%
\end{equation}
and therefore
\begin{equation}
\chi _b = \half b_{min}\Delta p_{b\perp} = \frac{\eta _b}{2\gamma _fv_f} \ll 1,\label{eq:bminQM23}%
\end{equation}
since $\gamma _f \gg 1$ and $\eta _b \sim v_f \sim 1$. Therefore,
the appropriate limiting expressions for the GWWM number spectra
are those listed in Eqs. (\ref{eq:NTph0})--(\ref{eq:NTWb0}) and
(\ref{eq:NLph0})--(\ref{eq:NLWb0}). By a judicious choice of $\eta
_b$, very good agreement can be found between the number spectrum
functions of the GWWM and those of other theories. Of course, the
GWWM agrees \emph{exactly} with the SWWM in the EM limit (by
construction), so the other theories here are the QWWM and EWM.
The general procedure for pinpointing an exact value of $\eta _b$
for a given application is to \emph{demand} that the GWWM $N_\mT $
and $N_\mL $ functions in the low boson energy (or, in the $x\to
0$) limit agree exactly with the same functions, in the same
limit, appearing in other theories. Then, a unique value of
$b_{min}$ becomes obvious, and consequently the corresponding
particular value of $\eta _b$ is identified. All values of $\eta
_b$ turn out to be $\sim 1$, as they should!

First consider the $N_\mT $ functions. The general form for the
low energy $N_\mT $ function in the GWWM is given in Eq.
(\ref{eq:NTGWWM4}). Following the above procedure, and (at the
risk of causing confusion) denoting the $N_\mT $ function found in
either of the other two theories as simply $N_\mT $ (for
simplicity),
\begin{subequations}
\begin{align}
\frac{N_0}{E_b}\, \ln \left[ \frac{2\mbox{\large e} ^{-(\gamma +v_f^2/2)}}{\chi _b} \right] &\equiv N_\mT  \label{eq:compp1}\\%
\frac{2\mbox{\large e} ^{-(\gamma +v_f^2/2)}}{\chi _b} &= \mbox{\Large e} ^{N_\mT E_b/N_0} \label{eq:compp2}\\%
{\chi _b} &= 2\mbox{\large e} ^{-(\gamma +v_f^2/2)}\, \mbox{\Large e} ^{-N_\mT E_b/N_0} \label{eq:compp3}\\%
\frac{\eta _b}{2\gamma _fv_f} &= 2\mbox{\large e} ^{-(\gamma +v_f^2/2)}\, \mbox{\Large e} ^{-N_\mT E_b/N_0} \qquad \mbox{(via Eq. (\ref{eq:bminQM23}))} \label{eq:compp4}\\%
\eta _b &= 4\gamma _fv_f\mbox{\large e} ^{-(\gamma +v_f^2/2)}\, \mbox{\Large e} ^{-N_\mT E_b/N_0}.\label{eq:compp5}%
\end{align}
\end{subequations}

When comparing to the QWWM, the correct $N_\mT $ function to use
is listed in Eq. (\ref{eq:NTQWWM0}). Plugging this function into
Eq. (\ref{eq:compp5}) yields
\begin{subequations}
\begin{align}
\eta _{\gamma} &= 4\gamma _fv_f\mbox{\large e} ^{-(\gamma +v_f^2/2)}\, \mbox{\Large e} ^{-\ln (\gamma _f\me ^{-1/2})} \label{eq:GWWMQWWMT1}\\%
&= 4\gamma _fv_f\mbox{\large e} ^{-(\gamma +v_f^2/2)}\, \left( \frac{\mbox{\large e} ^{1/2}}{\gamma _f}\right) \label{eq:GWWMQWWMT2}\\%
&= 4v_f\mbox{\large e} ^{-(\gamma +v_f^2/2-1/2)}\label{eq:GWWMQWWMT3}\\%
&= 4v_f\mbox{\large e} ^{-(\gamma -1/2\gamma _f^2)}\label{eq:GWWMQWWMT4}\\%
&\simeq 4\mbox{\large e} ^{-\gamma}\qquad \mbox{(via $v_f\simeq 1$ and $\gamma _f \gg 1$)}\label{eq:GWWMQWWMT5}\\%
&= 2.246\qquad \mbox{(to four significant figures).}\label{eq:GWWMQWWMT6}%
\end{align}
\end{subequations}
So $\eta _{\gamma}$ is found to be $\sim 1$, as it should be.

The $N_\mT $ function for massive bosons in the EWM in this same
limit is given in Eq. (\ref{eq:NTEWM03}). As shown, that equation
is actually the form for $N_\mT $ in the $E_b \to 0$ \emph{and}
$M_bB_{min} \ll 1$ limits. Use of this second limit here is
justified because it is equivalent to the $\chi _b\to 0$ (or $\chi
_b \ll 1$) limit being used in the GWWM, which can be seen by
noting that $\chi _b=b_{min}\Delta {p_{b\perp}}/2$ and $\Delta
p_{b\perp}/2\to m_b$ in the $E_b\to 0$ limit (via Eq.
(\ref{eq:chipbT3})). So, since $\chi _b=b_{min}m_b$ and $\chi
_b\ll 1$ is being considered in the GWWM, it seems perfectly
reasonable that the same limit, $M_bB_{min} \ll 1$, be used for
the EWM formula to which the GWWM is to be compared. It is also
the correct limit that is used in the literature
\cite{ref:Daws1,ref:Daws2,ref:Guni,ref:Kane,ref:Cahn1,ref:Alta}.
As a further simplification, the factor $(1-E_b/E_f)$
premultiplying Eq. (\ref{eq:NTEWM03}) will be discarded because it
tends to unity in the $E_b\to 0$ limit. Plugging in the resulting
expression for $N_\mT $ in Eq. (\ref{eq:compp5}) yields
\begin{subequations}
\begin{align}
\eta _b &= 4\gamma _fv_f\mbox{\large e} ^{-(\gamma +v_f^2/2)}\, \mbox{\Large e} ^{+\ln (M_bB_{min})} \label{eq:GWWMEWMT1}\\%
&= 4\gamma _fv_f\mbox{\large e} ^{-(\gamma +v_f^2/2)}\, (M_bB_{min}) .\label{eq:GWWMEWMT2}%
\end{align}
\end{subequations}
As discussed in the previous section, $B_{min}=1/(p
_{b\perp})_{max}$. Furthermore, for on-shell partons with
well-defined energies and momenta, $(p _{b\perp})_{max}$ is
naturally identified with $E_b$, so that $B_{min}=1/E_b$ in the
EWM. The parton theory posits that $x\equiv E_b/E_f=p_{b||}/p_f$
and $\vecp _{b\perp }=\veczero$ (on average), from which it can
easily be deduced that $M_b=xM_f$ \cite{ref:Halz}. Then, since
$E_b=xE_f=(M_b/M_f)(\gamma _fM_f)$, the identification $E_b=\gamma
_fM_b$ can be made. With this ansatz, the minimum impact parameter
in the EWM is found to be $B_{min}=1/\gamma _fM_b$. Eq.
(\ref{eq:GWWMEWMT2}) becomes
\begin{subequations}
\begin{align}
\eta _b &= 4\gamma _fv_f\mbox{\large e} ^{-(\gamma +v_f^2/2)}\, \left( \frac{M_b}{\gamma _fM_b}\right) \label{eq:GWWMEWMT3}\\%
&= 4v_f\mbox{\large e} ^{-(\gamma +v_f^2/2)} \label{eq:GWWMEWMT4}\\%
&\simeq 4\mbox{\large e} ^{-(\gamma +1/2)}\qquad \mbox{(via $v_f\simeq 1$ and $\gamma _f \gg 1$)}\label{eq:GWWMEWMT5}\\%
&= 1.362\qquad \mbox{(to four significant figures).}\label{eq:GWWMEWMT6}%
\end{align}
\end{subequations}
Therefore, $\eta _b\sim 1$ here as well, as it should.

Having now compared the $N_\mT $ functions among the various
theories, consider the $N_\mL $ functions. The $\chi _b\to 0$
limiting form for the $N_\mL $ function for massive equivalent
vector bosons in the GWWM is given in Eq. (\ref{eq:NLGWWM5}).
Since $\alpha _{\gamma}=0$ for photons, $N_\mL $ vanishes
identically in the GWWM, just as it does in the SWWM and QWWM. So,
the $N_\mL $ functions are always in perfect agreement for EM
interactions. To compare to the same function in the EWM (for
massive mediators), the $E_b\to 0$ and $E_f\simeq p_f$ limits are
taken. The resulting expression can be written
\begin{equation}
\lim _{E_b\to 0}\left\{ [N_\mL (E_b)]_{GWWM}\right\} \bigg| _{\chi _b \ll 1} = \frac{N_0}{2E_b}\, \frac{(1-E_b/E_f)}{[1+(1-\alpha _b)E_b/\alpha _bE_f]} .\label{eq:NLGWWM13}%
\end{equation}
The $N_\mL $ function for massive bosons in the EWM in this same
limit is given in Eq. (\ref{eq:NLEWM03}).
\begin{equation}
\lim _{E_b\to 0}\left\{ [N_\mL (E_b)]_{EWM}\right\} \bigg| _{M_bB_{min} \ll 1} = \frac{N_0}{2E_b}\, \left( 1-\frac{E_b}{E_f}\right) .\label{eq:NLEWM04}%
\end{equation}
Upon comparing these equations, it is apparent that agreement
between the GWWM and EWM longitudinal boson number spectra is only
achieved if the quantity $(1-\alpha _b)E_b/\alpha _bE_f$ appearing
in the denominator of Eq. (\ref{eq:NLGWWM13}) is insignificantly
small. As has been noted previously, $\alpha _b$ is typically
$\simeq 1$ and it is only the $E_b \ll E_f$ limit that is of
interest here, so this condition generally holds. In practice, the
actual numerical values of the corresponding $N_\mL $ functions in
the two theories only differ by a few percent!

Results from this section (for bosons radiating from electrons)
are shown in Figs. \ref{fig:frsp00} -- \ref{fig:mb20}. Fig.
\ref{fig:frsp00} shows the helicity-averaged frequency spectra
(cf. Eqs. (\ref{eq:dIdwdA1exp2}) -- (\ref{eq:dIdwdA3exp2})) for
the three WW pulses of photons radiating from a 500 GeV electron.
These functions generally depend on the boson energy $E_b$ as well
as the impact parameter $b$. To only display the energy
dependence, $b$ was set equal to the relevant minimum impact
parameter, $b_{min}$. For a given $E_b$, this choice for $b$ gives
the \emph{maximum} value that the frequency spectra will take.
Pulse 1, which represents transversely-polarized photons
travelling collinearly with the electron, is by far the most
dominant of the three pulses. Pulse 2, which is the fictitious
pulse travelling in a direction transverse to the electron's
velocity, is relatively negligible; it is shown amplified by a
factor of $\gamma _e^2$ so that it appears on the same graph as
Pulse 1. Pulse 3, which represents longitudinally-polarized
photons, does not appear at all on the graph because it vanishes
altogether. The number spectra for transversely-polarized photons,
as calculated via the GWWM (Eq. (\ref{eq:NTGWWM1})), or,
equivalently, the SWWM (Eq. (\ref{eq:NTSWWM1})), is compared to
the same function calculated via the QWWM (Eq. (\ref{eq:NTQWWM1}))
in Fig. \ref{fig:nt00A} and \ref{fig:nt00B}, for a particle
accelerator operating at beam energies of 500 and 1000 GeV,
respectively. For the 500 (1000) GeV electron case, the relative
differences between the functions rise from $0\%$ at low photon
energies to $169\%$ ($165\%$) at the highest possible photon
energies. The helicity-averaged frequency spectra, evaluated at
$b=b_{min}$, for the three pulses of equivalent $Z$ bosons ($W$
bosons) outside a 500 GeV electron are shown in Fig.
\ref{fig:frsp10} (Fig. \ref{fig:frsp20}). Like in the photon case,
Pulse 1 exceeds Pulse 2 by a factor of at least $\gamma _e^2$, but
unlike the photon case, Pulse 3, which describes
longitudinally-polarized $Z$ bosons ($W$ bosons), is not
completely negligible. Fig. \ref{fig:nt10A} (Fig. \ref{fig:nt20A})
compares the number spectra for transversely-polarized $Z$ bosons
($W$ bosons) radiating from a 500 GeV electron, as calculated via
the GWWM (Eq. (\ref{eq:NTGWWM1})), to the same function as
calculated via the EWM (Eq. (\ref{eq:NTEWM1})). Figs.
\ref{fig:nt10B} and \ref{fig:nt20B} show the same comparisons for
a collider operating at a beam energy of 1000 GeV. Relative
discrepancies rise from $0\%$ at low boson energies to $33\%$ at
high boson energies for both the $Z$ boson and $W$ boson, and both
the 500 and 1000 GeV electron, cases. Fig. \ref{fig:nl10A} (Fig.
\ref{fig:nl20A}) compares the number spectra for
longitudinally-polarized $Z$ bosons ($W$ bosons) radiating from a
500 GeV electron, as calculated via the GWWM (Eq.
(\ref{eq:NLGWWM1})), to the same function as calculated via the
EWM (Eq. (\ref{eq:NLEWM1})). Figs. \ref{fig:nl10B} and
\ref{fig:nl20B} show the same comparisons for a collider operating
at a beam energy of 1000 GeV. For the $Z$ boson case, relative
errors rise from $0\%$ at low boson energies to $7\%$ at high
boson energies, for both a 500 and 1000 GeV electron. For the $W$
boson case, relative errors remain at a steady $10^{-9}\, \%$ and
$10^{-8}\, \%$ for the 500 and 1000 GeV electrons, respectively,
as $x$ varies from $0$ to $1$. The mass of an equivalent $Z$ or
$W$ boson varies with the boson's energy according to Eq.
(\ref{eq:mbGWWM}). This function is shown in Fig. \ref{fig:mb10}
for $Z$ bosons and Fig. \ref{fig:mb20} for $W$ bosons.

To conclude, excellent agreement is found between the number
spectrum functions of the GWWM and other theories, for bosons
radiating from point particles. To achieve this agreement, the
free parameter $\eta _b$ (first introduced in Eq.
(\ref{eq:bminQM21})) appearing in the GWWM is always found to be
on the order of unity, as it is expected to be. When comparing the
$N_\mT $ for photons of the GWWM (i.e., the SWWM) to the same
function in the QWWM, $\eta _{\gamma }$ is found to be $4\me
^{-\gamma}=2.246$. And when comparing the $N_\mT $ for massive
bosons of the GWWM to the same function in the EWM, $\eta _b$ is
found to be $4\me ^{-(\gamma +1/2)}=1.362$. The $N_\mL $ function
for photons vanishes in both the GWWM (or the SWWM) and QWWM. The
expression for $N_\mL $ for massive bosons of the GWWM is found to
be equal to that in the EWM in the $E_b \to 0$ limit, which has
been pointed out previously is the limit of interest, and if
$\alpha _b\simeq 1$, which turns out to be almost always the case.

\section{Comparisons for Composite Particles} \label{sec:ComParticles} \indent

If the colliding particles are composite, the correct form for the
minimum impact parameter in the GWWM is complicated. If the
mediators are photons, Eq. (\ref{eq:bminGWWM21}) is used. As the
GWWM simplifies to the well documented SWWM in this limit, the
reader is referred to the literature on this subject. See
\cite{ref:Jack1}, \cite{ref:Jack2}, \cite{ref:Eich}, or any of the
many papers by Baur, Bertulani, Greiner, or Soff, to name a few.
In practice, the parameter $\eta _f$ appearing in Eq.
(\ref{eq:bminGWWM21}) can simply be set to $1$, as $R_1+R_2$ is
almost always (for applications to relativistic heavy ion
colliders) much greater than $\eta _f/\gamma _fm_fv_f$. The focus
thus turns to how the GWWM compares to the EWM when analyzing
nucleus-nucleus collisions. The references on this subject are
scarce, to say the least. One author who does calculate the number
spectra for massive bosons emitted from heavy ions is Papageorgiu
\cite{ref:Papa}. She uses the formulas of Kane et al., but in the
opposite limit as was considered for applications to collisions
between point particles \cite{ref:Kane}. That is, instead of
evaluating the general equations in the $(p _{b\perp})_{max}\gg
M_b$ limit, she evaluates them in the $(p _{b\perp})_{max}\ll M_b$
limit (cf. Eqs. (\ref{eq:NTEWMinf2})--(\ref{eq:NLEWMinf2})).
Furthermore, she identifies $(p _{b\perp})_{max}$ with $1/R$,
where $R$ is the radius of the parent nucleus, so the limit of
interest can be written $M_bB_{min} \gg 1$, where $B_{min}=R$.
Written this way, the number spectrum functions of interest are
Eqs. (\ref{eq:NTEWMinf3})--(\ref{eq:NLEWMinf3}). In view of the
fact that $M_bB_{min} \gg 1$, the factor $\ln
(1+1/M_b^2B_{min}^2)$ in Eq. (\ref{eq:NTEWMinf3}) reduces to
$1/M_b^2B_{min}^2$, and the factor $1/(1+M_b^2B_{min}^2)$ in Eq.
(\ref{eq:NLEWMinf3}) reduces to the same quantity,
$1/M_b^2B_{min}^2$. The equations can be written
\begin{subequations}
\begin{align}
\lim _{E_b\to 0}\left\{ [N_\mT (E_b)]_{EWM}\right\} \bigg| _{M_bB_{min} \gg 1} &= \frac{N_0}{2E_b}\, \frac{1}{M_b^2B_{min}^2} \label{eq:NTEWMinf4}\\%
\lim _{E_b\to 0}\left\{ [N_\mL (E_b)]_{EWM}\right\} \bigg| _{M_bB_{min} \gg 1} &= \frac{N_0}{2E_b}\, \frac{1}{M_b^2B_{min}^2} .\label{eq:NLEWMinf4}%
\end{align}
\end{subequations}
Clearly, in these limits, $N_\mT =N_\mL $, which is pointed out by
Papageorgiu. To compare the results of the GWWM with her results,
first note that the correct expression to use for $b_{min}$ for
such applications is given in Eq. (\ref{eq:bminGWWM22}). From this
equation and the definition of $\chi _b$, it is apparent that
\begin{equation}
\chi _b = \half b_{min}\Delta p_{b\perp} = \eta _b.\label{eq:bminGWWM23}%
\end{equation}
By the usual assumption that $\eta _b\sim 1$, it therefore follows
that $\chi _b\sim 1$. By demanding that $\chi _b$ not be $\ll 1$
or $\gg 1$, the Bessel functions appearing in the expressions for
$N_\mT $ and $N_\mL $ (cf. Eqs. (\ref{eq:NTGWWM1}) and
(\ref{eq:NLGWWM1})) do not simplify. But, if one considers the
limit of interest, $E_b\to 0$, and recalls Eq. (\ref{eq:chipbT3}),
one finds $\Delta p_{b\perp}/2\to m_b$, so that $b_{min}$ can be
written
\begin{equation}
b_{min} = \frac{\eta _b}{m_b},\label{eq:bminGWWM24}%
\end{equation}
which is the familiar relation relating the range of a force to
the mass of the associated messenger boson. This equation can be
rewritten as $m_bb_{min} = \eta _b$. It was just this quantity,
$M_bB_{min}$, that Papageorgiu assumed was $\gg 1$; in comparison
(and in contrast), again, it is assumed here that $m_bb_{min}\sim
1$. It is not too far-fetched, however, to allow $m_bb_{min} \gg
1$, or, equivalently, $\chi _b \gg 1$, here as well. A careful
examination of a table of values of the modified Bessel functions
appearing in the GWWM shows that the values of these functions
evaluated at $\chi _b=1$ differ from their values in the $\chi
_b\gg 1$ limits (cf. Eqs. (\ref{eq:K0gg1}) and (\ref{eq:K1gg1})),
also evaluated at $\chi _b=1$, by relative errors at only the 4\%
and 5\% levels, respectively \cite{ref:Abra}. Allowing for the
possibility that $\chi _b$ is somewhat greater than $1$ arrives at
even better agreement. For instance, those relative errors are
only 0.2\% and 0.4\%, respectively, when $\chi _b$ is set equal to
$5$; and, they are both about 0.1\% when $\chi _b$ is set equal to
$10$. For the sake of simplicity, then, it is assumed here that
$\chi _b \gg 1$ for applications of the method to nucleus-nucleus
collisions. The values of $\chi _b$ that are ultimately found to
work are between about $6$ and $9$, for \emph{all} nuclei. So
while this approximation is not in accord with the usual scenario,
that $\eta _b\simeq 1$, it greatly simplifies the analysis, and
arrives at very good results. With this limiting form for $\chi
_b$, the appropriate limiting expressions for the GWWM number
spectra are those listed in Eqs. (\ref{eq:NTGWWM12}) and
(\ref{eq:NLGWWM12}). As in the case of applications to pointlike
particles, the general procedure for finding an exact value of
$\eta _b$ is to \emph{demand} that the GWWM $N_\mT $ and $N_\mL $
functions in the low boson energy limit agree exactly with the
same functions, in the same limit, appearing in Papageorgiu's
EWM-based analysis.

Consider first the $N_\mT $ functions. As in the previous section,
equate the expression for the GWWM $N_\mT $ (found in Eq.
(\ref{eq:NTGWWM12})) to a generic $N_\mT $ function (to be
identified with Eq. (\ref{eq:NTEWMinf4}) below).
\begin{subequations}
\begin{align}
\piof \frac{N_0}{E_b}\, \mbox{\Large e} ^{-2\chi _b} &\equiv N_\mT  \label{eq:comcp1}\\%
\mbox{\Large e} ^{2\chi _b} &= \piof \frac{N_0}{N_\mT E_b} \label{eq:comcp2}\\%
2\chi _b &= \log{\left( \piof \frac{N_0}{N_\mT E_b} \right) } \label{eq:comcp3}\\%
\eta _b &= \half \, \log{\left( \piof \frac{N_0}{N_\mT E_b} \right) }\qquad \mbox{(via $\chi _b=\eta _b$).} \label{eq:comcp4}%
\end{align}
\end{subequations}
Now, substitute Eq. (\ref{eq:NTEWMinf4}) for $N_\mT $ here.
\begin{subequations}
\begin{align}
\eta _b &= \half\, \log{\left( \piot M_b^2B_{min}^2 \right) }\label{eq:comcp5}\\%
&= \log{\left( \rpiot M_bB_{min} \right) }\label{eq:comcp6}\\%
&= \log{\left( \rpiot M_bR \right) },\label{eq:comcp7}%
\end{align}
\end{subequations}
where $B_{min}$ has been rewritten as $R$ (the nuclear radius) in
the last line. An even more handy formula is found by substituting
the known value for $R$, which is given in Eq.
(\ref{eq:bminnuclei1}).
\begin{equation}
\eta _b = \log{\left( 1.504 M_b A^{1/3} \right) },\label{eq:comcp8}%
\end{equation}
where the constant $1.504$ is more precisely $1.2\sqrt{\mpi/2}$.
And, of course, $M_b$ is the on-shell value of the boson's mass,
and $A$ is the atomic mass of the nucleus that emits the boson.
With $m_W=80.42$ GeV and $m_Z=91.19$ GeV, and considering values
of $A$ to potentially range from $1$ (for hydrogen) to $238$ (for
uranium), $\eta _b$ is found to range from a minimum of $6.418$
(for $W$ bosons radiating from protons) to a maximum of $8.368$
(for $Z$ bosons radiating from uranium nuclei) \cite{ref:RPP}. So,
$\eta _b$ turns out to be closer (in order of magnitude) to $10$
than $1$. But, the use of the Heisenberg uncertainty principle is
such an inherently tricky business to begin with that these values
are not out of the realm of possibilities. Perhaps D. Griffiths
put it best when he said, ``In general, when you hear a physicist
invoke the uncertainty principle, keep a hand on your wallet"
\cite{ref:Grif}.

The $N_\mL $ functions are almost identical to the $N_\mT $
functions in the limits of interest. Just as Papageorgiu found
that $N_\mT \simeq N_\mL $, the $N_\mL $ of Eq.
(\ref{eq:NLGWWM12}) is equal to the $N_\mT $ of Eq.
(\ref{eq:NTGWWM12}) in the limit where $p_f\simeq E_f$, $E_b\ll
E_f$, and $(1-\alpha _b)E_b/\alpha _bE_f\ll 1$. The first two
limits are the usual ones of interest. The third approximation
holds in general, although there are exceptions (cf. Table
\ref{tab:alphab} and the discussion thereafter); see also Eq.
(\ref{eq:NLEWM04}) and the discussion following it. So $N_\mL
\simeq N_\mT $ in the GWWM, and since the $N_\mT $ function of the
GWWM is in excellent agreement with the $N_\mT $ of Papageorgiu's
EWM-based analysis (which, again, $\simeq N_\mL $ thereof), it is
concluded that the above choice for $\eta _b$ simultaneously
achieves good agreement between the $N_\mL $ function of the GWWM
and the $N_\mL $ of Papageorgiu's analysis.

Results from this section (for bosons radiating from a composite
particle) are shown in Figs. \ref{fig:frsp01} -- \ref{fig:mb21}.
Fig. \ref{fig:frsp01} shows the helicity-averaged frequency
spectra (cf. Eqs. (\ref{eq:dIdwdA1exp2}) --
(\ref{eq:dIdwdA3exp2})), evaluated at $b=b_{min}$, for the three
WW pulses of photons radiating from a lead ($^{208}Pb$) nucleus at
a relativistic heavy ion collider operating at a beam energy of
$3.4\mA$ TeV. Pulse 1, which represents transversely-polarized
photons travelling collinearly with the lead nucleus, is by far
the most dominant of the three pulses. Pulse 2, which is the
fictitious pulse travelling in a direction transverse to the
nucleus's velocity, is relatively negligible; it is shown
multiplied by a factor of $\gamma _f^2$ so that it appears on the
same graph as Pulse 1. Pulse 3, which represents
longitudinally-polarized photons, does not appear at all on the
graph because it vanishes altogether. The number spectra for
transversely-polarized photons, as calculated via the GWWM (Eq.
(\ref{eq:NTGWWM1})), or, equivalently, the SWWM (Eq.
(\ref{eq:NTSWWM1})), is compared to the same function calculated
via the version of the WWM developed by J\"{a}ckle and Pilkuhn
(Eq. (\ref{eq:NTJP})) in Figs. \ref{fig:nt01A} and
\ref{fig:nt01B}. The results shown in Fig. \ref{fig:nt01A} are
relevant to a collider operating at a beam energy of $2.76\mA$
TeV, and those shown in Fig. \ref{fig:nt01B} are relevant to beam
energies of $3.4\mA$ TeV, both energies of which will be
characteristic of the LHC collider at CERN that is currently being
upgraded \cite{ref:Norb2,ref:RPP}. The relative differences
between the functions are always about $10^{-5}\, \%$! The
helicity-averaged frequency spectra, evaluated at $b=b_{min}$, for
the three pulses of equivalent $Z$ bosons ($W$ bosons) outside a
lead nucleus at a $3.4\mA$ TeV relativistic heavy ion collider are
shown in Fig. \ref{fig:frsp11} (Fig. \ref{fig:frsp21}). Like in
the photon case, Pulse 1 exceeds Pulse 2 by a factor of at least
$\gamma _f^2$, but unlike the photon case, Pulse 3, which
describes longitudinally-polarized $Z$ bosons ($W$ bosons), is
nonnegligible, and, in fact, comparable to Pulse 1. Fig.
\ref{fig:nt11A} (Fig. \ref{fig:nt21A}) compares the number spectra
for transversely-polarized $Z$ bosons ($W$ bosons) radiating from
a lead nucleus at a $2.76\mA$ TeV relativistic heavy ion collider,
as calculated via the GWWM (Eq. (\ref{eq:NTGWWM1})), to the same
function as calculated via the EWM (Eq. (\ref{eq:NTEWM1})). Figs.
\ref{fig:nt11B} and \ref{fig:nt21B} are similar graphs, but
relevant to a $3.4\mA$ TeV collider. Relative differences are
always about $8.4\%$ ($8.5\%$) for all $Z$ boson ($W$ boson)
energies. Fig. \ref{fig:nl11A} (Fig. \ref{fig:nl21A}) compares the
number spectra for longitudinally-polarized $Z$ bosons ($W$
bosons) radiating from a lead nucleus at a $2.76\mA$ TeV
relativistic heavy ion collider, as calculated via the GWWM (Eq.
(\ref{eq:NLGWWM1})), to the same function as calculated via the
EWM (Eq. (\ref{eq:NLEWM1})). And Figs. \ref{fig:nl11B} and
\ref{fig:nl21B} are similar graphs, but relevant to a $3.4\mA$ TeV
collider. For both the $Z$ boson and $W$ boson cases, relative
errors are always about $2.7\%$. The mass of an equivalent $Z$ or
$W$ boson varies with the boson's energy according to Eq.
(\ref{eq:mbGWWM}). This function is shown in Fig. \ref{fig:mb11}
for $Z$ bosons and Fig. \ref{fig:mb21} for $W$ bosons.

In summary, splendid agreement is found between the number
spectrum functions of the GWWM and other theories, for bosons
radiating from composite particles. To achieve this agreement, the
free parameter $\eta _b$ (first introduced in Eq.
(\ref{eq:bminQM21})) appearing in the GWWM must be between $1$ and
$10$. For applications to EM interactions, the value of $\eta _b$
is actually completely inconsequential. When comparing the $N_\mT
$ for massive bosons of the GWWM to the same function in the EWM,
as applied to bosons radiating from a nucleus of atomic mass
number $A$, $\eta _b$ is found to be $\log{\left( 1.504 M_b
A^{1/3} \right) }$. This constant ranges from a minimum possible
value of $6.418$ (for $W$ bosons radiating from protons) to a
maximum possible value of $8.368$ (for $Z$ bosons radiating from
uranium nuclei) \cite{ref:RPP}. The $N_\mL $ function for photons
vanishes in both the GWWM (or the SWWM) and the version of the WWM
developed by J\"{a}ckle and Pilkuhn (Eq. (\ref{eq:NLJP})). The
expression for $N_\mL $ for massive bosons of the GWWM is found to
be equal to that in the EWM if, in addition to the above choice
for $\eta _b$, $(1-\alpha _b)E_b/\alpha _bE_f\ll 1$, or $\alpha
_b\simeq 1$, which, as has been pointed out previously, generally
holds true.

%% file: Summary.tex
%%%%%%%%%%%%%%%%%%
% Summary
%%%%%%%%%%%%%%%%%%

\chapter{Summary} \label{sec:Summary} \indent

In conclusion, an electroweak generalization of the semiclassical
Weizs\"{a}cker-\linebreak Williams Method was successfully
developed. In particular, the number spectrum functions describing
the distribution of massive and massless electroweak bosons
swarming about an ultrarelativistic fermion were derived. As in
the original method, the starting point for this derivation was
the equation of motion: Maxwell's equations for the massless
photon case and the Proca equation for the massive $W$ and $Z$
boson cases. The relevant equation was solved for a point charge
in relativistic motion. In all cases, the potentials and fields
surrounding the moving charge were found to be highly Lorentz
contracted in the plane transverse to the direction of motion. As
in the original method, these potentials and fields were then
approximated as pulses of plane wave radiation travelling
collinearly with the moving charge. Fourier transforms of all
quantities were taken to find the frequency spectra of these
pulses of ``equivalent bosons." An integration over the wavefront
area of the pulse and a division by the energy of the equivalent
boson then yielded the number spectrum functions. These
generalized functions differed in two ways from the original
function that only pertained to massless photons. One was the
charge of the fermion to which the boson couples and the other was
the mass of the boson. The charges for each of the three types of
electroweak interactions were derived from a knowledge of the
corresponding fermion 4-current in a very straightforward way. All
charges found were consistent with those appearing in the Standard
Model. Values of the boson masses were obtained in a much more
laborious manner. The main guiding principles behind this
calculation were the conservation of energy and momentum, Lorentz
invariance, and the concept of causality. The mass of a given
equivalent boson was found to depend in a very specific way on the
boson's energy. These equivalent bosons are interpreted as being
something analogous to off-shell SM boson states (i.e., virtual
particles) that are tailor-made to fit the semiclassical method
under consideration. The resulting number spectrum functions were
compared to similar functions found in other theories. In all
applications of the method, excellent agreement was obtained with
other, more reliable, theories, proving the generalized
Weizs\"{a}cker-Williams Method to be an accurate alternative to
these more exact theories.

%% file: AppendixA.tex
%%%%%%%%%%%%%%%%%
%  APPENDIX A
%%%%%%%%%%%%%%%%%

\singlespace
\setcounter{equation}{0}%
\renewcommand{\theequation}{A.\arabic{equation}}%

\begin{flushleft}
{\bf \Large Appendix A: Electroweak
4-Currents}\label{sec:AppendixA}
\end{flushleft}

\indent

In this section, a formula for the 4-current $J^\mu(\vecr ,t)$
that is common to all three types of EW interactions of
\emph{point charges} is derived. Suppressing the $\vecr$ and $t$
dependence for notational simplicity, the fermion 4-currents
appearing in the SM are generally of the form
\begin{equation}
J^\mu=g\Bar{\psi }\Tilde{\gamma }^\mu \Tilde{Q}\psi,
\label{eq:gencur}
\end{equation}
where $g$ is the relevant coupling constant, $\psi$ is a fermion
wave function (a solution to the Dirac equation (DE)), $\Bar{\psi
}$ is the Dirac adjoint of $\psi$, $\Tilde{\gamma }^\mu$ is a
Dirac matrix ( $\Tilde{}$ denotes matrix), and $\Tilde{Q}$ is the
relevant charge operator.

The particular representation of the Dirac matrices that will be
used here is the Weyl (or chiral) representation, where
\begin{equation}
\Tilde{\gamma }^0=
\begin{pmatrix}
 \;\;\; \Tilde{0} & -\Tilde{1}\\
-\Tilde{1} & \;\;\; \Tilde{0}\
\end{pmatrix}
\quad \mbox{and} \quad%
\Tilde{\gamma }^\mi =
\begin{pmatrix}
 \;\; \Tilde{0} & \Tilde{\sigma }^\mi\\
-\Tilde{\sigma }^\mi & \Tilde{0}\
\end{pmatrix}
\qquad \mbox{(Dirac matrices).}\label{eq:DirMat}
\end{equation}
This representation facilitates splitting the fermion wave
functions into definite chiral (right- and left-handed)
eigenstates, which is how they naturally appear in the SM. It is
to be understood that each element in Eq. (\ref{eq:DirMat}) is
actually a 2$\times$2 matrix. $\Tilde{1}$ is thus the 2$\times$2
unit matrix, $\Tilde{0}$ is the 2$\times$2 null matrix, and
$\Tilde{\sigma }^\mi$ is the $\mi ^{th}$ Pauli spin matrix. The
Pauli spin matrices are
\begin{eqnarray}
&\Tilde{\sigma}^1 =
\begin{pmatrix}
\;\,0 & 1\\
\;\,1 & \;0\
\end{pmatrix}
\mbox{,} \quad%
\Tilde{\sigma}^2=
\begin{pmatrix}
 \;\;\; 0 & -\mi\\
-\mi & \;\;\;\; 0\
\end{pmatrix}
\mbox{,} \quad \mbox{and} \quad%
\Tilde{\sigma}^3=
\begin{pmatrix}
\;\,1 & \; 0\\
\;\,0 & -1\
\end{pmatrix} \nonumber \\
&\hspace{3.3in} \mbox{(Pauli spin matrices),}\label{eq:PauMat}
\end{eqnarray}
where the elements this time are simply numbers. The following two
relations are useful and easy to verify.
\begin{equation}
\Tilde{\gamma }^0\Tilde{\gamma}^0 =
\begin{pmatrix}
\;\,\Tilde{1} & \Tilde{0}\\
\;\,\Tilde{0} & \Tilde{1}\
\end{pmatrix}
\qquad \mbox{and} \qquad%
\Tilde{\gamma }^0\Tilde{\gamma}^\mi =
\begin{pmatrix}
\;\,\Tilde{\sigma }^\mi &  \Tilde{0}\\
\;\,\Tilde{0} & -\Tilde{\sigma }^\mi\
\end{pmatrix}
.\label{eq:gamgamrelations}
\end{equation}
They will be made use of later.

Having now specified the Dirac matrices, the Dirac adjoint
$\Bar{\psi }$ of $\psi$ can be related in a simple way to the more
familiar Hermitian adjoint $\psi ^\dag$ of $\psi$:
\begin{equation}
\Bar{\psi }=\psi ^\dag \Tilde{\gamma }^0.\label{eq:DirAdjPsi}
\end{equation}
When all the whistles and bells are included, the fermion wave
functions $\psi$ more generally appear in the form
\begin{equation}
\psi ^\ms _\mc (x^\mu ,p^\mu )=N\, \phi^\ms _\mc (p^\mu )\,
\mbox{\Large
e} ^{-\mi p\cdot x}.\label{eq:genPsi}%
\end{equation}
$\ms$ labels the spin ($\ms =1\,(2)$ for spin-up (down)); $\mc$
labels the chirality ($\mc = \mR (\mL)$ for right (left) handed
states, and is suppressed for nonchiral states); $x^\mu=(t,\vecr
)$ is the fermion's 4-position; $p^\mu=(E,\vecp )$ is the
fermion's 4-momentum; $N$ is a (real) normalization constant to be
determined; and $\phi ^\ms _\mc (p^\mu )$ are 4-spinors which also
satisfy the DE (solutions to the DE are always arbitrary up to a
choice of normalization constant and overall phase factor). In a
general Lorentz frame (and in the Weyl representation), the
4-spinors for a \emph{massive} fermion are as follows.
\begin{equation}
\phi ^\ms _\mR (p^\mu )=\frac{1}{2\sqrt{m(E+m)}}\,
\begin{bmatrix}
(\Tilde{E}+\Tilde{m}+\Tilde{\sigma }\cdot \vecp)\Tilde{\chi }^\ms\\
\,\Tilde{0}\
\end{bmatrix}\label{eq:R4spinor}
\end{equation}
and
\begin{equation}
\phi ^\ms _\mL (p^\mu )=\frac{1}{2\sqrt{m(E+m)}}\,
\begin{bmatrix}
\Tilde{0}\\
\,(\Tilde{E}+\Tilde{m}-\Tilde{\sigma }\cdot \vecp)\Tilde{\chi
}^\ms\
\end{bmatrix},\label{eq:L4spinor}
\end{equation}
where $m$ is the mass of the fermion, $\Tilde{E}=\Tilde{1}E$,
$\Tilde{m}=\Tilde{1}m$, $\Tilde{\sigma }\cdot
\vecp=\Tilde{\sigma}^1p_x+\Tilde{\sigma}^2p_y+\Tilde{\sigma}^3p_z$,
and
\begin{equation}
\Tilde{\chi }^1=
\begin{pmatrix}
1\\
\;\,0\
\end{pmatrix}
\quad \mbox{and} \quad%
\Tilde{\chi }^2=
\begin{pmatrix}
0\\
\;\,1\
\end{pmatrix}
\label{eq:spinspinors}
\end{equation}
are the basis spinors corresponding to spin-up (down) states
\cite{ref:Ryde,ref:Halz}. For arbitrary spins $\ms$ and ${\ms }'$,
the relation

\begin{equation}
\chi ^{\ms \dag}\chi ^{{\ms }'} =\delta
_{ss'},\label{eq:chiorthog}
\end{equation}
where $\delta _{ss'}$ is the Kronecker delta ($\delta _{ss'}=1$ if
$s=s'$ and $\delta _{ss'}0$ if $s\ne s'$), is easily proved and
very useful. Eqs. (\ref{eq:R4spinor}) and (\ref{eq:L4spinor})
represent four independent solutions to the DE for massive
fermions. In the ultrarelativistic limit, where chirality is
identical with helicity, these four solutions uniquely correspond
to: a particle state with spin-up, a particle state with
spin-down, an antiparticle state with spin-up, and an antiparticle
state with spin-down. In a nonrelativistic limit, this description
(in terms of particle and spin states) is less cut-and-dry. A
nonchiral state $\phi ^\ms$ is simply the sum $\phi ^\ms _\mR+\phi
^\ms _\mL$:
\begin{equation}
\phi ^\ms (p^\mu )\equiv \phi ^\ms _\mR (p^\mu )+\phi ^\ms _\mL
(p^\mu ) =\frac{1}{2\sqrt{m(E+m)}}\,
\begin{bmatrix}
(\Tilde{E}+\Tilde{m}+\Tilde{\sigma }\cdot \vecp)\Tilde{\chi }^\ms\\
\;(\Tilde{E}+\Tilde{m}-\Tilde{\sigma }\cdot \vecp)\Tilde{\chi
}^\ms\
\end{bmatrix};\label{eq:nonchiral4spinor}
\end{equation}
it is also a solution to the DE. In the \emph{rest frame} limit
($E=m$ and $\vecp =\veczero$), it is easily deduced from Eqs.
(\ref{eq:R4spinor}) and (\ref{eq:L4spinor}) that
\begin{equation}
\phi ^\ms _\mR (p^\mu )=\oort
\begin{bmatrix}
\;\, \Tilde{\chi }^\ms\, \\
\;\, \Tilde{0}\, \
\end{bmatrix}
\quad \mbox{and} \quad%
\phi ^\ms _\mL (p^\mu)=\oort
\begin{bmatrix}
\Tilde{0}\\
\;\, \Tilde{\chi }^\ms\
\end{bmatrix}\qquad \mbox{(rest frame limit).}\label{eq:4spinorsrestframe}
\end{equation}
Therefore,
\begin{equation}
\phi ^\ms (p^\mu )\equiv \phi ^\ms _\mR (p^\mu )+\phi ^\ms _\mL
(p^\mu )=\oort
\begin{bmatrix}
\Tilde{\chi }^\ms\\
 \;\, \Tilde{\chi }^\ms\
\end{bmatrix}\qquad \mbox{(rest frame limit),}
\label{eq:4spinorsnonchiralrestframe}
\end{equation}
which is normalized to unity. In the \emph{massless} limit, it is
well-known that (for a \emph{particle}, as opposed to an
\emph{antiparticle}, state)
\begin{equation}
\phi ^\ms _\mR (p^\mu )=\oort
\begin{bmatrix}
\;\, \Tilde{\chi }^1\, \\
\;\, \Tilde{0}\, \
\end{bmatrix}
\quad \mbox{and} \quad%
\phi ^\ms _\mL (p^\mu)=\oort
\begin{bmatrix}
\Tilde{0}\\
\;\, \Tilde{\chi }^2\
\end{bmatrix}\qquad \mbox{(massless fermion),}\label{eq:massless4spinors}
\end{equation}
whence
\begin{equation}
\phi ^\ms (p^\mu )\equiv \phi ^\ms _\mR (p^\mu )+\phi ^\ms _\mL
(p^\mu )=\oort
\begin{bmatrix}
\Tilde{\chi }^1\\
 \;\, \Tilde{\chi }^2\
\end{bmatrix}\qquad \mbox{(massless fermion).}
\label{eq:nonchiralmassless4spinor}
\end{equation}
The $\phi ^\ms _\mR$ and $\phi ^\ms _\mL$ 4-spinors have the
property that they are eigenstates of the (normalized) helicity
operator, with eigenvalues $+1$ and $-1$, respectively. A copious
amount of experimental evidence indicates that neutrinos are
\emph{nearly} massless and are \emph{always} left-handed
\cite{ref:Grif,ref:Halz,ref:Ryde}. For this reason, the $\phi ^\ms
_\mL$ wave functions are used in the SM to represent these
fermions. The $\phi ^\ms _\mR$ wave functions are then reserved
for the antiparticle neutrino states. In short, the wave function
specifications listed above are merely a statement that massless
fermion \emph{particle} states are always left-handed, and
massless fermion \emph{antiparticle} states are always
right-handed.

The constant $N$ will now be determined for both massive and
massless fermion wave functions, in the context of the probability
4-current. The probability 4-current $J^{P\mu}_\mc$ associated
with a solution $\psi ^\ms _\mc$ of the DE is
\begin{equation}
J^{P\mu}_\mc=\Bar{\psi }^\ms _\mc \Tilde{\gamma }^\mu \psi ^\ms
_\mc \label{eq:probcurden}
\end{equation}
\cite{ref:Aitc,ref:Halz}. This 4-vector has the two main
properties required of such a 4-current: the time component, to be
interpreted as the probability density, is a scalar density that
is positive definite,
\begin{equation}
J^{P0}_\mc=\sum ^4 _{\mi =1}\left| \left( \psi ^\ms _\mc \right)
_\mi \right| ^2 >0; \label{eq:posdefofJP0}
\end{equation}
and $J^{P\mu}_\mc$ is conserved:
\begin{equation}
\partial^\mu J^P_{\mc \mu} =0\label{eq:consofprob}
\end{equation}
\cite{ref:Aitc}. Note that the summation index $\mi$ in Eq.
(\ref{eq:posdefofJP0}) runs over the four independent components
of $\psi ^\ms _\mc$ (equivalently, the four independent solutions
to the DE), which can be easily identified in Eqs.
(\ref{eq:R4spinor})--(\ref{eq:nonchiralmassless4spinor}). It is a
well known result (Noether's theorem) that if a 4-current is
conserved (in the above sense), there is a corresponding conserved
``charge" associated with it. This canonical charge is the
integral of the $0$ component of the 4-current over an infinite
spatial volume, evaluated on a hypersurface of constant time. For
the case at hand, this charge $Q^P_\mc$ is identified as the
probability of finding the particle \emph{somewhere} at some
instant in time, as it is merely the volume integral of the
probability density.
\begin{subequations}
\begin{align}
Q^P_\mc &=\lim_{V\to \infty}\int _V \dif ^3\vecr\, J^{P0}_\mc
\label{eq:totprob1}\\%
&=\lim_{V\to \infty}\int _V \dif ^3\vecr\, \Bar{\psi }^\ms _\mc
\Tilde{\gamma }^0\psi ^\ms _\mc , \label{eq:totprob2}\
\end{align}
\end{subequations}
where it is to be understood that the integral is to be carried
out on a hypersurface of constant time. Of interest to this study
are point particles. So the given particle is confined to a given
volume $V_0$, in the sense that $\psi ^\ms _\mc$ is to be taken to
vanish everywhere except inside $V_0$. Hence
\begin{subequations}
\begin{align}
Q^P_\mc &=\int _{V_0} \dif ^3\vecr\, \Bar{\psi }^\ms _\mc
\Tilde{\gamma }^0\psi ^\ms _\mc \label{eq:totprob3}\\%
&=\int _{V_0} \dif ^3\vecr\, N^2\Bar{\phi}^\ms _\mc \Tilde{\gamma }^0\phi ^\ms _\mc \label{eq:totprob4}\\%
&=\int _{V_0} \dif ^3\vecr\, N^2\phi ^{\ms \dag} _\mc
\Tilde{\gamma
}^0\Tilde{\gamma }^0\phi ^\ms _\mc \label{eq:totprob5}\\%
&=V_0N^2\phi ^{\ms \dag} _\mc \phi ^\ms _\mc.\label{eq:totprob6}\
\end{align}
\end{subequations}
This last step follows from the facts that $\Tilde{\gamma
}^0\Tilde{\gamma }^0=\Tilde{1}$, and $N$ and $\phi ^\ms _\mc$ are
independent of $\vecr$ (in any Lorentz frame). Since $Q^P_\mc$ is
Lorentz invariant (by Noether's theorem), it need only be
evaluated in the particle's own rest frame. Of course this step
can only be done in the case of massive fermions. For the massless
case, the following analysis can be performed in an arbitrary
Lorentz frame, and all results can be shown to agree with those
for the massive case. One need only keep in mind that $\ms =2$
($\ms =1$) for $R$ ($L$) chiral states (via Eq.
(\ref{eq:massless4spinors})). Assuming this to be the case, then,
the following will only be concerned with massive fermions. In the
rest frame of a massive fermion (recall Eqs.
(\ref{eq:4spinorsrestframe}) and
(\ref{eq:4spinorsnonchiralrestframe})),
\begin{subequations}
\begin{align}
\phi ^{\ms \dag} _\mR \phi ^\ms _\mR &=\half \left[ \Tilde{\chi
}^{\ms \dag}, \Tilde{0} \right]
\begin{bmatrix}
\;\;\Tilde{\chi }^\ms\;\\
\;\;\Tilde{0}\;\
\end{bmatrix}%
=\half \Tilde{\chi }^{\ms \dag}\Tilde{\chi }^\ms =\half
,\label{eq:Rnorm}\\%
\phi ^{\ms \dag} _\mL \phi ^\ms _\mL &=\half \left[ \Tilde{0},
\Tilde{\chi }^{\ms \dag} \right]
\begin{bmatrix}
\Tilde{0}\\
 \;\; \Tilde{\chi }^\ms\
\end{bmatrix}%
=\half \Tilde{\chi }^{\ms \dag}\Tilde{\chi }^\ms =\half
,\quad \mbox{and}\label{eq:Lnorm}\\%
\phi ^{\ms \dag} \phi ^\ms &=\half \left[ \Tilde{\chi }^{\ms
\dag}, \Tilde{\chi }^{\ms \dag} \right]
\begin{bmatrix}
\Tilde{\chi }^\ms\\
\; \Tilde{\chi }^\ms\
\end{bmatrix}%
=\half \left( \Tilde{\chi }^{\ms \dag}\Tilde{\chi }^\ms +
\Tilde{\chi }^{\ms \dag}\Tilde{\chi }^\ms \right)
=1.\label{eq:nonchiralnorm}\
\end{align}
\end{subequations}
So
\begin{equation}
Q^P_\mR = Q^P_\mL =\half V_0N^2 \label{eq:equalprobs}
\end{equation}
and
\begin{equation}
Q^P = Q^P_\mR + Q^P_\mL = V_0N^2 .\label{eq:totalprob}
\end{equation}
It is reasonable to interpret these results to mean that a massive
fermion is equally likely to be found in a right handed or left
handed state. In the case where the fermion is massless, this
statement is still true, since a fermion \emph{particle} state is
always left handed, while the \emph{antiparticle} state of the
same fermion is always right handed. Setting $Q^P \equiv 1$, which
says that (if found) there is a 100\% chance that the fermion will
be found in one of either of these two mutually exclusive states.
Then $Q^P_\mR = Q^P_\mL =1/2$ and
\begin{equation}
N = \frac{1}{\sqrt{V_0}}.\label{eq:Nspec}
\end{equation}
As stipulated, $N$ is independent of $\vecr$. Also, as mentioned
above, the same result can easily be obtained in the case where
the fermion is massless. For simplicity, all particles in this
thesis are to be regarded as pointlike. This constraint is made
quantitative by considering the limit of Eq. (\ref{eq:Nspec}) as
$V_0 \to 0$. In this limit, $1/V_0$ (centered on the particle with
4-position $x^\mu =(t,\vecr )$) becomes a Dirac delta function
$\delta [\vecr (t)]$, and
\begin{equation}
N \to \lim _{V_0 \to 0} \sqrt{\frac{1}{V_0}} = \sqrt{\delta [\vecr
(t)]}.\label{eq:Npointlike}
\end{equation}
The wave function becomes
\begin{equation}
\psi ^\ms _\mc (x^\mu ,p^\mu )=\sqrt{\delta [\vecr (t)]}\, \phi
^\ms _\mc (p^\mu )\, \mbox{\Large
e} ^{-\mi p\cdot x}.\label{eq:pointlikePsi}%
\end{equation}
It is this form that shall henceforth be adopted as the wave
function of \emph{any} particle (massive or not) appearing in this
thesis.

A word about the choice of normalization constant $N$ is in order.
The choice for $N$ specified in Eq. (\ref{eq:Nspec}) is referred
to as ``box normalization"; it is used extensively in
nonrelativistic quantum mechanics and occasionally in relativistic
quantum mechanics. \cite{ref:Grei,ref:Aitc,ref:Bran,ref:Mand}. For
example, a common problem encountered in nonrelativistic quantum
mechanics is that of determining the energy spectrum of a particle
confined to an infinite one dimensional square-well potential (a
``box"), say, centered at $x=0$ and having width $L$. In this
example, $L$ is a one dimensional version of the $V_0$ introduced
above. By construction, the particle's wave function (the momentum
eigenfunction) $\Psi (x)$ vanishes at the boundaries of the box
(i.e., $\Psi (-L/2)=\Psi (L/2)=0$), so all relevant integrals are
like those in Eqs. (\ref{eq:totprob3})--(\ref{eq:totprob6}), in
the sense that only the region of space within length $L$ is of
interest. In this particular example, $\Psi (x)$ is known to
generally be of the form
\begin{equation}
\Psi (x) = N\mbox{\Large e}^{\mi k_xx},\label{eq:PartinBox1}%
\end{equation}
for $|x|\le L/2$, and $\Psi (x)=0$ everywhere outside the box,
where $k_x$ is the particle's momentum, and $N$ is the
normalization constant to be determined. $N$ is deduced in the
same spirit as was done above: by normalizing $\Psi (x)$ so that
the probability of finding the particle between $x=-L/2$ and
$x=+L/2$ is exactly one. One finds (see, \cite{ref:Bran}, for
example)
\begin{subequations}
\begin{align}
1 &= \int _{-L/2}^{+L/2} \dif x\, {\Psi }^\ast(x) \Psi (x)\label{eq:PartinBox2}\\%
&= N^2 \int _{-L/2}^{+L/2} \dif x\, \mbox{\Large e}^{-\mi {k'}_xx} \mbox{\Large e}^{\mi k_xx} \label{eq:PartinBox3}\\%
&= N^2 \int _{-L/2}^{+L/2} \dif x\, \mbox{\Large e}^{\mi (k_x-{k'}_x)x} \label{eq:PartinBox4}\\%
&= N^2 \left( L\, \delta _{k_x{k'}_x} \right) ,\label{eq:PartinBox5}%
\end{align}
\end{subequations}
where $\delta _{k_x{k'}_x}$ is the Kronecker delta function. Since
$k_x={k'}_x$, it follows, therefore, that $N=1/\sqrt{L}$. If the
particle's wave function does not vanish on some given boundary,
but is still described by Eq. (\ref{eq:PartinBox1}), an
alternative normalization constant is deduced. In this case, the
above limits of integration are $\pm \infty$, instead of $\pm
L/2$, respectively. The appropriate orthonormality relation for
$\Psi (x)$ \emph{on the} $x=(-\infty,\, +\infty)$ \emph{interval}
is %
\begin{subequations}
\begin{align}
\delta (k_x-{k'}_x) &= \int _{-\infty }^{+\infty } \dif x\, {\Psi }^\ast(x) \Psi (x)\label{eq:PartinBox6}\\%
&= N^2 \int _{-\infty }^{+\infty } \dif x\, \mbox{\Large e}^{-\mi {k'}_xx} \mbox{\Large e}^{\mi k_xx} \label{eq:PartinBox7}\\%
&= N^2 \int _{-\infty }^{+\infty } \dif x\, \mbox{\Large e}^{\mi (k_x-{k'}_x)x} \label{eq:PartinBox8}\\%
&= N^2 \left[ 2\mpi \, \delta (k_x-{k'}_x) \right] ,\label{eq:PartinBox9}%
\end{align}
\end{subequations}
where $\delta (k_x-{k'}_x)$ is the usual Dirac delta function.
Thus, $N=1/\sqrt{2\mpi }$ for these cases. These prototypical
normalization constants are easily found to generalize to
$N=1/L^{3/2}$ (or $N=1/\sqrt{V_0}$, where $V_0=L^3$) and
$N=1/({2\mpi })^{3/2}$, respectively, in three dimensions. In
short, $N=1/\sqrt{V_0}$ is used in cases where the particle's wave
function is known to vanish on the boundary of some volume $V_0$,
and $N=1/({2\mpi })^{3/2}$ is used in cases where there is no
sharp discontinuity imposed on $\Psi (x)$. An example of where the
latter type of normalization constant might be used is in the wave
function describing a plane-wave wave packet freely propagating
through empty space. It is also used extensively in relativistic
quantum mechanics (in second quantization), where all operators
are expressed as infinite sums of plane wave states. In this
thesis, it is assumed \emph{for simplicity} that all fermions of
interest are \emph{very} localized. As such, the $N=1/\sqrt{V_0}$
is the more appropriate normalization constant to use. The obvious
question then arises as to what volume $V_0$ to use. To avoid this
whole issue, another simplifying assumption is made: the wave
functions are defined in the limit where $V_0 \to 0$, Eq.
(\ref{eq:Npointlike}). By taking this limit, the particles are
being envisioned as ideally pointlike, in the sense of having a
vanishing spatial dimension. Many authors use the box
normalization, $N=1/\sqrt{V_0}$, but few, if any, impose this
additional ``point particle constraint". The reason for this is
probably twofold. One is that, if particles \emph{are} being
treated as pointlike, the more obvious route of analysis is that
of second quantization, which uses the $N=1/({2\mpi })^{3/2}$
choice (as mentioned above), because quantum field theory is
invariably more accurate than such a classical approach as taken
here. Another reason might be that such a point particle
approximation is not always needed; it is not terribly difficult
(see, e.g., Sec. \ref{sec:Bmin}) to make reasonable approximations
about the density profile of an elementary particle. The delta
function approximation is done here merely for simplicity. In
particular, the fermion 4-potential $A^\mu$ depends on an integral
of the fermion 4-current $J^\mu$ (cf. Eq. (\ref{eq:PEsol1})). By
taking $N = \sqrt{\delta [\vecr (t)]}$ instead of
$N=1/\sqrt{V_0}$, a $J^\mu$ is obtained that depends on $\delta
[\vecr (t)]$ instead of some $1/V_0$, which \emph{considerably}
simplifies the expression: the delta function kills the integral
entirely, and the expression for $A^\mu$ reduces to a very compact
formula! In summary, the choice of box normalization (Eq.
\ref{eq:Nspec}) is merely one of at least two possible
conventional choices to make. The subsequent delta function
approximation made here (Eq. \ref{eq:Npointlike}) is not a
conventional step to take, but it is very reasonable, and
considerably simplifies the analysis (in particular, the
evaluation of Eq. (\ref{eq:PEsol1})).

It will prove useful later to have explicit expressions for the
two probability 4-currents $J^{P\mu }_\mR$ and $J^{P\mu }_\mL$.
So, at the risk of digressing for a while, attention is now turned
to this derivation. These 4-currents will first be derived in a
frame comoving with the particle. As mentioned before, this line
of attack cannot be performed for massless fermions, as there is
no such Lorentz frame for such particles. For those cases, the
analysis is to be carried out in an arbitrary Lorentz frame, and
all intermediate steps can be easily shown to be identical with
those arrived at in the massive fermion case. Also, two
constraints will be imposed on the system for simplicity. One is
that 4-momentum is to be conserved in going from the initial to
the final fermion states. The other is that the limit of the time
duration between the initial and final fermion states is to be
taken to go to zero. Under these conditions, Eq.
(\ref{eq:probcurden}), with Eq. (\ref{eq:pointlikePsi}) for $\psi
^\ms _\mc$, becomes
\begin{equation}
{J^{P\mu}}'_\mc=\delta [\vecrp (t')]\left[ \Bar{\phi }^\ms _\mc
({p^\mu }' )\mbox{\Large e} ^{+\mi mt'} \right] \Tilde{\gamma
}^\mu \left[ \phi ^\ms _\mc ({p^\mu }')\mbox{\Large e} ^{-\mi mt'}
\right] .\label{eq:pointprobcurden1}
\end{equation}
Primed quantities are those measured relative to the rest frame of
the fermion, so that ${p^\mu }'=(m,\veczero)$ and ${x^\mu
}'=(t',\veczero)$, where $t'$ is the fermion's proper time. Eq.
(\ref{eq:pointprobcurden1}) can be simplified a few steps further
before special cases must be considered.
\begin{subequations}
\begin{align}
{J^{P\mu}}'_\mc &=\delta [\vecrp (t')]\Bar{\phi }^\ms _\mc ({p^\mu
}' )\Tilde{\gamma }^\mu \phi ^\ms _\mc ({p^\mu
}')\label{eq:pointprobcurden2}\\%
&=\delta [\vecrp (t')]\phi ^{\ms \dag}_\mc ({p^\mu }'
)\Tilde{\gamma }^0 \Tilde{\gamma }^\mu \phi ^\ms _\mc ({p^\mu
}') .\label{eq:pointprobcurden3}%
\end{align}
\end{subequations}
Using Eqs. (\ref{eq:gamgamrelations}) and
(\ref{eq:4spinorsrestframe}), the $0$ components of the $R$, $L$,
and nonchiral probability 4-currents (in the particle's rest
frame) are as follows.
\begin{subequations}
\begin{align}
{J^{P0}}'_\mR &=\delta [\vecrp (t')]\half \left[ \Tilde{\chi
}^{\ms \dag}, \Tilde{0} \right]
\begin{bmatrix}
\;\;\Tilde{\chi }^\ms\;\\
\;\;\Tilde{0}\;\
\end{bmatrix}
\label{eq:JP0R1}\\%
&= \half \delta [\vecrp (t')].\label{eq:JP0R2}\\%
{J^{P0}}'_\mL &=\delta [\vecrp (t')]\half \left[ \Tilde{0},
\Tilde{\chi }^{\ms \dag} \right]
\begin{bmatrix}
\Tilde{0}\\
 \;\; \Tilde{\chi }^\ms\
\end{bmatrix}
\label{eq:JP0L1}\\%
&= \half \delta [\vecrp (t')].\label{eq:JP0L2}%
\end{align}
\end{subequations}
${J^{P0}}'$ is simply the sum of these two quantities:
\begin{subequations}
\begin{align}
{J^{P0}}'&={J^{P0}}'_\mR+{J^{P0}}'_\mL \label{eq:JP01}\\%
&=\half \delta [\vecrp (t')] +\half \delta [\vecrp (t')] \label{eq:JP02}\\%
&=\delta [\vecrp (t')] .\label{eq:JP03}%
\end{align}
\end{subequations}
The vector components can be found with the aid of Eq.
(\ref{eq:gamgamrelations}).
\begin{subequations}
\begin{align}
{J^{P\mi }}'_\mR &=\delta [\vecrp (t')]\half \left[ \Tilde{\chi
}^{\ms \dag}, \Tilde{0} \right]
\begin{pmatrix}
\;\,\Tilde{\sigma }^\mi &  \Tilde{0}\\
\;\,\Tilde{0} & -\Tilde{\sigma }^\mi\
\end{pmatrix}
\begin{bmatrix}
\;\;\Tilde{\chi }^\ms\;\\
\;\;\Tilde{0}\;\
\end{bmatrix}
\label{eq:JPiR1}\\%
&=\half \delta [\vecrp (t')]\left[ \Tilde{\chi }^{\ms \dag},
\Tilde{0} \right]
\begin{bmatrix}
\;\;\Tilde{\sigma }^\mi \Tilde{\chi }^\ms\;\\
\;\;\Tilde{0}\;\
\end{bmatrix}
\label{eq:JPiR2}\\%
&= \half \delta [\vecrp (t')] \left( \Tilde{\chi
}^{\ms \dag} \Tilde{\sigma }^\mi \Tilde{\chi }^\ms \right) .\label{eq:JPiR3}%
\end{align}
\end{subequations}
The only nonvanishing component here is ${J^{P3}}'_\mR$,
corresponding to the spin oriented along the $z$ axis (recall
$\Tilde{\chi }^s$ are eigenvectors of $\Tilde{S}^3=\Tilde{\sigma
}^3/2$, with eigenvalues $\pm 1/2$ for spin up (down)). Define
$\hats$ as the direction of the particle's spin, so that $\hats =
\pm \hatz$ is identified with $\ms = 1(2)$, or spin up (down).
Then
\begin{subequations}
\begin{align}
{\vecJ ^P}'_\mR &=\half \delta [\vecrp (t')] \hats \label{eq:JPiR4}\\%
&=\half \delta [\vecrp (t')] {\vecS
}',\label{eq:JPiR5}\\%
\end{align}
\end{subequations}
where ${\vecS }'$ is the vector component of the particle's
normalized 4-spin $s^\mu$ in its own rest frame. In the rest frame
of the particle, $s^\mu$ given as ${s^\mu}'=(0,{\vecS }')\equiv
(0,\hats)$ (i.e., $s^\mu$ is a 4-vector that reduces to the
particle's normalized spin $\hats$ in its own rest frame). A
similar result is found for ${\vecJ ^P}'_\mL$:
\begin{subequations}
\begin{align}
{J^{P\mi }}'_\mL &=\delta [\vecrp (t')]\half \left[ \Tilde{0}
,\Tilde{\chi }^{\ms \dag} \right]
\begin{pmatrix}
\;\,\Tilde{\sigma }^\mi &  \Tilde{0}\\
\;\,\Tilde{0} & -\Tilde{\sigma }^\mi\
\end{pmatrix}
\begin{bmatrix}
\Tilde{0}\\
 \;\; \Tilde{\chi }^\ms\
\end{bmatrix}
\label{eq:JPiL1}\\%
&=\half \delta [\vecrp (t')]\left[ \Tilde{0} ,\Tilde{\chi }^{\ms
\dag} \right]
\begin{bmatrix}
\;\;\;\Tilde{0}\\
\;\,-\Tilde{\sigma }^\mi \Tilde{\chi }^\ms\;\
\end{bmatrix}
\label{eq:JPiL2}\\%
&= -\half \delta [\vecrp (t')] \left( \Tilde{\chi }^{\ms \dag} \Tilde{\sigma }^\mi \Tilde{\chi }^\ms \right) .\label{eq:JPiL3}%
\end{align}
\end{subequations}
Thus
\begin{subequations}
\begin{align}
{\vecJ ^P}'_\mL &=-\half \delta [\vecrp (t')] \hats \label{eq:JPiL4}\\%
&=-\half \delta [\vecrp (t')] {\vecS }'.\label{eq:JPiL5}%
\end{align}
\end{subequations}
Finally,
\begin{subequations}
\begin{align}
{\vecJ ^P}'&={\vecJ ^P}'_\mR +{\vecJ ^P}'_\mL \label{eq:JPi1}\\%
&=\half \delta [\vecrp (t')] {\vecS }' -\half \delta [\vecrp (t')] {\vecS }' \label{eq:JPi2}\\%
&=\veczero .\label{eq:JPi3}%
\end{align}
\end{subequations}
The results of this paragraph can be summarized in a very simple
set of equations. Introduce the 4-velocity $u^\mu$ and normalized
4-spin $s^\mu$ of the particle. In the rest frame of the particle,
${u^\mu }' =(1,\veczero)$ and ${s^\mu }'=(0,\hats)$, where $\hats
=\pm \hatz$ for spin up (down), as mentioned above. For a frame in
which the particle moves with velocity $\vecv =v\hatz$, these
4-vectors transform (under basic Lorentz transformations) into
$u^\mu =\gamma (1,\vecv)$ and $s^\mu =\gamma (v\hats \cdot \hatz
,\hats)$. In the rest frame,
\begin{equation}
\renewcommand{\arraystretch}{1.3}
\left.
\begin{array}{r@{\, = \;}l}
{J ^{P\mu}}' _\mR  & \half \delta [\vecrp (t')] (1,\hats)=\half \delta [\vecrp (t')] ({u^\mu }'+{s^\mu }')\\
{J ^{P\mu}}' _\mL  & \half \delta [\vecrp (t')] (1,-\hats)=\half \delta [\vecrp (t')] ({u^\mu }'-{s^\mu }')\\
{J ^{P\mu}}' & \delta [\vecrp (t')] (1,\veczero)=\delta
[\vecrp (t')] {u^\mu }'\\
\end{array}
\right\} \qquad \mbox{(rest frame).} \label{eq:JPmurest}
\end{equation}
In a moving frame, the \emph{forms} of these 4-vector equations
are the same. Therefore
\begin{equation}
\renewcommand{\arraystretch}{1.3}
\left.
\begin{array}{r@{\, = \;}l}
J ^{P\mu} _\mR  & \half \delta [\vecr (t)] (u^\mu +s^\mu )\\
J ^{P\mu} _\mL  & \half \delta [\vecr (t)] (u^\mu -s^\mu )\\
J ^{P\mu} & \delta [\vecr (t)] u^\mu \\
\end{array}
\right\} \qquad \mbox{(arbitrary frame).} \label{eq:JPmumoving}
\end{equation}
As a double check, note that if the $0$ components of these
probability current 4-vectors are integrated over all space, the
correct total charges (total probabilities) are recovered. As
evaluated in the rest frame,
\begin{subequations}
\begin{align}
Q ^P_\mR &= \int \dif ^3\vecrp {J ^{P0}}' _\mR\label{eq:QPR1}\\%
&= \int \dif ^3\vecrp \left\{ \half \delta [\vecrp (t')] ({u^0}' +{s^0}' ) \right\} \label{eq:QPR2}\\%
&= \half \int \dif ^3\vecrp \delta [\vecrp (t')] (1+0) \label{eq:QPR3}\\%
&= \half \label{eq:QPR4}\\%
Q ^P_\mL &= \int \dif ^3\vecrp {J ^{P0}}' _\mL\label{eq:QPL1}\\%
&= \int \dif ^3\vecrp \left\{ \half \delta [\vecrp (t')] ({u^0}' -{s^0}' ) \right\} \label{eq:QPL2}\\%
&= \half \int \dif ^3\vecrp \delta [\vecrp (t')] (1-0) \label{eq:QPL3}\\%
&= \half \label{eq:QPL4}\\%
Q ^P &= \int \dif ^3\vecrp {J ^{P0}}' \label{eq:QP1}\\%
&= \int \dif ^3\vecrp \left\{ \delta [\vecrp (t')] \right\} \label{eq:QP2}\\%
&= \int \dif ^3\vecrp \delta [\vecrp (t')] \label{eq:QP3}\\%
&= 1 .\label{eq:QP4}%
\end{align}
\end{subequations}
Eq. (\ref{eq:JPmumoving}) will be used below to simplify a great
deal of the mathematics.

Referring back to Eq. (\ref{eq:gencur}), $\Tilde{\gamma }^\mu$,
$\psi$, and $\Bar{\psi }$ have so far been clearly specified; it
remains to specify $g$ and $\Tilde{Q}$. The SM is built upon the
SU(2)$_\mL \times$U(1)$_\mY$ symmetry of weak isospin and weak
hypercharge. Both of these symmetries have associated 4-currents.
There are actually three independent weak isospin 4-currents,
which are arranged as a 3-vector of 4-currents:
\begin{equation}
\vecJ ^\mu = g_W \Bar{\Psi }_\mL \Tilde{\gamma }^\mu \Tilde{\vecT }\Psi _\mL .\label{eq:weakisospincur}%
\end{equation}
$g_W$ is the weak isospin coupling constant, defined as
$g_W=e/\sin \theta _W$, where $e$ is the magnitude of the charge
on the electron and $\theta_W=28.74^\circ$ is the weak mixing (or
Weinberg) angle \cite{ref:RPP}. In the rationalized
Heaviside-Lorentz system of units being used here, $e=\sqrt{4\mpi
\alpha}=0.3028$ to four significant figures, where $\alpha
=7.297\times 10^{-3}\simeq 1/137$ is the fine structure constant
\cite{ref:RPP}. $\Tilde{\vecT }$ is a vector of weak isospin
charge operators: $\Tilde{\vecT }=(\Tilde{T}^1,\, \Tilde{T}^2,\,
\Tilde{T}^3)$, also referred to as the generator of SU(2)$_\mL$
transformations.
\begin{equation}
\Tilde{\vecT }=\half \Tilde{\vectau},\label{eq:weakisospincharges}%
\end{equation}
where $\Tilde{\vectau}=(\Tilde{\vectau }^1,\, \Tilde{\vectau
}^2,\, \Tilde{\vectau }^3)$ is the vector of Pauli spin matrices,
which were specified (under the more common name,
$\Tilde{\vecsig}$) in Eq. (\ref{eq:PauMat}). So,
\begin{subequations}
\begin{align}
&\Tilde{T}^1 =
\begin{pmatrix}
\;\,0 & \half \\
\;\,\half & \;0\
\end{pmatrix}
\mbox{,} \quad%
\Tilde{T}^2=
\begin{pmatrix}
 \;\;\; 0 & -\half \mi \\
-\half \mi & \;\;\;\; 0\
\end{pmatrix}
\mbox{,} \quad \mbox{and} \quad%
\Tilde{T}^3=
\begin{pmatrix}
\;\,\half & \; 0\\
\;\,0 & -\half \
\end{pmatrix} \nonumber \\
&\hspace{3.3in} \mbox{(weak isospin operators).}\label{eq:Tops}
\end{align}
\end{subequations}
$\Psi _\mL $ is a 2-spinor doublet of left-handed 4-spinor wave
functions (with the spin label suppressed), which has a total of
\emph{8} independent components. The two 4-spinors that comprise a
given doublet are members of the same generation of quarks or
leptons, whichever may be the case. They were previously denoted
as $\psi ^\ms _\mc$ (cf. Eq. (\ref{eq:pointlikePsi})). Suppressing
the spin $\ms$ labels for the moment, the $\Psi _\mL $ 2-spinors
will be either
\begin{equation}
\begin{bmatrix}
u_\mL \\
\;d_\mL \
\end{bmatrix}
\mbox{,} \quad%
\begin{bmatrix}
c_\mL \\
\;s_\mL \
\end{bmatrix}
\mbox{,} \quad \mbox{and} \quad%
\begin{bmatrix}
t_\mL \\
\;\,b_\mL \
\end{bmatrix}
\qquad \mbox{(left-handed quark
doublets),}\label{eq:quarkdoublets}
\end{equation}
if $\vecJ ^\mu$ is a 4-current of quarks, and one of the following
doublets:
\begin{equation}
\begin{bmatrix}
({\nu _e})_\mL \\
\;e_\mL \
\end{bmatrix}
\mbox{,} \quad%
\begin{bmatrix}
({\nu _\mu })_\mL \\
\;\mu _\mL \
\end{bmatrix}
\mbox{,} \quad \mbox{and} \quad%
\begin{bmatrix}
({\nu _\tau })_\mL \\
\;\tau _\mL \
\end{bmatrix}
\qquad \mbox{(left-handed lepton
doublets),}\label{eq:leptondoublets}
\end{equation}
if $\vecJ ^\mu$ is a 4-current of leptons. It is important to note
that weak isospin only couples left-handed chiral states to other
left-handed chiral states; it does not couple right-handed chiral
states to anything. Eq. (\ref{eq:weakisospincur}) is very often
written in the alternate form
\begin{equation}
\vecJ ^\mu = \half g_W \Bar{\Psi }\Tilde{\gamma }^\mu \left( \Tilde{1} - \Tilde{\gamma }^5 \right) \Tilde{\vecT }\Psi ,\label{eq:VAweakisospincur}%
\end{equation}
where
\begin{equation}
\Tilde{\gamma }^5=
\begin{pmatrix}
\;\,\Tilde{1} & \;\Tilde{0}\\
\;\,\Tilde{0} & -\Tilde{1}\
\end{pmatrix}
\qquad \mbox{(chirality operator)}\label{eq:DirMat5}
\end{equation}
is a fifth Dirac matrix (cf. Eq. (\ref{eq:DirMat})), called the
chirality operator. Note also the subtle technicality that, since
$\Psi _\mL $ has a total of 8 components, all Dirac matrices are
now 8$\times$8 matrices, formed by simply multiplying the
4$\times$4 versions by the 8$\times$8 identity matrix. Written
this way, the parity-violating nature of the weak interactions
becomes apparent. Under a parity operation, $\Tilde{\gamma }^\mu$
transforms as a vector, and $\Tilde{\gamma }^\mu \Tilde{\gamma
}^5$ transforms as an axial-vector. For this reason, the weak
interaction is commonly referred to as the ``V--A" interaction
\cite{ref:Grif}. The sum of such a vector current and an
axial-vector current is neither vector nor axial-vector in form.
Consequently, interactions involving this current are not
invariant under parity transformations; that is to say, parity is
not conserved in weak interactions. The weak hypercharge 4-current
$J ^{Y\mu }$ is simpler in this regard, in that it couples both
right- and left-handed chiral states.
\begin{equation}
J ^{Y\mu } = g_Y \Bar{\Psi }\Tilde{\gamma }^\mu \Tilde{Y}\Psi .\label{eq:weakhchargecur}%
\end{equation}
$g_Y$ is the weak hypercharge coupling constant, equal to $e/2\cos
\theta _W$. $\Tilde{Y}$ is the weak hypercharge operator, and
generator of U(1)$_\mY$ transformations. The wave function $\Psi$
that appears in this equation is nonchiral in nature; it can be
written as the sum of the right-handed and left-handed chiral
states: $\Psi = \Psi _\mR +\Psi _\mL $. Examples of the $\Psi _\mL
$ 2-spinor doublets are listed above; each is an eigenstate of the
$\Tilde{Y}$ operator. In contrast, the $\Psi _\mR $ states, which
can be listed in a way analogous to the $\Psi _\mL $ doublets
listed above, are \emph{not} eigenstates of $\Tilde{Y}$
--- no unique weak hypercharge eigenvalue can be assigned to a
right-handed doublet. However, the 4-spinor (right-handed) fermion
wave functions, a pair of which make up such a doublet, \emph{are}
eigenstates of this operator --- each has its own unique
hypercharge eigenvalue. For this reason, the left-handed states
are said to appear as ``isodoublets", and the right-handed states
are said to appear as ``isosinglets", in the SM. Now, these weak
isospin and hypercharge fermion 4-currents couple to bosons
fields. But, as written, they couple to ``nonphysical" boson
states, nonphysical meaning that they are not eigenstates of the
mass operator. The physical boson (mass) states are found by a
simple transformation of the nonphysical states, and the
corresponding 4-currents to which they couple are simultaneously
arrived at from the nonphysical 4-currents specified here by
similar transformations. The physical currents appearing in the SM
are:
\begin{subequations}
\begin{align}
J^{\gamma \mu }&=J^{Y\mu }\cos \theta _W+J^{3\mu }\sin \theta _W
\hspace{.242in} \mbox{(4-current to which the photon couples)} \label{eq:photoncurrent}\\%
J^{Z\mu }&=-J^{Y\mu }\sin \theta _W+J^{3\mu }\cos \theta _W \hspace{.01in} \mbox{(4-current to which the $Z$ boson couples)} \label{eq:Zcurrent}\\%
J^{+\mu }&=\oort \left( J^{1\mu }+\mi J^{2\mu }\right) \hspace{.47in} \mbox{(4-current to which the $W^-$ boson couples)}\label{eq:Wpcurrent}\\%
J^{-\mu }&=\oort \left( J^{1\mu }-\mi J^{2\mu }\right) \hspace{.47in} \mbox{(4-current to which the $W^+$ boson couples).}\label{eq:Wmcurrent}%
\end{align}
\end{subequations}
The relevant transformation operations are merely simple rotations
in parameter space.

In finding explicit expressions for the neutral electroweak
4-currents (i.e., $J^{\gamma \mu }$ and $J^{Z\mu }$), it is useful
to first solve for the $J^{Y\mu }$ and $J^{3\mu }$ currents,
separately. Denoting $T^3_\mL $ as the eigenvalue of the
$\Tilde{T}^3$ operator acting on $\Psi _\mL $, one finds
\begin{subequations}
\begin{align}
J^{3\mu }&=g_W\Bar{\Psi }_\mL \Tilde{\gamma }^\mu \Tilde{T}^3\Psi
_\mL \label{eq:J3mu1}\\%
&=g_WT^3_\mL \Bar{\Psi }_\mL \Tilde{\gamma }^\mu \Psi _\mL \label{eq:J3mu2}\\%
&=g_WT^3_\mL J^{P\mu } _\mL ,\label{eq:J3mu3}%
\end{align}
\end{subequations}
where $J^{P\mu }_\mL$ is the probability 4-current (cf. Eq.
(\ref{eq:probcurden})) associated with $\Psi _\mL $. Eq.
(\ref{eq:JPmumoving}) can now be used to simplify. Eq.
(\ref{eq:J3mu3}) becomes
\begin{subequations}
\begin{align}
J^{3\mu }&=g_WT^3_\mL \left\{ \half \delta [\vecr (t)] (u^\mu -s^\mu ) \right\} \label{eq:J3mu4}\\%
&=\delta [\vecr (t)] q^{3\mu},\label{eq:J3mu5}%
\end{align}
\end{subequations}
where
\begin{equation}
q^{3\mu} \equiv \frac{1}{2}g_WT^3_\mL (u^\mu -s^\mu
)\label{eq:3charge4vector}%
\end{equation}
is a new 4-vector, called the ``weak isospin 4-charge". Similarly,
denoting $Y_\mR $ ($Y_\mL $) as the eigenvalue of $\Tilde{Y}$
acting on $\Psi _\mR $ ($\Psi _\mL $),
\begin{subequations}
\begin{align}
J^{Y\mu }&=g_Y\Bar{\Psi }\Tilde{\gamma }^\mu \Tilde{Y}\Psi\label{eq:JYmu1}\\%
&=g_Y\left( \Bar{\Psi }_\mR +\Bar{\Psi }_\mL \right) \Tilde{\gamma
}^\mu\Tilde{Y}\left( \Psi _\mR +\Psi _\mL \right) \label{eq:JYmu2}\\%
&=g_Y\left( \Bar{\Psi }_\mR +\Bar{\Psi }_\mL \right) \Tilde{\gamma
}^\mu\left( \Tilde{Y}\Psi _\mR +\Tilde{Y}\Psi _\mL \right) \label{eq:JYmu3}\\%
&=g_Y\left( \Bar{\Psi }_\mR +\Bar{\Psi }_\mL \right) \Tilde{\gamma
}^\mu\left( Y_\mR \Psi _\mR +Y_\mL \Psi _\mL \right) \label{eq:JYmu4}\\%
&=g_Y\left( \Bar{\Psi }_\mR \Tilde{\gamma }^\mu Y_\mR \Psi _\mR  +
\Bar{\Psi }_\mR \Tilde{\gamma }^\mu Y_\mL \Psi _\mL  +\Bar{\Psi
}_\mL \Tilde{\gamma }^\mu Y_\mR \Psi _\mR  +\Bar{\Psi }_\mL
\Tilde{\gamma
}^\mu Y_\mL \Psi _\mL  \right) \label{eq:JYmu5}\\%
&=g_Y\left( \Bar{\Psi }_\mR \Tilde{\gamma }^\mu Y_\mR \Psi _\mR
+\Bar{\Psi
}_\mL \Tilde{\gamma }^\mu Y_\mL \Psi _\mL  \right) .\label{eq:JYmu6}%
\end{align}
\end{subequations}
This last step is not so obvious. That $\Bar{\Psi }_\mR
\Tilde{\gamma }^\mu Y_\mL \Psi _\mL =0$ and $\Bar{\Psi }_\mL
\Tilde{\gamma }^\mu Y_\mR \Psi _\mR =0$ can be shown using a fair
amount of matrix manipulations (see Griffiths, \cite{ref:Grif},
for example). The details are not very illuminating, and are not
relevant for the purposes of this Appendix, so will be left out.
Eq. (\ref{eq:JYmu6}) can be simplified one step further:
\begin{equation}
J^{Y\mu }=g_YY_\mR J^{P\mu }_\mR  +g_YY_\mL J^{P\mu }_\mL ,\label{eq:JYmu7}%
\end{equation}
where $J^{P\mu }_\mR $ and $J^{P\mu }_\mL $ are the probability
4-currents (cf. Eq. (\ref{eq:probcurden}) again) associated with
$\Psi _\mR $ and $\Psi _\mL $, respectively. Making use of Eq.
(\ref{eq:JPmumoving}) again, this equation simplifies to
\begin{subequations}
\begin{align}
J^{Y\mu }&=g_YY_\mR \left\{ \half \delta [\vecr (t)] (u^\mu +s^\mu
)\right\} +g_YY_\mL \left\{ \half \delta [\vecr (t)] (u^\mu -s^\mu
)\right\} \label{eq:JYmu8}\\%
&=\delta [\vecr (t)] \half g_Y\left[ Y_\mR (u^\mu +s^\mu )+Y_\mL
(u^\mu
-s^\mu )\right] \label{eq:JYmu9}\\%
&=\delta [\vecr (t)] \half g_Y\left[ (Y_\mR +Y_\mL )u^\mu
+(Y_\mR -Y_\mL )s^\mu )\right] \label{eq:JYmu10}\\%
&=\delta [\vecr (t)] q^{Y\mu} ,\label{eq:JYmu11}%
\end{align}
\end{subequations}
where
\begin{equation}
q^{Y\mu} \equiv \half g_Y\left[ (Y_\mR +Y_\mL )u^\mu
+(Y_\mR -Y_\mL )s^\mu \right] \label{eq:Ycharge4vector}%
\end{equation}
is the ``weak hypercharge 4-charge". Eqs. (\ref{eq:J3mu5}) and
(\ref{eq:JYmu11}) are great simplifications of the two currents
that go into making up the two neutral electroweak currents of the
SM. Note that neither $J^{1\mu }$ nor $J^{2\mu }$ can be expressed
in this way, because $\Psi _\mL $ does not satisfy
particle-preserving eigenvalue equations for $\Tilde{T}^1$ and
$\Tilde{T}^2$ as it does for $\Tilde{T}^3$. This result follows
from the fact that $\Tilde{T}^3$ is the only of the three
components of $\Tilde{\vecT }$ that is diagonal. Nevertheless,
expressions for charged electroweak 4-currents can be derived from
$J^{1\mu }$ and $J^{2\mu }$.

First consider the current that couples to the photon.
\begin{subequations}
\begin{align}
J^{\gamma \mu }&=J^{Y\mu }\cos \theta _W+J^{3\mu }\sin \theta _W\label{eq:photcur1}\\%
&=\left\{ \delta [\vecr (t)] q^{Y\mu} \right\} \cos \theta _W+\left\{ \delta [\vecr (t)] q^{3\mu}\right\} \sin \theta _W\label{eq:photcur2}\\%
&=\delta [\vecr (t)] \left( q^{Y\mu}\cos \theta _W+q^{3\mu} \sin \theta _W \right) \label{eq:photcur3}\\%
&=\delta [\vecr (t)] \left\{ \half g_Y\left[ (Y_\mR +Y_\mL )u^\mu +(Y_\mR -Y_\mL )s^\mu )\right] \cos \theta _W+\right. \nonumber\\ & \quad \left. +\half g_W\left[ T^3_\mL u^\mu-T^3_\mL s^\mu \right] \sin \theta _W\right\} \label{eq:photcur4}\\%
&=\delta [\vecr (t)] \left\{ \half \left( \frac{e}{2\cos \theta _W}\right) \left[ (Y_\mR +Y_\mL )u^\mu+(Y_\mR -Y_\mL )s^\mu )\right] \cos \theta _W+\right. \nonumber\\ & \quad \left. +\left( \frac{e}{2\sin \theta _W} \right) \left[ T^3_\mL u^\mu-T^3_\mL s^\mu \right] \sin \theta _W\right\} \label{eq:photcur5}\\%
&=\delta [\vecr (t)] \frac{e}{2} \left\{ \half \left[ (Y_\mR +Y_\mL )u^\mu+(Y_\mR -Y_\mL )s^\mu )\right] +\left[ T^3_\mL u^\mu-T^3_\mL s^\mu \right] \right\} \label{eq:photcur6}\\%
&=\delta [\vecr (t)] \frac{e}{2} \left\{ \left[ \half Y_\mR + \left( T^3_\mL +\half Y_\mL \right) \right] u^\mu+\left[ \half Y_\mR -\left( T^3_\mL +\half Y_\mL \right) \right] s^\mu \right\} .\label{eq:photcur7}%
\end{align}
\end{subequations}
A weak interaction analog of the the Gell-Mann--Nishijima formula,
\begin{equation}
Y_\mc =2(Q^\gamma _\mc -T^3_\mc )\qquad \mbox{(Gell-Mann--Nishijima formula),}\label{eq:GMN2}%
\end{equation}
where $\mc$ denotes a generic chiral states ($\mc=\mR,\, \mL$),
can be used to simplify. As applied to $\mR$ states, this equation
reads
\begin{subequations}
\begin{align}
Y_\mR &=2(Q^\gamma _\mR -T^3_\mR )\label{eq:GMNR1}\\%
&=2Q^\gamma _\mR \qquad \mbox{(because $T^3_\mR =0$ always)},\label{eq:GMNR2}%
\end{align}
\end{subequations}
since none of the isosinglet $\mR$ states that appear in the SM
are eigenstates of the $\Tilde{T^3}$ operator. For the $\mL$
states,
\begin{equation}
Y_\mL =2(Q^\gamma _\mL -T^3_\mL ).\label{eq:GMNL}%
\end{equation}
Note also that the electric charge of an $\mR$ state is identical
to that of an $\mL$ state, so $Q^\gamma _\mR=Q^\gamma _\mL \equiv
Q^\gamma$. With Eqs. (\ref{eq:GMNR2}) and (\ref{eq:GMNL}), and
this relation between $Q^\gamma _\mR$ and $Q^\gamma _\mL$, Eq.
(\ref{eq:photcur7}) becomes
\begin{subequations}
\begin{align}
J^{\gamma \mu }&=\delta [\vecr (t)] \frac{e}{2} \left\{ \left[ Q^\gamma _\mR+ \left( Q^\gamma _\mL \right) \right] u^\mu+\left[ Q^\gamma _\mR-\left( Q^\gamma _\mL \right) \right] s^\mu \right\} \label{eq:photcur8}\\%
&=\delta [\vecr (t)] \frac{e}{2} \left\{ \left[ 2Q^\gamma \right] u^\mu+\left[ 0 \right] s^\mu \right\} \label{eq:photcur9}\\%
&=\delta [\vecr (t)] \left( Q^\gamma e \right) u^\mu , \label{eq:photcur10}%
\end{align}
\end{subequations}
which is a familiar result from electrodynamics \cite{ref:Wang}.

The current that couples to the $Z$ boson can be found in a
similar way.
\begin{subequations}
\begin{align}
J^{Z \mu }&=-J^{Y\mu }\sin \theta _W+J^{3\mu }\cos \theta _W\label{eq:Zcur1}\\%
&=-\left\{ \delta [\vecr (t)] q^{Y\mu} \right\} \sin \theta _W+\left\{ \delta [\vecr (t)] q^{3\mu}\right\} \cos \theta _W\label{eq:Zcur2}\\%
&=\delta [\vecr (t)] \left( -q^{Y\mu}\sin \theta _W+q^{3\mu} \cos \theta _W \right) \label{eq:Zcur3}\\%
&=\delta [\vecr (t)] \left\{ -\half g_Y\left[ (Y_\mR +Y_\mL )u^\mu +(Y_\mR -Y_\mL )s^\mu )\right] \sin \theta _W+\right. \nonumber\\ & \quad \left. +\frac{1}{2}g_W\left[ T^3_\mL u^\mu-T^3_\mL s^\mu \right] \cos \theta _W\right\} \label{eq:Zcur4}\\%
&=\delta [\vecr (t)] \left\{ -\half \left( \frac{e}{2\cos \theta _W} \right) \left[ (Y_\mR +Y_\mL )u^\mu+(Y_\mR -Y_\mL )s^\mu )\right] \sin \theta _W+\right. \nonumber\\ & \quad \left. +\left( \frac{e}{2\sin \theta _W} \right) \left[ T^3_\mL u^\mu-T^3_\mL s^\mu \right] \cos \theta _W\right\} \label{eq:Zcur5}\\%
&=\delta [\vecr (t)] \frac{(e/\sin \theta _W\cos \theta _W)}{2}\left\{ -\half \left[ (Y_\mR +Y_\mL )u^\mu+(Y_\mR -Y_\mL )s^\mu )\right] \sin ^2\theta _W+\right. \nonumber\\ & \quad \left. +\left[ T^3_\mL u^\mu-T^3_\mL s^\mu \right] \cos ^2\theta _W\right\} \label{eq:Zcur6}\\%
&=\delta [\vecr (t)] \frac{g_Z}{2}\left\{ \left[ T^3_\mL -T^3_\mL \sin ^2\theta _W-\half \left( Y_\mR +Y_\mL \right) \sin ^2\theta _W \right] u^\mu+\right. \nonumber\\ & \quad \left. +\left[ -T^3_\mL +T^3_\mL \sin ^2\theta _W-\half \left( Y_\mR -Y_\mL \right) \sin ^2\theta _W \right] s^\mu \right\} ,\nonumber \\ &\quad \: \mbox{where $g_Z\equiv  \mbox{\Large $\frac{e}{\sin \theta _W\cos \theta _W}$}$}\label{eq:Zcur7}\\%
&=\delta [\vecr (t)] \frac{g_Z}{2}\left\{ \left[ T^3_\mL -\left( T^3_\mL +\half \left\{ Y_\mR +Y_\mL \right\} \right) \sin ^2\theta _W \right] u^\mu+\right. \nonumber\\ & \quad \left. +\left[ -T^3_\mL +\left( T^3_\mL -\half \left\{ Y_\mR -Y_\mL \right\} \right) \sin ^2\theta _W \right] s^\mu \right\} .\label{eq:Zcur8}%
\end{align}
\end{subequations}
The sum of, and difference between, Eqs. (\ref{eq:GMNR2}) and
(\ref{eq:GMNL}) can be used to simplify. These relations are
easily found to be equivalent to
\begin{equation}
2Q^\gamma = T^3_\mL+\half \left( Y_\mR +Y_\mL \right) \label{eq:GMNsum}%
\end{equation}
and
\begin{equation}
0 = T^3_\mL-\half \left( Y_\mR -Y_\mL \right) ,\label{eq:GMNdiff}%
\end{equation}
respectively. Eq. (\ref{eq:Zcur8}) simplifies to
\begin{equation}
J^{Z \mu }=\delta [\vecr (t)] \frac{g_Z}{2}\left[ \left( T^3_\mL -2Q^\gamma \sin ^2\theta _W \right) u^\mu +\left( -T^3_\mL  \right) s^\mu \right] .\label{eq:Zcur9}%
\end{equation}
Even greater simplification can be achieved for the $J^{Z \mu }$
current by making the following definitions.
\begin{subequations}
\begin{align}
q_V &\equiv \half g_Z(T^3_\mL -2Q^\gamma \sin^2\theta_W)\label{eq:qV}\\%
q_A &\equiv -\half g_ZT^3_\mL ,\label{eq:qA}%
\end{align}
\end{subequations}
where
\begin{equation}
g_Z \equiv \frac{e}{\sin \theta _W\cos \theta _W},\label{eq:gZ}%
\end{equation}
as already introduced above (cf. Eq. (\ref{eq:Zcur7})). Then $J^{Z
\mu }$ can be written
\begin{equation}
J^{Z \mu }=\delta [\vecr (t)] q^\mu,\label{eq:JZmu}%
\end{equation}
where
\begin{equation}
q^\mu \equiv q_Vu^\mu +q_As^\mu \label{eq:qmu}%
\end{equation}
is the ``4-charge" of the fermion. It is interesting that a clear
V--A structure to the weak interaction is evident when $J^{Z \mu
}$ is written in this way, just as it was when $\vecJ ^{\mu }$ was
written in terms of $\Tilde{\gamma }^{\mu }$ and $\Tilde{\gamma
}^5$ (recall Eq. (\ref{eq:VAweakisospincur})). $u^\mu$ transforms
as a vector (hence the subscript V on its coefficient), and
$s^\mu$ transforms as an axial-vector (hence the subscript A on
its coefficient), under parity transformations. So $J^{Z \mu }$
has a mixed vector--axial-vector form, just like $\vecJ ^\mu $
(cf. Eq. (\ref{eq:VAweakisospincur}) again). In contrast, the
electromagnetic current $J^{\gamma \mu}$ is a purely vector
quantity: $J^{\gamma \mu}=\delta [\vecr (t)] q^\mu = \delta [\vecr
(t)] (q_Vu^\mu +q_As^\mu )$, where $q_V=Q^\gamma e$ and $q_A=0$.
Because the coefficient of $s^\mu$ is zero, $J^{\gamma \mu}$
transforms in exactly the same way under parity transformations as
$u^\mu$ does --- as a vector!

The specifications of the charged weak 4-currents, $J^{+ \mu}$ and
$J^{- \mu}$, require a bit more care. There are two caveats to
keep in mind this time. One is that charged weak interactions
change one flavor of quark or lepton into another. For example, in
the reaction $u \to d + W^+$, an up quark is transmuted into a
down quark, by way of $W^+$ emission. In these situations, the
idea of one particular particle as being the source of the current
loses its meaning. Carr makes a note of this observation in his
development of a classical description of lepton neutral current
forces \cite{ref:Carr}. The SM avoids this picture by using weak
isospin doublets (representing \emph{pairs} of such particles)
instead of wave functions representing single fermions (recall
Eqs. (\ref{eq:quarkdoublets}) and (\ref{eq:leptondoublets})). The
other caveat is that neutrinos are massless, so this subtlety must
not be overlooked when repeating the analysis that was carried out
for neutral weak 4-currents. To address the former caveat, the
4-currents will be considered in full matrix (and weak isodoublet)
notation. The latter caveat will be addressed when the need
arises.

The charged weak 4-currents will be considered simultaneously. In
place of $J^{+\mu }$ and $J^{-\mu }$, as separate quantities, the
symbol $J^{\pm \mu}$ will be used. Eqs. (\ref{eq:Wpcurrent}) and
(\ref{eq:Wmcurrent}), and (\ref{eq:Tops}) yield
\begin{subequations}
\begin{align}
J^{\pm \mu } &= \oort \left( J^{1\mu }\pm \mi J^{2\mu }\right) \label{eq:Jpmmu1}\\%
&= \oort g_W\Bar{\Psi }_\mL \Tilde{\gamma }^\mu \left( \Tilde{T}^1\pm \mi \Tilde{T}^2 \right) \Psi _\mL \label{eq:Jpmmu2}\\%
&= \oort g_W\Psi ^\dag _\mL \left( \Tilde{\gamma }^0 \Tilde{\gamma
}^\mu \right) \left[
\begin{pmatrix}
\;\,0 & \half \\
\;\,\half & \;0\
\end{pmatrix}
\pm \mi
\begin{pmatrix}
 \;\;\; 0 & -\half \mi \\
-\half \mi & \;\;\;\; 0\
\end{pmatrix}
\right] \Psi _\mL \label{eq:Jpmmu3}\\%
&=\oort g_W\Psi ^\dag _\mL \left( \Tilde{\gamma }^0 \Tilde{\gamma
}^\mu \right) \left[ \half
\begin{pmatrix}
\;\,0 & (1\pm 1) \\
\;\,(1\mp 1) & \;0\
\end{pmatrix}
\right] \Psi _\mL \label{eq:Jpmmu4}\\%
&=\halfort g_W\Psi ^\dag _\mL \left( \Tilde{\gamma }^0
\Tilde{\gamma }^\mu \right)
\begin{pmatrix}
\;\,0 & (1\pm 1) \\
\;\,(1\mp 1) & \;0\
\end{pmatrix}
\Psi _\mL \label{eq:Jpmmu5}\\%
&=\halfort g_W\left[ \psi ^\dag _{1\mL }, \psi ^\dag _{2\mL
}\right] \left( \Tilde{\gamma }^0 \Tilde{\gamma }^\mu \right)
\begin{pmatrix}
\;\,0 & (1\pm 1) \\
\;\,(1\mp 1) & \;0\
\end{pmatrix}
\begin{bmatrix}
\psi _{1\mL }\\
\;\psi _{2\mL }\
\end{bmatrix}
\label{eq:Jpmmu6}\\%
&=\halfort g_W\left[ \psi ^\dag _{1\mL }, \psi ^\dag _{2\mL
}\right] \left( \Tilde{\gamma }^0 \Tilde{\gamma }^\mu \right)
\begin{bmatrix}
(1\pm 1) \psi _{2\mL }\\
\;\,(1\mp 1)\psi _{1\mL }\
\end{bmatrix}
,\label{eq:Jpmmu7}%
\end{align}
\end{subequations}
where $\psi _{1\mL }$ and $\psi _{2\mL }$ are the two 4-spinor
components (upper and lower, respectively) of the weak isospin $L$
doublet (cf. Eqs. (\ref{eq:quarkdoublets}),
(\ref{eq:leptondoublets}) and (\ref{eq:pointlikePsi})).

The $J^{\pm 0}$ components are
\begin{subequations}
\begin{align}
J^{\pm 0} &=\halfort g_W\left( \psi ^\dag _{1\mL }, \psi ^\dag
_{2\mL }\right) \left[ \Tilde{\gamma }^0 \Tilde{\gamma }^0 \right]
\begin{bmatrix}
(1\pm 1) \psi _{2\mL }\\
\;\,(1\mp 1)\psi _{1\mL }\
\end{bmatrix}
\label{eq:Jpm01}\\%
&=\halfort g_W\left[ \psi ^\dag _{1\mL }, \psi ^\dag _{2\mL
}\right]
\begin{bmatrix}
(1\pm 1) \psi _{2\mL }\\
\;\,(1\mp 1)\psi _{1\mL }\
\end{bmatrix}
\qquad \mbox{(via Eq. (\ref{eq:gamgamrelations}))}\label{eq:Jpm02}\\%
&=\halfort g_W\left[ (1\pm 1) \psi ^\dag _{1\mL } \psi _{2\mL } +
(1\mp 1) \psi ^\dag _{2\mL } \psi _{1\mL } \right]
.\label{eq:Jpm03}%
\end{align}
\end{subequations}
Thus
\begin{subequations}
\begin{align}
J^{+0} &= \oort g_W\psi ^\dag _{1\mL } \psi _{2\mL } \label{eq:Jp01}\\%
J^{-0} &= \oort g_W\psi ^\dag _{2\mL } \psi _{1\mL } .\label{eq:Jm01}%
\end{align}
\end{subequations}
$J^{+0}$ ($J^{-0}$) corresponds to $\psi _{2\mL }$ ($\psi _{1\mL
}$) transforming into $\psi _{1\mL }$ ($\psi _{2\mL }$). In order
to preserve the idea of the 4-current as being associated with one
particular particle (as Carr \cite{ref:Carr} pointed out was a
desirable feature), as opposed to the more abstract weak
isodoublet quantity, consider the following scenario. First,
consider the $J^{+0}$ component. Obviously (recall Eqs.
(\ref{eq:quarkdoublets})--(\ref{eq:leptondoublets})) a $W^-$ is
emitted from $\psi _{2\mL }$ and $\psi _{2\mL }$ simultaneously
changes into $\psi _{1\mL }$. To preserve the flavor (identity) of
the fermion, consider the process whereby the $W^-$ is
subsequently absorbed by $\psi _{1\mL }$ immediately after it is
emitted by $\psi _{2\mL }$. Alternatively, the $W^-$ that is
emitted has vanishing energy and momentum. Let the appearance and
immediate disappearance of $\psi _{1\mL }$ occur over an
infinitesimal period of time, and within an infinitesimal volume
of space --- the usual ``virtual particle" scenario. This limiting
procedure allows for $\psi _{2\mL }$ to retain its identity during
the ``charge measurement process" to an arbitrary degree of
accuracy. A similar procedure can be used to identify the $J^{-0}$
component, which would be the charge density of $\psi _{1\mL }$.
To continue the analysis, assumptions must be made about the
masses of $\psi _{1\mL }$ and $\psi _{2\mL }$. First consider the
simpler cases where both $\psi _{1\mL }$ and $\psi _{2\mL }$ are
massive. It is also easier to evaluate the quantities in the rest
frame of the incident particle; Lorentz boosts can always be made
to generalize to arbitrary frames. If $\psi _{2\mL }$ ($\psi
_{1\mL }$) is the incident particle, $\psi _{1\mL }$ ($\psi _{2\mL
}$) will only make an appearance during an arbitrarily short
interval of time. Hence, $\psi _{1\mL }$ ($\psi _{2\mL }$) is also
at rest. These rest-frame wave functions were specified in Eq.
(\ref{eq:pointlikePsi}), using Eq. (\ref{eq:4spinorsrestframe})
for $\phi ^\ms _\mc ({p^\mu}')$ (recall primes denote the rest
frame of a fermion). ${J^{+0}}'$ works out as follows.
\begin{subequations}
\begin{align}
{J^{+0}}' &= \left. \oort g_W\psi ^\dag _{1\mL } \psi _{2\mL }
\right| _{\mbox{$\psi _{1\mL }$\,and\,$\psi _{2\mL }$\,rest frame
at
${t_1}'={t_2}'\equiv t'$}}\label{eq:Jp02}\\%
&= \oort g_W\left[ \sqrt{\delta [\vecrp (t')]}\, \oort \left(
\Tilde{0}, \Tilde{\chi }^{\ms _1 \dag} \right) \mbox{\Large e}
^{+\mi m_1 t'} \right] \times \nonumber \\ & \times \left[
\sqrt{\delta [\vecrp (t')]}\, \oort
\begin{pmatrix}
\Tilde{0}\\
 \;\; \Tilde{\chi }^{\ms _2}\
\end{pmatrix}
\mbox{\Large e} ^{-\mi m_2 t'} \right] \label{eq:Jp03}\\%
&= \delta [\vecrp (t')]\halfort g_W \left[ \Tilde{0}, \Tilde{\chi
}^{\ms _1 \dag} \right]
\begin{bmatrix}
\Tilde{0}\\
 \;\; \Tilde{\chi }^{\ms _2}\
\end{bmatrix}
\mbox{\Large e} ^{-\mi (m_2-m_1) t'}\label{eq:Jp04}\\%
&= \delta [\vecrp (t')]\halfort g_W (\Tilde{\chi }^{\ms _1 \dag} \Tilde{\chi }^{\ms _2} )\mbox{\Large e} ^{-\mi (m_2-m_1) t'}\label{eq:Jp05}\\%
&= \delta [\vecrp (t')]\halfort g_W (\delta_{\ms _1 \ms _2} )\mbox{\Large e} ^{-\mi (m_2-m_1) t'}\qquad \mbox{(via Eq. (\ref{eq:chiorthog}))}.\label{eq:Jp06}%
\end{align}
\end{subequations}
Note that in the SM, $\psi _{2\mL }$ is \emph{always} massive, but
$\psi _{1\mL }$ can be either massive or massless (cf. Eqs.
(\ref{eq:quarkdoublets})--(\ref{eq:leptondoublets})). If $\psi
_{1\mL }$ were massless (i.e., if a charged lepton current is
under consideration), one would use Eq.
(\ref{eq:massless4spinors}), instead of Eq.
(\ref{eq:4spinorsrestframe}), in the formula for $\psi _{1\mL }$
in Eq. (\ref{eq:Jp03}) above. In the end, it would merely amount
to using the particular value $\ms _1=2$. Also, since particle 1
would be travelling at the speed of light, say in the $\hatz$
direction, ${\vecp _1}'={E_1}'\hatz$; thus the $\psi _{1\mL }$
phase factor would be $\mbox{\Large e} ^{+\mi {E_1}'(t'-z')}$
instead of $\mbox{\Large e} ^{+\mi m_1t'}$. With these two
modifications, the ${J^{+0}}'$ would read
\begin{equation}
{J^{+0}}'=\delta [\vecrp (t')]\halfort g_W (\delta_{2 \ms _2}
)\mbox{\Large e} ^{-\mi [(m_2-{E_1}')t'+{E_1}'z']
}\label{eq:Jp06massless}%
\end{equation}
In either case, an SU(2)$\times$U(1) local gauge transformation
can be made to eliminate the overall phase factor of ${J^{+0}}'$.
This transformation entails transforming \emph{all} particle wave
functions $\Psi$ in the SM lagrangian, simultaneously, by the same
factor. In particular, the left handed isodoublets $\Psi _\mL$
transform as $\Psi _\mL \to {\Psi _\mL}'=\mbox{\Large e} ^{\mi
(\vecalph \cdot \Tilde{\vecT }+ \beta \Tilde{Y})}\Psi _\mL$, and
the right handed isosinglets $\Psi _\mR$ transform as $\Psi _\mR
\to {\Psi _\mR}'=\mbox{\Large e} ^{\mi \beta \Tilde{Y}}\Psi _\mR$,
where $\vecalph =\vecalph (x^\mu )$ and $\beta =\beta (x^\mu )$
are arbitrary functions of space-time \cite{ref:Halz}. Since one
is completely free to choose the $\vecalph$ and $\beta$ functions,
they can be chosen in such a way that the argument of the
exponential function vanishes. In the case where particle 1 is
massive, the SU(2)$\times$U(1) gauge freedom of the SM is being
used here to reparameterize the rest-frame time coordinate so that
$t'=0$ at this boson emission and immediate reabsorption event. If
particle 1 is massless, this choice of gauge amounts to
reparameterizing particle 2's time coordinate in such a way that
$t'={E_1}'z'/(m_2-{E_1}')$ at this event. With these appropriate
choices, Eq. (\ref{eq:Jp06}) becomes
\begin{equation}
{J^{+0}}'=\delta [\vecrp (t')]\halfort g_W (\delta_{\ms _1 \ms _2}),\label{eq:Jp07}%
\end{equation}
where $\ms _1$ is simply set equal to 2 if particle 1 is massless.
Recalling Eqs. (\ref{eq:Jp01}) and (\ref{eq:Jm01}), ${J^{-0}}'$ is
arrived at from ${J^{+0}}'$ by simply swapping indices 1 and 2:
\begin{equation}
{J^{-0}}'=\delta [\vecrp (t')]\halfort g_W (\delta_{\ms _2 \ms _1}).\label{eq:Jm02}%
\end{equation}
Or, letting $\ms _1 =\ms _2$ be assumed, one can write
\begin{equation}
{J^{\pm 0}}'=\delta [\vecrp (t')] \halfort g_W.\label{eq:Jpm04}%
\end{equation}
Note, then, that this expression for ${J^{\pm 0}}'$ can be used
for \emph{any} of the SM fermions. The possibility that particle 2
(the incident particle) is massive while particle 1 is massless
was discussed above. If particle 2 is massless and particle 1 is
massive, the ${J^{+0}}'$ 4-current is merely arrived at from the
above derivation of ${J^{+0}}'$ by swapping indices 1 and 2 in Eq.
(\ref{eq:Jp01}), and letting the primes refer to the rest frame of
particle 1 instead of to that of particle 2.

Referring back to Eq. (\ref{eq:Jpmmu7}), the $J ^{\pm \mi}$
components are
\begin{subequations}
\begin{align}
J^{\pm \mi} &=\halfort g_W\left( \psi ^\dag _{1\mL }, \psi ^\dag
_{2\mL }\right) \left[ \Tilde{\gamma }^0 \Tilde{\gamma }^\mi
\right]
\begin{bmatrix}
(1\pm 1) \psi _{2\mL }\\
\;\,(1\mp 1)\psi _{1\mL }\
\end{bmatrix}
\label{eq:Jpmi1}\\%
&=\halfort g_W\left[ \psi ^\dag _{1\mL }, \psi ^\dag _{2\mL
}\right]
\begin{pmatrix}
\;\,\Tilde{\sigma }^\mi &  \Tilde{0}\\
\;\,\Tilde{0} & -\Tilde{\sigma }^\mi\
\end{pmatrix}
\begin{bmatrix}
(1\pm 1) \psi _{2\mL }\\
\;\,(1\mp 1)\psi _{1\mL }\
\end{bmatrix}
\qquad \mbox{(via Eq. (\ref{eq:gamgamrelations}))}\label{eq:Jpmi2}\\%
&=\halfort g_W\left[ \psi ^\dag _{1\mL }, \psi ^\dag _{2\mL
}\right]
\begin{bmatrix}
\;\,(1\pm 1) \Tilde{\sigma }^\mi \psi _{2\mL }\\
-(1\mp 1)\Tilde{\sigma }^\mi \psi _{1\mL }\
\end{bmatrix}
.\label{eq:Jpmi3}%
\end{align}
\end{subequations}
Thus
\begin{subequations}
\begin{align}
J^{+\mi } &= \oort g_W\psi ^\dag _{1\mL } \Tilde{\sigma }^\mi \psi _{2\mL } \label{eq:Jpi1}\\%
J^{-\mi } &= -\oort g_W\psi ^\dag _{2\mL } \Tilde{\sigma }^\mi \psi _{1\mL } .\label{eq:Jmi1}%
\end{align}
\end{subequations}
As pointed out previously (immediately following
(\ref{eq:DirMat5})), because the $\psi$ functions here have 4
components each, the $\Tilde{\sigma }^\mi$ matrices are actually
4-dimensional generalizations of the usual 2$\times$2 Pauli spin
matrices, formed by multiplying the 2$\times$2 versions by the
4$\times$4 identity matrix. First consider the $J^{+\mi }$
3-current in the rest frame of an incident \emph{massive} particle
$\psi _{2\mL }$.
\begin{subequations}
\begin{align}
{J^{+\mi}}' &= \left. \oort g_W\psi ^\dag _{1\mL } \Tilde{\sigma }^\mi \psi _{2\mL } \right| _{\mbox{$\psi _{1\mL }$\,and\,$\psi _{2\mL }$\,rest frame at ${t_1}'={t_2}'\equiv t'$}}\label{eq:Jpi2}\\%
&= \oort g_W\left[ \sqrt{\delta [\vecrp (t')]}\, \oort \left(
\Tilde{0}, \Tilde{\chi }^{\ms _1 \dag} \right) \mbox{\Large e}
^{+\mi m_1 t'} \right] \times \nonumber \\ & \quad \times
\begin{pmatrix}
\;\,\Tilde{\sigma }^\mi &  \Tilde{0}\\
\;\,\Tilde{0} & \Tilde{\sigma }^\mi\
\end{pmatrix}
\left[ \sqrt{\delta [\vecrp (t')]}\, \oort
\begin{pmatrix}
\Tilde{0}\\
 \;\; \Tilde{\chi }^{\ms _2}\
\end{pmatrix}
\mbox{\Large e} ^{-\mi m_2 t'} \right] \label{eq:Jpi3}\\%
&= \delta [\vecrp (t')]\halfort g_W \left[ \Tilde{0}, \Tilde{\chi
}^{\ms _1 \dag} \right]
\begin{bmatrix}
\,\Tilde{0}\\
\;\,\Tilde{\sigma }^\mi \Tilde{\chi }^{\ms _2}\
\end{bmatrix}
\mbox{\Large e} ^{-\mi (m_2-m_1) t'}\label{eq:Jpi4}\\%
&= \delta [\vecrp (t')]\halfort g_W (\Tilde{\chi }^{\ms _1 \dag}
\Tilde{\sigma }^\mi \Tilde{\chi }^{\ms _2})\mbox{\Large e}
^{-\mi (m_2-m_1) t'}.\label{eq:Jpi5}%
\end{align}
\end{subequations}
At this point, it is convenient to express the ${J^+}'$ 3-current
as a general 3-vector, instead of in terms of its components. One
finds
\begin{subequations}
\begin{align}
{\vecJ ^+}' &= \delta [\vecrp (t')]\halfort g_W (\Tilde{\chi
}^{\ms _1 \dag} \Tilde{\vecsig } \Tilde{\chi }^{\ms
_2})\mbox{\Large e} ^{-\mi (m_2-m_1) t'}\label{eq:Jpi6}\\%
&= \delta [\vecrp (t')]\halfort g_W (\Tilde{\chi }^{\ms _1 \dag}
\hats \Tilde{\chi }^{\ms _2})\mbox{\Large e}
^{-\mi (m_2-m_1) t'}\label{eq:Jpi7}\\%
&= \delta [\vecrp (t')]\halfort g_W (\delta_{\ms _1 \ms _2} \hats
)\mbox{\Large e} ^{-\mi (m_2-m_1) t'}\qquad
\mbox{(via Eq. (\ref{eq:chiorthog}))},\label{eq:Jpi8}%
\end{align}
\end{subequations}
where $\hats$ is the unit vector that points in the direction of
the spin of $\psi _{2\mL }$, as before. If particle 1 were
massless, the only changes would be that $\ms _1=2$ and the phase
factor for $\psi _{1\mL }$ would be $\mbox{\Large e} ^{+\mi
{E_1}'(t'-z')}$ instead of $\mbox{\Large e} ^{+\mi m_1t'}$, where
${E_1}'$ ($z'$) is the energy (z-coordinate) of particle 1 in the
rest frame of particle 2, as introduced above. The above equation
would then read
\begin{equation}
{\vecJ ^+}'=\delta [\vecrp (t')]\halfort g_W (\delta_{2\ms _2}
\hats )\mbox{\Large e} ^{-\mi [(m_2-{E_1}')t'+{E_1}'z']}.\label{eq:Jpi8massless}%
\end{equation}
In either case, a local SU(2)$\times$U(1) gauge transformation is
performed again to redefine the zero of particle 2's rest-frame
time $t'$, so that the argument of the exponential function
vanishes. The final form of ${\vecJ ^+}'$ is thus
\begin{equation}
{\vecJ ^+}'=\delta [\vecrp (t')] \halfort g_W (\delta_{\ms _1 \ms _2} \hats ),\label{eq:Jpi9}%
\end{equation}
where $\ms _1$ must be set to 2 if particle 1 is massless. An
inspection of Eqs. (\ref{eq:Jpi1}) and (\ref{eq:Jmi1}) yields (by
merely swapping indices 1 and 2, and adding a $-$ sign in the
above equation)
\begin{equation}
{\vecJ ^-}'=-\delta [\vecrp (t')] \halfort g_W (\delta_{\ms _2 \ms _1} \hats ).\label{eq:Jmi2}%
\end{equation}
In short, with $\ms _2 =\ms _1$ assumed, one can write
\begin{equation}
{\vecJ ^\pm }'=\pm \delta [\vecrp (t')] \halfort g_W \hats.\label{eq:Jpmi4}%
\end{equation}
As with the $J^{\pm 0}$ component, this formula is correct for
\emph{all} SM fermions, whether massless or not. The case where
particle 2 is massive while particle 1 is massless was discussed
above. If particle 2 is massless and particle 1 is massive,
${\vecJ ^\pm }'$ is found by swapping indices 1 and 2 in Eq.
(\ref{eq:Jmi2}), and letting the primes refer to the rest frame of
particle 1 instead of to that of particle 2.

Eqs. (\ref{eq:Jpm04}) and (\ref{eq:Jpmi4}) yield the scalar and
vector components, respectively, of the charged 4-current ${J^{\pm
\mu }}'$. One can write
\begin{subequations}
\begin{align}
{J ^{\pm \mu}}' &= \left( {J ^{\pm 0}}',{\vecJ ^\pm }' \right)
\label{eq:Jpmmu8}\\%
&= \left( \left\{ \delta [\vecrp (t')] \halfort g_W \right\} ,
\left\{ \pm \delta [\vecrp (t')] \halfort g_W \hats \right\}
\right) \label{eq:Jpmmu9}\\%
&= \delta [\vecrp (t')] \halfort g_W \left( 1, \pm \hats \right) .
\label{eq:Jpmmu10}\\%
&= \delta [\vecrp (t')] \halfort g_W \left( 1, \pm \vecSp \right)
.\label{eq:Jpmmu11}\\%
&= \delta [\vecrp (t')] \halfort g_W \left(
{u^\mu }' \pm {s^\mu }' \right) .\label{eq:Jpmmu12}%
\end{align}
\end{subequations}
Since $J ^{\pm \mu}$ is a 4-vector, the expression in an arbitrary
Lorentz frame can be gotten from the above equation by a simple
Lorentz boost. Since the \emph{form} of a 4-vector expression
remains invariant under a Lorentz boost, in a frame in which the
incident particle is travelling at speed $v$, one easily finds
\begin{equation}
J ^{\pm \mu} =\delta [\vecr (t)] \halfort g_W \left(
u^\mu \pm s^\mu \right) .\label{eq:Jpmmu13}%
\end{equation}
To conform to the way the EM and neutral weak 4-currents were
previously expressed (Eqs. (\ref{eq:photcur10}) and
(\ref{eq:JZmu}), respectively), this formula is now recast into
the following form:
\begin{equation}
J ^{\pm \mu} =\delta [\vecr (t)] q^{\pm \mu},\label{eq:Jpmmu14}%
\end{equation}
where
\begin{equation}
q ^{\pm \mu} =\halfort g_W \left(
u^\mu \pm s^\mu \right),\label{eq:qpmmu}%
\end{equation}
is the 4-charge associated with the $J ^{\pm \mu}$ current,
respectively. Recalling the discussion following Eqs.
(\ref{eq:Jp01}) and (\ref{eq:Jm01}), the $J ^{\pm \mu}$ is to be
identified with the \emph{emission} of a $W^\mp$ boson,
respectively. So it might be said that a fermion's 4-charge $q
^{\pm \mu}$ is the 4-charge to which the $W^\mp$ boson couples.
Note the subtle point that, with this terminology, the $J ^{\pm
\mu}$ is `identified with' the $W^\mp$ boson, and \emph{not} the
$W^\pm$ boson.

All of these electroweak currents share a common form:
$J^\mu(\vecr ,t) =\delta [\vecr (t)] q^\mu$, where $q^\mu =
q_Vu^\mu + q_As^\mu$ is the corresponding 4-charge. The only
difference among the four currents is the particular pair of
values that $q_V$ and $q_A$ assume. In summary,
\begin{equation}
\left.
\begin{array}{r@{\, = \;}l}
q_V & Q^\gamma e\\
q_A & 0\\
\end{array}
\right\} \qquad \mbox{(charges to which the $\gamma$
couples)}\label{eq:chEMA}
\end{equation}
for electromagnetic interactions,
\begin{equation}
\left.
\begin{array}{r@{\, = \;}l}
q_V & \half g_Z(T^3_\mL-2Q^\gamma \sin^2\theta_W)\\
q_A & -\half g_ZT^3_\mL\\
\end{array}
\right\} \qquad \mbox{(charges to which the $Z$ boson
couples)}\label{eq:chNEWA}
\end{equation}
for neutral weak interactions, and
\begin{equation}
\left.
\begin{array}{r@{\, = \;}l}
q_V & \mbox{\large $\halfort$}g_W\\
q_A & \mp \mbox{\large $\halfort$}g_W\\
\end{array}
\right\} \qquad \mbox{(charges to which the $W^\pm$ boson
couples)}\label{eq:chCEWA}
\end{equation}
for charged weak interactions. The constants appearing in these
equations are as follows. $e=\sqrt{4\mpi \alpha}=0.3028$ is the
electromagnetic coupling constant, where $\alpha =7.297\times
10^{-3}\simeq 1/137$ is the fine structure constant
\cite{ref:RPP}. $g_Z=e/\sin\theta_W\cos\theta_W=0.7183$ is the
neutral weak coupling constant, where $\theta_W=28.74^\circ$ is
the weak mixing (or Weinberg) angle \cite{ref:RPP}. And
$g_W=e/\sin\theta_W=0.6298$ is the charged weak coupling constant
\cite{ref:RPP}. The charge quantum numbers appearing in the
electromagnetic and neutral weak 4-charges are different for
different particles. A table of values for all SM particles is
most informative. For completeness, a table of values of all of
the charge quantum numbers discussed in this Appendix is provided
here.
\renewcommand{\arraystretch}{1.4}
\begin{table}
\renewcommand{\tablename}{TABLE}
\caption{\label{tab:chargesA} CHARGE QUANTUM NUMBERS OF VARIOUS SM
FERMIONS}
\medskip
\begin{tabular*}{5in}{@{\extracolsep{\fill}}p{1.85in}ccccccc@{\hspace{1em}}@{\extracolsep{1em}}}
\hline\hline
SM Fermion & $Q^\gamma _\mL$ & $Q^\gamma _\mR$ & $Q^\gamma$ & $T^3_\mL$ & $T^3_\mR$ & $Y_\mL$ & $Y_\mR$\\
\hline
$(\nu _e)_\mL $, $(\nu _\mu )_\mL $, $(\nu _\tau )_\mL $ & \hspace{4pt} $0$ & \hspace{4pt} $0$ & \hspace{4pt} $0$ & \hspace{4pt} $\half$ & \hspace{4pt} $0$ & $-1$ & \hspace{4pt} $0$\\
$e^-_\mL $, $\mu^-_\mL $, $\tau^-_\mL $ & $-1$ & $-1$ & $-1$ & $-\half$ & \hspace{4pt} $0$ & $-1$ & \hspace{4pt} $0$\\
$(\nu _e)_\mR $, $(\nu _\mu )_\mR $, $(\nu _\tau )_\mR $ & \hspace{4pt} $0$ & \hspace{4pt} $0$ & \hspace{4pt} $0$ & \hspace{4pt} $0$ & \hspace{4pt} $0$ & \hspace{4pt} $0$ & \hspace{4pt} $0$\\
$e^-_\mR $, $\mu^-_\mR $, $\tau^-_\mR $ & $-1$ & $-1$ & $-1$ & \hspace{4pt} $0$ & \hspace{4pt} $0$ & \hspace{4pt} $0$ & $-2$\\
$u_\mL $, $c_\mL $, $t_\mL $ & \hspace{4pt} $\frac{2}{3}$ & \hspace{4pt} $\frac{2}{3}$ & \hspace{4pt} $\frac{2}{3}$ & \hspace{4pt} $\half$ & \hspace{4pt} $0$ & \hspace{4pt} $\frac{1}{3}$ & \hspace{4pt} $0$\\
$d_\mL $, $s_\mL $, $b_\mL $ & $-\frac{1}{3}$ & $-\frac{1}{3}$ & $-\frac{1}{3}$ & $-\half$ & \hspace{4pt} $0$ & \hspace{4pt} $\frac{1}{3}$ & \hspace{4pt} $0$\\
$u_\mR $, $c_\mR $, $t_\mR $ & \hspace{4pt} $\frac{2}{3}$ & \hspace{4pt} $\frac{2}{3}$ & \hspace{4pt} $\frac{2}{3}$ & \hspace{4pt} $0$ & \hspace{4pt} $0$ & \hspace{4pt} $0$ & \hspace{4pt} $\frac{4}{3}$\\
$d_\mR $, $s_\mR $, $b_\mR $ & $-\frac{1}{3}$ & $-\frac{1}{3}$ & $-\frac{1}{3}$ & \hspace{4pt} $0$ & \hspace{4pt} $0$ & \hspace{4pt} $0$ & $-\frac{2}{3}$\\
\hline\hline
\end{tabular*}
\end{table}

The vector and axial-vector charges, $q_V$ and $q_A$,
respectively, that were introduced in this appendix are by no
means a new invention. Many authors have used these charges, in
one form or another, in their presentation of electroweak
interactions (esp in the context of the V--A interaction). The
reader is referred to \cite{ref:Guni,ref:Grif,ref:Halz}, and
\cite{ref:Barg1}, to name only a few good references that make use
of these ``weak charges". In addition, there is a great number of
published papers on tests of atomic parity violation, which make
extensive reference to what is called \emph{the} ``weak charge",
$Q_W$. A close examination of the equations reveal that $Q_W$ is
identically the vector charge $q_V$ to which the $Z$ boson
couples, introduced here. The idea is that when an electron
interacts with a nucleus, it can do so via either photon or $Z$
boson exchange; the latter type of interaction violates parity. By
precisely measuring the parity-violating (electric dipole) term,
one can infer the weak charge, which can then be used to place
useful constraints on the SM. Some references on these exciting
experiments are \cite{ref:Bouc1} (seminal work),
\cite{ref:Bouc2,ref:Bouc3,ref:Comm,ref:Klap,ref:Rent,ref:Pesk2},
and \cite{ref:Barg2}.

%% file: AppendixB.tex
%%%%%%%%%%%%%%%%%
%  APPENDIX B
%%%%%%%%%%%%%%%%%

\singlespace
\setcounter{equation}{0}%
\renewcommand{\theequation}{B.\arabic{equation}}%

\begin{flushleft}
{\bf \Large Appendix B: Helicity and
Chirality}\label{sec:AppendixB}
\end{flushleft}

\indent

The purpose of this appendix is to define helicity and chirality,
and to show that they are one and the same in the high energy
limit. In this thesis, helicity $\lambda$ (actually
\emph{normalized} helicity) appears in the various expressions for
the charges of the fermions under consideration. The final
expressions for the number spectra and equivalent boson masses are
all averaged over all possible helicity states of the parent
fermion. Chirality manifests itself in the particular kinds of
particles under consideration. All SM fermions appear in either a
``left-handed" $\mL$ or a ``right-handed" $\mR$ chiral state.

In short, the helicity of a particle is the projection of the
particle's spin in the direction of the particle's motion. For
convenience, the particle is taken to be travelling in the $\hatz$
direction. If the particle's spin is denoted $\vecS =\pm S\hatz$,
the normalized helicity is then given as $\lambda =\vecS \cdot
\hatz /S =S_z/S$. As written, $\lambda$ is an eigenvalue of the
normalized helicity operator $\Lambda ^{\mu \nu}$ that acts on a
given wave function of interest. The exact expression that
$\Lambda ^{\mu \nu}$ takes depends on the type of particle the
wave function is representing. In this thesis, the particles are
generally either SM fermions (spin-one-half) or vector bosons
(spin-one). Formulas for nuclei are built up from the expressions
for the SM quarks by superposition (of the potentials and fields).
The equation of motion describing the dynamics of a fermion is the
Dirac equation, and the wave function representing the fermion is
a 4-spinor (cf. Appendix A). Vector bosons obey the Proca
equation, and are represented by 4-vectors (cf. Sec.
\ref{sec:PolarizationFourVector}). Of interest for either case is
the $\tilde{S_z}$ operator (the\; $\tilde{}$\; symbol here denotes
an operator), as noted above. $\tilde{S_z}$ is the generator of
rotations about the $z$ axis, and is related to the relevant
rotation matrix $\tilde{R} _z$ according to
\begin{equation}
\tilde{R} _z=\mbox{\Large e}^{\mi \tilde{S_z} \theta},\label{eq:Rz}%
\end{equation}
where $\theta$ is the parameter of the rotation group
corresponding to the generator $\tilde{S_z}$ of rotations
\cite{ref:Ryde}. An equation that serves to define the helicity
operator follows from Eq. (\ref{eq:Rz}) by solving for
$\tilde{S_z}$:
\begin{equation}
\tilde{S} _z=\frac{1}{\mi} \left. \frac{\dif \tilde{R} _z}{\dif \theta} \right| _{\theta=0}.\label{eq:Sz}%
\end{equation}
An excellent discussion of how 4-spinors and 4-vectors transform
under rotations is presented in \cite{ref:Ryde}. The important
results are that the $\tilde{R} _z$ operator in 4-spinor space is
\begin{equation}
\tilde{R} _z=
\begin{pmatrix}
\mbox{\Large e}^{\mi \tilde{\sigma} _z \theta /2} & \, \tilde{0} \\
\, \tilde{0} & \mbox{\Large e}^{\mi \tilde{\sigma} _z \theta/2} \\
\end{pmatrix}
\qquad \mbox{(rotation operator for
4-spinors),}\label{eq:Rz4spinors}
\end{equation}
where $\tilde{\sigma} _z$ is the 3$^{\mathrm{rd}}$ Pauli spin
matrix (i.e., $\tilde{\sigma }^3$) (cf. Eq. (\ref{eq:PauMat})) and
$\tilde{0}$ is the 2$\times$2 null matrix, and
\begin{equation}
\tilde{R} _z=
\begin{pmatrix}
1 & \;\;\; 0 & \; 0 & \;\;\; 0\\
0 & \;\;\,\, \cos {\theta} & \sin {\theta} & \;\;\; 0\\
0 & -\sin {\theta} & \cos {\theta} & \;\;\; 0\\
0 & \;\;\; 0 & \; 0 & \;\;\; 1
\end{pmatrix}
\qquad \mbox{(rotation operator for
4-vectors).}\label{eq:Rz4vectors}
\end{equation}
By Eq. (\ref{eq:Sz}), the corresponding helicity operators are
thus
\begin{equation}
\tilde{S} _z=
\begin{pmatrix}
\half \tilde{\sigma} _z & \, \tilde{0} \\
\, \tilde{0} & \half \tilde{\sigma} _z \\
\end{pmatrix}
\qquad \mbox{(helicity operator for
4-spinors),}\label{eq:Sz4spinors}
\end{equation}
and
\begin{equation}
\tilde{S} _z=
\begin{pmatrix}
0 & \;\;\; 0 & \;\;\; 0 & \;\;\; 0\\
0 & \;\;\; 0 & -\mi & \;\;\; 0\\
0 & \;\;\; \mi & \;\;\; 0 & \;\;\; 0\\
0 & \;\;\; 0 & \;\;\; 0 & \;\;\; 0
\end{pmatrix}
\qquad \mbox{(helicity operator for
4-vectors).}\label{eq:Sz4vectors}
\end{equation}
The \emph{normalized} helicity operators are gotten from the above
equations by diving through by the particles total spin. In the
former case, this factor is $1/2$ because fermions have a total
spin $S=1/2$. Vector particles have $S=1$, so this factor is
merely $1$ in the latter case. Thus (renaming these matrices as
$\Lambda ^{\mu \nu}$ for clarity),
\begin{equation}
\Lambda ^{\mu \nu}=
\begin{pmatrix}
\tilde{\sigma} _z & \, \tilde{0} \\
\, \tilde{0} & \tilde{\sigma} _z \\
\end{pmatrix}
\qquad \mbox{(normalized helicity operator for
4-spinors),}\label{eq:nSz4spinors}
\end{equation}
and
\begin{equation}
\Lambda ^{\mu \nu}=
\begin{pmatrix}
0 & \;\;\; 0 & \;\;\; 0 & \;\;\; 0\\
0 & \;\;\; 0 & -\mi & \;\;\; 0\\
0 & \;\;\; \mi & \;\;\; 0 & \;\;\; 0\\
0 & \;\;\; 0 & \;\;\; 0 & \;\;\; 0
\end{pmatrix}
\qquad \mbox{(normalized helicity operator for
4-vectors).}\label{eq:nSz4vectors}
\end{equation}
The second of these equations was exactly the $\Lambda ^{\mu \nu}$
stated in Eq. (\ref{eq:HelMat}). It was pointed out in Sec.
\ref{sec:PolarizationFourVector} that the vector boson states are
eigenstates of this operator. Longitudinal boson states have
helicity (eigenvalue) $\lambda=0$, and transverse boson states
have either $\lambda =+1$ or $\lambda =-1$. The SM fermions are
eigenstates of the first of these operators. To see this, recall
the fermion wave functions introduced in Appendix A (cf. Eq.
(\ref{eq:pointlikePsi})):
\begin{equation}
\psi ^\ms _\mc (x^\mu ,p^\mu )=\sqrt{\delta [\vecr (t)]}\, \phi
^\ms _\mc (p^\mu )\, \mbox{\Large
e} ^{-\mi p\cdot x},\label{eq:pointlikePsiB}%
\end{equation}
where
\begin{equation}
\phi ^\ms _\mR (p^\mu )=\frac{1}{2\sqrt{m(E+m)}}\,
\begin{bmatrix}
(\Tilde{E}+\Tilde{m}+\Tilde{\sigma }\cdot \vecp)\Tilde{\chi }^\ms\\
\,\Tilde{0}\
\end{bmatrix}\label{eq:R4spinorB}
\end{equation}
for right-handed chiral states, and
\begin{equation}
\phi ^\ms _\mL (p^\mu )=\frac{1}{2\sqrt{m(E+m)}}\,
\begin{bmatrix}
\Tilde{0}\\
\,(\Tilde{E}+\Tilde{m}-\Tilde{\sigma }\cdot \vecp)\Tilde{\chi
}^\ms\
\end{bmatrix}\label{eq:L4spinorB}
\end{equation}
for left-handed states. It can easily be verified (since the
$\Tilde{\chi }^\ms$ 2-spinors are eigenvalues of the
$\tilde{\sigma} _z$ operator) that the $\psi ^\ms _\mc$ wave
functions satisfy eigenvalue equations of the form
\begin{equation}
\Lambda ^{\mu \nu}\, \psi ^\ms _\mc =\lambda\, \psi ^\ms _\mc
.\label{eq:eigeqpsimassive}
\end{equation}
$\lambda$ is either $\pm 1$, depending on whether $s=1$ or $2$
(i.e., depending on whether the fermion is in a spin-up or a
spin-down state). It is important to note that a given chiral
state $\mc$ can have either $\lambda=+1$ or $\lambda=-1$. This is
\emph{not} the case when the particle is travelling at or near the
speed of light. In that limit, $\mR$ states always have $\lambda
=+1$, and $\mL$ states always have $\lambda =-1$. To see this,
recall the fermion wave functions in the massless (or $v \to 1$)
limit. They are simply Eq. (\ref{eq:pointlikePsiB}), with the
expressions in Eq. (\ref{eq:massless4spinors}),
\begin{equation}
\phi ^\ms _\mR (p^\mu )=\oort
\begin{bmatrix}
\;\, \Tilde{\chi }^1\, \\
\;\, \Tilde{0}\, \
\end{bmatrix}
\quad \mbox{and} \quad%
\phi ^\ms _\mL (p^\mu)=\oort
\begin{bmatrix}
\Tilde{0}\\
\;\, \Tilde{\chi }^2\
\end{bmatrix}\qquad \mbox{(massless fermion),}\label{eq:massless4spinorsB}
\end{equation}
used for the $\phi ^\ms _\mc (p^\mu )$ functions. Since
$\Tilde{\chi }^1$ ($\Tilde{\chi }^2$) is an eigenstate of the
$\tilde{\sigma} _z$ operator with eigenvalue $+1$ ($-1$), the
eigenvalue equations for these massless fermion states read
\begin{equation}
\Lambda ^{\mu \nu}\, \psi ^\ms _\mc =\lambda\, \psi ^\ms _\mc
,\label{eq:eigeqpsimassless}
\end{equation}
where (as stated above) $\lambda=+1$ ($-1$) for $\mc =\mR$ ($\mL$)
states.

Chirality is a bit easier to work with, as it does not depend on
the velocity of the particle under consideration. However, the
exact form of the chirality operator depends on the chosen
representation. In the Weyl representation, as has been used
throughout this thesis, this operator takes the form
\begin{equation}
\Tilde{\gamma }^5=
\begin{pmatrix}
\;\,\Tilde{1} & \;\Tilde{0}\\
\;\,\Tilde{0} & -\Tilde{1}\
\end{pmatrix}
\qquad \mbox{(chirality operator).}\label{eq:DirMat5B}
\end{equation}
It was introduced in Appendix A, in the context of explaining the
meaning of the ``V--A" interaction. In the usual 4$\times$4
representation, $\Tilde{\gamma }^5$ acts on the 4-spinor fermion
wave functions discussed above and in Appendix A. The eigenvalue
equation for $\mR$ states reads
\begin{subequations}
\begin{align}
\Tilde{\gamma }^5\psi ^\ms _\mR &=
\begin{pmatrix}
\;\,\Tilde{1} & \;\Tilde{0}\\
\;\,\Tilde{0} & -\Tilde{1}\
\end{pmatrix}\,%
\left\{ \sqrt{\delta [\vecr (t)]}\, \phi ^\ms _\mR (p^\mu )\,
\mbox{\Large e} ^{-\mi p\cdot x} \right\} \label{eq:eigeqGam5R1}\\%
&=\frac{\sqrt{\delta [\vecr (t)]}}{2\sqrt{m(E+m)}}\, \left\{
\begin{pmatrix}
\;\,\Tilde{1} & \;\Tilde{0}\\
\;\,\Tilde{0} & -\Tilde{1}\
\end{pmatrix}\,%
\begin{bmatrix}
(\Tilde{E}+\Tilde{m}+\Tilde{\sigma }\cdot \vecp)\Tilde{\chi }^\ms\\
\,\Tilde{0}\
\end{bmatrix} \right\} \,
\mbox{\Large e} ^{-\mi p\cdot x} \label{eq:eigeqGam5R2}\\%
&=\frac{\sqrt{\delta [\vecr (t)]}}{2\sqrt{m(E+m)}}\, \left\{ (+1)\,%
\begin{bmatrix}
(\Tilde{E}+\Tilde{m}+\Tilde{\sigma }\cdot \vecp)\Tilde{\chi }^\ms\\
\,\Tilde{0}\
\end{bmatrix} \right\} \,
\mbox{\Large e} ^{-\mi p\cdot x} \label{eq:eigeqGam5R3}\\%
&=(+1)\,%
\left\{ \sqrt{\delta [\vecr (t)]}\, \phi ^\ms _\mR (p^\mu )\,
\mbox{\Large e} ^{-\mi p\cdot x} \right\} \label{eq:eigeqGam5R4}\\%
&= \mc \, \psi ^\ms _\mR ,\label{eq:eigeqGam5R5}%
\end{align}
\end{subequations}
where $\mc =+1$. And that for the $\mL$ states reads
\begin{subequations}
\begin{align}
\Tilde{\gamma }^5\psi ^\ms _\mL &=
\begin{pmatrix}
\;\,\Tilde{1} & \;\Tilde{0}\\
\;\,\Tilde{0} & -\Tilde{1}\
\end{pmatrix}\,%
\left\{ \sqrt{\delta [\vecr (t)]}\, \phi ^\ms _\mL (p^\mu )\,
\mbox{\Large e} ^{-\mi p\cdot x} \right\} \label{eq:eigeqGam5L1}\\%
&=\frac{\sqrt{\delta [\vecr (t)]}}{2\sqrt{m(E+m)}}\, \left\{
\begin{pmatrix}
\;\,\Tilde{1} & \;\Tilde{0}\\
\;\,\Tilde{0} & -\Tilde{1}\
\end{pmatrix}\,%
\begin{bmatrix}
\Tilde{0}\\
\,(\Tilde{E}+\Tilde{m}-\Tilde{\sigma }\cdot \vecp)\Tilde{\chi
}^\ms\
\end{bmatrix} \right\} \,
\mbox{\Large e} ^{-\mi p\cdot x} \label{eq:eigeqGam5L2}\\%
&=\frac{\sqrt{\delta [\vecr (t)]}}{2\sqrt{m(E+m)}}\, \left\{ (-1)\,%
\begin{bmatrix}
\Tilde{0}\\
\,(\Tilde{E}+\Tilde{m}-\Tilde{\sigma }\cdot \vecp)\Tilde{\chi
}^\ms\
\end{bmatrix} \right\} \,
\mbox{\Large e} ^{-\mi p\cdot x} \label{eq:eigeqGam5L3}\\%
&=(-1)\,%
\left\{ \sqrt{\delta [\vecr (t)]}\, \phi ^\ms _\mL (p^\mu )\,
\mbox{\Large e} ^{-\mi p\cdot x} \right\} \label{eq:eigeqGam5L4}\\%
&= \mc \, \psi ^\ms _\mL ,\label{eq:eigeqGam5L5}%
\end{align}
\end{subequations}
where $\mc =-1$. In summary, $\mc =+1$ for $\mR$ chiral states,
and $\mc =-1$ for $\mL$ chiral states, regardless of velocity.

Comparing the $\lambda$ eigenvalues to the $\mc$ eigenvalues, an
interesting observation can be made. At nonrelativistic velocities
(i.e., \emph{not} in the massless, or $v \to 1$, limit), $\lambda$
and $\mc$ are unrelated. But if the fermion is travelling at or
near the speed of light (in the ultrarelativistic limit),
$\lambda$ is identical to $\mc$. In particular, in that limit,
$\mR$ chiral states have $\lambda = \mc = +1$, and $\mL$ chiral
states have $\lambda = \mc = -1$. So in that limit (and
\emph{only} in that limit), helicity can be identified with
chirality. The terms $\mR$ and $\mL$ chiral states are merely an
analogy to polarized light, in the sense that in classical ED, EM
plane waves with positive (negative) helicity are said to be left
(right) circularly polarized \cite{ref:Aitc,ref:Jack1}.

%% file: Biblio.tex
%%%%%%%%%%%%%%%%%%
% Bibliography
%%%%%%%%%%%%%%%%%%

\singlespace

%% file: Figures.tex
%%%%%%%%%%%%%%%%%
%  Figures
%%%%%%%%%%%%%%%%%

\begin{landscape}

\newpage
\begin{figure}[t]
\begin{center}
\epsfig{file=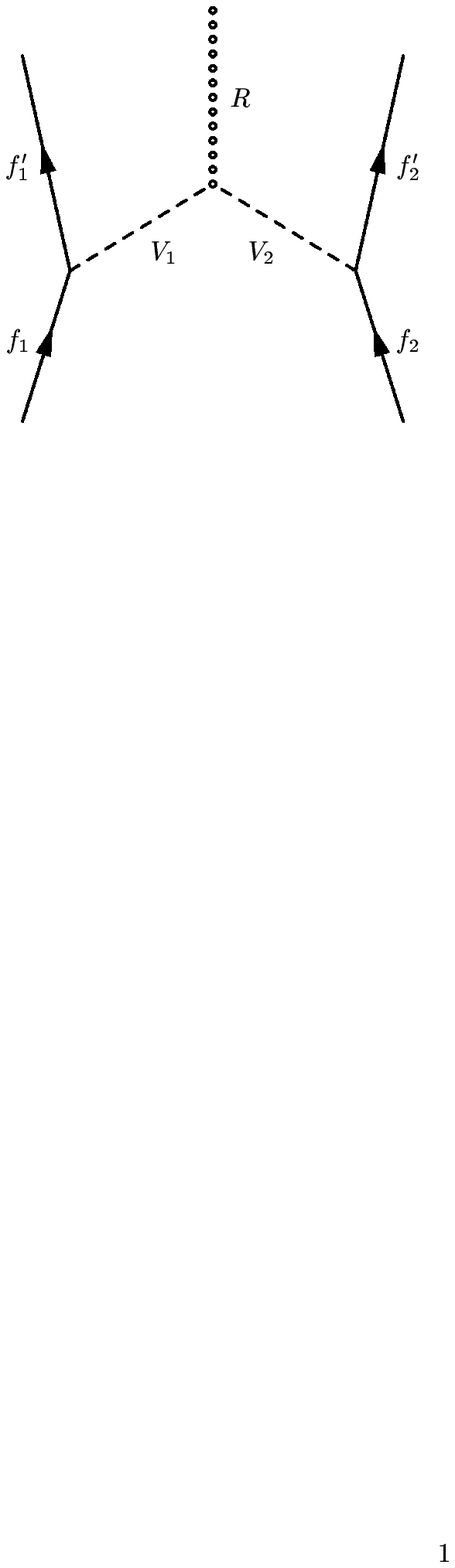,width=4.0in}
\end{center}
\renewcommand{\figurename}{FIG.}
\caption{\label{fig:Rprod} Resonance $R$ production via vector
boson ($V=W$ or $Z$ boson) fusion in a peripheral collision of two
fermions. The reaction is precluded by conservation of energy if
the mass of $R$ is less than the sum of the masses of the bosons.}
\end{figure}

\clearpage

\newpage
\begin{figure}[t]
\begin{center}
\epsfig{file=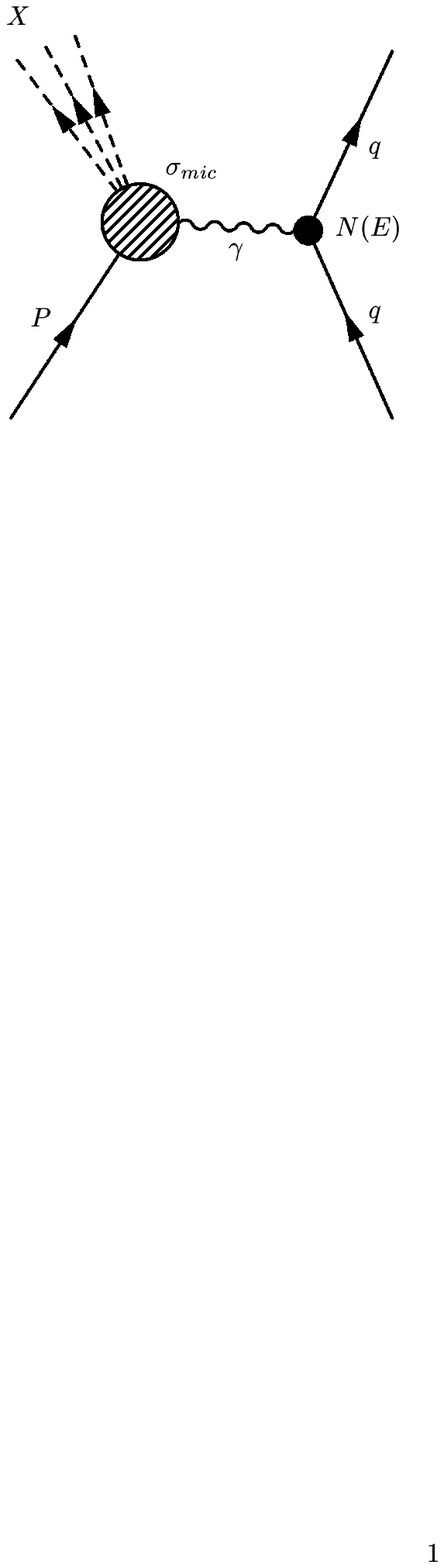,width=4.0in}
\end{center}
\renewcommand{\figurename}{FIG.}
\caption{\label{fig:Xprod} A peripheral collision between an
incident particle $q$ and a target particle $P$, by way of a
one-photon exchange. At or near the distance of closest approach,
$q$ emits the photon $\gamma$, which then subsequently interacts
with $P$ and produces some (arbitrary) final state $X$. The total
cross section $\sigma_{mac}$ for this reaction can be written
$\sigma_{mac}=\int \dif E\, N(E)\, \sigma_{mic}$. In a crude
sense, $N(E)$ gives the probability that $q$ emits the photon
$\gamma$ at energy $E$, and $\sigma_{mic}$ is the probability that
$\gamma$ then collides with $P$ and produces $X$; the integral
runs over all allowable values of $E$.}
\end{figure}

\clearpage

\newpage
\begin{figure}[t]
\begin{center}
\epsfig{file=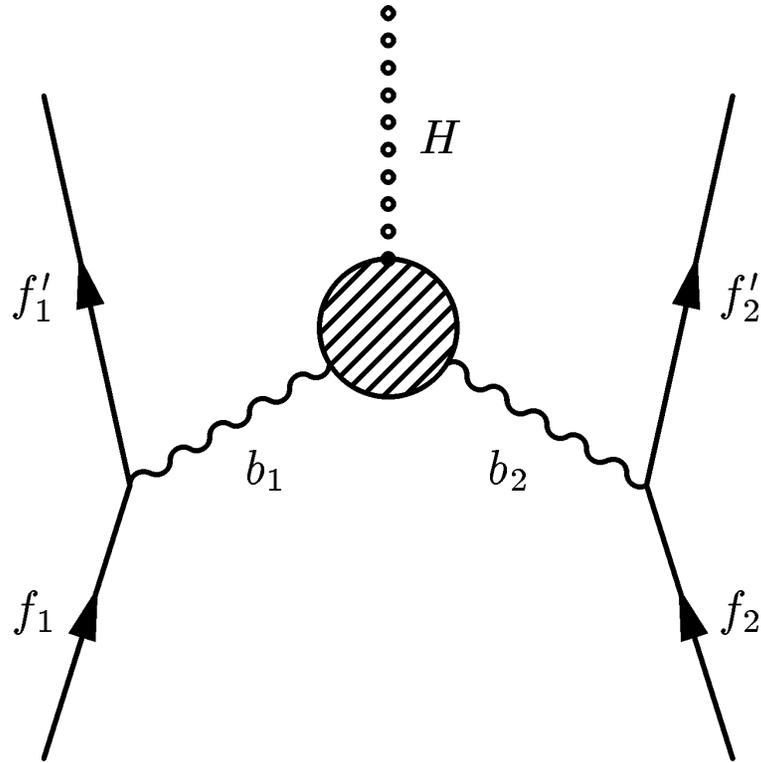,width=4.0in}
\end{center}
\renewcommand{\figurename}{FIG.}
\caption{\label{fig:Hprod} Standard Model Higgs boson $H$
production via vector boson ($b=\gamma$, $W$ or $Z$ boson) fusion
in a peripheral collision of two fermions. The shaded region at
the $bbH$ vertex just indicates that the $bb\rightarrow H$
production mechanism is in all generality not a tree level
process.}
\end{figure}

\newpage
\begin{figure}[t]
\begin{center}
\epsfig{file=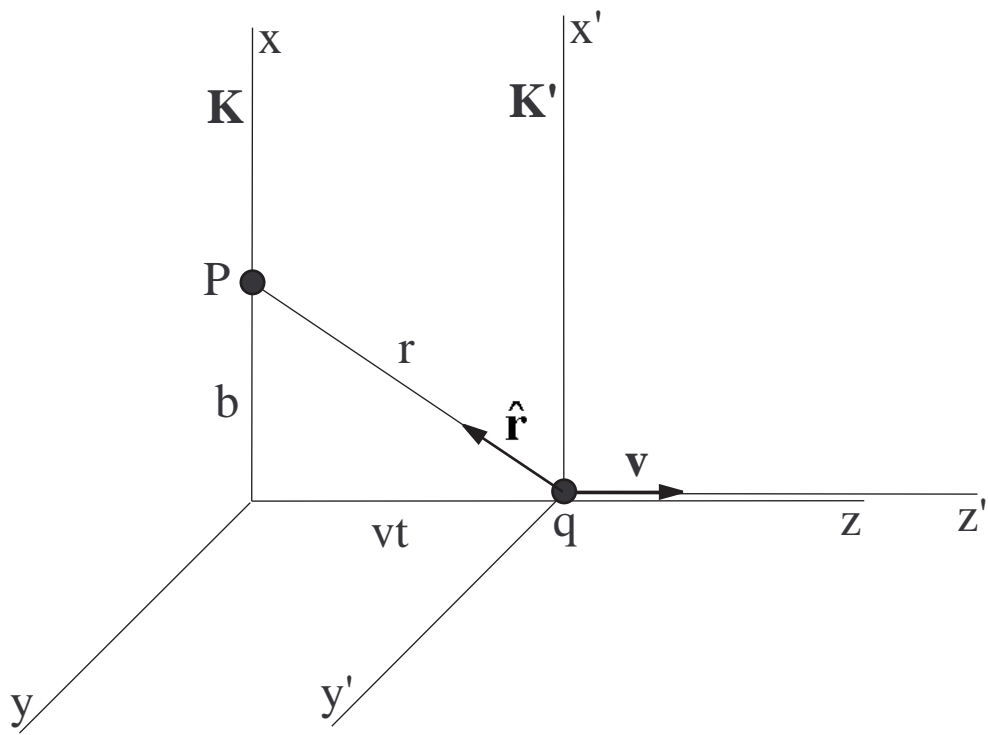,width=5.25in}
\end{center}
\renewcommand{\figurename}{FIG.}
\caption{\label{fig:LTWWM} A particle $q$ at rest at the origin of
frame $K'$ moves in the $z/z'$ direction past point $P$ in frame
$K$ with velocity $\vecv$. Relative to the origin of $K$, $P$ is
located at coordinates $(b,\, 0,\, 0)$ and the coordinates of $q$
are $(0,\, 0,\, vt)$.}
\end{figure}

\newpage
\begin{figure}[t]
\begin{center}
\epsfig{file=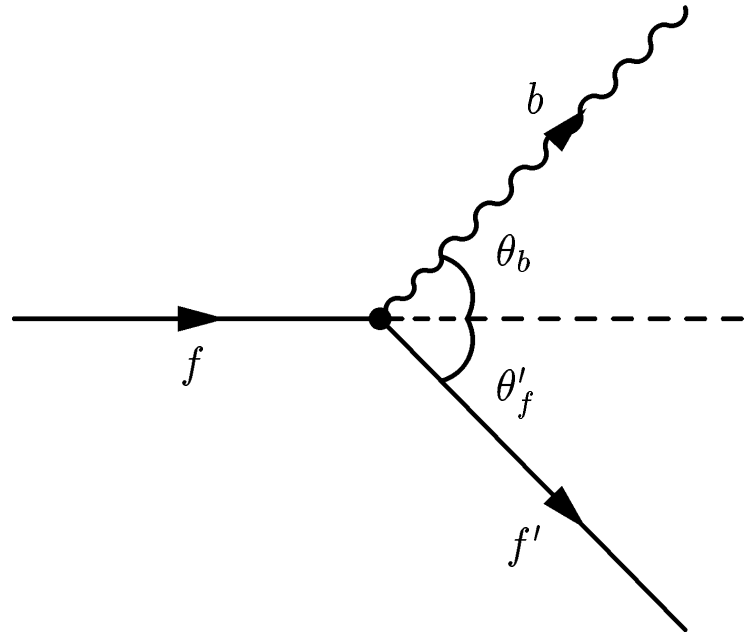,width=4.0in}
\end{center}
\renewcommand{\figurename}{FIG.}
\caption{\label{fig:fbf} A fermion $f$ that is incident from the
left emits a boson $b$ into angle $\theta _b$ with respect to the
original direction of motion; the final state fermion $f'$ is
similarly scattered into an angle ${\theta _f}'$.}
\end{figure}

\newpage
\begin{figure}[t]
\begin{center}
\epsfig{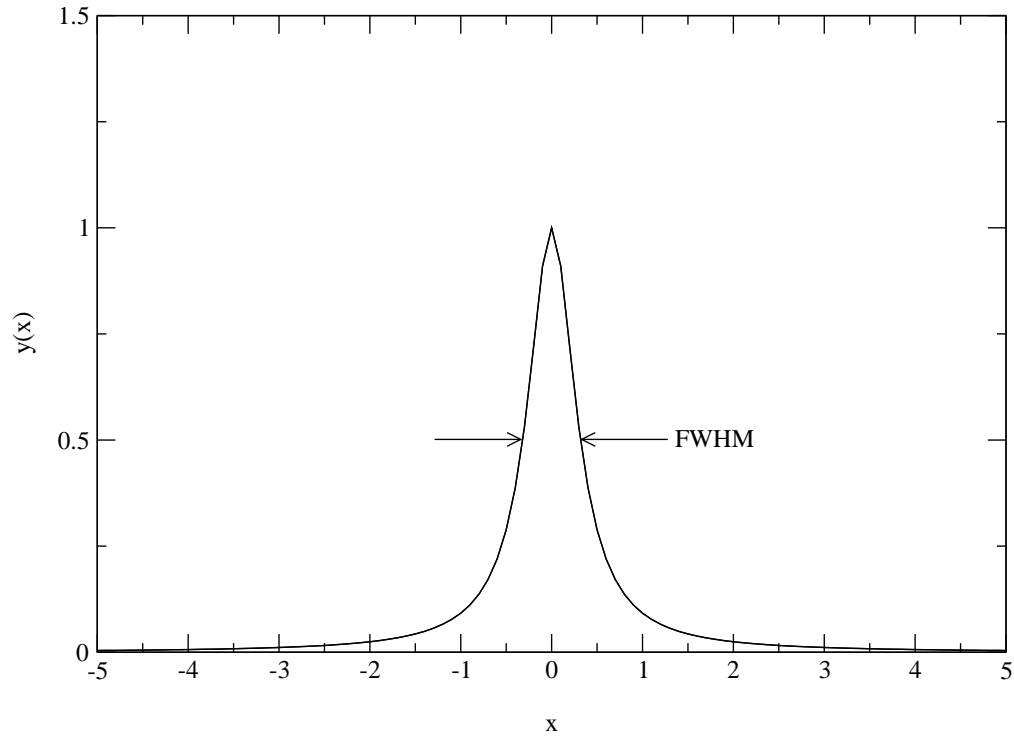}
\end{center}
\renewcommand{\figurename}{FIG.}
\caption{\label{fig:BWcurve} A Breit-Wigner (or Lorentzian) curve.
The function describing this curve is generally of the form
$y(x)=(\Delta x/2\mpi)/[(x-x_0)^2+(\Delta x/2)^2]$, where $x_0$ is
the $x$-coordinate of the peak and $\Delta x$ is the full width at
half maximum (FWHM). The amplitude (greatest $y$ value) is always
given by $2/\mpi \Delta x$. The curve shown above was constructed
to be centered at $x=0$ and have a normalized amplitude, so that
$x_0=0$ and $\Delta x=2/\mpi$.}
\end{figure}

\clearpage

\newpage
\begin{figure}[t]
\begin{center}
\epsfig{file=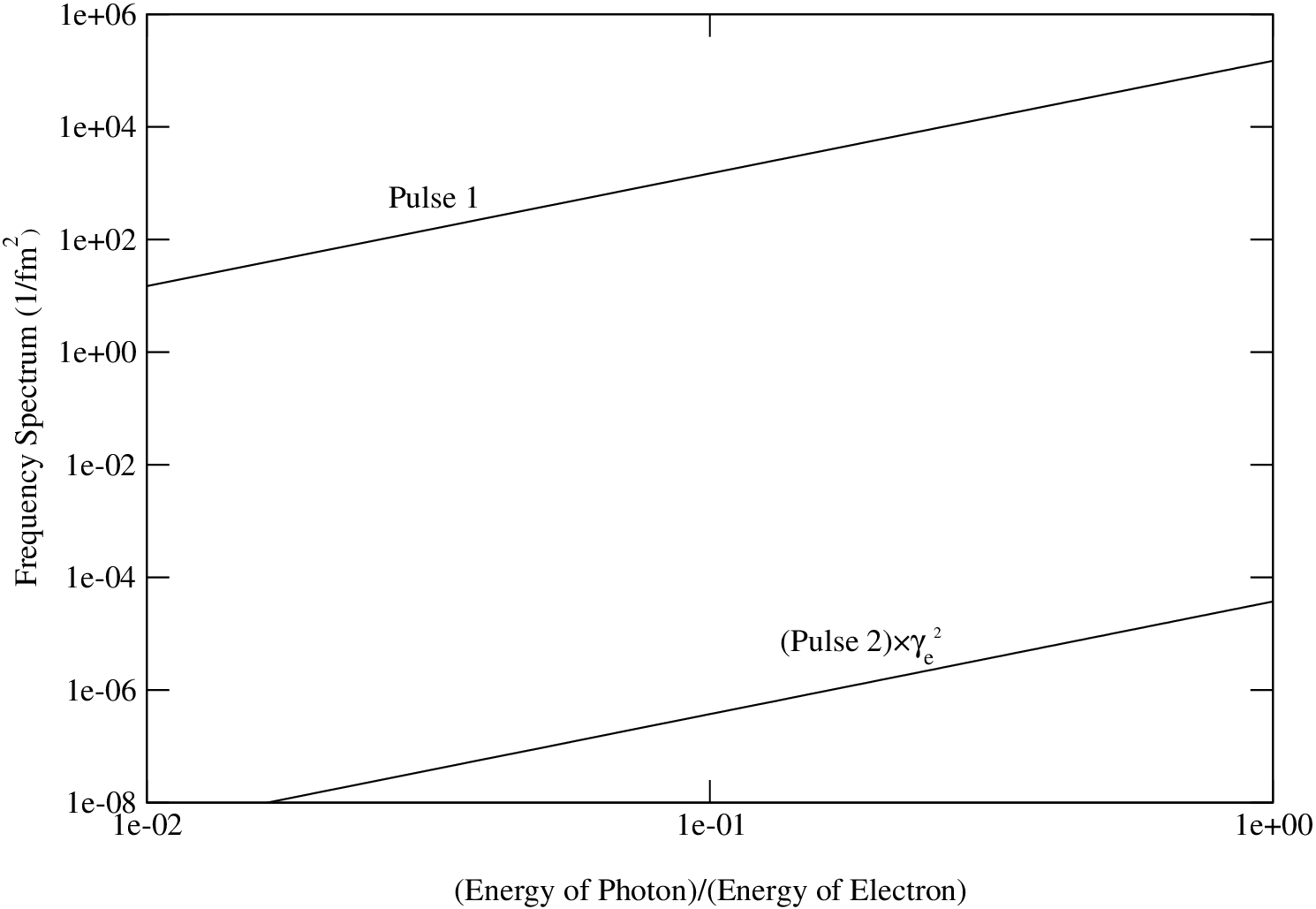,width=5.25in}
\end{center}
\renewcommand{\figurename}{FIG.}
\caption{\label{fig:frsp00} Comparison of helicity-averaged
frequency spectra of the three pulses of equivalent photons
outside a 500 GeV electron. The helicity-averaged frequency
spectrum, $\langle \dif^2I/\dif \omega\, \dif A \rangle$,
evaluated at the minimum impact parameter $b_{min}$, (cf. Eqs.
(\ref{eq:dIdwdA1exp2}) -- (\ref{eq:dIdwdA3exp2})) is plotted on
the y-axis and the Feynman scaling variable, $x=E_{\gamma}/E_e$,
is plotted on the x-axis. Because the spectrum for Pulse 2,
corresponding to transversely-polarized photons, is typically
suppressed by a factor of $\gamma _e^2$ (which can be quite large)
relative to that for Pulse 1, which also corresponds to
transversely-polarized photons, it is shown here amplified by
$\gamma _e^2$. Pulse 3, which corresponds to
longitudinally-polarized photons, does not reveal itself on the
graph because it vanishes everywhere, on account of the fact that
longitudinally-polarized photons simply do not occur in nature.}
\end{figure}

\newpage
\begin{figure}[t]
\begin{center}
\epsfig{file=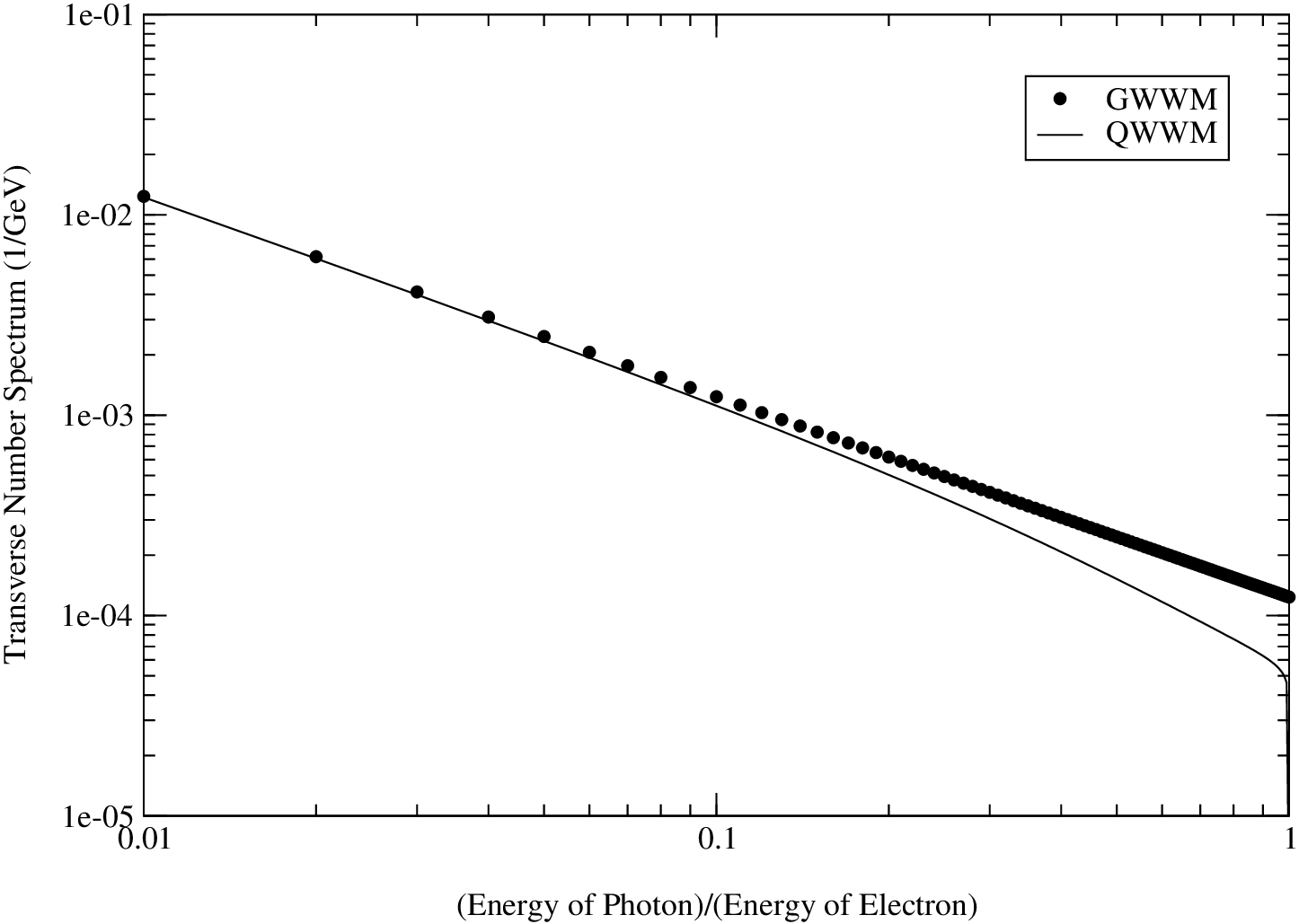,width=5.25in}
\end{center}
\renewcommand{\figurename}{FIG.}
\caption{\label{fig:nt00A} Comparison of number spectra for
transversely-polarized photons radiating from a 500 GeV electron.
Note that the number spectra for longitudinally-polarized photons
always vanishes because photons are never found in longitudinal
polarization states. $N_\mT $ is plotted on the y-axis and the
Feynman scaling variable, $x=E_{\gamma}/E_e$, is plotted on the
x-axis. The dotted curve shows the results of the GWWM (i.e., the
SWWM) and the solid curve shows the predictions of the QWWM.
Relative errors rise from $0\%$ at $x=0$ to $169\%$ at $x=0.99$;
the $N_\mT $ in the EWM drops rapidly to zero beyond $x=0.99$.
There is a slight dependence of the results on $E_e$, and a
moderate dependence on $b_{min}$.}
\end{figure}

\newpage
\begin{figure}[t]
\begin{center}
\epsfig{file=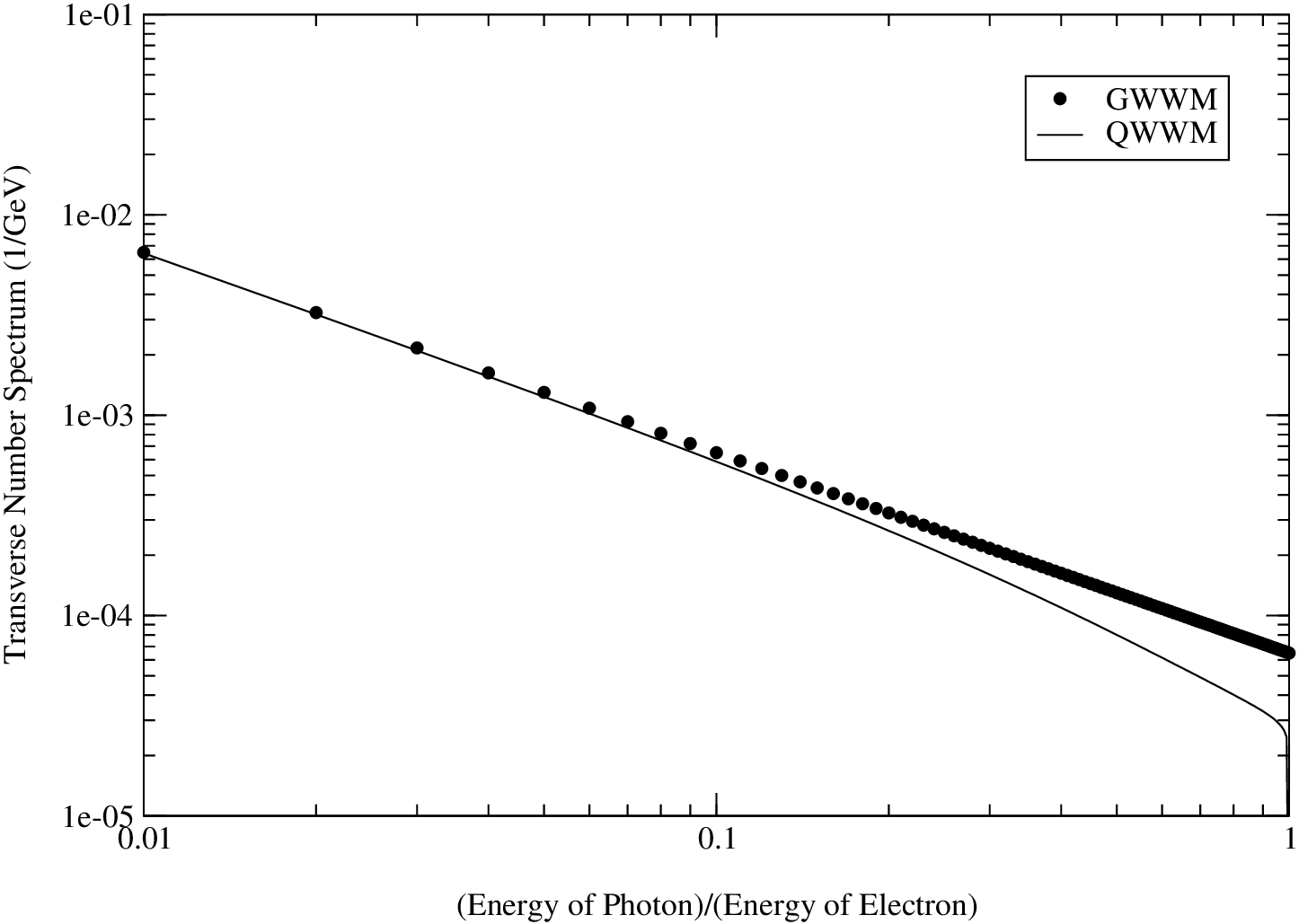,width=5.25in}
\end{center}
\renewcommand{\figurename}{FIG.}
\caption{\label{fig:nt00B} Comparison of number spectra for
transversely-polarized photons radiating from a 1000 GeV electron.
Note that the number spectra for longitudinally-polarized photons
always vanishes because photons are never found in longitudinal
polarization states. $N_\mT $ is plotted on the y-axis and the
Feynman scaling variable, $x=E_{\gamma}/E_e$, is plotted on the
x-axis. The dotted curve shows the results of the GWWM (i.e., the
SWWM) and the solid curve shows the predictions of the QWWM.
Relative errors rise from $0\%$ at $x=0$ to $165\%$ at $x=0.99$;
the $N_\mT $ in the EWM drops rapidly to zero beyond $x=0.99$.
There is a slight dependence of the results on $E_e$, and a
moderate dependence on $b_{min}$.}
\end{figure}

\clearpage

\newpage
\begin{figure}[t]
\begin{center}
\epsfig{file=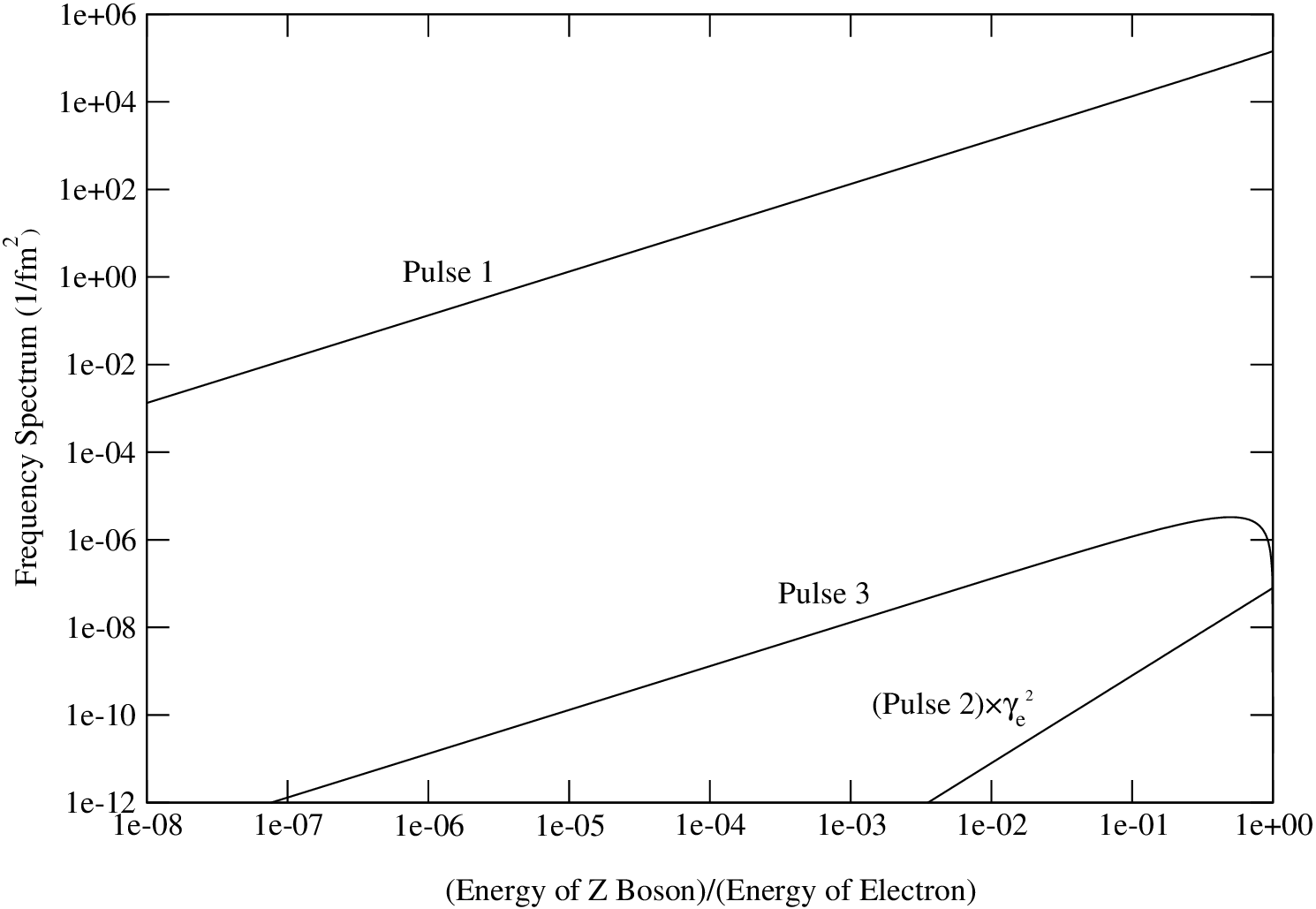,width=5.25in}
\end{center}
\renewcommand{\figurename}{FIG.}
\caption{\label{fig:frsp10} Comparison of helicity-averaged
frequency spectra of the three pulses of equivalent $Z$ bosons
outside a 500 GeV electron. The helicity-averaged frequency
spectrum, $\langle \dif^2I/\dif \omega\, \dif A \rangle$,
evaluated at the minimum impact parameter $b_{min}$, (cf. Eqs.
(\ref{eq:dIdwdA1exp2}) -- (\ref{eq:dIdwdA3exp2})) is plotted on
the y-axis and the Feynman scaling variable, $x=E_Z/E_e$, is
plotted on the x-axis. Because the spectrum for Pulse 2,
corresponding to transversely-polarized $Z$ bosons, is typically
suppressed by a factor of $\gamma _e^2$ (which can be quite large)
relative to that for Pulse 1, which also corresponds to
transversely-polarized $Z$ bosons, it is shown here amplified by
$\gamma _e^2$. Pulse 3 corresponds to the longitudinally-polarized
$Z$ bosons.}
\end{figure}

\newpage
\begin{figure}[t]
\begin{center}
\epsfig{file=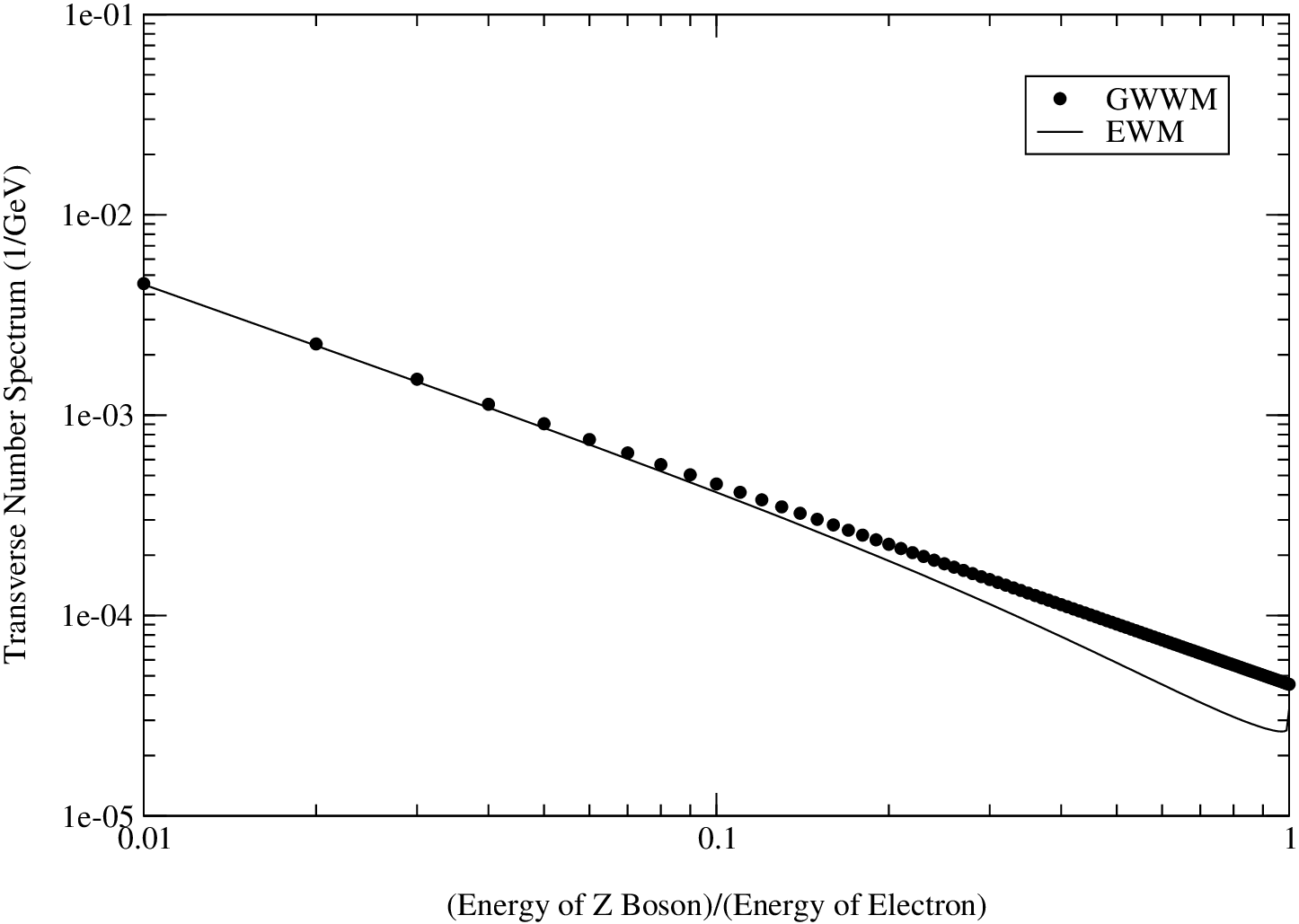,width=5.25in}
\end{center}
\renewcommand{\figurename}{FIG.}
\caption{\label{fig:nt10A} Comparison of number spectra for
transversely-polarized $Z$ bosons radiating from a 500 GeV
electron. $N_\mT $ is plotted on the y-axis and the Feynman
scaling variable, $x=E_Z/E_e$, is plotted on the x-axis. The
dotted curve shows the results of the GWWM and the solid curve
shows the predictions of the EWM. Relative errors rise from $0\%$
at $x=0$ to $33\%$ at $x=1$. There does not seem to be any
dependence of the results on $E_e$, but there is a strong
dependence on $b_{min}$.}
\end{figure}

\newpage
\begin{figure}[t]
\begin{center}
\epsfig{file=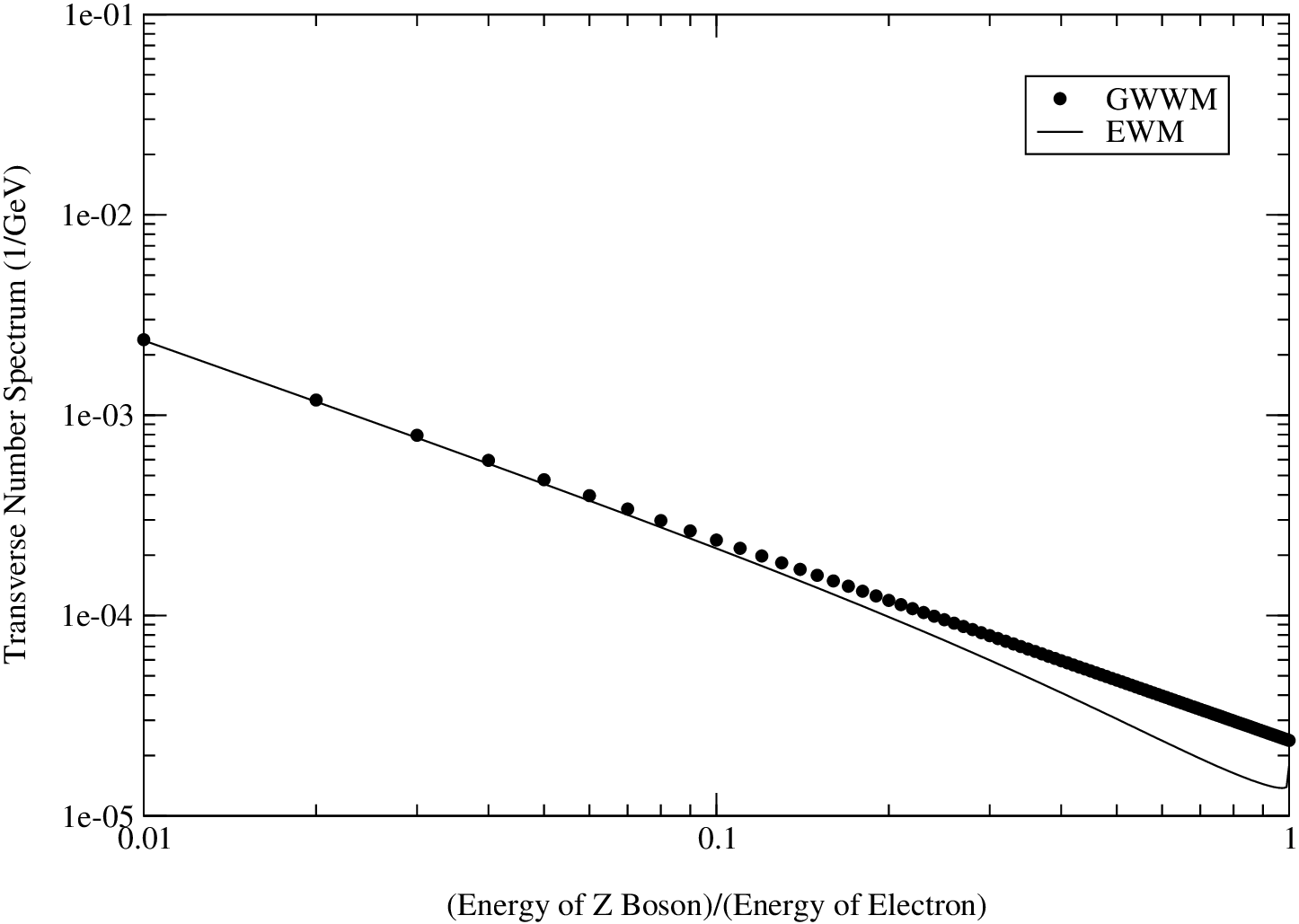,width=5.25in}
\end{center}
\renewcommand{\figurename}{FIG.}
\caption{\label{fig:nt10B} Comparison of number spectra for
transversely-polarized $Z$ bosons radiating from a 1000 GeV
electron. $N_\mT $ is plotted on the y-axis and the Feynman
scaling variable, $x=E_Z/E_e$, is plotted on the x-axis. The
dotted curve shows the results of the GWWM and the solid curve
shows the predictions of the EWM. Relative errors rise from $0\%$
at $x=0$ to $33\%$ at $x=1$. There does not seem to be any
dependence of the results on $E_e$, but there is a strong
dependence on $b_{min}$.}
\end{figure}

\newpage
\begin{figure}[t]
\begin{center}
\epsfig{file=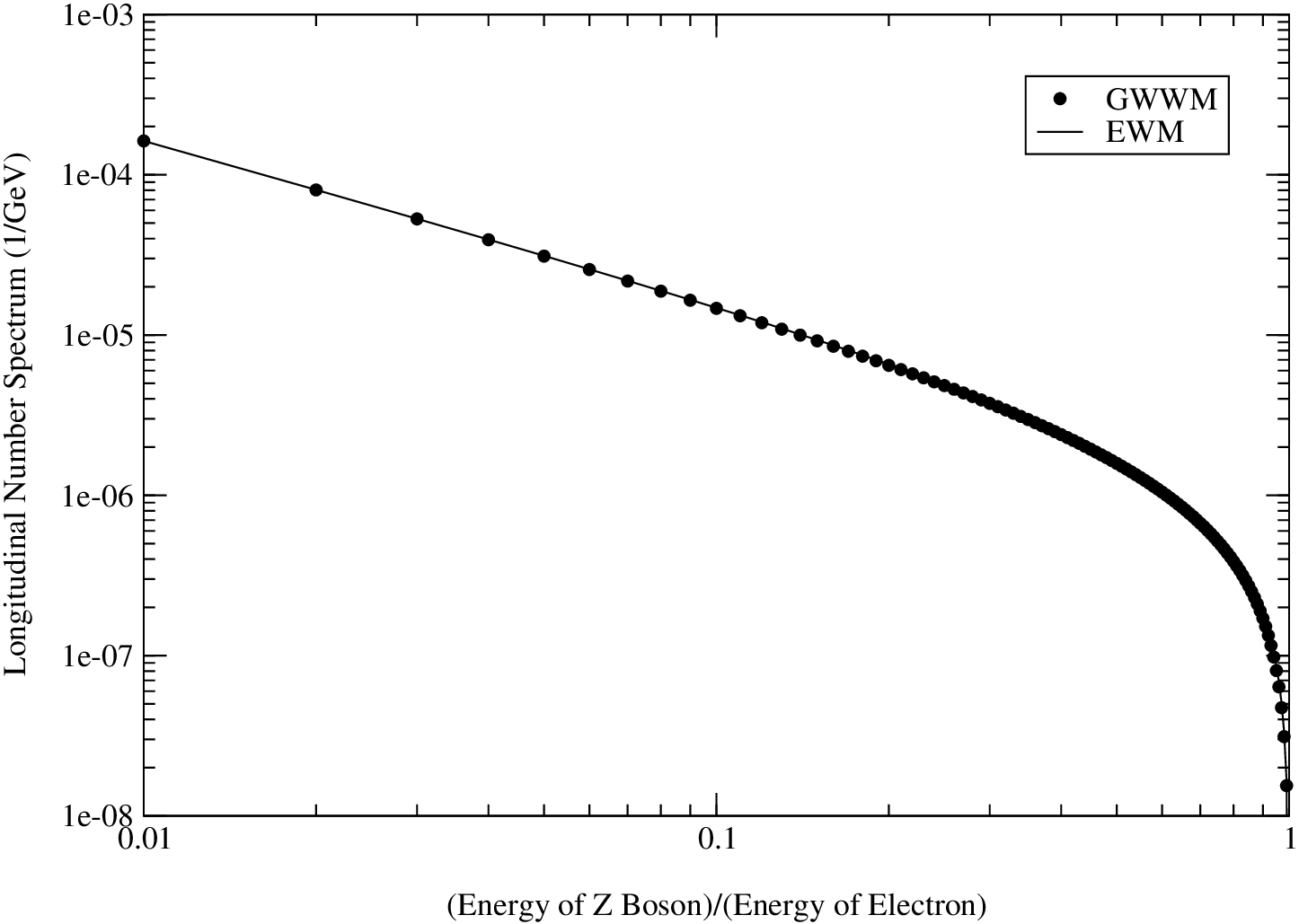,width=5.25in}
\end{center}
\renewcommand{\figurename}{FIG.}
\caption{\label{fig:nl10A} Comparison of number spectra for
longitudinally-polarized $Z$ bosons radiating from a 500 GeV
electron. $N_\mL $ is plotted on the y-axis and the Feynman
scaling variable, $x=E_Z/E_e$, is plotted on the x-axis. The solid
curve shows the predictions of the EWM and the dotted curve shows
the results of the GWWM. Relative errors rise from 0\% at $x=0$ to
$7\%$ at $x=1$. There does not seem to be any dependence of the
results on $E_e$, and only a slight dependence on $b_{min}$.}
\end{figure}

\newpage
\begin{figure}[t]
\begin{center}
\epsfig{file=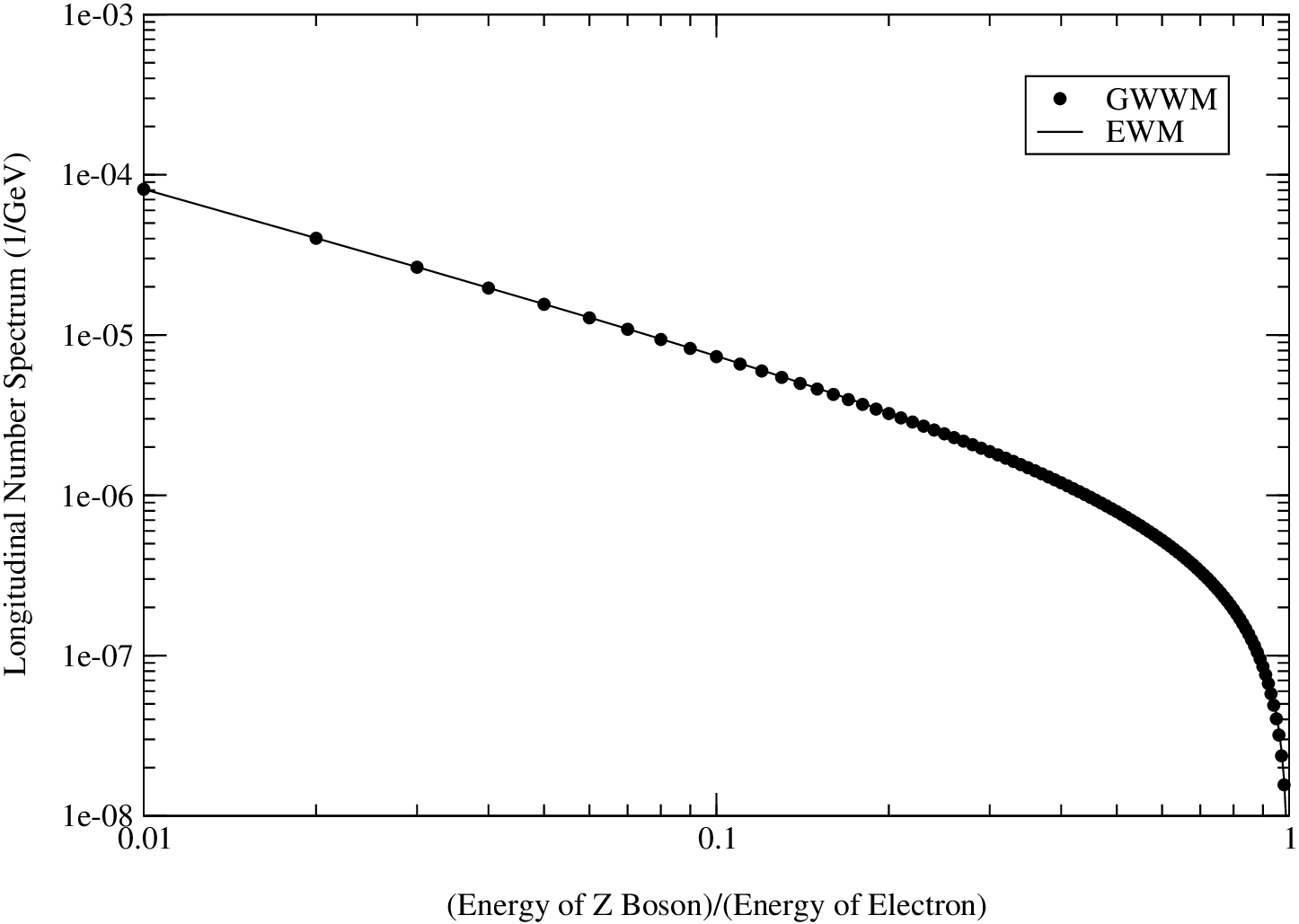,width=5.25in}
\end{center}
\renewcommand{\figurename}{FIG.}
\caption{\label{fig:nl10B} Comparison of number spectra for
longitudinally-polarized $Z$ bosons radiating from a 1000 GeV
electron. $N_\mL $ is plotted on the y-axis and the Feynman
scaling variable, $x=E_Z/E_e$, is plotted on the x-axis. The solid
curve shows the predictions of the EWM and the dotted curve shows
the results of the GWWM. Relative errors rise from 0\% at $x=0$ to
$7\%$ at $x=1$. There does not seem to be any dependence of the
results on $E_e$, and only a slight dependence on $b_{min}$.}
\end{figure}

\newpage
\begin{figure}[t]
\begin{center}
\epsfig{file=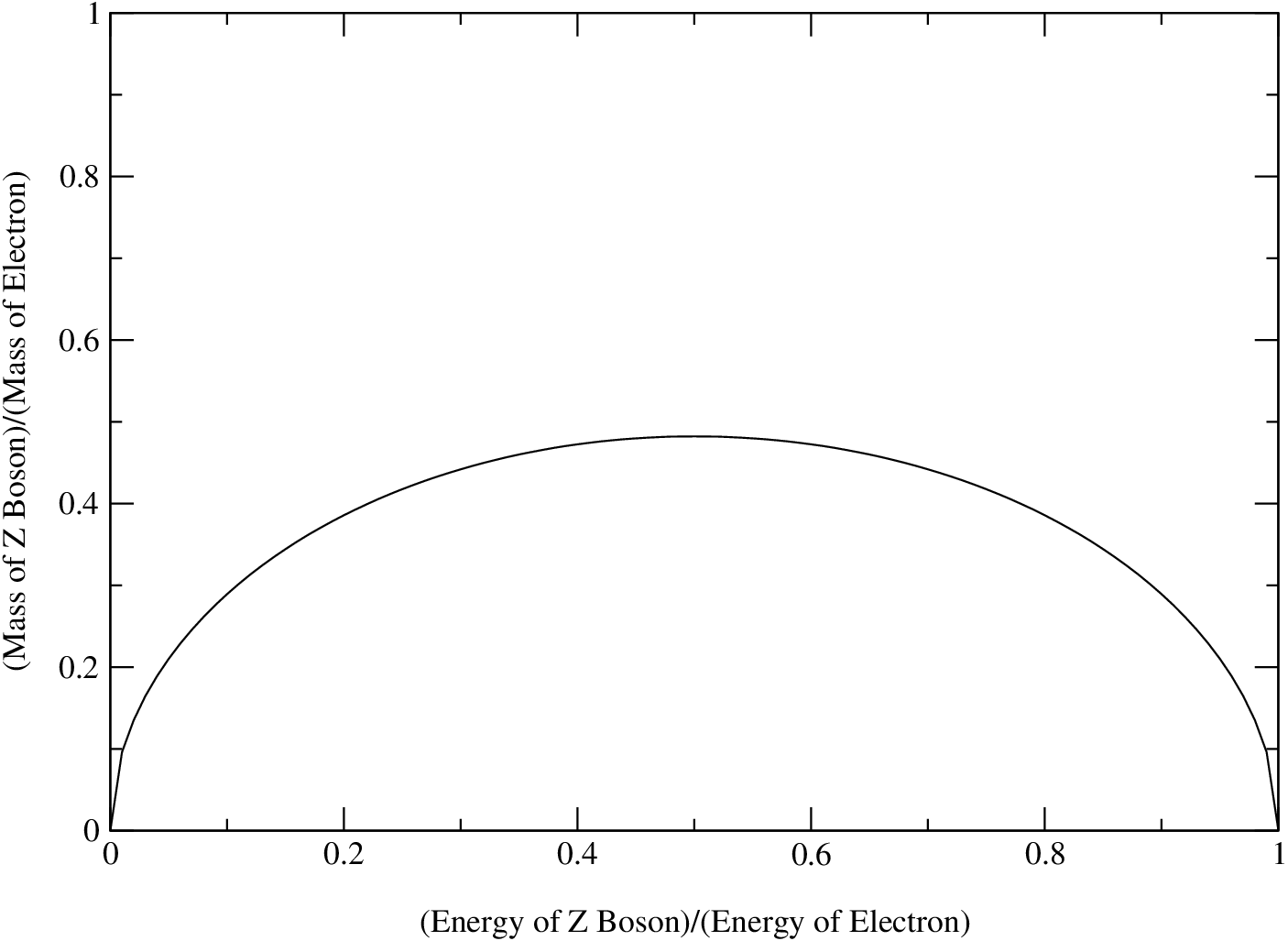,width=5.25in}
\end{center}
\renewcommand{\figurename}{FIG.}
\caption{\label{fig:mb10} The mass $m_Z$ of an equivalent $Z$
boson emitted from an electron. The ratio of $m_Z$ to the mass
$m_e$ of the electron is plotted on the y-axis and the Feynman
scaling variable, $x=E_Z/E_e$, is plotted on the x-axis. $m_Z$
vanishes at $E_Z=0$ and $E_Z=E_e-m_e$, and peaks to a maximum
value of $m_e\sqrt{0.9300}/2$ at $E_ev_e/2$, where $v_e$ is the
speed of the electron.}
\end{figure}

\clearpage

\newpage
\begin{figure}[t]
\begin{center}
\epsfig{file=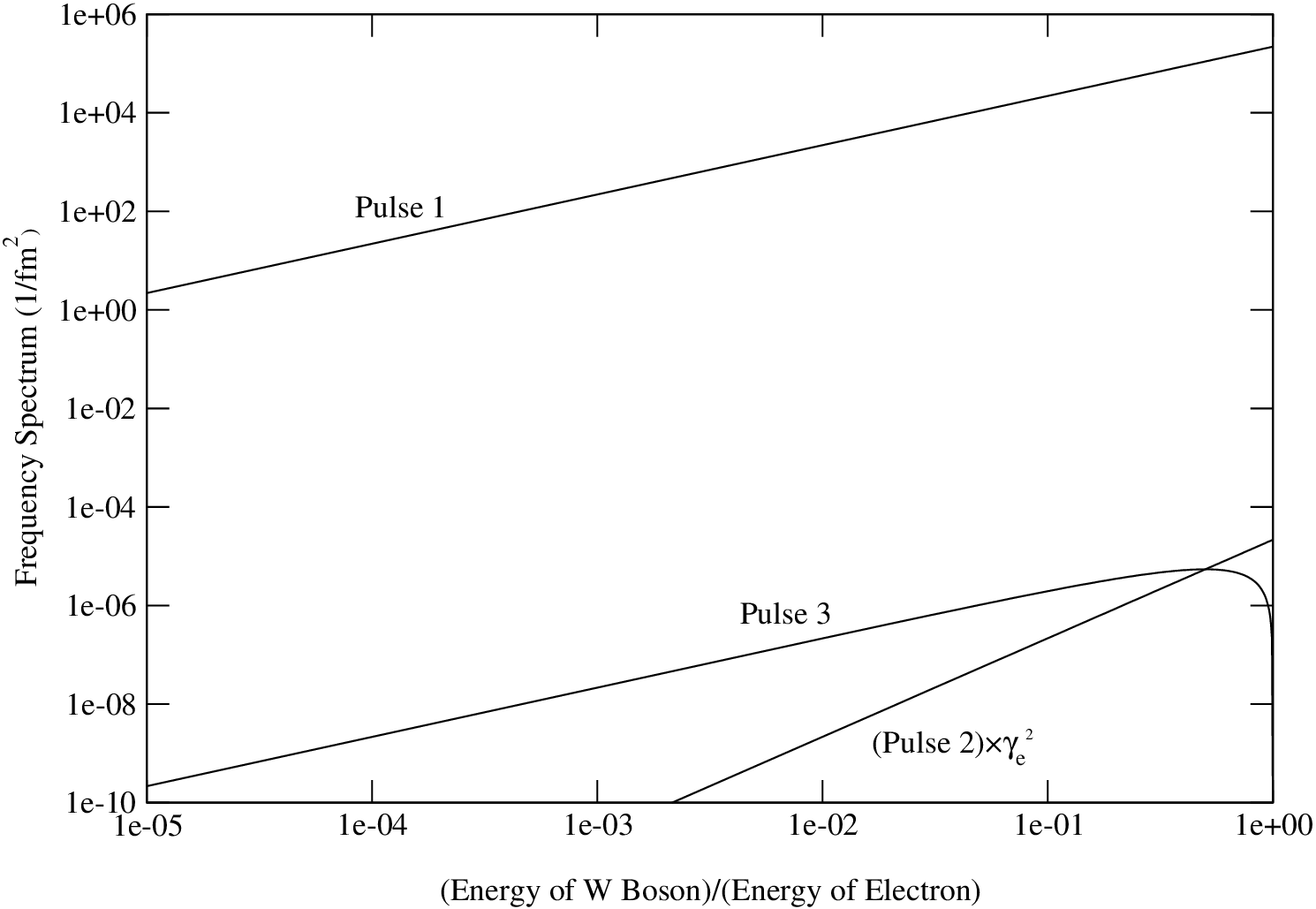,width=5.25in}
\end{center}
\renewcommand{\figurename}{FIG.}
\caption{\label{fig:frsp20} Comparison of helicity-averaged
frequency spectra of the three pulses of equivalent $W$ bosons
outside a 500 GeV electron. The helicity-averaged frequency
spectrum, $\langle \dif^2I/\dif \omega\, \dif A \rangle$,
evaluated at the minimum impact parameter $b_{min}$, (cf. Eqs.
(\ref{eq:dIdwdA1exp2}) -- (\ref{eq:dIdwdA3exp2})) is plotted on
the y-axis and the Feynman scaling variable, $x=E_W/E_e$, is
plotted on the x-axis. Because the spectrum for Pulse 2,
corresponding to transversely-polarized $W$ bosons, is typically
suppressed by a factor of $\gamma _e^2$ (which can be quite large)
relative to that for Pulse 1, which also corresponds to
transversely-polarized $W$ bosons, it is shown here amplified by
$\gamma _e^2$. Pulse 3 corresponds to the longitudinally-polarized
$W$ bosons.}
\end{figure}

\newpage
\begin{figure}[t]
\begin{center}
\epsfig{file=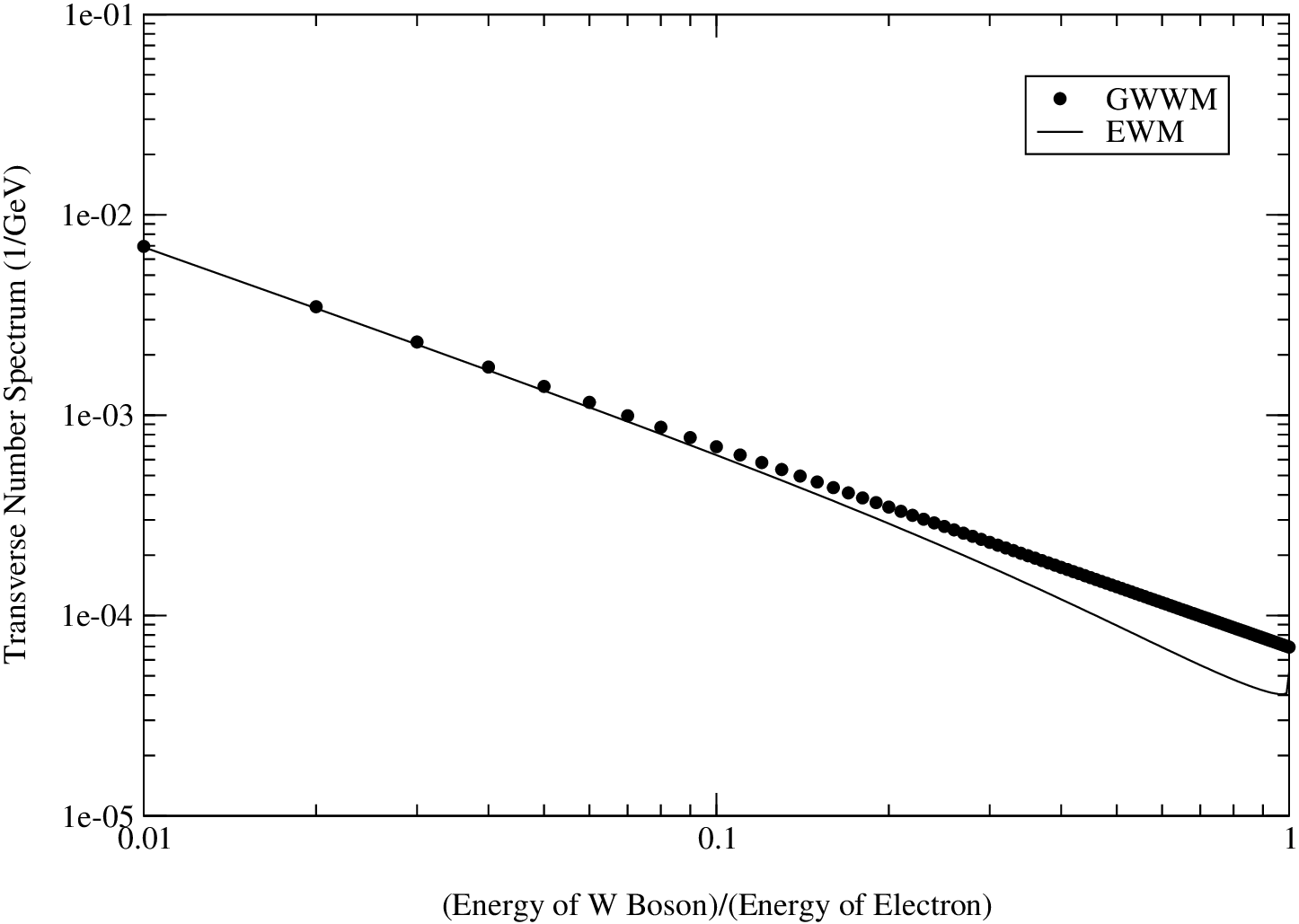,width=5.25in}
\end{center}
\renewcommand{\figurename}{FIG.}
\caption{\label{fig:nt20A} Comparison of number spectra for
transversely-polarized $W$ bosons radiating from a 500 GeV
electron. $N_\mT $ is plotted on the y-axis and the Feynman
scaling variable, $x=E_W/E_e$, is plotted on the x-axis. The
dotted curve shows the results of the GWWM and the solid curve
shows the predictions of the EWM. Relative errors rise from $0\%$
at $x=0$ to $33\%$ at $x=1$. There does not seem to be any
dependence of the results on $E_e$, but there is a strong
dependence on $b_{min}$.}
\end{figure}

\newpage
\begin{figure}[t]
\begin{center}
\epsfig{file=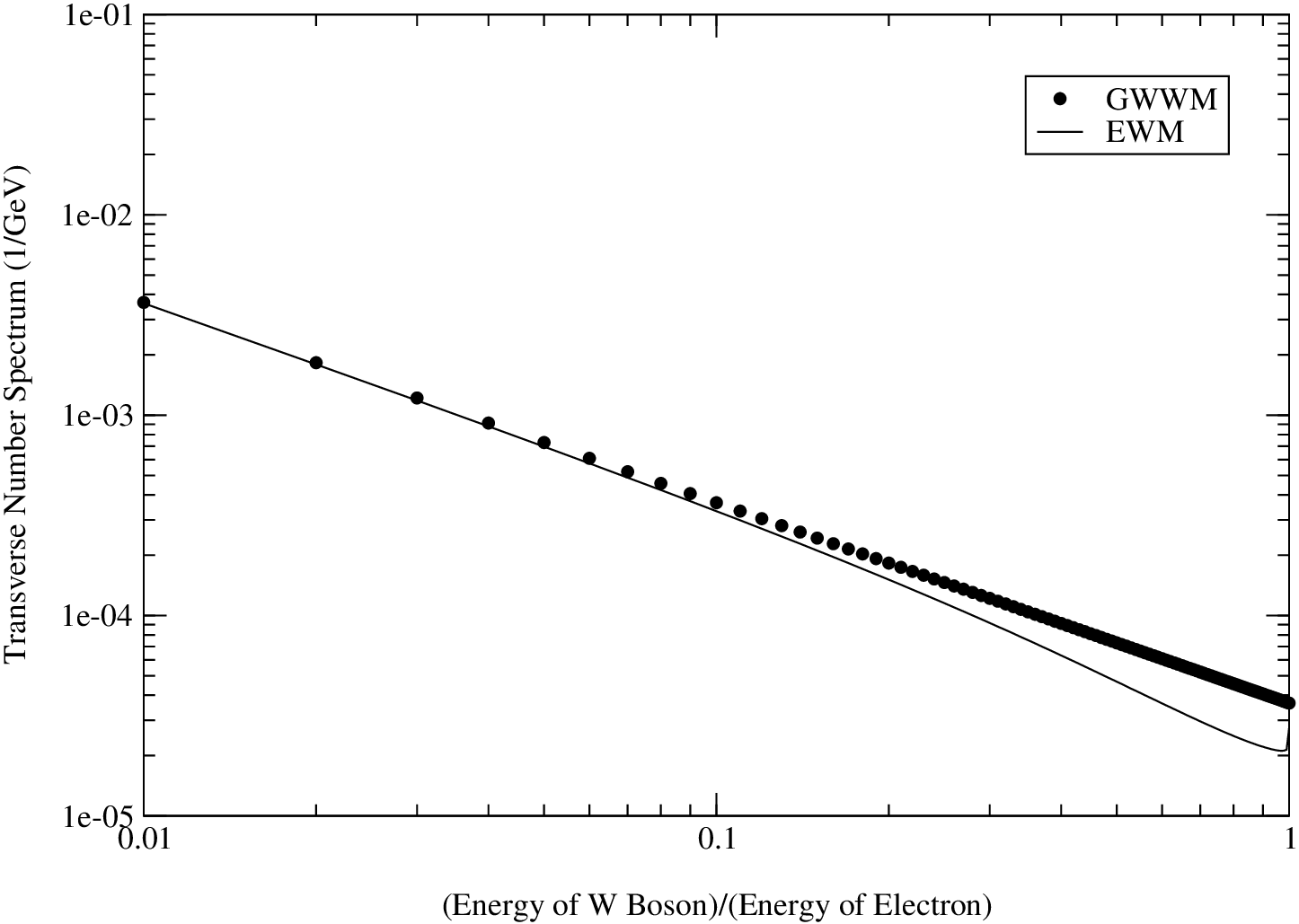,width=5.25in}
\end{center}
\renewcommand{\figurename}{FIG.}
\caption{\label{fig:nt20B} Comparison of number spectra for
transversely-polarized $W$ bosons radiating from a 1000 GeV
electron. $N_\mT $ is plotted on the y-axis and the Feynman
scaling variable, $x=E_W/E_e$, is plotted on the x-axis. The
dotted curve shows the results of the GWWM and the solid curve
shows the predictions of the EWM. Relative errors rise from $0\%$
at $x=0$ to $33\%$ at $x=1$. There does not seem to be any
dependence of the results on $E_e$, but there is a strong
dependence on $b_{min}$.}
\end{figure}

\newpage
\begin{figure}[t]
\begin{center}
\epsfig{file=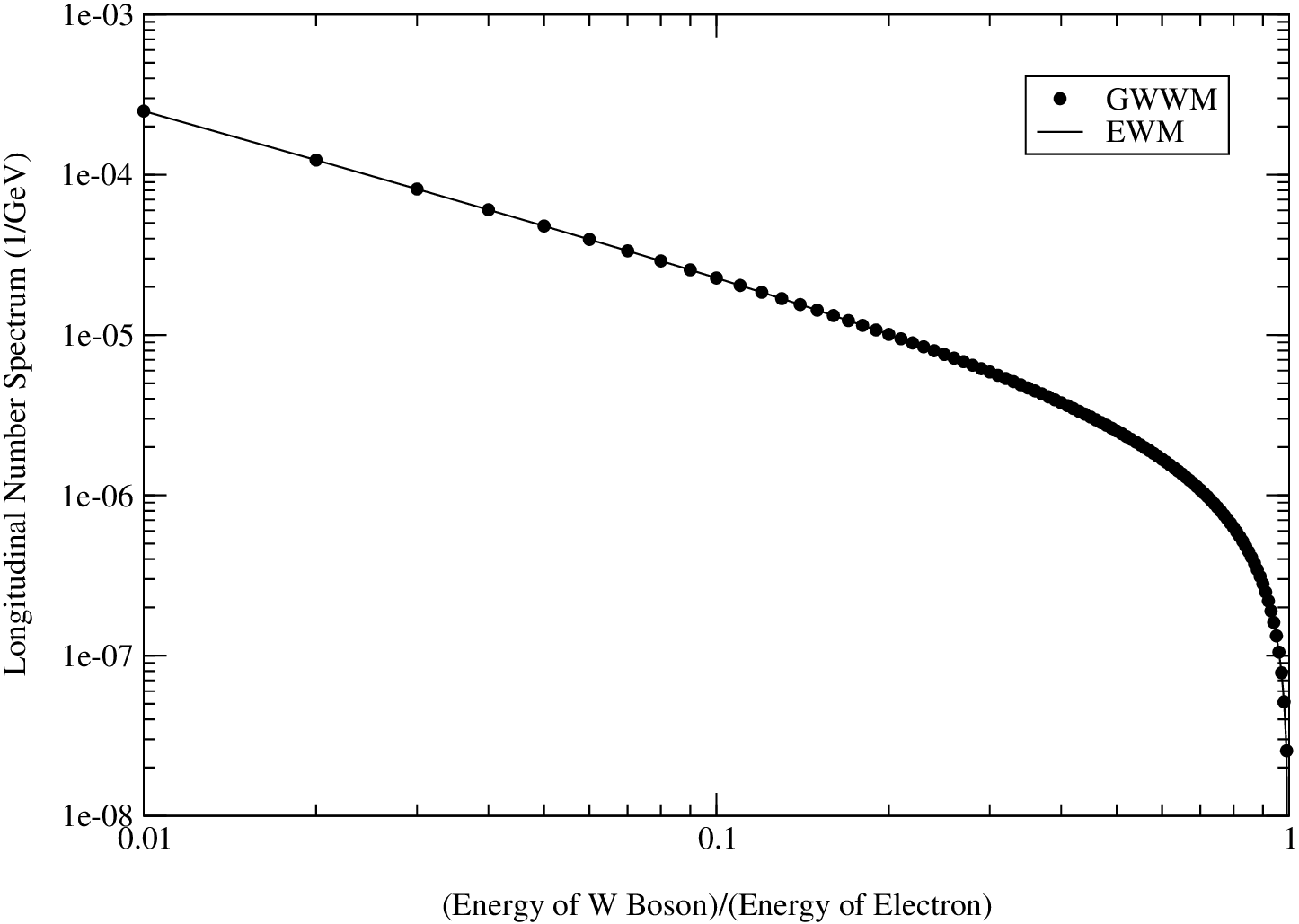,width=5.25in}
\end{center}
\renewcommand{\figurename}{FIG.}
\caption{\label{fig:nl20A} Comparison of number spectra for
longitudinally-polarized $W$ bosons radiating from a 500 GeV
electron. $N_\mL $ is plotted on the y-axis and the Feynman
scaling variable, $x=E_W/E_e$, is plotted on the x-axis. The solid
curve shows the predictions of the EWM and the dotted curve shows
the results of the GWWM. Relative errors are always on the order
of magnitude of $10^{-9}\, \%$, from $x=0$ to $x=1$. There is a
slight dependence of the results on $E_e$ and $b_{min}$.}
\end{figure}

\newpage
\begin{figure}[t]
\begin{center}
\epsfig{file=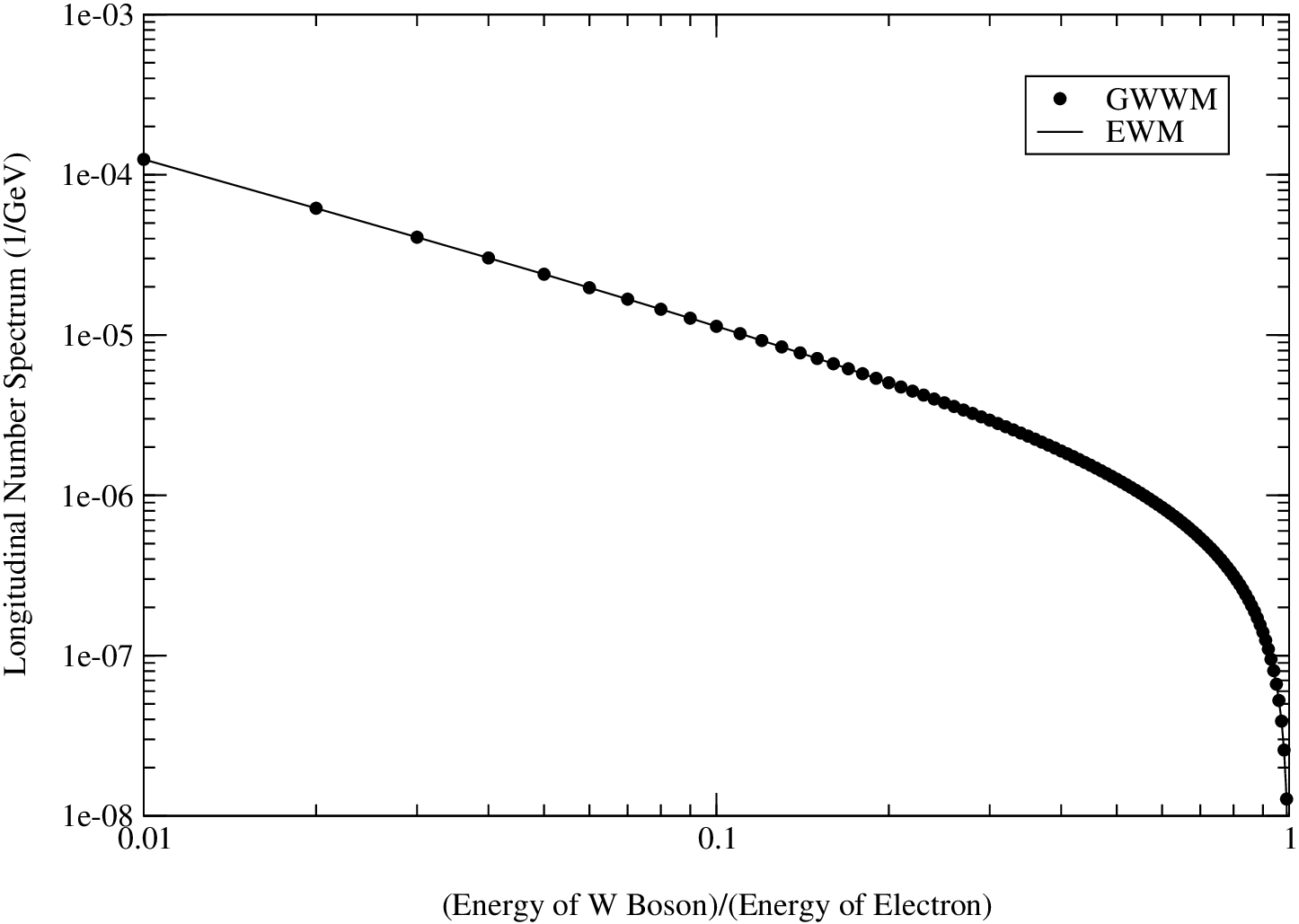,width=5.25in}
\end{center}
\renewcommand{\figurename}{FIG.}
\caption{\label{fig:nl20B} Comparison of number spectra for
longitudinally-polarized $W$ bosons radiating from a 1000 GeV
electron. $N_\mL $ is plotted on the y-axis and the Feynman
scaling variable, $x=E_W/E_e$, is plotted on the x-axis. The solid
curve shows the predictions of the EWM and the dotted curve shows
the results of the GWWM. Relative errors are always on the order
of magnitude of $10^{-8}\, \%$, from $x=0$ to $x=1$. There is a
slight dependence of the results on $E_e$ and $b_{min}$.}
\end{figure}

\newpage
\begin{figure}[t]
\begin{center}
\epsfig{file=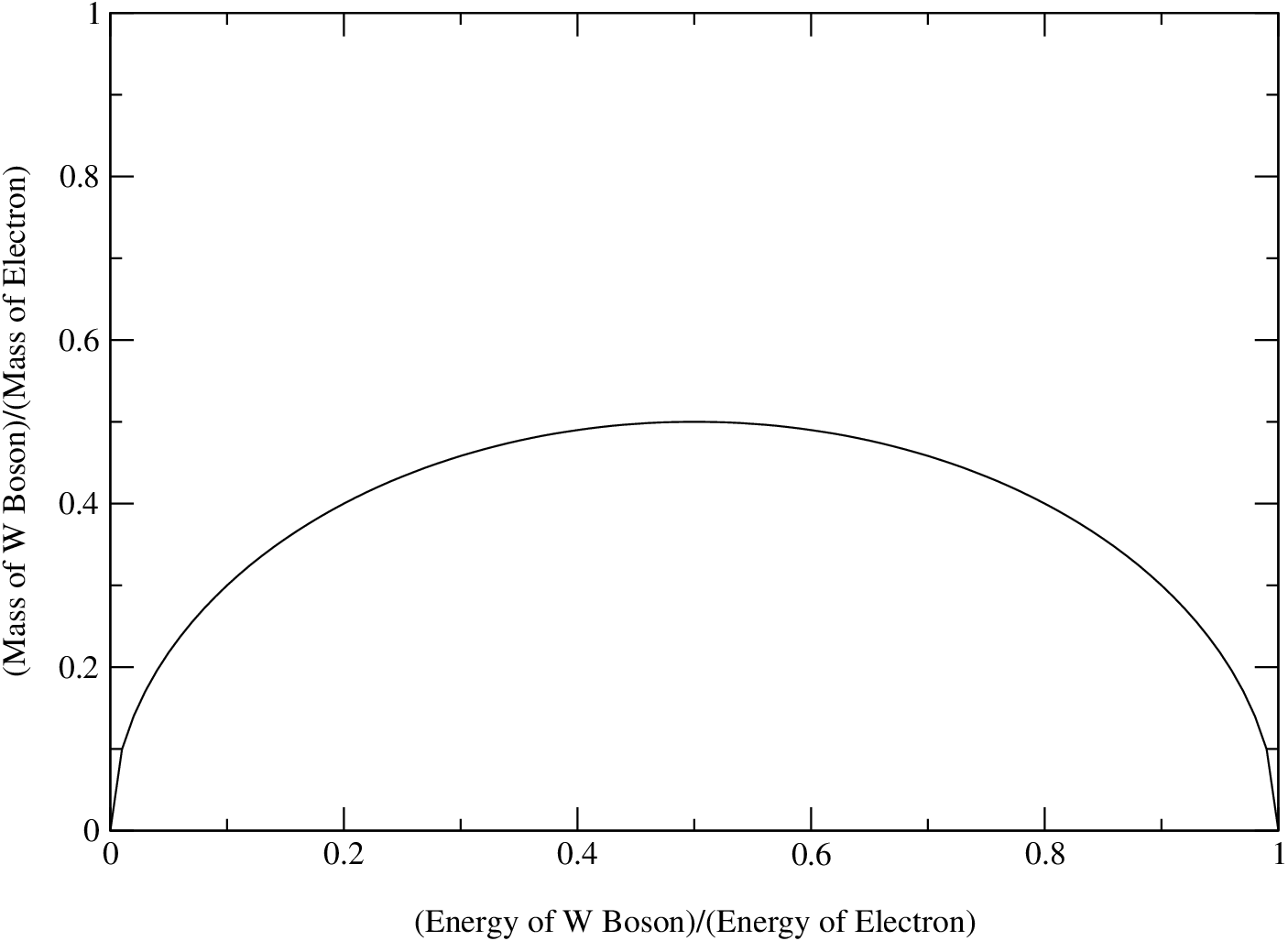,width=5.25in}
\end{center}
\renewcommand{\figurename}{FIG.}
\caption{\label{fig:mb20} The mass $m_W$ of an equivalent $W$
boson emitted from an electron. The ratio of $m_W$ to the mass
$m_e$ of the electron is plotted on the y-axis and the Feynman
scaling variable, $x=E_W/E_e$, is plotted on the x-axis. $m_W$
vanishes at $E_W=0$ and $E_W=E_e-m_e$, and peaks to a maximum
value of $m_e/2$ at $E_ev_e/2$, where $v_e$ is the speed of the
electron.}
\end{figure}

\clearpage

\newpage
\begin{figure}[t]
\begin{center}
\epsfig{file=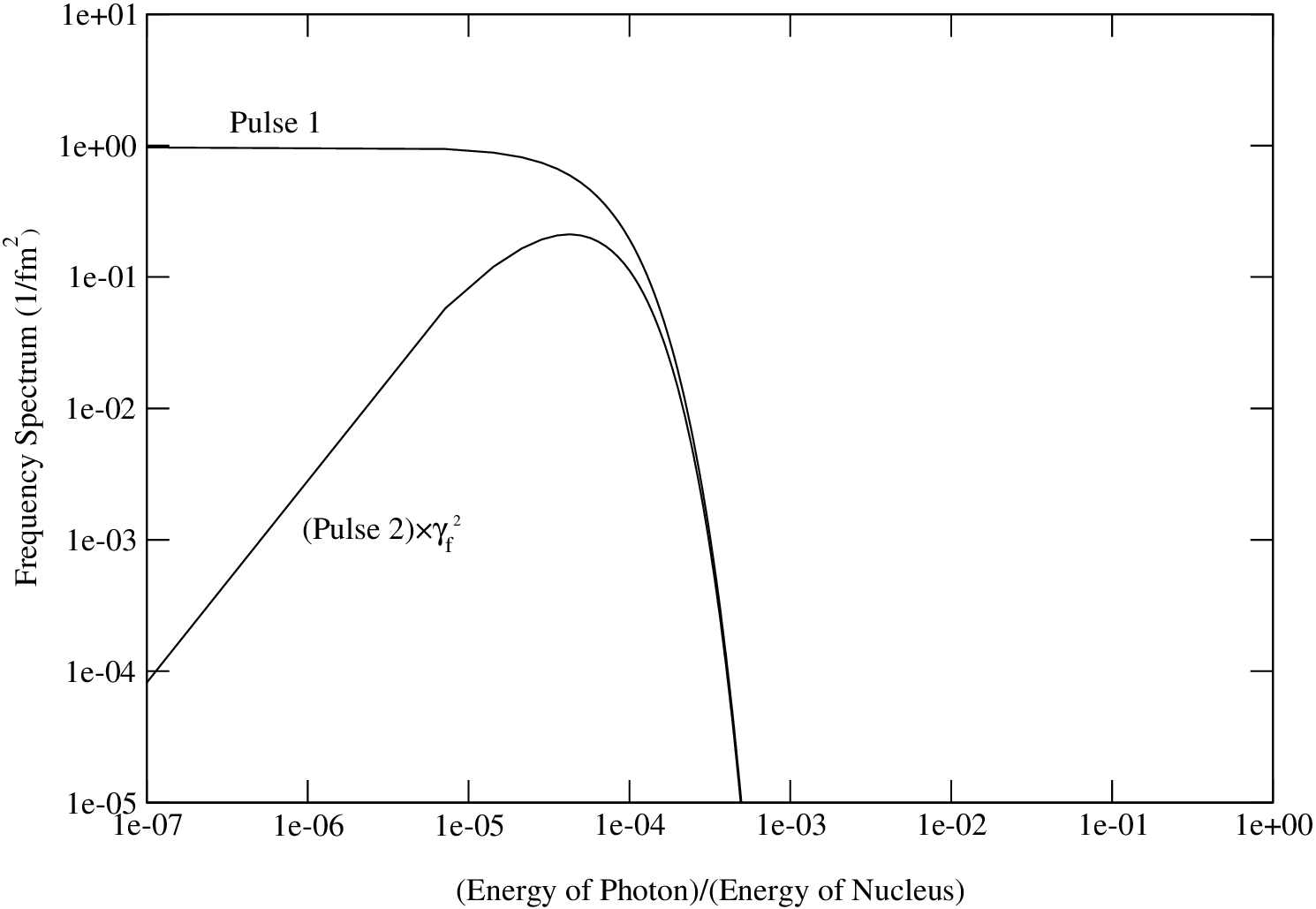,width=5.25in}
\end{center}
\renewcommand{\figurename}{FIG.}
\caption{\label{fig:frsp01} Comparison of helicity-averaged
frequency spectra of the three pulses of equivalent photons
outside a lead ($^{208}Pb$) nucleus at a relativistic heavy ion
collider operating at a beam energy of $3.4\mA$ TeV. The
helicity-averaged frequency spectrum, $\langle \dif^2I/\dif
\omega\, \dif A \rangle$, evaluated at the minimum impact
parameter $b_{min}$, (cf. Eqs. (\ref{eq:dIdwdA1exp2}) --
(\ref{eq:dIdwdA3exp2})) is plotted on the y-axis and the Feynman
scaling variable, $x=E_{\gamma}/E_f$, is plotted on the x-axis.
Because the spectrum for Pulse 2, corresponding to
transversely-polarized photons, is typically suppressed by a
factor of $\gamma _f^2$ (which can be quite large) relative to
that for Pulse 1, which also corresponds to transversely-polarized
photons, it is shown here amplified by $\gamma _f^2$. Pulse 3,
which corresponds to longitudinally-polarized photons, does not
reveal itself on the graph because it vanishes everywhere, on
account of the fact that longitudinally-polarized photons simply
do not occur in nature.}
\end{figure}

\newpage
\begin{figure}[t]
\begin{center}
\epsfig{file=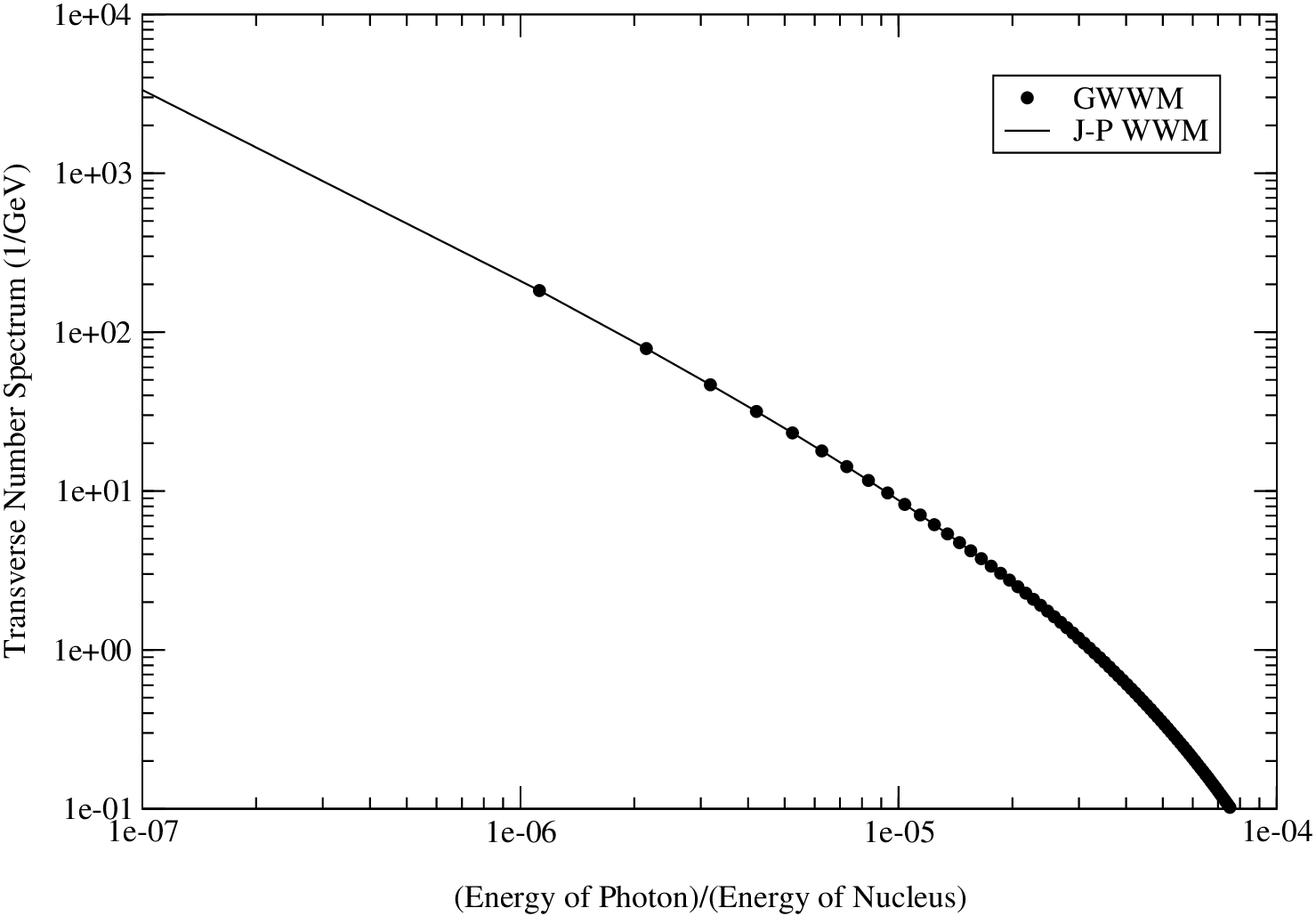,width=5.25in}
\end{center}
\renewcommand{\figurename}{FIG.}
\caption{\label{fig:nt01A} Comparison of number spectra for
transversely-polarized photons radiating from a lead ($^{208}Pb$)
nucleus at a relativistic heavy ion collider operating at a beam
energy of $2.76\mA$ TeV. Note that the number spectra for
longitudinally-polarized photons always vanishes because photons
are never found in longitudinal polarization states. $N_\mT $ is
plotted on the y-axis and the Feynman scaling variable,
$x=E_{\gamma}/E_{nuc}$, is plotted on the x-axis. The dotted curve
shows the results of the GWWM, which are also identically those of
the SWWM, and the solid curve shows the predictions of the
semiclassical version of the WWM developed by J\"{a}ckle and
Pilkuhn \cite{ref:JaPi}. Relative errors between the GWWM and the
J\"{a}ckle-Pilkuhn WWM are always on the order of magnitude of
$10^{-5}\, \%$, from $x=0$ to $x=1$. There does not seem to be any
dependence of the results on $E_{nuc}$, but there is a strong
dependence on $b_{min}$.}
\end{figure}

\newpage
\begin{figure}[t]
\begin{center}
\epsfig{file=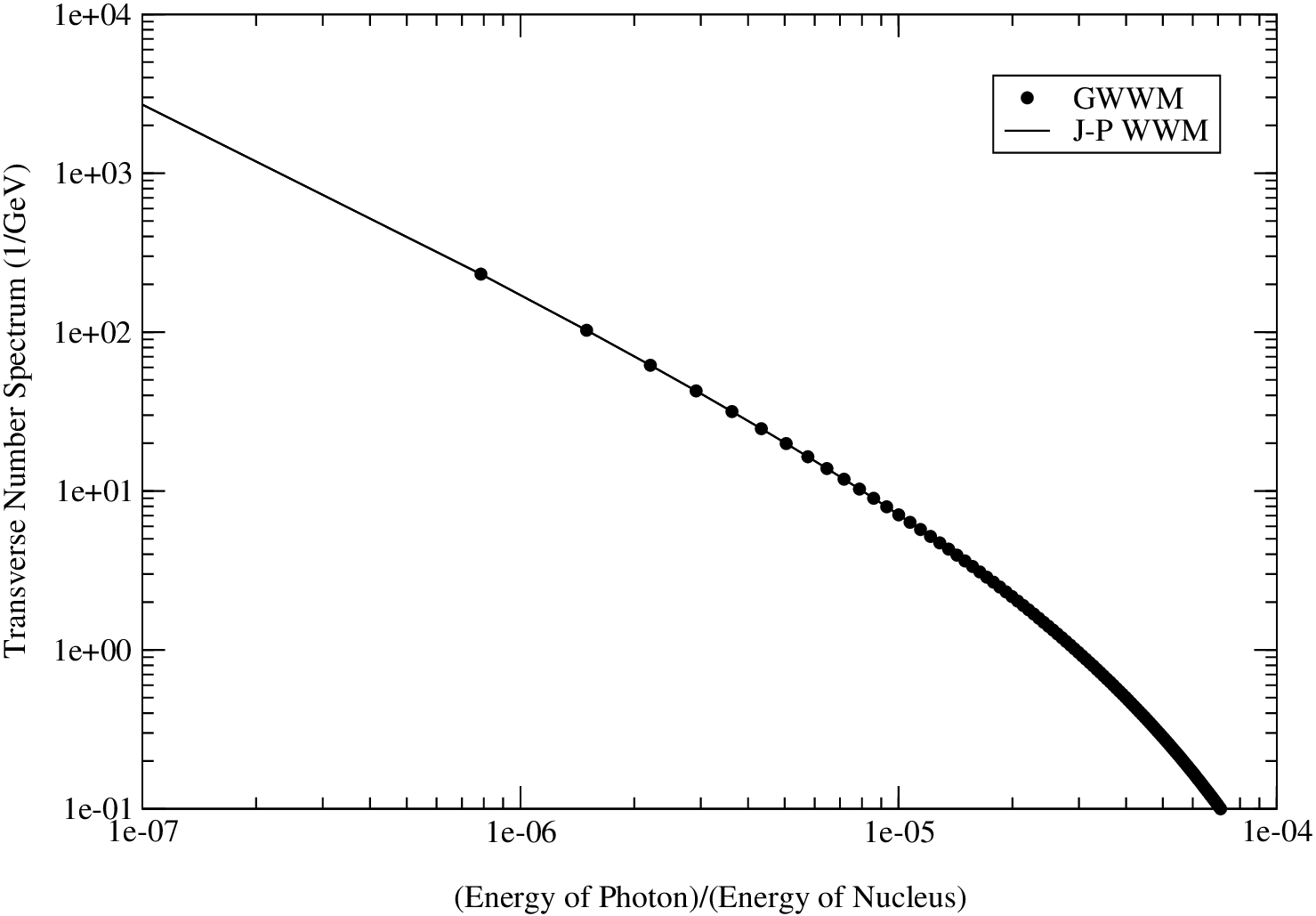,width=5.25in}
\end{center}
\renewcommand{\figurename}{FIG.}
\caption{\label{fig:nt01B} Comparison of number spectra for
transversely-polarized photons radiating from a lead ($^{208}Pb$)
nucleus at a relativistic heavy ion collider operating at a beam
energy of $3.4\mA$ TeV. Note that the number spectra for
longitudinally-polarized photons always vanishes because photons
are never found in longitudinal polarization states. $N_\mT $ is
plotted on the y-axis and the Feynman scaling variable,
$x=E_{\gamma}/E_{nuc}$, is plotted on the x-axis. The dotted curve
shows the results of the GWWM, which are also identically those of
the SWWM, and the solid curve shows the predictions of the
semiclassical version of the WWM developed by J\"{a}ckle and
Pilkuhn \cite{ref:JaPi}. Relative errors between the GWWM and the
J\"{a}ckle-Pilkuhn WWM are always on the order of magnitude of
$10^{-5}\, \%$, from $x=0$ to $x=1$. There does not seem to be any
dependence of the results on $E_{nuc}$, but there is a strong
dependence on $b_{min}$.}
\end{figure}

\clearpage

\newpage
\begin{figure}[t]
\begin{center}
\epsfig{file=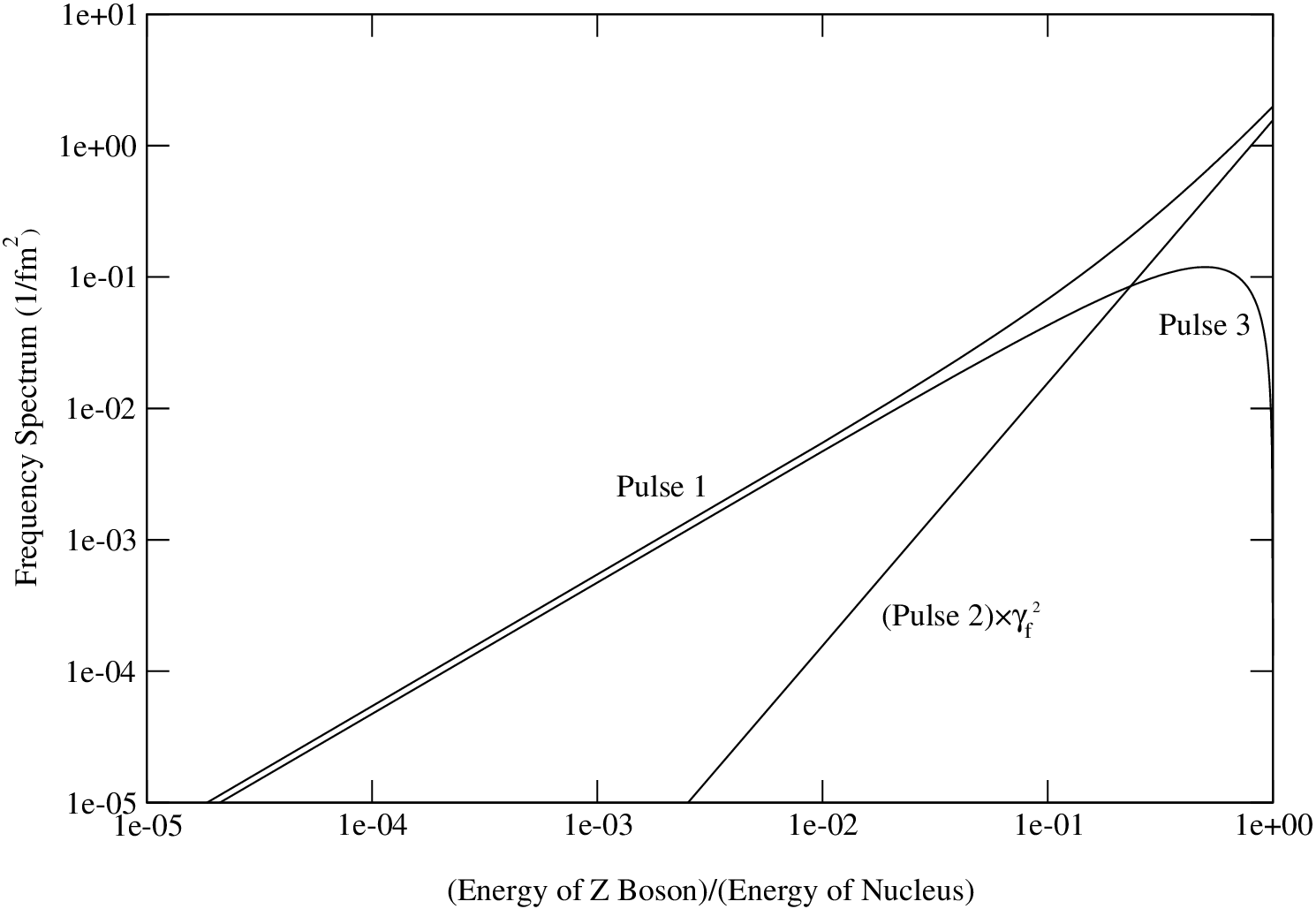,width=5.25in}
\end{center}
\renewcommand{\figurename}{FIG.}
\caption{\label{fig:frsp11} Comparison of helicity-averaged
frequency spectra of the three pulses of equivalent $Z$ bosons
outside a lead ($^{208}Pb$) nucleus at a relativistic heavy ion
collider operating at a beam energy of $3.4\mA$ TeV. The
helicity-averaged frequency spectrum, $\langle \dif^2I/\dif
\omega\, \dif A \rangle$, evaluated at the minimum impact
parameter $b_{min}$, (cf. Eqs. (\ref{eq:dIdwdA1exp2}) --
(\ref{eq:dIdwdA3exp2})) is plotted on the y-axis and the Feynman
scaling variable, $x=E_Z/E_f$, is plotted on the x-axis. Because
the spectrum for Pulse 2, corresponding to transversely-polarized
$Z$ bosons, is typically suppressed by a factor of $\gamma _f^2$
(which can be quite large) relative to that for Pulse 1, which
also corresponds to transversely-polarized $Z$ bosons, it is shown
here amplified by $\gamma _f^2$. Pulse 3 corresponds to the
longitudinally-polarized $Z$ bosons.}
\end{figure}

\newpage
\begin{figure}[t]
\begin{center}
\epsfig{file=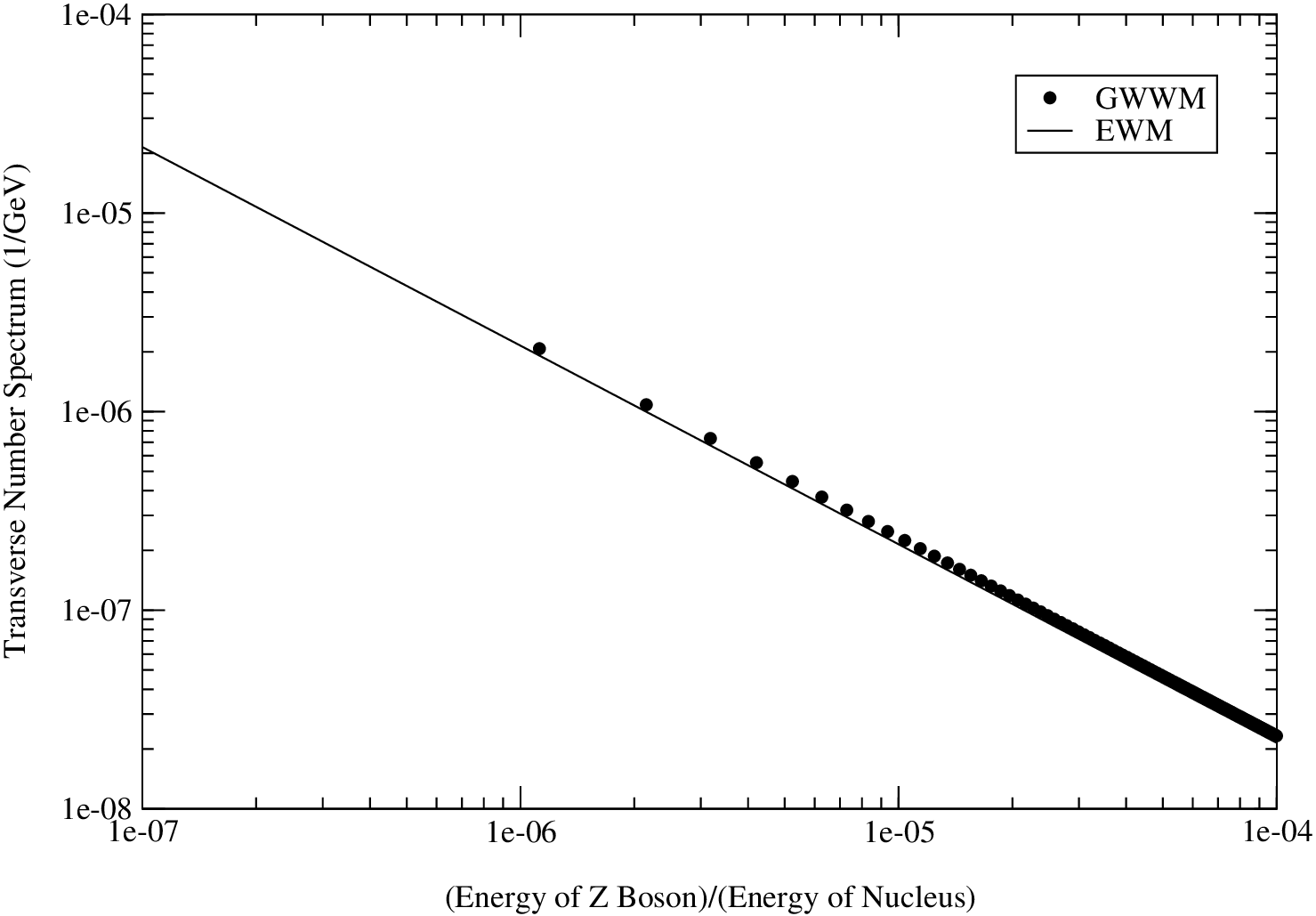,width=5.25in}
\end{center}
\renewcommand{\figurename}{FIG.}
\caption{\label{fig:nt11A} Comparison of number spectra for
transversely-polarized $Z$ bosons radiating from a lead
($^{208}Pb$) nucleus at a relativistic heavy ion collider
operating at a beam energy of $2.76\mA$ TeV. $N_\mT $ is plotted
on the y-axis and the Feynman scaling variable, $x=E_Z/E_{nuc}$,
is plotted on the x-axis. The dotted curve shows the results of
the GWWM and the solid curve shows the predictions of the EWM.
Relative errors are always about $8.4\%$, from $x=0$ to $x=1$.
There does not seem to be any dependence of the results on
$E_{nuc}$, but there is a strong dependence on $b_{min}$.}
\end{figure}

\newpage
\begin{figure}[t]
\begin{center}
\epsfig{file=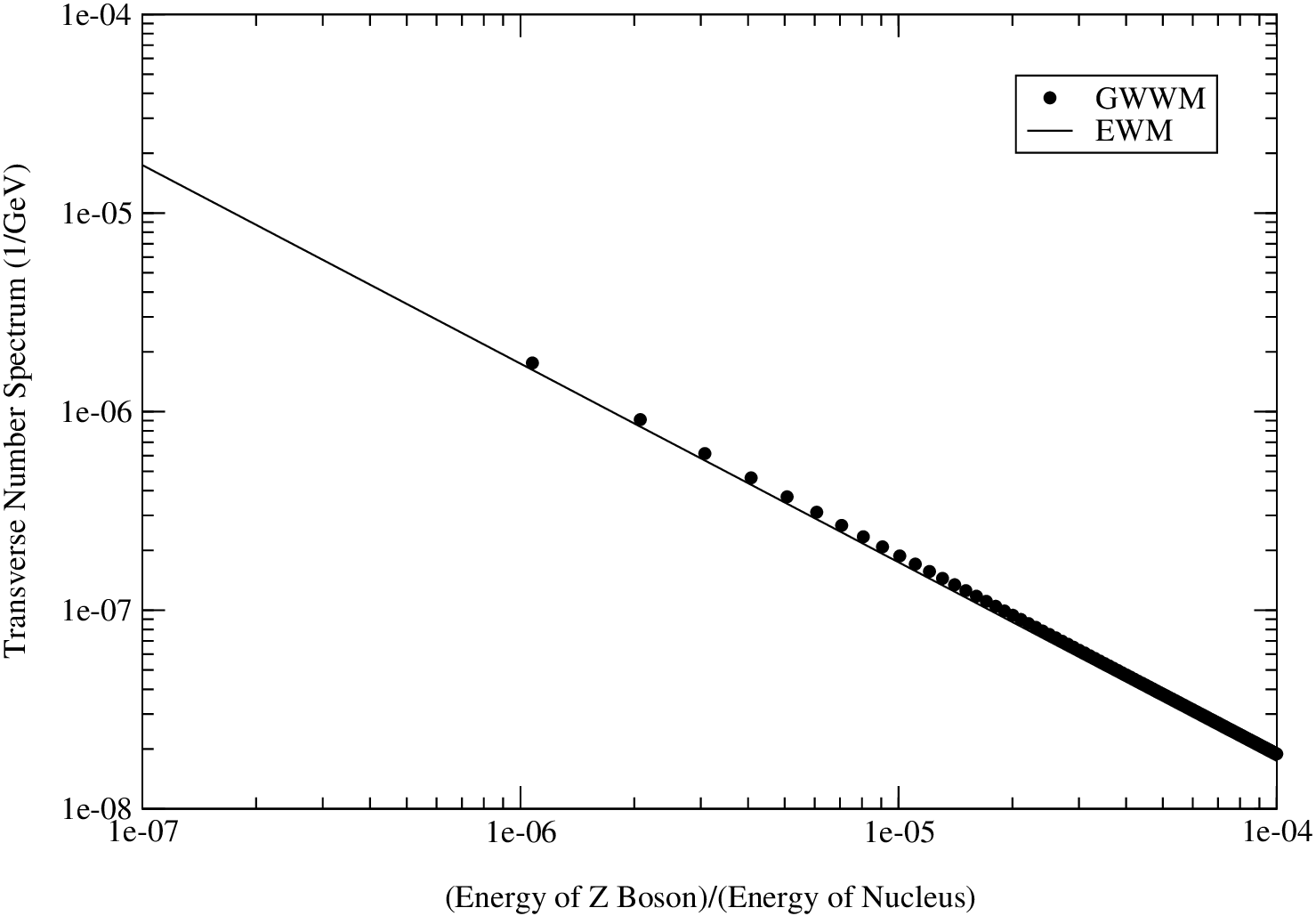,width=5.25in}
\end{center}
\renewcommand{\figurename}{FIG.}
\caption{\label{fig:nt11B} Comparison of number spectra for
transversely-polarized $Z$ bosons radiating from a lead
($^{208}Pb$) nucleus at a relativistic heavy ion collider
operating at a beam energy of $3.4\mA$ TeV. $N_\mT $ is plotted on
the y-axis and the Feynman scaling variable, $x=E_Z/E_{nuc}$, is
plotted on the x-axis. The dotted curve shows the results of the
GWWM and the solid curve shows the predictions of the EWM.
Relative errors are always about $8.4\%$, from $x=0$ to $x=1$.
There does not seem to be any dependence of the results on
$E_{nuc}$, but there is a strong dependence on $b_{min}$.}
\end{figure}

\newpage
\begin{figure}[t]
\begin{center}
\epsfig{file=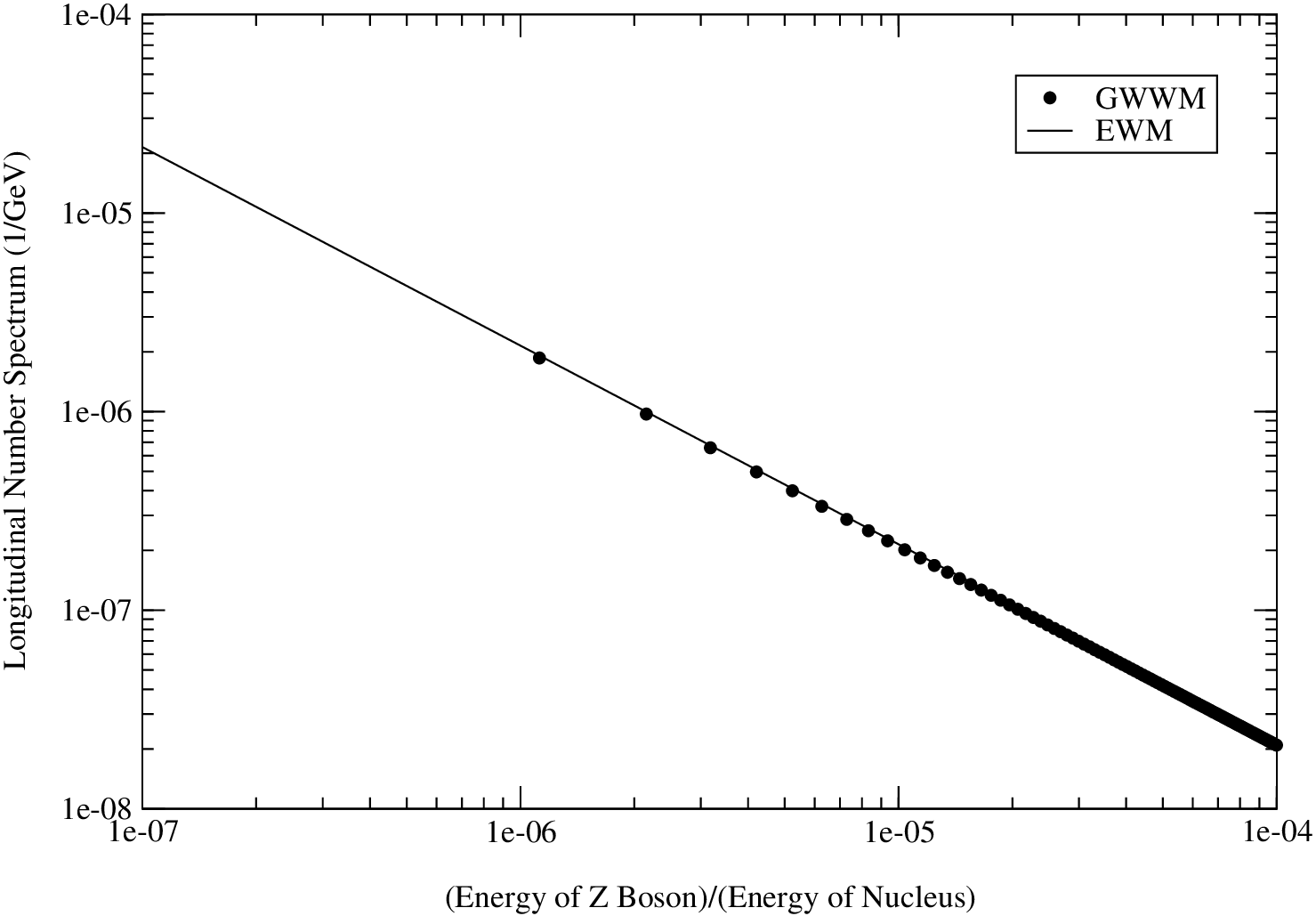,width=5.25in}
\end{center}
\renewcommand{\figurename}{FIG.}
\caption{\label{fig:nl11A} Comparison of number spectra for
longitudinally-polarized $Z$ bosons radiating from a lead
($^{208}Pb$) nucleus at a relativistic heavy ion collider
operating at a beam energy of $2.76\mA$ TeV. $N_\mL $ is plotted
on the y-axis and the Feynman scaling variable, $x=E_Z/E_{nuc}$,
is plotted on the x-axis. The solid curve shows the predictions of
the EWM and the dotted curve shows the results of the GWWM.
Relative errors are always about $2.7\%$, from $x=0$ to $x=1$.
There does not seem to be any dependence of the results on
$E_{nuc}$, but there is a strong dependence on $b_{min}$.}
\end{figure}

\newpage
\begin{figure}[t]
\begin{center}
\epsfig{file=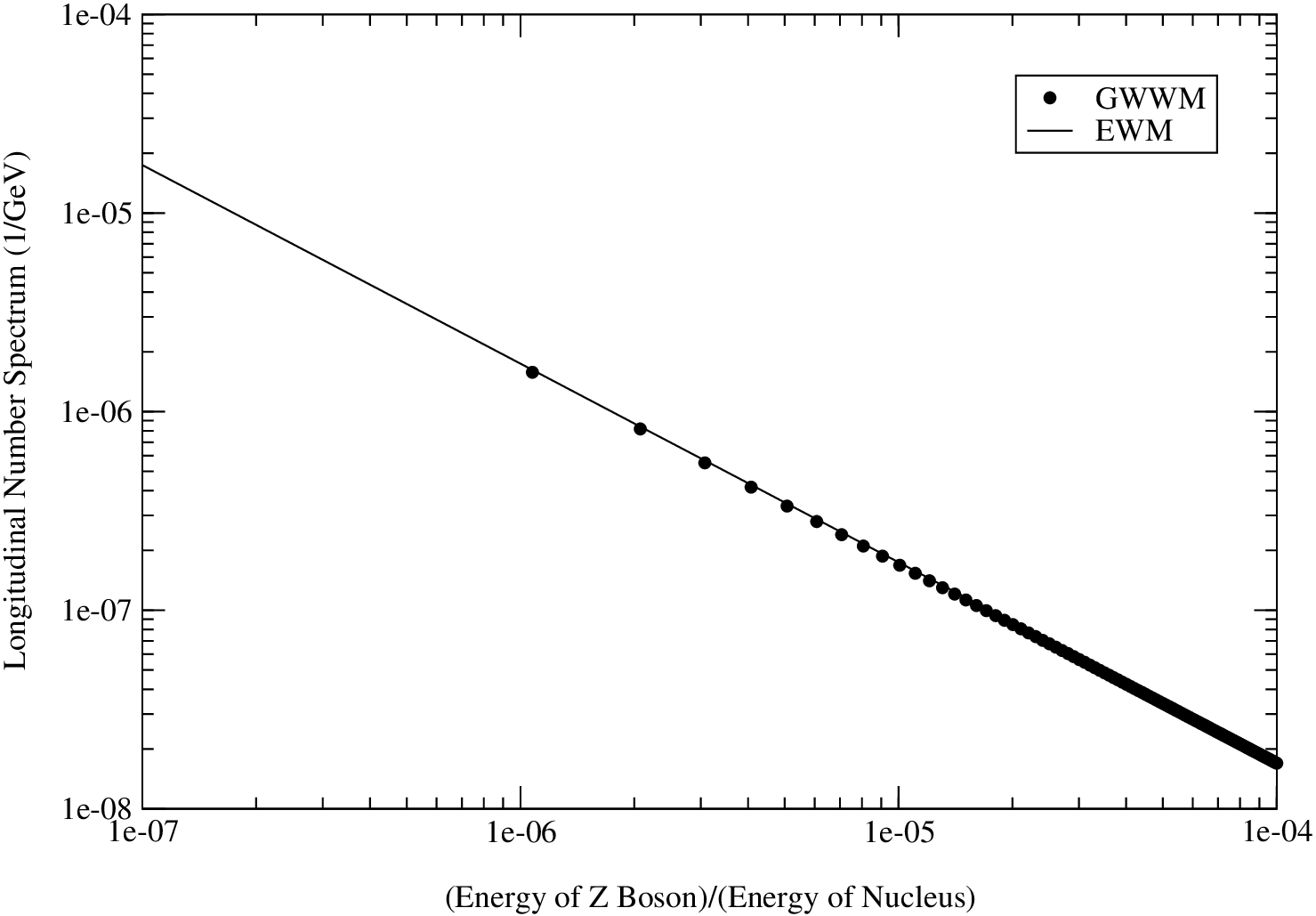,width=5.25in}
\end{center}
\renewcommand{\figurename}{FIG.}
\caption{\label{fig:nl11B} Comparison of number spectra for
longitudinally-polarized $Z$ bosons radiating from a lead
($^{208}Pb$) nucleus at a relativistic heavy ion collider
operating at a beam energy of $3.4\mA$ TeV. $N_\mL $ is plotted on
the y-axis and the Feynman scaling variable, $x=E_Z/E_{nuc}$, is
plotted on the x-axis. The solid curve shows the predictions of
the EWM and the dotted curve shows the results of the GWWM.
Relative errors are always about $2.7\%$, from $x=0$ to $x=1$.
There does not seem to be any dependence of the results on
$E_{nuc}$, but there is a strong dependence on $b_{min}$.}
\end{figure}

\newpage
\begin{figure}[t]
\begin{center}
\epsfig{file=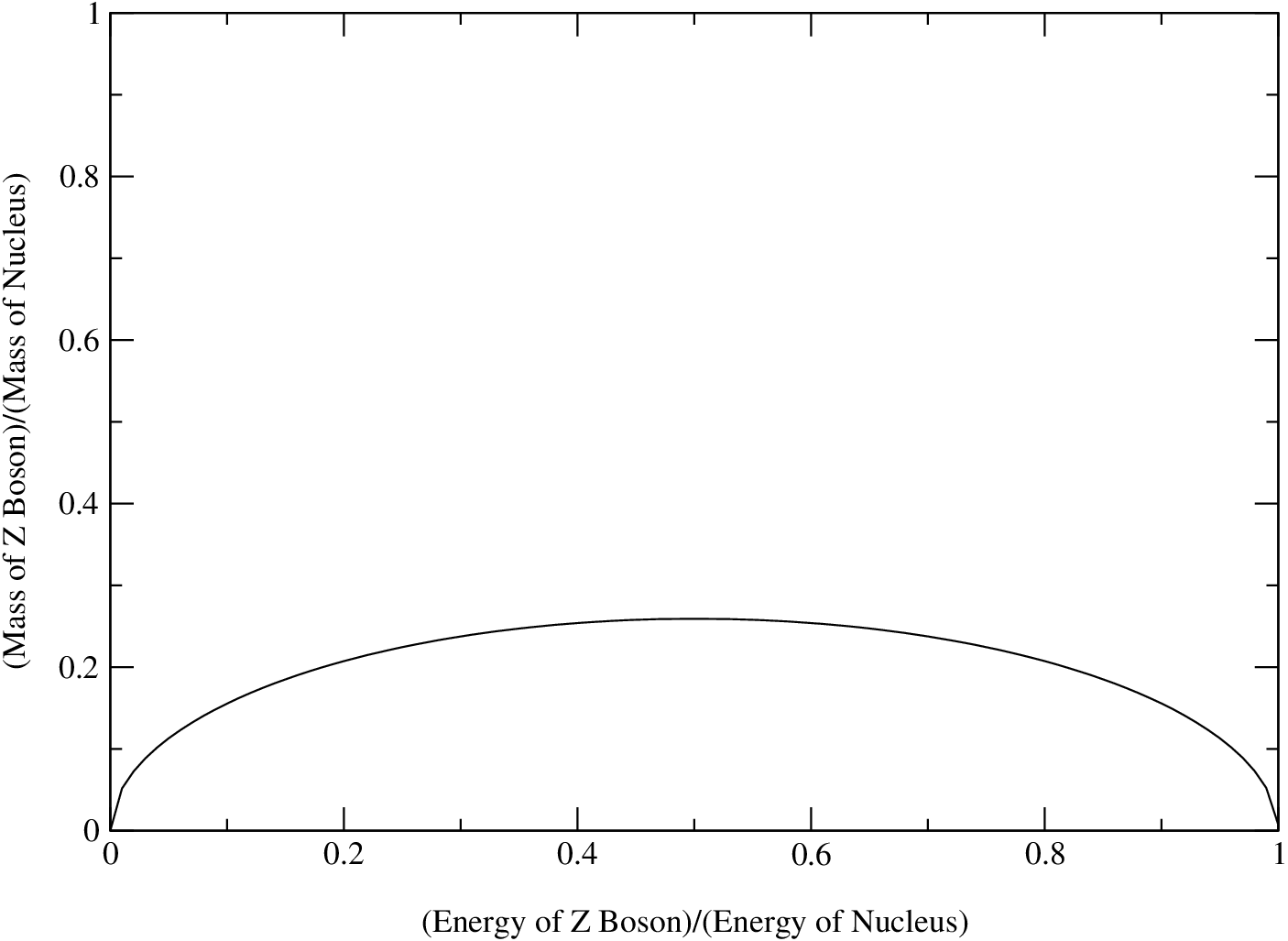,width=5.25in}
\end{center}
\renewcommand{\figurename}{FIG.}
\caption{\label{fig:mb11} The mass $m_Z$ of an equivalent $Z$
boson emitted from a lead ($^{208}Pb$) nucleus. The ratio of $m_Z$
to the mass $m_f$ of the nucleus is plotted on the y-axis and the
Feynman scaling variable, $x=E_Z/E_f$, is plotted on the x-axis.
$m_Z$ vanishes at $E_Z=0$ and $E_Z=E_f-m_f$, and peaks to a
maximum value of $m_f\sqrt{0.2686}/2$ at $E_fv_f/2$, where $v_f$
is the speed of the nucleus.}
\end{figure}

\clearpage

\newpage
\begin{figure}[t]
\begin{center}
\epsfig{file=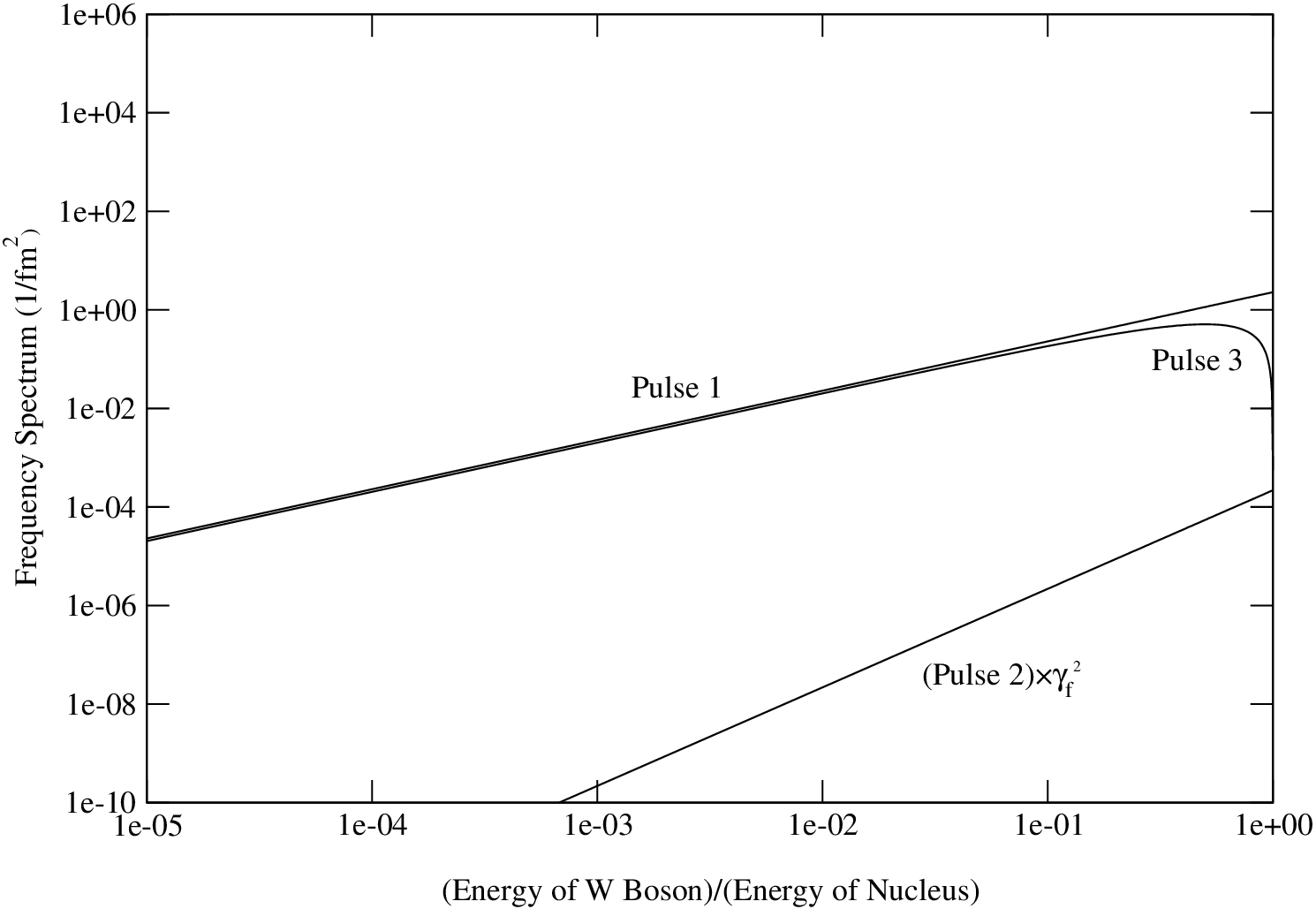,width=5.25in}
\end{center}
\renewcommand{\figurename}{FIG.}
\caption{\label{fig:frsp21} Comparison of helicity-averaged
frequency spectra of the three pulses of equivalent $W$ bosons
outside a lead ($^{208}Pb$) nucleus at a relativistic heavy ion
collider operating at a beam energy of $3.4\mA$ TeV. The
helicity-averaged frequency spectrum, $\langle \dif^2I/\dif
\omega\, \dif A \rangle$, evaluated at the minimum impact
parameter $b_{min}$, (cf. Eqs. (\ref{eq:dIdwdA1exp2}) --
(\ref{eq:dIdwdA3exp2})) is plotted on the y-axis and the Feynman
scaling variable, $x=E_W/E_f$, is plotted on the x-axis. Because
the spectrum for Pulse 2, corresponding to transversely-polarized
$W$ bosons, is typically suppressed by a factor of $\gamma _f^2$
(which can be quite large) relative to that for Pulse 1, which
also corresponds to transversely-polarized $Z$ bosons, it is shown
here amplified by $\gamma _f^2$. Pulse 3 corresponds to the
longitudinally-polarized $W$ bosons.}
\end{figure}

\newpage
\begin{figure}[t]
\begin{center}
\epsfig{file=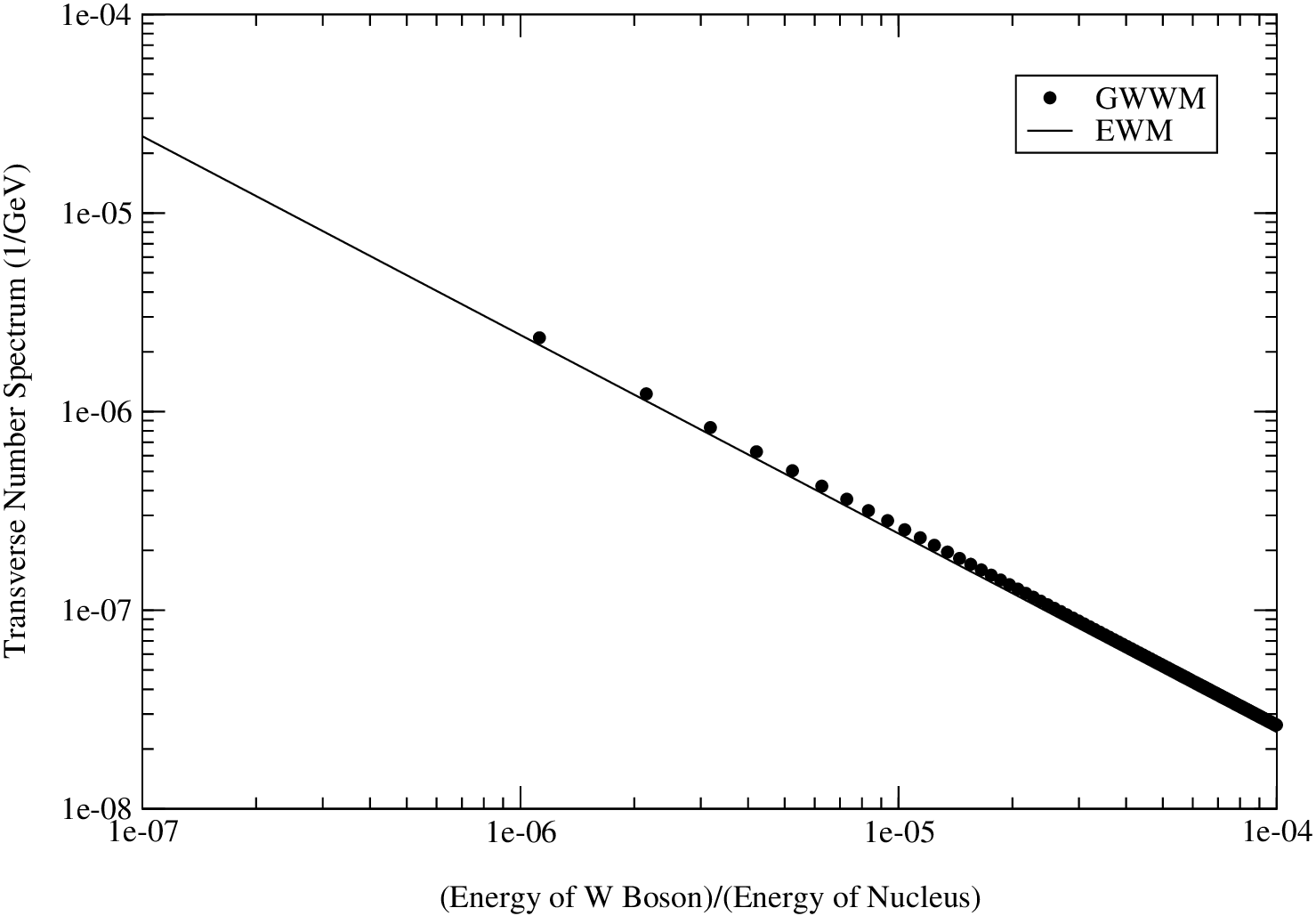,width=5.25in}
\end{center}
\renewcommand{\figurename}{FIG.}
\caption{\label{fig:nt21A} Comparison of number spectra for
transversely-polarized $W$ bosons radiating from a lead
($^{208}Pb$) nucleus at a relativistic heavy ion collider
operating at a beam energy of $2.76\mA$ TeV. $N_\mT $ is plotted
on the y-axis and the Feynman scaling variable, $x=E_W/E_{nuc}$,
is plotted on the x-axis. The dotted curve shows the results of
the GWWM and the solid curve shows the predictions of the EWM.
Relative errors are always about $8.5\%$, from $x=0$ to $x=1$.
There does not seem to be any dependence of the results on
$E_{nuc}$, but there is a strong dependence on $b_{min}$.}
\end{figure}

\newpage
\begin{figure}[t]
\begin{center}
\epsfig{file=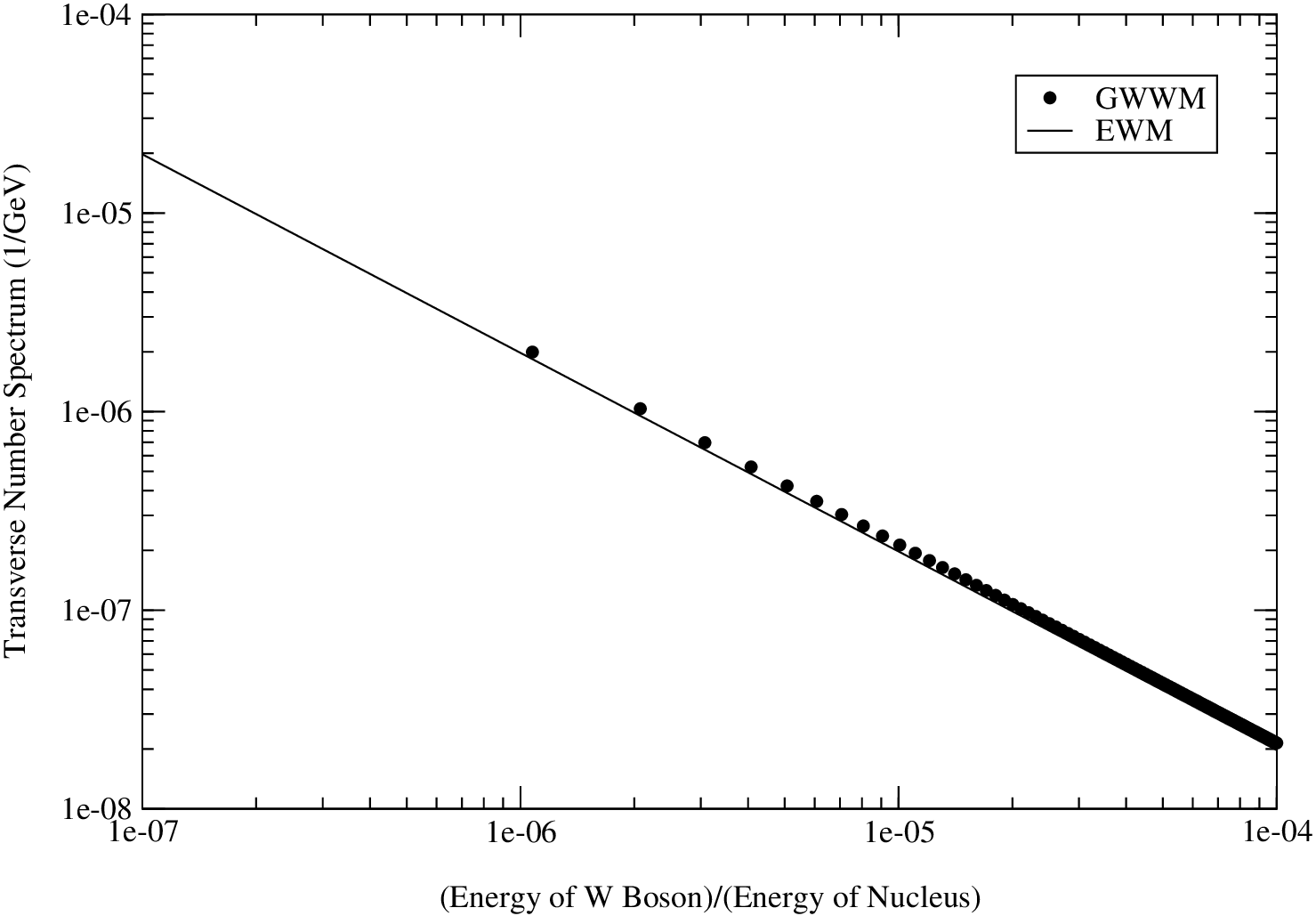,width=5.25in}
\end{center}
\renewcommand{\figurename}{FIG.}
\caption{\label{fig:nt21B} Comparison of number spectra for
transversely-polarized $W$ bosons radiating from a lead
($^{208}Pb$) nucleus at a relativistic heavy ion collider
operating at a beam energy of $3.4\mA$ TeV. $N_\mT $ is plotted on
the y-axis and the Feynman scaling variable, $x=E_W/E_{nuc}$, is
plotted on the x-axis. The dotted curve shows the results of the
GWWM and the solid curve shows the predictions of the EWM.
Relative errors are always about $8.5\%$, from $x=0$ to $x=1$.
There does not seem to be any dependence of the results on
$E_{nuc}$, but there is a strong dependence on $b_{min}$.}
\end{figure}

\newpage
\begin{figure}[t]
\begin{center}
\epsfig{file=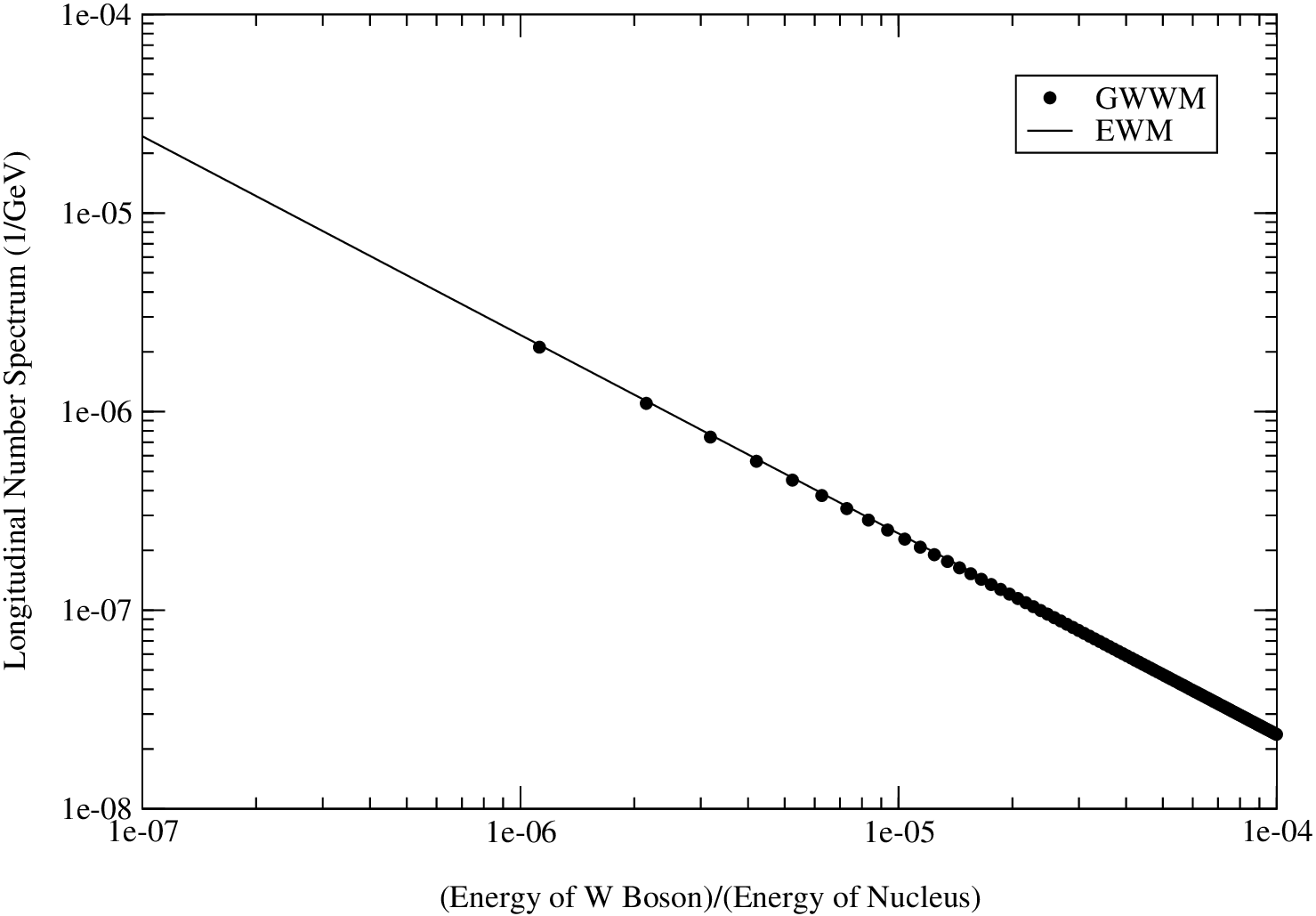,width=5.25in}
\end{center}
\renewcommand{\figurename}{FIG.}
\caption{\label{fig:nl21A} Comparison of number spectra for
longitudinally-polarized $W$ bosons radiating from a lead
($^{208}Pb$) nucleus at a relativistic heavy ion collider
operating at a beam energy of $2.76\mA$ TeV. $N_\mL $ is plotted
on the y-axis and the Feynman scaling variable, $x=E_W/E_{nuc}$,
is plotted on the x-axis. The solid curve shows the predictions of
the EWM and the dotted curve shows the results of the GWWM.
Relative errors are always about $2.7\%$, from $x=0$ to $x=1$.
There does not seem to be any dependence of the results on
$E_{nuc}$, but there is a strong dependence on $b_{min}$.}
\end{figure}

\newpage
\begin{figure}[t]
\begin{center}
\epsfig{file=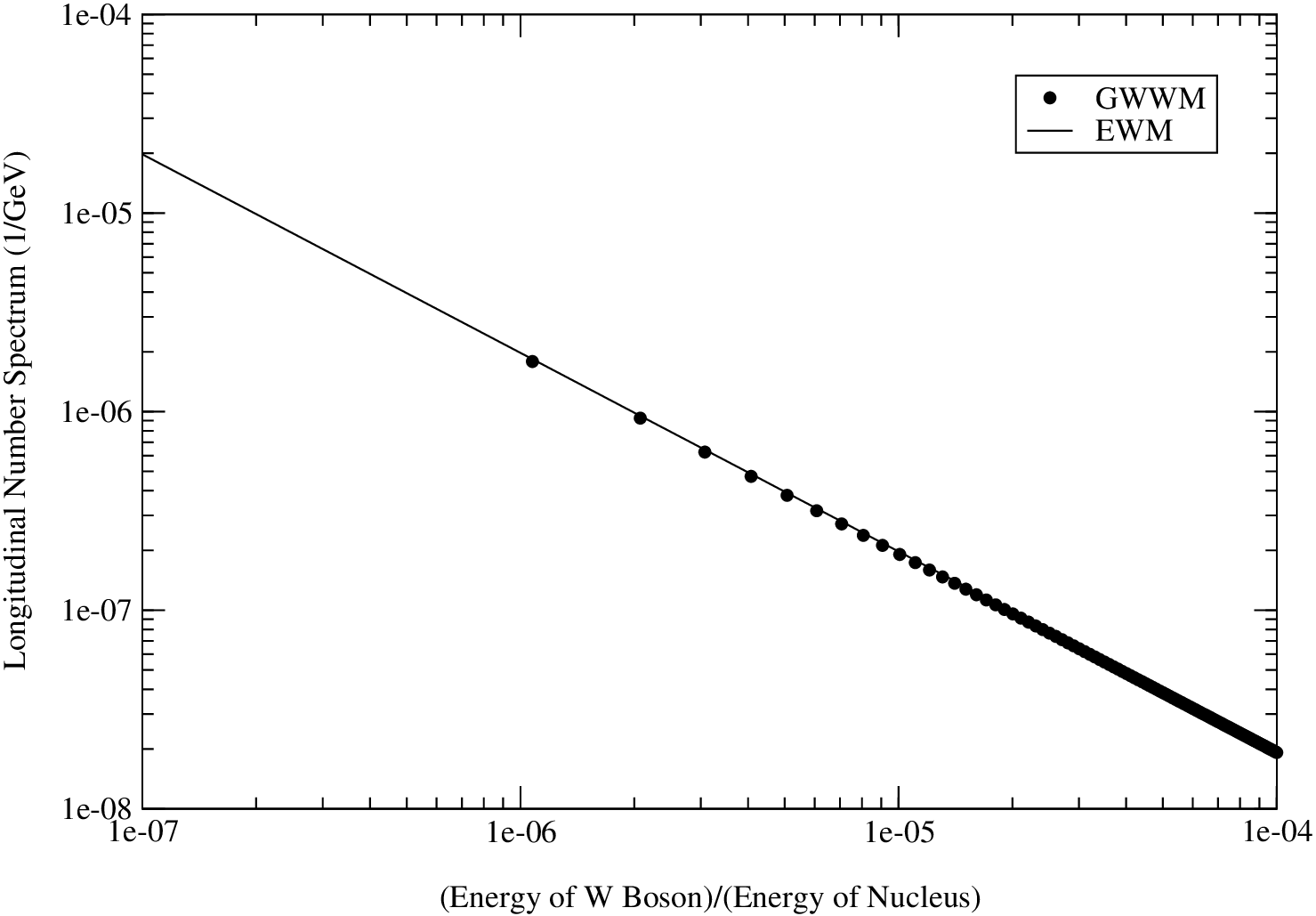,width=5.25in}
\end{center}
\renewcommand{\figurename}{FIG.}
\caption{\label{fig:nl21B} Comparison of number spectra for
longitudinally-polarized $W$ bosons radiating from a lead
($^{208}Pb$) nucleus at a relativistic heavy ion collider
operating at a beam energy of $3.4\mA$ TeV. $N_\mL $ is plotted on
the y-axis and the Feynman scaling variable, $x=E_W/E_{nuc}$, is
plotted on the x-axis. The solid curve shows the predictions of
the EWM and the dotted curve shows the results of the GWWM.
Relative errors are always about $2.7\%$, from $x=0$ to $x=1$.
There does not seem to be any dependence of the results on
$E_{nuc}$, but there is a strong dependence on $b_{min}$.}
\end{figure}

\newpage
\begin{figure}[t]
\begin{center}
\epsfig{file=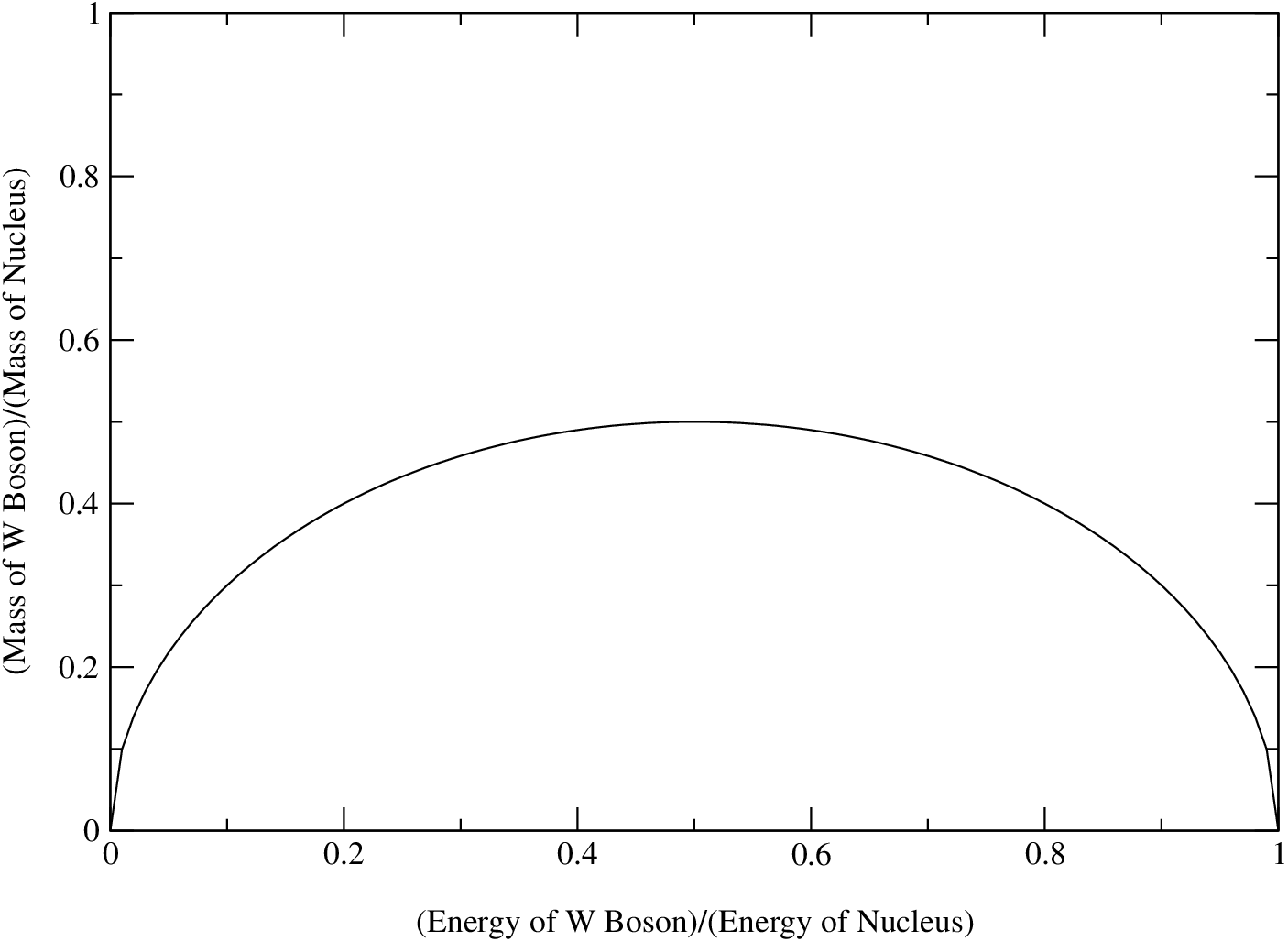,width=5.25in}
\end{center}
\renewcommand{\figurename}{FIG.}
\caption{\label{fig:mb21} The mass $m_W$ of an equivalent $W$
boson emitted from a lead ($^{208}Pb$) nucleus. The ratio of $m_W$
to the mass $m_f$ of the nucleus is plotted on the y-axis and the
Feynman scaling variable, $x=E_W/E_f$, is plotted on the x-axis.
$m_W$ vanishes at $E_W=0$ and $E_W=E_f-m_f$, and peaks to a
maximum value of $m_f/2$ at $E_fv_f/2$, where $v_f$ is the speed
of the nucleus.}
\end{figure}

\end{landscape}